\documentclass[twoside,12pt]{article}
\usepackage{epsfig,amsmath}
\usepackage{amssymb}

\usepackage[ colorlinks=true, pdfstartview=FitV, linkcolor=black, citecolor=blue, urlcolor=blue]{hyperref} %comment out if you prefer colorless version.
\usepackage{color} %comment out if you prefer colorless version.
\usepackage{tensor}
\usepackage{bm}
\usepackage{booktabs}
\usepackage[utf8]{inputenc}

\newcommand{\vect}[1]{\mathbf{#1}}
\newcommand{\vp}{\vect{p}}

\newcommand{\vx}{\vect{x}}

\newcommand{\vzero}{\vect{0}}

\newcommand{\vE}{\vect{E}}
\newcommand{\vB}{\vect{B}}

\newcommand{\Nc}{N_c}
\newcommand{\Nf}{N_f}
\newcommand{\Cf}{C_F}
\newcommand{\nf}{n_F}

\newcommand{\Cem}{C_{\text{em}}}

\newcommand{\re}{{\text{Re}}}

\newcommand{\comment}[1]{}

\newcommand{\be}{\begin{equation}}
\newcommand{\ee}{\end{equation}}
\newcommand{\bea}{\begin{eqnarray}}
\newcommand{\eea}{\end{eqnarray}}

\newcommand{\cev}[1]{\reflectbox{\ensuremath{\vec{\reflectbox{\ensuremath{#1}}}}}}
\def\slash#1{\not\!#1}
\def\slashb#1{\not\!\!#1}

\begin{document}

\title{ \vspace{1cm} Recent Progress in QCD Condensate Evaluations \\
and Sum Rules} 

\author{P.\ Gubler,$^{1}$  
D.\ Satow$^2$\\ 
\\
$^1$ Advanced Science Research Center, Japan Atomic Energy Agency, \\
Tokai, Ibaraki 319-1195, Japan, \\
gubler@post.j-parc.jp \\
$^2$ Arithmer Inc.,
1-6-1, Roppongi, \\
Minato-ku, Tokyo 106-6040, Japan
}

\maketitle

\begin{abstract} 
We review the recent status of the QCD sum rule approach to study the properties of hadrons in vacuum and in hot or dense matter. 
Special focus is laid on the progress made in the evaluation of the QCD condensates, which are the input of all QCD sum rule calculations, 
and for which much new information has become available through high precision lattice QCD calculations, chiral perturbation theory and 
experimental measurements. 
Furthermore, we critically examine common analysis methods for QCD sum rules and contrast them with potential alternative 
strategies. 
The status of QCD sum rule studies investigating 
the modification of hadrons at finite density 
as well as recent derivations of exact sum rules applicable to finite temperature spectral functions, 
are also reviewed. 
\end{abstract}
\eject
\tableofcontents

%%%%%%%%%%%%%%%%%%%%
\section{Introduction}
The QCD sum rule (QCDSR) method, formulated and proposed in the seminal papers of Shifman, Vainshtein and Zakharov \cite{Shifman:1978bx,Shifman:1978by} 
in the late seventies (for earlier attempts, see also Refs.\,\cite{Novikov:1976tn,Novikov:1977cm,Novikov:1977dq}), is today still being frequently used as a tool to 
compute hadronic properties from QCD\footnote{Similar sum rules were formulated even before by other authors in Refs.\,\cite{Sakurai:1973rh,Shankar:1977ap}.}. 
Initially, its main purpose was to compute basic observables such as ground state masses or magnetic moments of hadrons. 
Such calculations were rather successful \cite{Reinders:1984sr} (see, however, Ref.\,\cite{Novikov:1981xi} for a discussion about exceptional channels), which 
led to the firm establishment of the method in the hadron physics/QCD community. 

QCDSRs rely on several approximations and assumptions such as the truncation of the operator product expansion (OPE) or the pole dominance of the 
sum rules, as will be discussed in detail in Section \ref{sec:Formalism}. These approximations typically limit the precision of QCDSR predictions to 
about 10\,\% to 20\,\%. 
Nevertheless, even with the advancement of lattice QCD, which is by now able to precisely compute many hadronic observables with physical pion masses 
and up to four active flavors \cite{Aoki:2013ldr}, QCDSR still have a role to play. Typical settings and problems for which QCDSR can be relevant even 
today are the following. 
1) QCDSR provide non-trivial relations between hadronic observables and the QCD vacuum (condensates). Especially interesting in this context is the 
relation between hadronic 
properties and the spontaneous breaking of chiral symmetry.  
2) The behavior of hadrons at finite density can be studied in QCDSR at least up to densities of the order of 
normal nuclear matter density \cite{Hatsuda:1991ez}. The status of such works will be discussed in Section \ref{sec:HadronsDensity}. 
In lattice QCD such calculations are presently still not possible because of the sign problem, which prevents efficient important sampling techniques to work. 
3) QCDSRs often do not require heavy numerical analyses and can hence be used for first exploratory studies to obtain a rough idea on what the final result will look like. 
This can lead to important hints, for instance for more precise lattice QCD studies. 
4) QCDSRs can provide constraints on certain integrals (moments) of hadronic spectral functions (see for example Section \ref{sec:exact-sumrule} of this review for a derivation of 
such sum rules at finite temperature). These can be used either for checks for spectral functions computed from 
hadronic models, for determining condensate values in case the spectral function itself is known, or for constraining parameters in spectral fits of lattice QCD data. 
5) QCDSR studies of exotic hadrons are possible and indeed have become rather popular in recent years \cite{Nielsen:2009uh}. Care is, however, needed 
as for states with more than three quarks, the OPE convergence often becomes problematic and the continuum contribution 
to the sum rules tends to be significant. 

The goal of this review is to summarize some of the recent progress in the field of QCDSRs. 
As this method by now already has a rather long history, a large number of reviews have been written over the years  
\cite{Reinders:1984sr,Nielsen:2009uh,Novikov:1983gd,Cohen:1994wm,Leinweber:1995fn,Shifman:1998rb,Colangelo:2000dp,Narison:2007spa,
Ayala:2016vnt,Dominguez:2018zzi}. Hence, to avoid too many redundancies, we will only touch briefly upon the QCDSR derivation and its basic features, but instead discuss 
novel developments in more detail that have roughly occurred during the last decade. We will particularly focus on up-to-date estimates of the QCD condensates in vacuum, finite temperature, 
finite density and in a constant and homogeneous magnetic field, taking into account the latest results from lattice QCD and chiral perturbation theory. 
Non-scalar condensates, which become non-zero only in a hot, dense or magnetic medium will also be reviewed and updated estimates 
for them will be given wherever possible. 
We will furthermore describe advancements in analysis techniques, using alternative forms of sum rules (in contrast to the 
most frequently employed Borel sum rules) and the maximum entropy method, which can be used to extract the 
spectral function from the sum rules without relying on any strong assumption about its form \cite{Gubler:2013moa}. 

As a disclaimer for the reader, let us note that 
no attempt to discuss all possible applications of QCDSRs and to review the corresponding recent literature, will be made in this article.  
Considering the large number of QCDSR related papers that appear on the arXiv weekly if not daily, this would clearly go beyond the intended scope for this review and 
the ability and time of the authors. We will, however, review recent works studying the modification of hadrons in nuclear matter, 
as these will potentially have a large impact on related experimental studies planned at various 
experimental facilities such as FAIR, NICA, HIAF and J-PARC. As a second application, we will outline the derivation of exact sum rules at finite temperature, 
discuss their properties and provide specific sum rules 
for the energy-momentum tensor and vector current correlators. 
These can be useful either to constrain fits of 
spectral functions to lattice data or to determine certain combinations of condensate values or hydrodynamic transport coefficients. 

This review article is organized as follows. In Section \ref{sec:Formalism}, a brief introduction of the basic QCDSR features, such as 
the dispersion relation and the OPE is given and followed by a detailed discussion about our present knowledge of QCD condensates in vacuum, 
at finite density, temperature and in a constant and homogeneous magnetic field in Section \ref{sec:QCD.condensates}. 
In Section \ref{sec:analysis.strategies}, traditional and more advanced analysis techniques for practical 
QCDSR studies are reviewed. 
Section \ref{sec:HadronsDensity} discusses applications of QCDSR to studies of hadronic spectral functions in dense matter. 
In Section \ref{sec:exact-sumrule}, the derivations of several exact sum rules are reviewed and their potential 
applications discussed. 
Finally, Section \ref{sec:SummandOutl} gives a short summary and outlook. 
In Appendix \ref{sec:Appendix1}, specific OPE expressions for various correlators needed for the derivation of the exact sum rules 
in Section \ref{sec:exact-sumrule} are provided.

%%%%%%%%%%%%%%%%%%%%
\section{Formalism of QCD sum rules \label{sec:Formalism}} %derivation of QCD sum rule, OPE, 
In this section, we will introduce the QCDSR method, its basic idea and concrete implementation. 
Following partly Ref.\,\cite{Gubler:2013moa}, 
we will also examine the inputs and tools required for this method, the operator product expansion (OPE) and 
the QCD condensates arising from the non-trivial vacuum of QCD.  
We will furthermore discuss how QCDSRs can 
be generalized to the case of non-zero temperature, density or magnetic field, especially how the QCD condensates are modified in 
hot, dense or magnetic matter and how new Lorentz-symmetry-violating condensates are generated. Finally, we will 
review how information about physical states can be extracted from the sum rules. In particular, we will  
critically assess the ``pole + continuum" assumption, which is routinely used in QCDSR studies, but 
is not necessarily universally applicable for all channels and becomes particularly questionable for finite density and/or temperature 
and/or magnetic field spectra. 

\subsection{Basics}
The method of QCD sum rules relies in essence on two basic concepts: the analyticity of the two-point function (correlator) of an interpolating 
field and asymptotic freedom of QCD. 
As will be shown in more detail below, the 
former allows one to derive dispersion relations that relate the deep Euclidean region of the correlator with an 
integral over its imaginary part (the spectral function) in the physical (positive) energy region. 
The latter, asymptotic freedom, then makes it possible to systematically compute the correlator in the deep Euclidean region 
using the OPE, which incorporates both perturbative and non-perturbative aspects into the calculation and 
becomes exact in the high-energy limit. The OPE gives rise to an expansion of non-perturbative expectation values of operators 
with increasing mass dimension and corresponding Wilson coefficients that are used to describe the short-distance dynamics 
of the correlator and can be obtained perturbatively. 
One is then left with equations that relate certain integrals of the spectral function (or in other words, sums of contributions of physical states, 
hence the name ``sum rules") with the result of the OPE. 
The high-energy part of the spectral function is furthermore often substituted by the analytically continued OPE expression, making use of the 
quark-hadron duality. Integrals that only involve the low-energy part of the spectral function can thus be derived from QCD via the OPE. 
Let us discuss each step outlined above more explicitly. 

\subsubsection{The dispersion relation}  
Variations of the dispersion relation derived here are used in many branches of physics \cite{Toll:1956cya,Pickering:1985gf,SheikBahae:2005}. 
In some fields, they are referred to as Kramers-Kroenig relations \cite{deL.Kronig:26,Kramers:1927}. 
First, we define the correlator as 
\be
\Pi(q^2) = i \int d^4x e^{iqx} \langle 0|T[J(x) J^{\dagger}(0) ] |0\rangle . 
\label{eq:correlator}
\ee
Here, $J(x)$ is a general operator that in principle 
can have Dirac or Lorentz indices, in which case $\Pi(q^2)$ 
becomes a matrix. For simplicity these non-essential complications are ignored here. The symbol $\langle 0|$ denotes the 
non-trivial QCD vacuum, but can be generalized for instance to the ground state of nuclear matter as will be 
done later. Furthermore, when considering sum rules at finite temperature, the retarded correlator 
should be used instead of the above time-ordered one, 
because it has suitable analytic properties when regarded as function of $q_0 = \omega$ \cite{Hatsuda:1992bv,Abrikosov:QFT} (see also Section \ref{sec:exact-sumrule}). 

\begin{figure}[tb]
\begin{center}
\begin{minipage}[t]{8 cm}
\vspace{0.5 cm}
\hspace{-1.5 cm}
\includegraphics[width=11cm,bb=0 0 360 252]{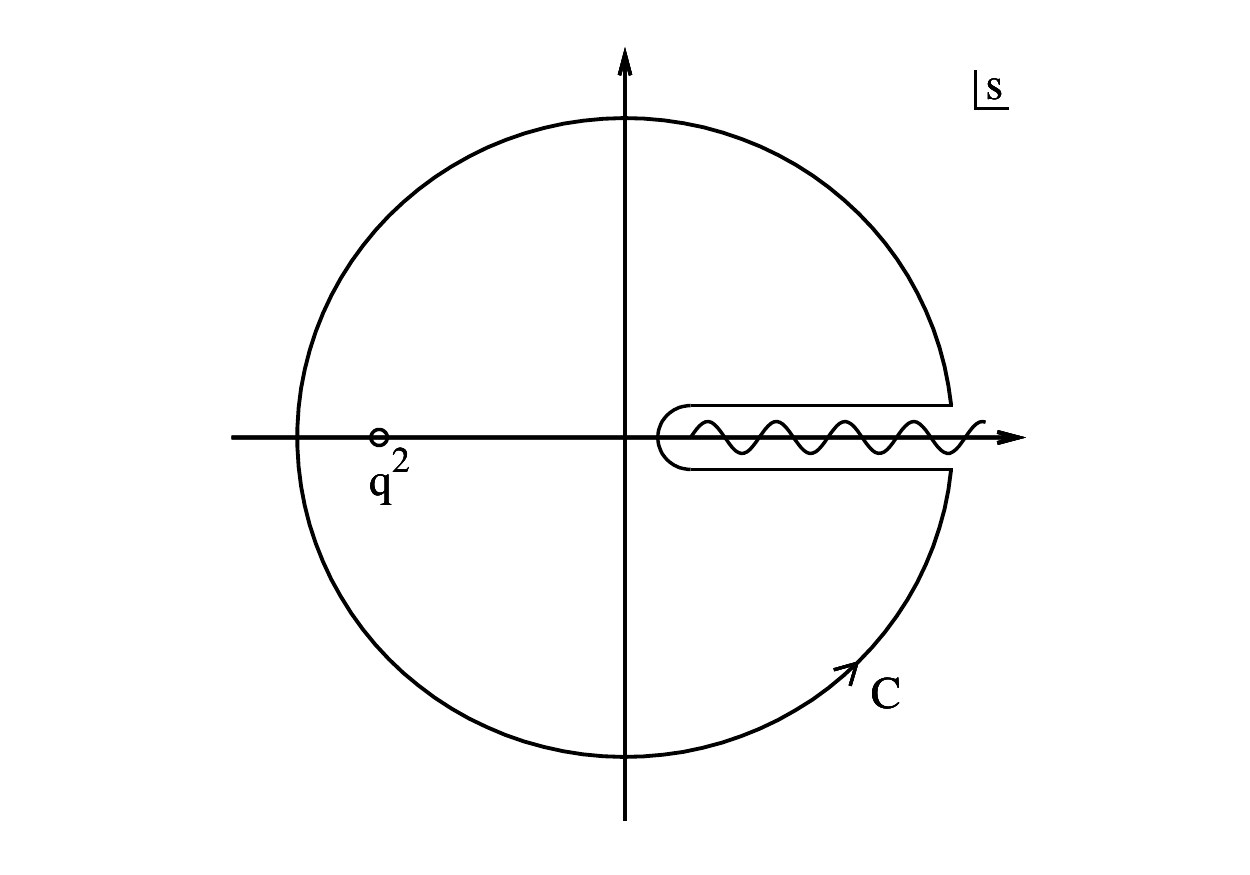}
%\vspace{1 cm}
\end{minipage}
\begin{minipage}[t]{16.5 cm}
\caption{The contour $C$ on the complex energy of the variable $s$, used in Eq.\,(\ref{eq:derive.correlator.1}). 
The wavy line represents the location of potential poles and cuts of the correlator $\Pi(s)$. 
Taken from Fig.\,A.1 of Ref.\,\cite{Gubler:2013moa}. 
\label{fig:contour.for.trd.sr}}
\end{minipage}
\end{center}
\end{figure}

The function $\Pi(q^2)$ is known to be analytic on the whole complex $q^2$ plane except the positive real 
axis, where it can have poles and cuts, which correspond to the physical states that are generated by the operator 
$J^{\dagger}(0)$. Making use of this analyticity, we employ the Cauchy theorem to obtain 
\begin{align}
\Pi(q^2) &= \frac{1}{2\pi i} \oint_C ds \frac{\Pi(s)}{s - q^2} \nonumber \\
&= \frac{1}{2\pi i} \oint_{|s| = R} ds \frac{\Pi(s)}{s - q^2} + \frac{1}{2\pi i} \int_0^{R} ds \frac{\Pi(s+i\epsilon) - \Pi(s-i\epsilon)}{s - q^2}.  
\label{eq:derive.correlator.1}
\end{align}
Here, $R$ denotes the radius of the large circle in Fig.\,\ref{fig:contour.for.trd.sr}. 
%%%%%%%%%%%%%%%%%%%%%%%%%%%%%%%%%%%%%%%%%%%%%%%%%%%%%%%%%%%%%%%%%%%%%%%%%%%%%%%%%%%%%%%%%%%%%%%%%%%%%%%%%%%%%%%%%%%%
%%%%%%%%%%%%%%%%%%%%%%%%%%%%%%%%%%%%%%%%%%%%%%%%%%%%%%%%%%%%%%%%%%%%%%%%%%%%%%%%%%%%%%%%%%%%%%%%%%%%%%%%%%%%%%%%%%%%
Next, we take $R$ to infinity, which means that the first term in Eq.\,(\ref{eq:derive.correlator.1}) vanishes if $\Pi(s)$ decreases fast enough 
at $|s| = R \to \infty$. 
As will be demonstrated in the next paragraph, this is not the case in 
many practical situations and therefore subtraction terms have to be 
introduced. 
We will here 
for simplicity assume that the first term indeed vanishes for $|s| = R \to \infty$. 
The second term can be cast into a simple form by the Schwarz reflection principle, which gives  
\be
\Pi(s+i\epsilon) - \Pi(s-i\epsilon) = 2i \mathrm{Im} \Pi(s+i\epsilon). 
\label{eq:derive.correlator.2}
\ee
We hence have derived the dispersion relation as 
\be
\Pi(q^2) = \frac{1}{\pi} \int_0^{\infty} ds \frac{\mathrm{Im} \Pi(s+i\epsilon)}{s - q^2} = \frac{1}{\pi} \int_0^{\infty} ds \frac{\rho(s)}{s - q^2},  
\label{eq:derive.correlator.3}
\ee
where we have defined $\mathrm{Im} \Pi(s+i\epsilon) = \rho(s)$. 

Let us for a moment return to the case where the first term in Eq.\,(\ref{eq:derive.correlator.1}) does not vanish for $|s| = R \to \infty$ 
and/or the integral on the right hand side of Eq.\,(\ref{eq:derive.correlator.3}) diverges, 
in which case subtraction terms have to be introduced to tame the divergence. 
If this divergence is logarithmic, one only needs one subtraction term, 
 \begin{align}
\widetilde{\Pi}(q^2) & \equiv \Pi(q^2) - \Pi(0) \nonumber \\
& = \frac{q^2}{\pi} \int_0^{\infty} ds \frac{\rho(s)}{s(s - q^2)}. 
\label{eq:derive.correlator.4}
\end{align}
The same prescription can be applied arbitrary many times by subtracting the Taylor expansion of 
$\Pi(q^2)$ around $q^2 = 0$ term by term, by which power like divergences of any order can be 
eliminated, which suffices for all practical applications in QCD. Note that $\Pi(0)$ is a divergent constant  
in the above example, which, however, does not play any important role in the formulation 
of the final form of the sum rules. Indeed, applying the Borel transform to Eq.\,(\ref{eq:derive.correlator.4}), 
this constant (or any positive power of $q^2$) vanishes. In fact, the correlator is in any case only well 
defined modulo power terms of $q^2$ (see Refs.\,\cite{Romatschke:2009ng,CaronHuot:2009ns}). 
We conclude this section by noting that 
the discussion preceding Eq.\,(\ref{eq:derive.correlator.3}) is not the only path  
to derive a dispersion relation. As will be seen later in Section \ref{deriv.of.exaxt.sr}, the derivation of exact sum rules at 
finite temperature can be done using a somewhat different method. 

\subsubsection{The quark-hadron duality \label{sec:duality}}
One more concept often mentioned in relation to the derivation of QCD sum rules is the so-called quark-hadron duality. 
We refer the interested reader to Refs.\,\cite{Shifman:2000jv,Hofmann:2003qf} for more detailed discussions and here only give a brief description. 
The quark-hadron duality was first proposed in Ref.\,\cite{Poggio:1975af} and says that a hadronic and experimentally 
measurable spectral function $\rho(s)$ appropriately averaged over a certain energy range can be described by the 
corresponding expression calculated from QCD and its degrees of freedom, quarks and gluons. More precisely, one 
sometimes distinguishes between a local and global quark-hadron duality \cite{Hofmann:2003qf}. The former refers to the case where the non-energy-averaged  
hadronic spectral function agrees with its QCD counterpart within uncertainties. 
At low energies, this local duality is often strongly violated due to the sharp resonance peaks which cannot be 
accurately described by perturbative QCD. On the other hand, at high energies, where hadronic resonances are wide and 
overlapping, the local duality is often satisfied rather well. In practical QCD sum rule analyses, one 
makes use of this and approximates the spectral function above a certain threshold $s_{th}$ by its QCD expression 
[see Section \ref{sec:conventional.analysis} and especially Eq.\,(\ref{eq:pole.plus.continuum})]. 
The global quark-hadron duality in contrast refers to the (approximate) equality between an integrated hadronic spectral function 
and the integral of the same quantity computed from QCD. Specifically, considering Eq.\,(\ref{eq:derive.correlator.3}), this 
corresponds to the statement that $\Pi(q^2)$ on the left-hand side for sufficiently large $Q^2 = -q^2$ is equal to 
the integral of $\rho(s)/[\pi(s - q^2)]$ on the right-hand side, where $\rho(s)$ is the hadronic spectral function. 

\subsection{The operator product expansion \label{sec:OPE}}
Here we discuss the second technique used to derive the sum rules, the operator product expansion (OPE). 
Originally proposed by Wilson \cite{Wilson:1969zs}, it can in position space be summarized as 
\be
\hat{A}(x) \hat{B}(y) \xrightarrow{x \to y} \sum_n C_n(x-y) \hat{O}_n \Bigl(\frac{x+y}{2}\Bigr).
\label{eq:OPE.1}
\ee  
Here, $\hat{A}(x)$ and $\hat{B}(y)$ are arbitrary operators defined at positions $x$ and $y$. 
The essence of the above equation is that if $x$ is sufficiently close to $y$, the product of $\hat{A}(x)$ and $\hat{B}(y)$ can be expanded in 
a series of local operators $\hat{O}_n$ defined somewhere in between $x$ and $y$ [we could just as well have written $\hat{O}_n(x)$ or $\hat{O}_n(y)$ 
instead of $\hat{O}_n \bigl(\frac{x+y}{2}\bigr)$], with corresponding coefficients $C_n(x-y)$, which depend only on the distance 
between $x$ and $y$ and are simply C-numbers. The $C_n(x-y)$ are called Wilson coefficients, which are governed by the short distance dynamics of $x-y$ and 
can therefore due to asymptotic freedom of QCD be calculated perturbatively if the distance $x - y$ is small enough. 
Potential contact terms, which are proportional to $\delta^{(4)}(x-y)$ or its derivative, are neglected in all the OPE expressions of our manuscript. 
This causes no problem because such terms do not appear in the final form of the sum rule after the Borel transform. 

After taking the expectation value with respect to some general state $|\Omega\rangle$ (which can be the vacuum, the thermal ensemble or the 
ground state of nuclear matter) it is usually assumed that the expectation values of the local operators $\hat{O}_n$ are position independent. 
Thus, computing the Fourier transform of Eq.\,(\ref{eq:OPE.1}) sandwiched between $|\Omega\rangle$, we obtain 
\be
i \int d^4 x e^{iq(x-y)} \langle \Omega| \hat{A}(x) \hat{B}(y) |\Omega\rangle \xrightarrow{|q^2| \to \infty} \sum_n C_n(q) \langle \Omega| \hat{O}_n |\Omega\rangle, 
\label{eq:OPE.2}
\ee
where $C_n(q)$ denotes the Fourier transform (times $i$) of $C_n(x-y)$. 
Using dimensional analysis, one can easily determine the functional forms of $C_n(x-y)$ and $C_n(q)$. 
In the short distance or large energy limit where the OPE is applicable, low energy scales such as light quark masses can 
be ignored, such that $x-y$ or $q$ are the only dimensional quantities that can appear in $C_n(x-y)$ and $C_n(q)$ 
(this is not necessarily true for channels involving heavy quarks $c$ or $b$, where the simple arguments 
given here have to be modified). 
Assuming the mass dimensions of $\hat{A}(x)$, $\hat{B}(y)$ and $\hat{O}_n$ to be $d_A$, $d_B$ and $d_n$, 
we get for $C_n(x-y)$, 
\be
C_n(x-y) \xrightarrow{x \to y} \Biggl[\frac{1}{(x-y)^2} \Biggr]^{(d_A + d_B - d_n)/2}, 
\label{eq:OPE.3}
\ee
and for $C_n(q)$, 
\be
C_n(q) \xrightarrow{|q^2| \to \infty} q^{d_A + d_B - d_n - 4}. 
\label{eq:OPE.4}
\ee
In the last equation, we have ignored potential logarithmic factors of $\log(-q^2/\mu^2)$ ($\mu^2$: renormalization scale), which 
occur for $d_A + d_B - d_n - 4 \geq 0$, but are not important for the discussion here. 
As we see in Eq.\,(\ref{eq:OPE.4}), operators $\hat{O}_n$ with the smallest values of $d_n$ dominate the expansion if 
$q^2$ is large enough. 
The operators $\hat{O}_n$ are generally constructed from quark fields (which have mass dimension 3/2), 
gluon field strengths (mass dimension 2) 
and covariant derivatives (mass dimension 1). 
If the state $|\Omega\rangle$ corresponds to the vacuum ($| 0 \rangle$), only Gauge- and Lorentz-invariant 
operators can have non-zero expectation values. Up to mass dimension 6, these are 
\begin{align}
\mathrm{dimension}\, 0:\,\, & 1, \nonumber \\
\mathrm{dimension}\, 3:\,\, & \langle 0 | \bar{q} q |0 \rangle, \nonumber \\
\mathrm{dimension}\, 4:\,\, & \langle 0 | G^a_{\mu \nu} G^{a\mu\nu} |0 \rangle, \nonumber \\
\mathrm{dimension}\, 5:\,\, & \langle 0 | \bar{q} \sigma_{\mu\nu} t^a G^{a\mu\nu} q |0 \rangle, \nonumber \\
\mathrm{dimension}\, 6:\,\, & \langle 0 | \bar{q} q \bar{q} q |0 \rangle,\, \langle 0 | \bar{q} \gamma_5 q \bar{q} \gamma_5 q |0 \rangle, \label{eq:condensates.vacuum} \\
& \langle 0 | \bar{q} t^a q \bar{q} t^a q |0 \rangle,\, \langle 0 | \bar{q} \gamma^{\mu} t^a q \bar{q} \gamma_{\mu} t^a q |0 \rangle,\,\dots \nonumber \\
& \langle 0 | f^{abc} G^{a\nu}_{\mu} G^{b\lambda}_{\nu} G^{c\mu}_{\lambda} |0 \rangle,  \nonumber \\
\dots & \nonumber
\end{align}
Here, $t^a = \lambda^a/2$, $\lambda^a$ being the Gell-Mann matrices, 
while $f^{abc}$ stands for the structure constants of the $SU(3)$ (color) group. 
We have in Eq.\,(\ref{eq:condensates.vacuum}) for simplicity only considered one species of 
quarks, which is denoted as $q$. 
In the above list we have not included the gauge non-invariant gluon condensate of dimension 2, 
$\langle 0 | A^a_{\mu} A^{a\mu} |0 \rangle$. The potential existence and relevance of this condensate has generated a fairly large body of work 
(see for instance Refs.\,\cite{Lavelle:1988eg,Lavelle:1992yh,Boucaud:2000nd,Gubarev:2000eu,Gubarev:2000nz,Kondo:2001nq}), but is nevertheless far less established 
than those given in Eq.\,(\ref{eq:condensates.vacuum}) and is usually not considered in present-day QCD sum rule studies. 
At dimension 6, we have shown only a few representative examples of all possible four-quark condensates, of which some can be related by 
Fierz-transformations \cite{Thomas:2007gx}. For the gluonic condensates at dimension 6, one can in fact construct one more operator 
with two covariant derivatives and two gluon fields, which however can be rewritten as a four-quark condensate by the use of the equation of motion. 
With the exception of the four-quark condensates, the above list is therefore complete up to dimension 6.  

Once one starts to consider the case of finite temperature, density or magnetic field, more condensates can be constructed because Lorentz symmetry gets 
partly broken by these external fields. 
For the case of finite temperature and density, 
the most simple way to do this is to define a normalized four-vector $u^{\mu}$ ($u^2 = 1$) with spatial components that correspond to the velocity of the hot or dense medium 
and to then assemble all possible combinations of quark fields, gluon field strengths, covariant derivatives and $u^{\mu}$ as before. 
In this derivation, one usually considers the medium to be colorless and invariant with respect to parity and time reversal, which we 
will assume as well in the discussions of this review. 
The details of this procedure have been discussed for instance in Refs.\,\cite{Jin:1992id,Zschocke:2011aa}. Here, we just reproduce the final findings, which are 
\begin{align}
\mathrm{dimension}\, 3:\,\, & \langle \Omega | \bar{q} \gamma^{\mu} q | \Omega \rangle, \nonumber \\
\mathrm{dimension}\, 4:\,\, & \langle \Omega | \mathcal{ST} \bar{q} \gamma^{\mu} iD^{\nu} q | \Omega \rangle, \langle \Omega | \bar{q} iD^{\mu} q | \Omega \rangle, 
\langle \Omega |\mathcal{ST} G_{\alpha}^{a\mu} G^{a\nu\alpha} | \Omega \rangle, \nonumber \\
\mathrm{dimension}\, 5:\,\, & \langle \Omega |\mathcal{ST} \bar{q} iD^{\mu} iD^{\nu} q | \Omega \rangle, 
\langle \Omega |\mathcal{ST} \bar{q} \gamma^{\alpha} iD^{\mu} iD^{\nu} q | \Omega \rangle, 
\langle \Omega | \bar{q} \gamma^{\mu} \sigma_{\alpha \beta} G^{a \alpha \beta} t^a q | \Omega \rangle,  \label{eq:condensates.medium} \\
\mathrm{dimension}\, 6:\,\, & \langle \Omega |\mathcal{ST} \bar{q} \gamma^{\mu} t^a q \bar{q} \gamma^{\nu} t^a q | \Omega \rangle, 
\langle \Omega |\mathcal{ST} \bar{q} \gamma^{\mu} iD^{\nu} iD^{\alpha} iD^{\beta} q | \Omega \rangle,\,\dots \nonumber \\
& \langle \Omega |\mathcal{ST} G^a_{\alpha \beta} iD^{\mu} iD^{\mu} G^{a\alpha \beta} | \Omega \rangle, 
\langle \Omega |\mathcal{ST} G^{a\alpha}_{\delta} iD^{\beta} iD^{\mu} G^{a\nu \delta} | \Omega \rangle,\,\dots \nonumber \\
\dots & \nonumber
\end{align}
Here the letters $\mathcal{ST}$ stand for the operation of making the Lorentz indices symmetric and traceless, 
\begin{align}
\mathcal{ST} O^{\mu \nu} =&\,\, \frac{1}{2} (O^{\mu \nu} + O^{\nu \mu}) - \frac{1}{4} g^{\mu \nu} \tensor{O}{_{\alpha}^{\alpha}}, \label{eq:symmetric.traceless.1} \\
\mathcal{ST} O^{\mu \nu \alpha} =&\,\, \frac{1}{6} (O^{\mu \nu \alpha} + O^{\mu \alpha \nu} + O^{\nu \mu \alpha} + O^{\nu \alpha \mu} + O^{\alpha \mu \nu} + O^{\alpha \nu \mu}) \nonumber \\
& + \frac{1}{3}(g^{\mu \nu} A^{\alpha} + g^{\mu \alpha} A^{\nu} + g^{\nu \alpha} A^{\mu}), \label{eq:symmetric.traceless.2} \\ 
\mathcal{ST} O^{\mu \nu \alpha \beta} =&\,\, \frac{1}{24}(O^{\mu \nu \alpha \beta}+ \text{23 other orderings of }\mu \nu \alpha \beta) \nonumber \\ 
& + \frac{1}{6}(g^{\mu \nu} B^{\alpha \beta} + g^{\mu \alpha} B^{\nu \beta} + g^{\mu \beta} B^{\nu \alpha} + 
g^{\nu \alpha} B^{\mu \beta} + g^{\nu \beta} B^{\mu \alpha} + g^{\alpha \beta} B^{\mu \nu}) \nonumber \\
& + \frac{1}{3}C(g^{\mu \nu} g^{\alpha \beta} + g^{\mu \alpha} g^{\nu \beta} + g^{\mu \beta} g^{\nu \alpha}). 
\label{eq:symmetric.traceless}
\end{align}
$A^{\mu}$ can easily be obtained from the tracelessness condition of $\mathcal{ST} O^{\mu \nu \alpha}$, 
\begin{align}
A^{\mu} = - \frac{1}{6}(\tensor{O}{_{\alpha}^{\alpha}^{\mu}} + \tensor{O}{_{\alpha}^{\mu}^{\alpha}} + \tensor{O}{^{\mu}_{\alpha}^{\alpha}}).
\end{align}
In Eq.\,(\ref{eq:symmetric.traceless}), we define $B^{\alpha \beta}$ to be symmetric and traceless. From the tracelessness condition of $\mathcal{ST} O^{\mu \nu \alpha \beta}$, we then have  
\begin{align}
B^{\alpha \beta} =&\,\, \frac{1}{16} \Bigl[ 
(\tensor{O}{_{\delta}^{\delta}_{\sigma}^{\sigma}} + \tensor{O}{_{\delta \sigma}^{\delta \sigma}} 
+ \tensor{O}{_{\delta \sigma}^{\sigma \delta}}) g^{\alpha \beta} \nonumber \\ 
&\qquad -( \tensor{O}{_{\delta}^{\delta \alpha \beta}} + \tensor{O}{_{\delta}^{\delta \beta \alpha}} + \tensor{O}{_{\delta}^{\alpha \delta \beta}} 
+ \tensor{O}{_{\delta}^{\beta \delta \alpha}} + \tensor{O}{_{\delta}^{\alpha \beta \delta}} + \tensor{O}{_{\delta}^{\beta \alpha \delta}} \nonumber \\
&\qquad + \tensor{O}{^{\alpha}_{\delta}^{\delta \beta}} + \tensor{O}{^{\beta}_{\delta}^{\delta \alpha}} + \tensor{O}{^{\alpha}_{\delta}^{\beta \delta}} 
+ \tensor{O}{^{\beta}_{\delta}^{\alpha \delta}} + \tensor{O}{^{\alpha \beta}_{\delta}^{\delta}} +  \tensor{O}{^{\beta \alpha}_{\delta}^{\delta}})
\Bigr], \\ 
C =&\,\,-\frac{1}{24}(\tensor{O}{_{\delta}^{\delta}_{\sigma}^{\sigma}} + \tensor{O}{_{\delta \sigma}^{\delta \sigma}} 
+ \tensor{O}{_{\delta \sigma}^{\sigma \delta}}). 
\label{eq:symmetric.traceless.3}
\end{align}
It is noteworthy that the $0$ component of the Lorentz violating dimension 3 condensate is just $\langle \Omega | q^{\dagger} q | \Omega \rangle$, which is nothing but the 
quark number density of the state $| \Omega \rangle$. 
Furthermore, the first and third condensates on the second line of Eq.\,(\ref{eq:condensates.medium}) are proportional to the quark and gluon components of the energy momentum 
tensor. 

We do not provide the complete set of independent operators of dimension 6 in Eq.\,(\ref{eq:condensates.medium}), but again only a few representative examples. 
For the complete list of operators appearing in the vector channel OPE, see Ref.\,\cite{Kim:2017nyg}. A recent discussion about the independent 
Lorentz violating gluonic operators of dimension 6 and a calculation of their anomalous dimensions is given in Ref.\,\cite{Kim:2015ywa}. 
Moreover, the non-scalar condensates appearing in a magnetic field generally have a different structure. They will be discussed in Section\,\ref{sec:conds.in.magn.field}. 

Finally, we consider the renormalization group (RG) effect on the OPE. 
The expectation values of the operators in Eq.\,(\ref{eq:OPE.2}) make sense only when the energy scale is specified at which the operators 
and their corresponding Wilson coefficients are evaluated.
In the present case, $q$ is the natural choice for the scale, which we take to be large in the derivation of the sum rule. 
On the other hand, the expectation values obtained from, say, lattice QCD, are evaluated at a finite energy scale such as $T$. 
Such expectation values evaluated at different scales are related by RG equations. 
The perturbative RG equation provides scaling properties additional to the canonical dimension $d_n$, 
\begin{align}
\label{eq:RG-log}
{\cal O}_n(q)\sim \left(\frac{\ln(q_0/\Lambda_{\text{QCD}})}{\ln(q/\Lambda_{\text {QCD}})}\right)^{a_n}{\cal O}_n(q_0), 
\end{align}
where $a_n$ is proportional to the anomalous dimension of the operator ${\cal O}_n$. 
Furthermore, a general operator may mix with other operators of the same dimension due to the RG effect. 
This point is not always taken into account in the conventional sum rule analysis, in which the finite UV cutoff is introduced so that the effect of the anomalous dimension is negligible.
However, for the exact sum rules to be reviewed in Sec.\,\ref{sec:exact-sumrule}, we will consider the infinite energy limit, in which this effect 
has to be taken into account. 

This RG effect actually generates a very useful byproduct, particularly handy for finite temperature calculations \cite{CaronHuot:2009ns}. 
The correlator from the OPE is at finite $T$ usually calculated in imaginary time. To obtain the retarded Green function, 
which plays a central role for the derivation of the exact sum rules, 
one needs to do an analytic continuation to real time. 
The logarithmic factor coming from the RG scaling/mixing of Eq.\,(\ref{eq:RG-log}) gives a constant imaginary contribution after such an analytic continuation, 
which means that the OPE can predict the spectral function at high energy.
As the other parts in the OPE [$C_n(q)$] have polynomial (and possibly logarithmic) dependence on $q$, the resultant spectral function has the same $q$ dependence, 
This structure in the spectral function is called UV tail. 
The explicit form of the UV tail in the vector channel is given in Appendix \ref{sec:Appendix1}. 

\subsubsection{Status of higher order Wilson coefficient computations}
Over the years, higher order $\alpha_s$ terms of Wilson coefficients have been computed for many channels. We will give a short overview of these calculations here. 
Reviewing the numerous purely perturbative computations, which in principle correspond to Wilson coefficients of the identity operator, would however go beyond the scope of this review. 
With the exception of a number of exotic channels, we will therefore only consider terms involving condensates of at least mass dimension 3. 

\paragraph{Mesonic correlators}\mbox{}\\
The most detailed information about NLO and NNLO $\alpha_s$ terms is available for two-quark mesonic channels. 
Let us first consider currents with two light quarks. 
For the vector and axial-vector channels, the NLO $\alpha_s$ corrections of the dimension 3 quark condensate (which appears at linear order in the quark mass $m_q$) were computed for 
the first time in Ref.\,\cite{Pascual:1981jr}. 
For the same vector and axial-vector channels, 
$\alpha_s$ and $\alpha_s^2$ terms of the dimension 3 quark condensate and the dimension 4 gluon condensate were calculated in Ref.\,\cite{Chetyrkin:1985kn}
The same terms were computed similarly for the vector, scalar and pseudoscalar channels in Ref.\,\cite{Surguladze:1990sp} (see also Ref.\,\cite{Surguladze:1995yp}). 
For all the above channels, the LO quark condensate terms are of order $\mathcal{O}(1)$ while those for the gluon condensate are of order $\mathcal{O}(\alpha_s)$. 
The mixed condensate 
$\langle 0 | \bar{q} \sigma_{\mu\nu} t^a G^{a\mu\nu} q |0 \rangle$, whose Wilson coefficient is proportional to $m_q$ and the strong coupling $g$, at LO is known to vanish 
for the vector current correlator \cite{Shifman:1978bx}. It 
would hence be useful to calculate the respective NLO term, especially in the phenomenologically important vector channel. To our knowledge, this has presently not yet been done for 
any channel. 
The NLO corrections in the four-quark operator Wilson coefficients for the vector and axial-vector channels were obtained in Ref.\,\cite{Lanin:1986zs} (this reference is unfortunately rather difficult to find online, 
the corresponding results are however reproduced and further discussed in Refs.\,\cite{Braaten:1991qm,Adam:1993uu}). 

Next, we discuss $\alpha_s$ corrections for heavy-light quark current correlators, about which much less is known and presently only NLO 
terms for the quark condensate $\langle \overline{q} q \rangle$ have been computed. This was first done in Ref.\,\cite{Jamin:2001fw} for the 
pseudoscalar channel. Later, in Ref.\,\cite{Gelhausen:2013wia} the same $\alpha_s$ correction was also calculated for the vector channel. 
The appendix of Ref.\,\cite{Gelhausen:2013wia} is especially useful, as it gives explicit OPE expressions of pseudoscalar and vector 
channels both before and after the Borel transform. Furthermore, results for the scalar and axial-vector channels 
are available in Ref.\,\cite{Wang:2015uya}. 

Finally, we turn to meson current correlators with two heavy quarks (quarkonia), which have only gluonic operators in their OPE, as heavy quark condensates can 
be recast as gluonic condensates with the help of the heavy quark expansion. In principle, light quark operators can also contribute,  
but appear only at order $\mathcal{O}(\alpha_s^2)$ and will therefore not be discussed here. 
The NLO corrections to the Wilson coefficient of the dimension 4 gluon condensate for the scalar, pseudoscalar, vector and 
axial-vector channels were obtained in Ref.\,\cite{Broadhurst:1994qj}. These are the only NLO results available so far for quarkonium correlators. 

For mesons containing four or more quarks, NLO calculations have up to now only been carried out for purely perturbative terms. 
For a large number of light tetraquark channels, this was done in Ref.\,\cite{Groote:2014pva}. 

\paragraph{Baryonic correlators}\mbox{}\\
For baryonic channels, only a few NLO terms have been obtained so far. 
The first attempt to compute $\alpha_s$ corrections to the dimension 3 chiral condensate term were made in Ref.\,\cite{Krasnikov:1982ea}. 
Later, more terms in more channels (NLO terms of the dimension 3 chiral condensate, the dimension 5 mixed condensate and dimension 6 four-quark condensates for both the 
proton and $\Delta$ channels), were calculated in Ref.\,\cite{Chung:1984gr}, which were however not fully consistent with those of 
Ref.\,\cite{Krasnikov:1982ea}. The results of Ref.\,\cite{Krasnikov:1982ea} and parts of Ref.\,\cite{Chung:1984gr} were further 
corrected in Refs.\,\cite{Jamin:1987gq,Ovchinnikov:1991mu}. More recently, perturbative corrections to the Wilson coefficient 
of the dimension 3 vector condensate (which vanishes in vacuum, but is non-zero at finite density) were obtained in Ref.\,\cite{Groote:2008hz}. 

For exotic baryons with at least five valence quarks or anti-quarks, no NLO corrections to non-perturbative condensate terms have so 
far been computed. Purely perturbative NLO $\alpha_s$ terms were however obtained in Refs.\,\cite{Groote:2006sy,Groote:2011my} for light quark pentraquark correlators. 

\section{The QCD condensates \label{sec:QCD.condensates}} 
It has long been known that non-perturbative quantum fluctuations generate condensates, which break chiral or dilatation symmetries. 
These symmetries are present in the Lagrangian of massless QCD, but are not reflected in the hadronic spectrum. 
Nevertheless, with a complete and non-perturbative understanding of QCD still missing, many features of 
these condensates are not yet well understood and established. Until not long ago, the QCD condensates were for instance thought of as properties of 
the QCD vacuum, while it was recently claimed in Ref.\,\cite{Brodsky:2010xf} that they are in fact properties of hadrons themselves. 
This led to a vigorous debate about the true nature of the condensates (see for example Refs.\,\cite{Reinhardt:2012xs,Brodsky:2012ku,Lee:2012ga}). 
We will in this section not go into the intricate details of this debate, but pragmatically focus on what is presently known about the individual condensate values 
and about their modifications in extreme environments. 

\subsection{Vacuum}
The vacuum condensate that is presently by far best known and understood is the quark (or chiral) condensate averaged over the 
lightest $u$ and $d$ quarks: $\langle 0 | \overline{q}q | 0 \rangle = (\langle 0 | \overline{u}u | 0 \rangle + \langle 0 | \overline{d}d | 0 \rangle)/2$. 
It is an order parameter of chiral symmetry breaking 
i.e. its value being non-zero means that this symmetry is spontaneously broken in the vacuum. 
Earliest estimates of the quark condensates have been obtained based on the Gell-Mann-Oakes-Renner relation \cite{GellMann:1968rz}, 
\begin{equation}
f_{\pi}^2 m_{\pi}^2 = -2 m_q \langle 0 | \overline{q}q | 0 \rangle. 
\label{eq:GellMann.Oakes}
\end{equation}
Here, $f_{\pi}$ and $m_{\pi}$ are the pion decay constant and mass, which can be measured experimentally, while $m_q$ is the averaged $u$ and $d$ quark mass. 
This relation is however not exact because Eq.\,(\ref{eq:GellMann.Oakes}) is only the leading order result of 
the chiral expansion and receives corrections due to non-zero quark masses \cite{Gasser:1983yg,Jamin:2002ev}. 
Nowadays, lattice QCD is able to compute the chiral condensate at the physical point with good precision and with most (if not all) 
systematic uncertainties under control. The Flavour Lattice Averaging Group (FLAG) \cite{Aoki:2016frl} 
presently (November 2018) gives an averaged value 
of 
\begin{equation}
\langle 0 | \overline{q}q | 0 \rangle = -[272(5)\,\mathrm{MeV}]^3 \hspace{0.5cm} \cite{Bazavov:2010yq,Borsanyi:2012zv,Durr:2013goa,Boyle:2015exm,Cossu:2016eqs}
\label{eq:quark.cond.value}
\end{equation}
for $N_f = 2 + 1$ flavors in the $\overline{\mathrm{MS}}$ scheme at a renormalization scale of 2 GeV (see their webpage for updates). 

The strange quark condensate $\langle 0 | \overline{s}s | 0 \rangle$ is much less well determined. 
Old QCDSR analyses studying the energy levels and splittings of baryons led to a value of 
$\langle 0 | \overline{s}s | 0 \rangle / \langle 0 | \overline{q}q | 0 \rangle = 0.8 \pm 0.1$ \cite{Reinders:1984sr}. 
From the lattice, there are to our knowledge at present only two publicly available results, which read  
\begin{align}
\langle 0 | \overline{s}s | 0 \rangle &= -[290(15)\,\mathrm{MeV}]^3  \hspace{0.5cm} (N_f = 2 + 1 + 1) \hspace{0.5cm}  \cite{McNeile:2012xh}, \label{eq:strange.quark.cond.value} \\
\langle 0 | \overline{s}s | 0 \rangle &= -[296(11)\,\mathrm{MeV}]^3  \hspace{0.5cm} (N_f = 2 + 1) \hspace{0.5cm}  \cite{Davies:2018hmw}. \label{eq:strange.quark.cond.value.2}
\end{align}
Both are given in the $\overline{\mathrm{MS}}$ scheme at a renormalization scale of 2 GeV. 
In Ref.\,\cite{McNeile:2012xh}, similar values were obtained for both $\langle 0 | \overline{s}s | 0 \rangle$ 
and 
$\langle 0 | \overline{q}q | 0 \rangle$: $\langle 0 | \overline{s}s | 0 \rangle / \langle 0 | \overline{q}q | 0 \rangle \simeq  1.08(16)$.
The tendency of this result does not agree with the above-mentioned older estimate of Ref.\,\cite{Reinders:1984sr}, 
which is smaller than 1 and is still widely used in practice. 
It would therefore be helpful to have further independent lattice computations that could check the reliability 
of Eqs.\,(\ref{eq:strange.quark.cond.value}) and (\ref{eq:strange.quark.cond.value.2}). 

The gluon condensate is usually defined as a product with the strong coupling constant, 
which is a scale-independent quantity: 
$\langle 0 | \frac{\alpha_s}{\pi} G^a_{\mu \nu} G^{a\mu\nu} |0 \rangle \equiv \langle 0 | \frac{\alpha_s}{\pi} G^2 |0 \rangle$. 
A first estimate of its value was obtained in Refs.\,\cite{Shifman:1978bx,Shifman:1978by} from an analysis of 
charmonium sum rules, for which the gluon condensate is the leading order non-perturbative power correction. 
Their value 
\begin{equation}
\langle 0 | \frac{\alpha_s}{\pi} G^2 | 0 \rangle =  (0.012 \pm 0.004)\,\mathrm{GeV}^4
\label{eq:gluon.cond.value}
\end{equation}
is frequently used even in current QCDSR studies, simply because no significant progress 
in its determination has since been made and no later estimate can beyond any doubt claim to be more reliable. 
Over the years, estimates have been given that are a few times larger \cite{Marrow:1987mx} or smaller \cite{Ioffe:2010zz}, 
which shows that the systematic uncertainties in the determination of this condensate are still large. 
For further details and references, we refer the reader to Table 1 of Ref.\,\cite{Narison:2018dcr} 
for a compilation of available gluon condensate estimates. 

It is, however, worth discussing here some recent progress in computing the gluon condensate on the lattice. 
At first sight this 
seems to be a relatively straightforward task as the operator $G^a_{\mu \nu} G^{a\mu\nu}$ is directly related 
to the plaquette in a lattice QCD computation. 
Attempts in this direction were accordingly made already in the very early days of lattice QCD calculations \cite{DiGiacomo:1981lcx,Kripfganz:1981ri}. 
The situation has, however, turned out to be more complicated than initially expected, 
because one in principle needs to subtract a perturbative contribution 
from the lattice result to obtain the purely non-perturbative value of the gluon condensate. 
The way one defines (and truncates) this perturbative part will therefore change the final value of 
the gluon condensate obtained in the calculation. Recently, the technique of the numerical stochastic perturbation theory was used to compute 
the corresponding perturbative series to high orders (up to $\alpha_s^{35}$!), after which it was subtracted from the respective 
lattice observable. 
For more detailed discussions about this issue, see Refs.\,\cite{Horsley:2012ra,Bali:2014sja}. 
The final values obtained for the gluon condensate in this approach are 
\begin{align}
\langle 0 | \frac{\alpha_s}{\pi} G^2 | 0 \rangle & = 0.028(3)\,\mathrm{GeV}^4  \hspace{0.3cm} (\alpha_s^{20})  \hspace{0.3cm} \cite{Horsley:2012ra}, \\
\langle 0 | \frac{\alpha_s}{\pi} G^2 | 0 \rangle & = 0.077\,\mathrm{GeV}^4 \hspace{0.3cm}
\Bigl[
\delta \langle 0 | \frac{\alpha_s}{\pi} G^2 | 0 \rangle = 0.087\,\mathrm{GeV}^4 \Bigr]  \hspace{0.3cm} (\alpha_s^{35})  \hspace{0.3cm} \cite{Bali:2014sja}. 
\label{eq:gluon.cond.value.lattice}
\end{align}
Here, $\delta \langle 0 | \frac{\alpha_s}{\pi} G^2 | 0 \rangle$ is an estimate of the uncertainty due to the truncation prescription of 
the perturbative series. The contents of the round brackets indicate the highest perturbative order taken into account. 
The lattice results tend to be considerably larger than the phenomenological estimate of Eq.\,(\ref{eq:gluon.cond.value}), but likewise 
have large systematic uncertainties due to the needed subtraction of the perturbative part. 
In all, it can be concluded from the above discussion that the gluon condensate values presently are not much more than order 
of magnitude estimates with large uncertainties. 

As a final remark, let us here mention the non-local generalization of $\langle 0 | \frac{\alpha_s}{\pi} G^2 | 0 \rangle$, 
which is a result of a resummation of covariant derivatives between the two gluonic operators. For a detailed discussion, see Ref.\,\cite{Hoang:2002nt}. 
 
Next, we consider $\langle 0 | \bar{q} \sigma_{\mu\nu} t^a G^{a\mu\nu} q |0 \rangle \equiv 
\langle 0 | \bar{q} \sigma G q |0 \rangle$ the mixed quark-gluon condensate of dimension 5. 
Its value is usually given in combination with the strong coupling constant $g$ and the dimension 3 chiral condensate, 
\begin{equation}
m_0^2 \equiv \frac{\langle 0 | \bar{q} g \sigma G q |0 \rangle}{\langle 0 | \bar{q} q |0 \rangle}.  
\label{eq:mixed.cond.para}
\end{equation}
Here, condensates containing $q$ again stand for the average over 
$u$ and $d$ quark condensates. 
Information about $m_0^2$ was extracted already long time ago from sum rules of the nucleon channel \cite{Belyaev:1982sa},  
\begin{equation}
m_0^2 = (0.8 \pm 0.2) \, \mathrm{GeV}^2. 
\label{eq:mixed.cond.value}
\end{equation}
The above value is still most frequently employed in the contemporary QCD sum rule literature. Other estimates for $m_0^2$ 
(or $\langle 0 | \bar{q} g \sigma G q |0 \rangle$) have been given in the global color symmetry model \cite{Zong:2002mh}, 
the field correlator method \cite{DiGiacomo:2004ff},
Dyson-Schwinger equations \cite{Lu:2010zze}, 
an effective quark-quark interaction model \cite{Meissner:1997ks}, 
the instanton liquid model \cite{Polyakov:1996kh,Nam:2006ng}
and holographic QCD \cite{Kim:2008ff}. To obtain an estimate of  $m_0^2$ that 
is more reliable than Eq.\,(\ref{eq:mixed.cond.value}), a precise lattice QCD computation 
would certainly be most helpful. 
Two lattice calculations were in fact already performed more than 15 years ago \cite{Doi:2002wk,Chiu:2003iw}. 
Ref.\,\cite{Doi:2002wk} obtained a result, that is significantly larger than Eq.\,(\ref{eq:mixed.cond.value}), $m_0^2 \simeq 2.5\,\mathrm{GeV}^2$, 
while Ref.\,\cite{Chiu:2003iw} reported a value consistent with Eq.\,(\ref{eq:mixed.cond.value}), $m_0^2 = 0.98(2)\,\mathrm{GeV}^2$. 
The lattice results hence have clearly not yet converged and updated calculations would be desirable. One possible 
problem for the above lattice studies is the potential mixing of $\bar{q} \sigma G q$ with lower dimensional operators, which 
can occur on the lattice, but was not taken into account in Refs.\,\cite{Doi:2002wk,Chiu:2003iw}. This issue needs to be 
carefully handled in any future lattice calculation. 

The strange mixed quark-gluon condensate $\langle 0 | \bar{s} \sigma G s|0 \rangle$ is parametrized in a similar way, 
\begin{equation}
m_1^2 \equiv \frac{\langle 0 | \bar{s} g \sigma G s |0 \rangle}{\langle 0 | \bar{s} s |0 \rangle},   
\label{eq:strange.mixed.cond.para.1}
\end{equation}
or, alternatively by the ratio with the $u$ and $d$ counterpart, 
\begin{equation}
R \equiv \frac{\langle 0 | \bar{s} g \sigma G s |0 \rangle}{\langle 0 | \bar{q} g \sigma G q |0 \rangle}.    
\label{eq:strange.mixed.cond.para.2}
\end{equation}
For $R$, a number of estimates have been given during the years \cite{Khatsimovsky:1987bb,Beneke:1992ba,Aladashvili:1995zj,Braun:2004vf}, 
which can roughly be summarized in the following range
\begin{equation}
R =  0.9 \pm 0.2. 
\label{eq:strange.mixed.cond.value}
\end{equation}
Note, however, that Ref.\,\cite{Nam:2006ng} obtains a value that is considerably smaller ($R \simeq 0.5$). 
This translates to 
\begin{align}
m_1^2 & = R m_0^2 \Bigl( \frac{\langle 0 | \overline{s}s | 0 \rangle}{\langle 0 | \overline{q}q | 0 \rangle} \Bigr)^{-1} \nonumber \\
& = 0.8 \pm 0.3\,\mathrm{GeV}^2, 
\label{eq:strange.mixed.cond.value.2}
\end{align}
where we have used Eqs.\,(\ref{eq:mixed.cond.value}), (\ref{eq:strange.mixed.cond.value}) and 
$\langle 0 | \overline{s}s | 0 \rangle / \langle 0 | \overline{q}q | 0 \rangle = 0.95 \pm 0.15$, which combines QCDSRs and 
lattice calculations for this last quantity. 
For $\langle 0 | \bar{s} \sigma G s|0 \rangle$, no lattice QCD calculation has yet been performed, which 
hopefully will be done in the future. 

At dimension 6, there is one condensate constructed only from gluon fields, 
$\langle 0 |g^3 f^{abc} G^{a\nu}_{\mu} G^{b\lambda}_{\nu} G^{c\mu}_{\lambda} |0 \rangle$, 
where, as for the dimension 4 gluon condensate, appropriate powers of the strong coupling constant are 
multiplied. 
The value of this quantity is not well known, with only one available estimate based on the dilute instanton 
gas model \cite{Novikov:1979ux}, 
\begin{equation}
\langle 0 |g^3 f^{abc} G^{a\nu}_{\mu} G^{b\lambda}_{\nu} G^{c\mu}_{\lambda} |0 \rangle = 
\frac{48 \pi^2}{5} \frac{1}{\rho_c^2} \langle 0 | \frac{\alpha_s}{\pi} G^2 | 0 \rangle, 
\label{eq:three.gluon.condensate}
\end{equation}
where $\rho_c$ is the instanton radius. We here use $\rho_c \simeq  0.3\,\mathrm{fm}$, which is based on an estimate from the instanton liquid model, for which 
the instanton density is fitted to the dimension 4 gluon condensate value, which fixes $\rho_c$ \cite{Shuryak:1981ff,Shuryak:1994rr,Shuryak:1995pv} and 
lattice QCD measurements \cite{Michael:1994uu,Chu:1994vi}. With Eq.\,(\ref{eq:gluon.cond.value}), 
one gets 
\begin{equation}
\langle 0 |g^3 f^{abc} G^{a\nu}_{\mu} G^{b\lambda}_{\nu} G^{c\mu}_{\lambda} |0 \rangle  \simeq 
0.045\,\mathrm{GeV}^6. 
\label{eq:three.gluon.condensate.value}
\end{equation}
It would certainly be useful to test the above estimate in an independent lattice QCD calculation, which  
was already tried in Ref.\,\cite{Panagopoulos:1989zn} some time ago. However, here again the problem 
of mixing with lower dimensional operators occurs, which has to be treated with care. 

At dimension 6 there are furthermore a large number of four-quark condensates that can have 
a non-zero value in vacuum. These condensates have attracted some interest because of a proposed scenario, in which  
the chiral symmetry could be broken by non-zero four-quark condensates, while the more common 
``two-quark" condensate $\langle 0 | \overline{q}q | 0 \rangle$ vanishes \cite{Kogan:1998zc,Kanazawa:2015kca}. 
Generally, the four-quark condensates can be given as 
\begin{equation}
\langle 0 |\overline{q}^{i}_{\alpha} \overline{q}^{k}_{\beta} q^{l}_{\gamma} q^{m}_{\delta} |0 \rangle,   
\label{eq:four.quark.condensate}
\end{equation}
for which the color indices ($i$, $k$, \dots) and the spinor indices ($\alpha$, $\beta$, \dots) have to 
be contracted to give a color and Lorentz singlet. This can be done in various ways, which 
leads to multiple independent condensates, of which some are given in Eq.\,(\ref{eq:condensates.vacuum}) for illustration. 
None of these four-quark condensates are however well constrained in any meaningful way. 
The only method presently known to obtain a concrete numerical value for them is the so-called vacuum 
saturation approximation (also sometimes referred to as factorization), which reads \cite{Shifman:1978bx} 
\begin{equation}
\langle 0 |\overline{q}^{i}_{\alpha} \overline{q}^{k}_{\beta} q^{l}_{\gamma} q^{m}_{\delta} |0 \rangle \simeq 
\frac{1}{144} \Bigl( 
\delta^{im} \delta^{kl} \delta_{\alpha \delta} \delta_{\beta \gamma} - \delta^{il} \delta^{km} \delta_{\alpha \gamma} \delta_{\beta \delta}
\Bigr) \langle 0 | \overline{q}q | 0 \rangle^2.
\label{eq:four.quark.condensate.factorization}
\end{equation}
The idea behind this approximation is to insert a complete set of states between the two $\overline{q}$ and $q$ quarks 
and to then assume that the vacuum contribution dominates the sum of states, such that one ends up with the squared chiral 
condensate $\langle 0 | \overline{q}q | 0 \rangle^2$. This approximation was shown to be valid in the large $N_c$ limit \cite{Novikov:1983jt}, 
but it is not known to what degree it is violated in real QCD with $N_c = 3$. 
To take into account the 
violation of this approximation, the symbol $\kappa$ is frequently introduced and multiplied to the right-hand side of Eq.\,(\ref{eq:four.quark.condensate.factorization}). 
The case $\kappa = 1$ thus stands for the vacuum saturation approximation, while values different from 1 parametrize 
its violation. 
During the years a number of values have been obtained, which depend on the studied channel and also on the 
flavor content of $q$. The proposed estimates range from close to 1 \cite{Ioffe:2010zz} to $2 \sim 3$ \cite{Narison:2007spa} and 
even up to $\sim 6$ \cite{Leinweber:1995fn}. For the case of $s$ quarks, a value of $\sim 7$ was reported from an 
analysis of finite energy sum rules in the $\phi$ meson channel \cite{Gubler:2015yna,Gubler:2016itj}. 

Condensates with mass dimensions larger than 6 can play an important role in sum rules derived from interpolating fields with 
three or more quarks, where the convergence of the OPE is usually slower. As it was discussed in Ref.\,\cite{Ioffe:1981kw}, 
the leading order OPE terms are composed of a number of loops (if the interpolating field has $n$ quarks, the number of loops is $n -1$ 
for the leading order OPE term at leading order in $\alpha_s$). These loops are numerically suppressed due to their momentum 
integrals. Going to higher order OPE terms, some of these loops are cut, hence less numerically suppressed and therefore 
enhanced compared to the leading order terms. 
As a general rule of thumb, one thus should compute the OPE up to the point where all loops are cut, to achieve satisfactory 
OPE convergence. For interpolating fields with $n$ quarks, one hence can expect terms up to $~\langle 0| \overline{q}q| 0 \rangle^{n-1}$ 
[that is, terms with mass dimension $3(n - 1)$] to give a significant contribution to the OPE. 
For baryonic currents with three (five) quark fields, one should therefore at least take into account terms up to dimension 6 (12), 
while for tetraquark current, one needs terms up to dimension 9. 
To evaluate condensates with dimensions larger than 6, usually some sort of vacuum saturation approximation similar to 
Eq.\,(\ref{eq:four.quark.condensate.factorization}) is used. 
Results based on this approximation should, however, be treated with care, as their systematic uncertainties are large as we 
have seen for the four-quark condensates above. The OPE hence becomes less reliable as the number of quark fields in the 
interpolating fields are increased. This means that QCD sum rule studies of exotics such as tetraquarks or pentaquarks 
have considerably larger systematic uncertainties and are less reliable than those of quark-antiquark mesons or three quark 
baryons. 

\subsection{Hot, dense or magnetic medium}
In this Section, we will discuss the evaluation of QCD condensates in a hot, dense or magnetic medium. We will not only consider 
the modification of condensates that are non-zero already in vacuum, but also of the Lorentz violating condensates of Eq.\,(\ref{eq:condensates.medium}), 
which only appear at finite temperature or density, and similar ones that appear in a magnetic field. 
The OPE in essence divides the correlator into a low-energy part that involves the condensates and a high-energy part that 
is treated perturbatively as Wilson coefficients. For most applications, it therefore only makes sense to consider the condensates at relatively low temperatures 
and densities ($T \lesssim \Lambda_{QCD} \sim T_c$, $\rho \lesssim \Lambda_{QCD}^3 \sim \rho_0$, where $T_c$ is the critical temperature of the hadron - quark-gluon 
plasma phase transition and $\rho_0$ the normal nuclear matter density) because only here the 
division of scales remains valid and condensates can be treated as low-energy objects. 
We will in the following discuss the evaluation of condensates at finite temperature, density and a magnetic field separately. 
At low temperatures and densities, both effects can be combined as independent superpositions, 
as it was done for instance in Ref.\,\cite{Zschocke:2002mn}. 

\subsubsection{Condensates at finite temperature \label{conds_T}}
The study of the thermal behavior of condensates has quite a long history, 
several theoretical approaches being at our disposal for this task. At low temperatures below $T_c$, 
the hadron resonance gas (HRG) model and/or chiral perturbation 
theory, which consider the effect of a hot pion (and, if needed, other hadrons) gas, can be applied. At very high temperatures much above 
$T_c$, on the other hand, perturbative QCD and hard thermal loop (HTL) approaches can be used. 
While HTL methods cannot be employed to calculate the QCD condensates directly, they can be 
of use to compute thermodynamic quantities such as energy density and pressure, which in turn are needed 
to estimate the gluon condensate behavior at finite temperature. 
Furthermore, lattice QCD 
in recent years has become increasingly powerful in simulating hot QCD for realistic pion masses and is nowadays the most 
precise tool to study condensates at finite temperature\footnote{Alternative methods to estimate the temperature dependences 
of the condensates have been proposed in the literature. Especially, approaches which make use of QCD sum rules by introducing a 
temperature dependence for the continuum threshold parameter $s_{th}$ (see Sec.\,\ref{sec:conventional.analysis}), are frequently discussed. 
The temperature dependences of the condensates are in such approaches related to the behavior of the threshold parameters. 
For more details, see for instance Refs.\,\cite{Ayala:2016vnt,Hofmann:1999nn}}. 
We will in this section review recent progress especially of lattice QCD in evaluating the various condensates that 
are used in QCDSRs, starting from those with the lowest dimension. 

Lattice QCD has so far 
mostly been used to study scalar condensates of low dimensions (see the following two Subsections). Therefore, 
one often considers a free and dilute gas of pions and, if needed, kaons and the $\eta$ meson in QCDSR studies. 
Condensates in this model are expressed as \cite{Hatsuda:1992bv} 
\begin{align}
\langle \mathcal{O} \rangle_{T} = \langle 0 | \mathcal{O} | 0 \rangle 
+ \sum_{a = 1}^{3} \int \frac{d^3 \bm{k}}{2E(\bm{k}) (2 \pi)^3} \langle \pi^{a}(\bm{k}) | \mathcal{O} | \pi^{a}(\bm{k}) \rangle 
n_{\mathrm{B}}[E(\bm{k})/T], 
\label{eq:HRG.model}
\end{align} 
with $E(\bm{k}) = \sqrt{\bm{k}^2 + m_{\pi}^2}$ and $n_{\mathrm{B}}(x) = (e^x - 1)^{-1}$. 
Here and throughout the rest of this review, 
the thermal expectation value $\langle \mathcal{O} \rangle_{T}$ is defined as 
\begin{align}
\langle \mathcal{O} \rangle_{T} \equiv \frac{\mathrm{Tr}(\mathcal{O} e^{-H/T})}{ \mathrm{Tr}(e^{-H/T})}. 
\end{align} 
Furthermore, the normalization 
\begin{align}
\langle \pi^{a}(\bm{k}) | \pi^{b}(\bm{p}) \rangle = 2 E(\bm{k}) (2 \pi)^3 \delta^{ab} \delta^3(\bm{k} - \bm{p})
\label{eq:pi.normalization}
\end{align}
for the pionic states is used. Clearly, this model is only applicable for sufficiently low temperatures below $T_c$, 
where pions are the dominant thermal excitations. We will assess the range of validity of this approximation in the 
following Subsection which discusses the chiral condensate of dimension 3, as for this quantity reliable lattice 
QCD data are available for a wide range of temperatures. 

\paragraph{Condensates of dimension 3}\mbox{}\\
At dimension 3, we consider the chiral condensate, which is naturally important for understanding 
what phase of chiral symmetry is realized at what temperature. Therefore, it has been studied intensively 
in chiral perturbation theory \cite{Gerber:1988tt} and later in lattice QCD. 
We will here not attempt to give a full account of past works, 
but just give an overview of state-of-the-art lattice QCD studies about the behavior of 
$\langle \overline{q} q \rangle_{T}$ and $\langle \overline{s} s \rangle_{T}$ 
at finite temperature. 

Computing the chiral condensate as a function of temperature in full QCD with several active flavors, 
realistic quark masses and even taking the continuum limit is by now an achievable task. 
In recent years, two groups, the BMW collaboration and the HotQCD 
collaboration have provided such results, of which some will be reproduced here. 
The chiral condensate on the lattice generally requires both multiplicative and additive renormalizations. 
One convenient way of removing such renormalization artifacts is to consider a renormalization group invariant quantity involving the chiral 
condensate and furthermore 
to subtract the vacuum part from the condensate at finite temperature. 

\begin{figure}[tb]
\begin{center}
\begin{minipage}[t]{8 cm}
\vspace{1.7 cm}
\hspace{-1.0 cm}
\includegraphics[width=9.5cm,bb=0 0 300 235]{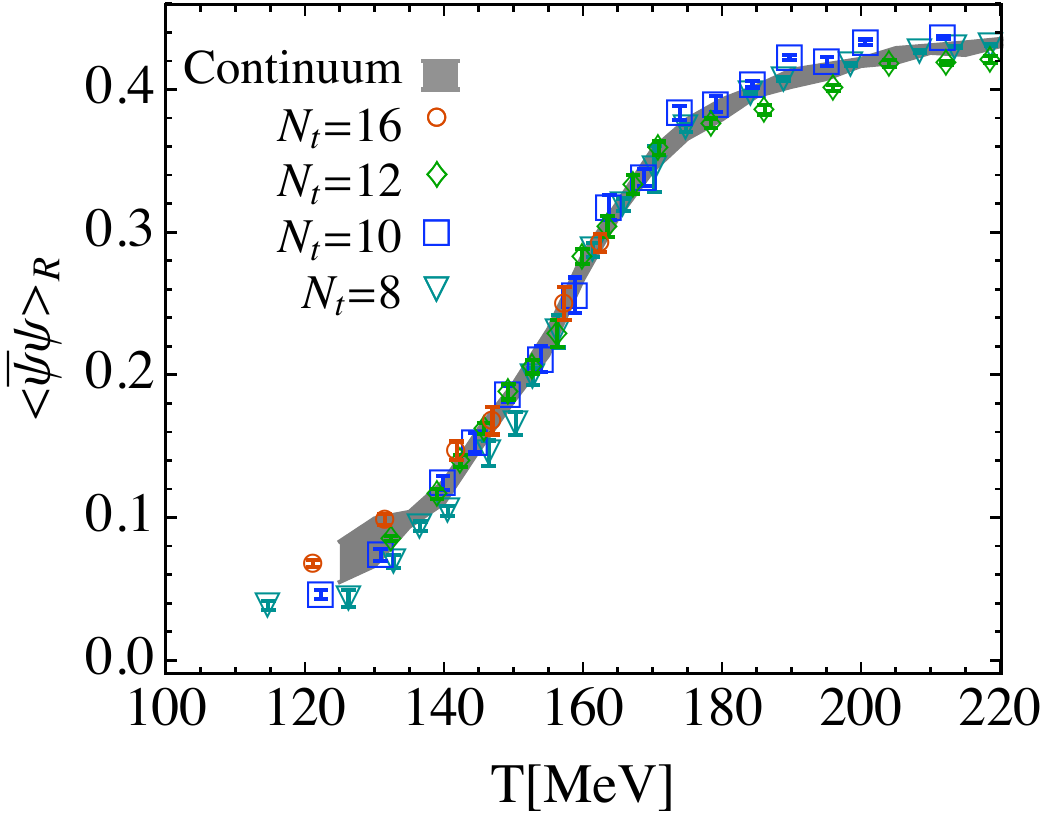}
%\vspace{0.0 cm}
\end{minipage}
\begin{minipage}[t]{16.5 cm}
\caption{The quantity $\langle \overline{\psi} \psi \rangle_R$ defined in Eq.\,(\ref{eq:cond.BMW}) 
as a function of temperature for different $N_{t}$, which are 
the number of lattice sites in the imaginary time direction. 
The gray band corresponds to the continuum limit. Taken from the left plot in Fig. 4 of Ref.\,\cite{Borsanyi:2010bp}.  
\label{fig:chiral.cond.BMW}}
\end{minipage}
\end{center}
\end{figure}

The BMW collaboration for this purpose introduced $\langle \overline{\psi} \psi \rangle_R$ \cite{Borsanyi:2010bp}, 
\begin{equation}
\langle \overline{\psi} \psi \rangle_R = - \Bigl[ \langle \overline{\psi} \psi \rangle_{l,T} - \langle \overline{\psi} \psi \rangle_{l,0} \Bigr] \frac{m_l}{X^4}, \hspace{0.5cm} \mathrm{(}l=u\mathrm{, }\,d\mathrm{)}, 
\label{eq:cond.BMW}
\end{equation}
where $X$ is an arbitrary quantity with dimension of mass. Here, we have kept the original notation used in Ref.\,\cite{Borsanyi:2010bp}, where 
the chiral condensate is defined with an opposite sign compared to our conventions. Hence, for instance, $\langle \overline{\psi} \psi \rangle_{l,0} > 0$. 
The results of Ref.\,\cite{Borsanyi:2010bp} are shown in Fig.\,\ref{fig:chiral.cond.BMW} including different 
lattice sizes with varying discretizations and the continuum limit (gray band). It is seen that the results for all 
discretizations 
lie close to each other and that hence the continuum limit 
can be safely taken.  
%%%%%%%%%%%%%%%%%%%%%%%%%%%%%%%%%%%%%%%%%%%%%%%%%%%%%%%%%%%%%%%%%%%%%%%%%%%%%%%%%%%%%%%%%%%%%%%%%%%%%%%%%%%%%%%%%%%%
%%%%%%%%%%%%%%%%%%%%%%%%%%%%%%%%%%%%%%%%%%%%%%%%%%%%%%%%%%%%%%%%%%%%%%%%%%%%%%%%%%%%%%%%%%%%%%%%%%%%%%%%%%%%%%%%%%%%

\begin{figure}[tb]
\begin{center}
\begin{minipage}[t]{16.5 cm}
\vspace{0.5 cm}
%\hspace{0.5 cm}
%\epsfig{file=Delta_R_l,scale=0.7}
%\epsfig{file=Delta_R_s,scale=0.7}
\includegraphics[width=8.5cm,bb=0 0 360 252]{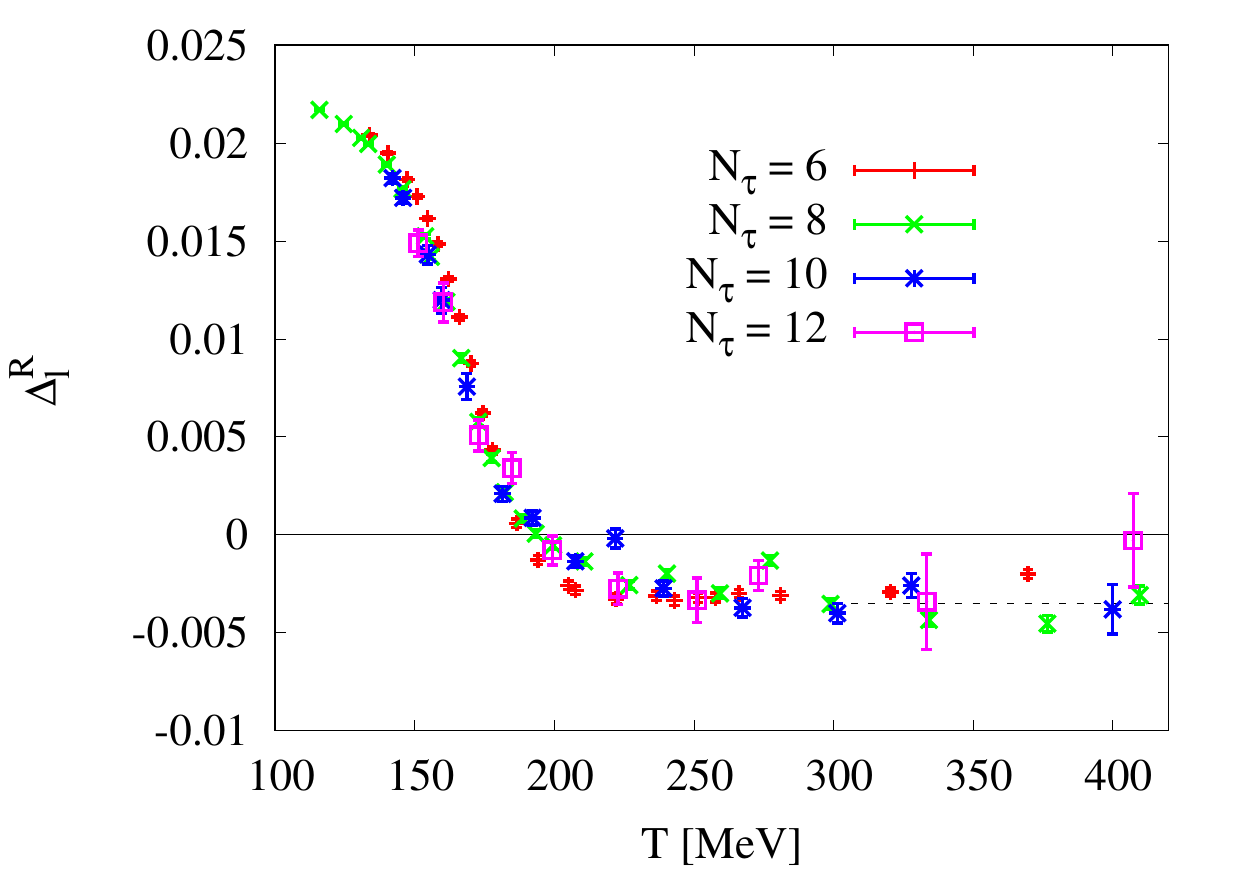}
\includegraphics[width=8.5cm,bb=0 0 360 252]{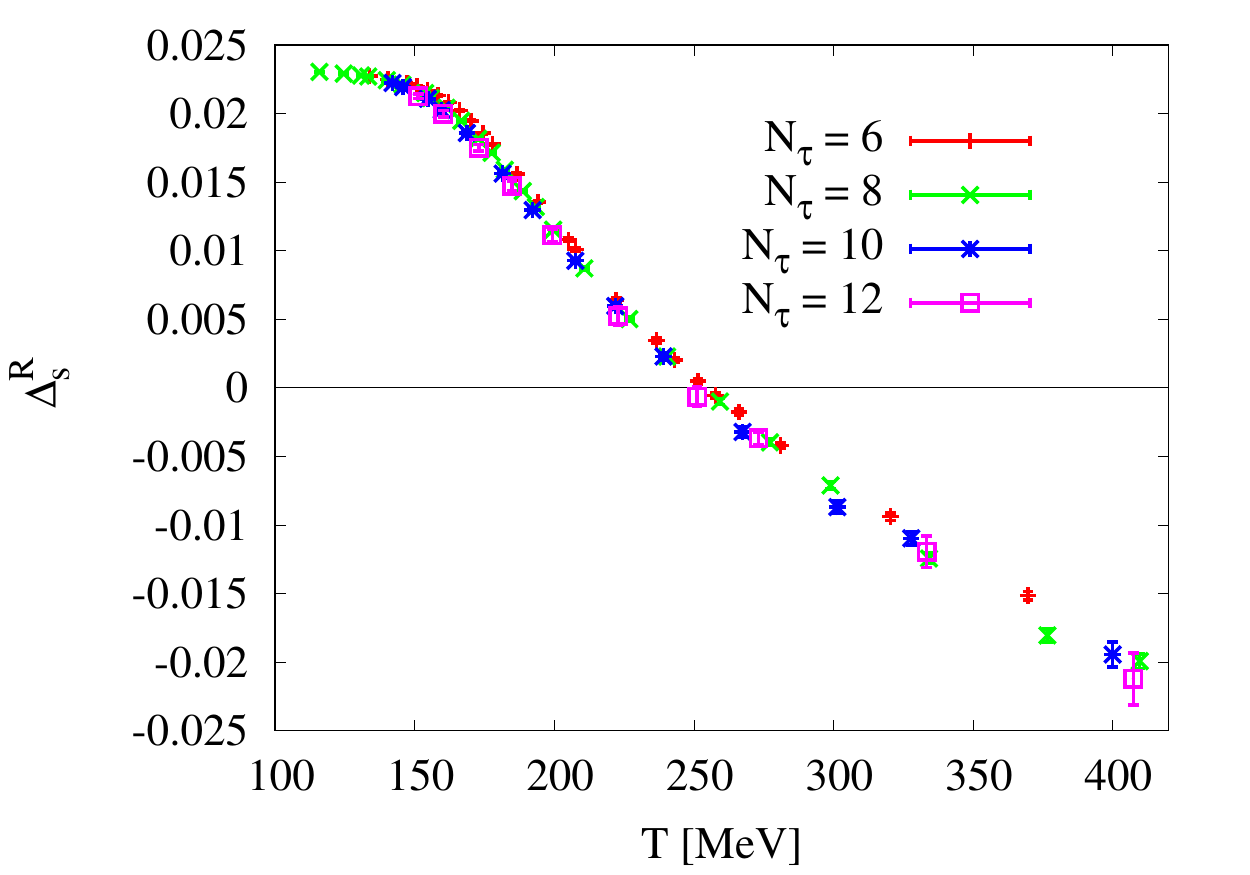}
%\vspace{1 cm}
\end{minipage}
\begin{minipage}[t]{16.5 cm}
\caption{The quantity $\Delta^R_q$ defined in Eq.\,(\ref{eq:cond.HotQCD}) for $q = l = u$, $d$ (left plot) 
and $q = s$ (right plot) for various $N_t$, which are the number of lattice sites in the imaginary time direction. 
Adapted from Fig.\,8 of Ref.\,\cite{Bazavov:2011nk}. Numerical data used to compute the data points are taken from Ref.\,\cite{Bazavov:2014pvz}. 
The dashed black line represents the result of a constant fit to the data points above 300 MeV.  
\label{fig:chiral.cond.HotQCD}}
\end{minipage}
\end{center}
\end{figure}

The HotQCD collaboration on the other hand introduced the similar quantity $\Delta^R_q$ \cite{Bazavov:2011nk}, 
\begin{equation}
\Delta^R_q = d + 2 m_s r_1^4 \Bigl[ \langle \overline{\psi} \psi \rangle_{q,T} - \langle \overline{\psi} \psi \rangle_{q,0} \Bigr], 
\label{eq:cond.HotQCD}
\end{equation}
where $q$ either represents $u$, $d$ quarks or the $s$ quark. Here, the same sign convention as in Eq.\,(\ref{eq:cond.BMW}) is 
employed. The artificial parameter $d$ is determined such that $\Delta^R_q$ approximately vanishes in the high temperature limit. 
In Ref.\,\cite{Bazavov:2011nk} it was obtained as $d = 0.0232244$. Finally, $r_1$ is a parameter determined from the slope of the static 
quark anti-quark potential evaluated on the lattice, which is used to convert lattice units into physical units. 
In Refs.\cite{Bazavov:2011nk,Bazavov:2014pvz}, $r_1 = 0.3106\,\mathrm{fm}$ was used. 
We show the results given numerically in Ref.\,\cite{Bazavov:2014pvz} for $\Delta^R_l$ ($l = u$, $d$) and $\Delta^R_s$ in 
Fig.\,\ref{fig:chiral.cond.HotQCD}. As for the BMW results, $\Delta^R_l$ and $\Delta^R_s$ do not much depend on the 
number of lattice sites $N_{\tau}$ in the imaginary time direction and can hence assumed to be already close to the continuum limit. 
%%%%%%%%%%%%%%%%%%%%%%%%%%%%%%%%%%%%%%%%%%%%%%%%%%%%%%%%%%%%%%%%%%%%%%%%%%%%%%%%%%%%%%%%%%%%%%%%%%%%%%%%%%%%%%%%%%%%
%%%%%%%%%%%%%%%%%%%%%%%%%%%%%%%%%%%%%%%%%%%%%%%%%%%%%%%%%%%%%%%%%%%%%%%%%%%%%%%%%%%%%%%%%%%%%%%%%%%%%%%%%%%%%%%%%%%%

\begin{figure}[tb]
\begin{center}
\begin{minipage}[t]{16.5 cm}
\vspace{1 cm}
%\hspace{0.5 cm}
%\epsfig{file=Delta_divided_compare_l_4,scale=0.7}
%\epsfig{file=Delta_divided_s_3,scale=0.7}
\includegraphics[width=8.5cm,bb=0 0 360 252]{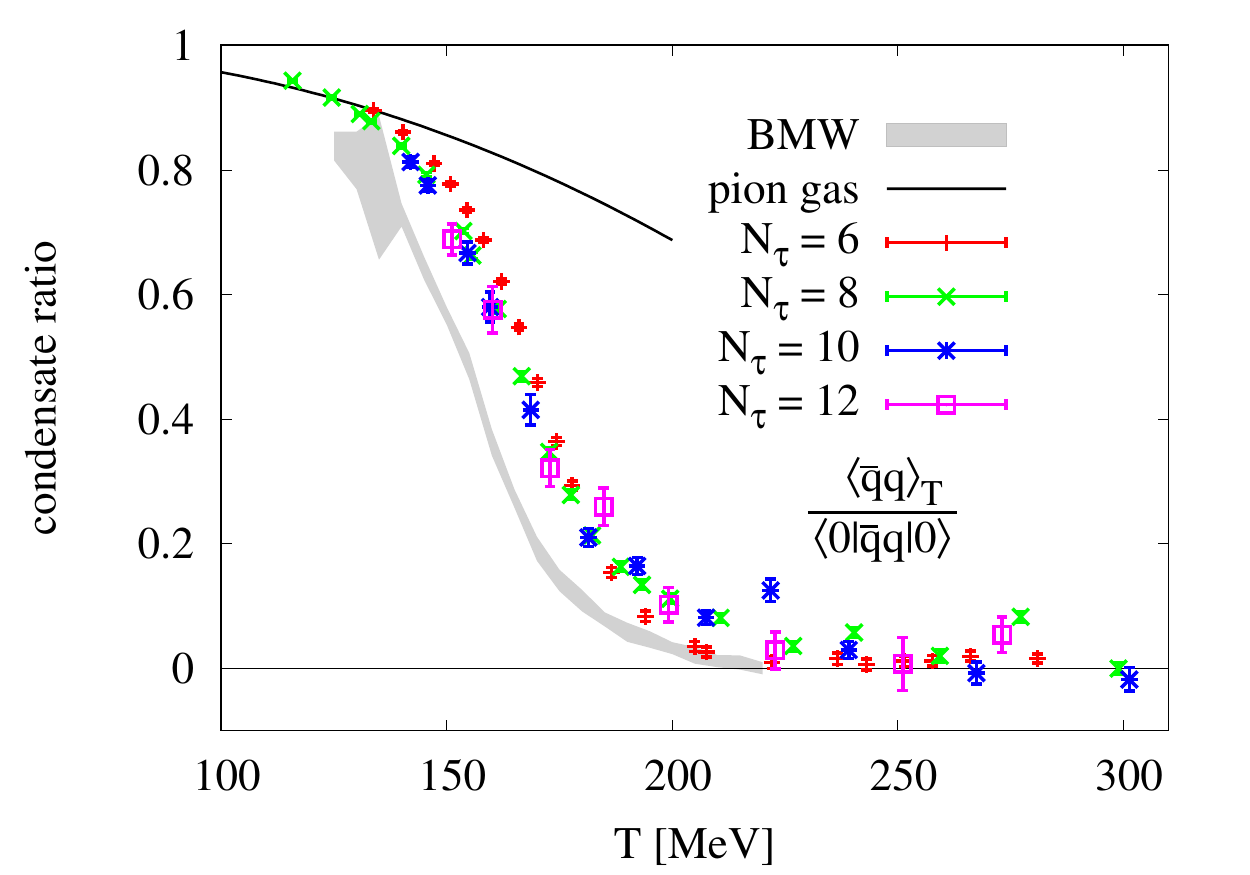}
\includegraphics[width=8.5cm,bb=0 0 360 252]{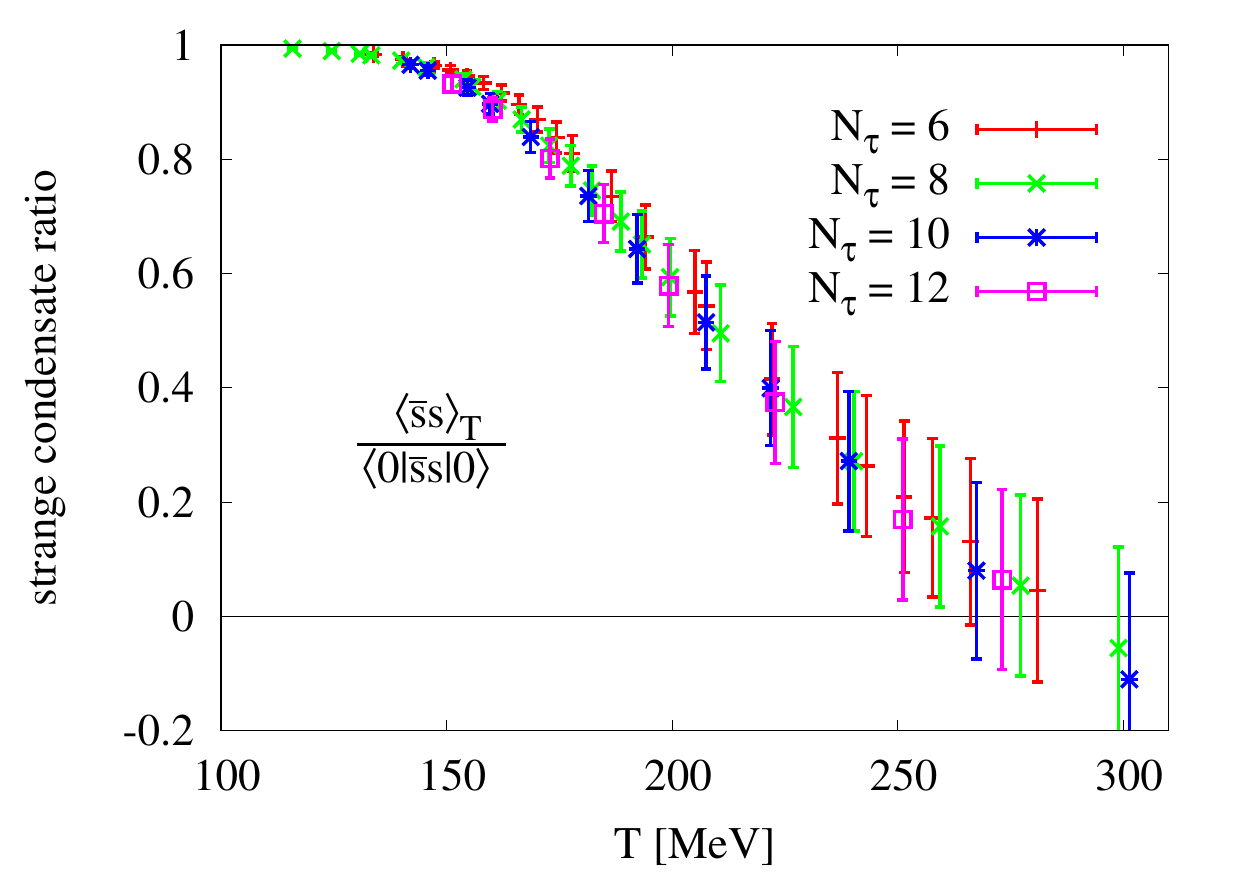}
%\vspace{1 cm}
\end{minipage}
\begin{minipage}[t]{16.5 cm}
\caption{The ratio $\langle \overline{q} q \rangle_{T} / \langle 0| \overline{q} q| 0 \rangle$, with $q = u$, $d$ (left plot) and 
$\langle \overline{s} s \rangle_{T} / \langle 0| \overline{s} s| 0 \rangle$ (right plot) 
for various $N_t$, which are the number of lattice sites in the imaginary time direction. 
For $\langle T| \overline{q} q| T \rangle / \langle 0| \overline{q} q| 0 \rangle$, 
Eqs.\,(\ref{eq:BMW.formula}) and (\ref{eq:HotQCD.formula}) and the corresponding data provided by the BMW \cite{Borsanyi:2010bp} and HotQCD \cite{Bazavov:2014pvz} 
collaborations were used. For $\langle \overline{s} s \rangle_{T} / \langle 0| \overline{s} s| 0 \rangle$, the data for $\Delta^R_s$ together with 
the values of $\langle 0| \overline{s} s| 0 \rangle$, $m_s$ and $r_1$ (see text) were employed.  
\label{fig:chiral.cond.ratio}}
\end{minipage}
\end{center}
\end{figure}

For applying these lattice findings to actual QCDSR calculations, it is helpful to convert them into quantities that 
are easier to use. For the $u$ and $q$ quark condensates, it seen both in 
Figs.\,\ref{fig:chiral.cond.BMW} and \ref{fig:chiral.cond.HotQCD} that $\langle \overline{\psi} \psi \rangle_R$ and $\Delta^R_l$ 
approach a constant at high temperatures. Assuming that the condensate completely vanishes in this temperature region, 
one can convert both $\langle \overline{\psi} \psi \rangle_R$ and $\Delta^R_l$ into $\langle \overline{q} q  \rangle_{T} / \langle 0| \overline{q} q| 0 \rangle$. 
Specifically, we have 
\begin{align}
\frac{\langle \overline{q} q  \rangle_{T}}{\langle 0| \overline{q} q| 0 \rangle} & = 
1 - \frac{\langle \overline{\psi} \psi \rangle_R(T)}{\langle \overline{\psi} \psi \rangle_R(\infty)} \hspace{0.5cm} \text{(BMW collaboration)}, \label{eq:BMW.formula}\\
\frac{\langle \overline{q} q  \rangle_{T}}{\langle 0| \overline{q} q| 0 \rangle} & = 1 - \frac{d - \Delta^R_l(T)}{d - \Delta^R_l(\infty)} \hspace{0.5cm} \text{(HotQCD collaboration)}.
\label{eq:HotQCD.formula}
\end{align}
For $\langle \overline{\psi} \psi \rangle_R(\infty)$, we use the largest temperature data point provided by the BMW collaboration, while for $\Delta^R_l(\infty)$ we use 
a fit to all data above 300 MeV given in Ref.\,\cite{Bazavov:2014pvz}. The result of this fit is indicated by the dashed line in the left plot of Fig.\,\ref{fig:chiral.cond.HotQCD}. 
Values of $\langle \overline{q} q \rangle_{T} / \langle 0| \overline{q} q| 0 \rangle$ from both collaborations are shown and compared in the left plot of Fig.\,\ref{fig:chiral.cond.ratio}. 
%%%%%%%%%%%%%%%%%%%%%%%%%%%%%%%%%%%%%%%%%%%%%%%%%%%%%%%%%%%%%%%%%%%%%%%%%%%%%%%%%%%%%%%%%%%%%%%%%%%%%%%%%%%%%%%%%%%%
%%%%%%%%%%%%%%%%%%%%%%%%%%%%%%%%%%%%%%%%%%%%%%%%%%%%%%%%%%%%%%%%%%%%%%%%%%%%%%%%%%%%%%%%%%%%%%%%%%%%%%%%%%%%%%%%%%%%
The BMW (continuum limit) result is shown by the gray band, while the data points are from HotQCD. Both findings agree qualitatively, 
even though there is still a small ($\sim$ 10 MeV) discrepancy seen in the temperature at which the condensate drops most steeply. 
This shows that some systematic uncertainties that go beyond the errors shown in Fig.\,\ref{fig:chiral.cond.ratio} still remain, likely 
related to the continuum extrapolation \cite{Borsanyi:2010bp} and the setting of the scale, which are, however, reasonably well under 
control. If needed, one can extrapolate the above results to lower temperatures by a simple pion gas model \cite{Hatsuda:1992bv,Gerber:1988tt}, 
as described in the following paragraph. 

We next compare the lattice QCD results to those of the pion gas model and examine up the what 
temperatures it is able to describe the lattice data reasonably well. In this model, the chiral condensate at finite temperature 
can with the help of PCAC and current algebra be given as \cite{Hatsuda:1992bv,Gasser:1986vb} 
\begin{align}
\frac{\langle \overline{q} q  \rangle_{T}}{\langle 0| \overline{q} q| 0 \rangle} & = 
1 -  \frac{T^2}{8 f_{\pi}^2} B_1 \Bigl(\frac{m_{\pi}}{T} \Bigr), 
\label{eq:chiral.cond.pion.gas}
\end{align}
where we have defined 
\begin{align}
B_n(x) = \frac{1}{\zeta(2n) \Gamma(2n)} \int_x^{\infty}dy y^{2(n-1)} \frac{\sqrt{y^2 - x^2}}{e^y - 1}.  
\label{eq:chiral.cond.pion.gas.2}
\end{align}
The corresponding curve is shown as a solid black line in the left plot of Fig.\,\ref{fig:chiral.cond.ratio}, for which 
we have used $f_{\pi} = 93$ MeV and $m_{\pi} = 140$ MeV. 
Comparing this curve to the lattice data, it is observed that the pion gas model remains approximately valid up to temperatures 
of about 140 MeV but quickly breaks down for higher temperatures. This gives a rough idea about the 
reliability of this model. 
To improve the consistency with lattice data, one 
could try to improve it by adding other hadron species and 
further artificial terms\footnote{``Artificial terms'' here are terms that 
have no apparent physical interpretation [unlike the second term on the right hand side of 
Eq.\,(\ref{eq:chiral.cond.pion.gas})], but are introduced to get better agreement with lattice QCD data. 
In Ref.\,\cite{Hohler:2013eba}, for instance, a term $-\alpha T^{10}$ was added to the right hand side of Eq.\,(\ref{eq:chiral.cond.pion.gas}) 
for this purpose.}. Doing this, it is possible to extend its range 
of applicability to temperatures slightly above $T_c$ (see for instance Ref.\,\cite{Hohler:2013eba}). 

For the strange quark condensate, more input is needed as $\Delta^R_s$, shown on the right plot of Fig.\,\ref{fig:chiral.cond.HotQCD} 
does not approach any constant value even for temperatures larger than those shown. We therefore use the value given in 
Eq.\,(\ref{eq:strange.quark.cond.value}) and $m_s = 96 \pm 6$ MeV \cite{Patrignani:2016xqp} (for which we have symmetrized the upper and lower error for simplicity). 
With these values and $r_1$, given earlier, we can obtain $\langle \overline{s} s \rangle_{T} / \langle 0| \overline{s} s| 0 \rangle$ from 
$\Delta^R_s$. The result is shown in the right plot of Fig.\,\ref{fig:chiral.cond.ratio}. In contrast to the $u$ and $d$ condensate, 
the strange quark condensate does not decrease suddenly around $T_c$, but shows only a gently decreasing behavior, approaching 
zero at temperatures above around $2 T_c$. Such a qualitative difference between the $u$, $d$ and $s$ condensates was already 
predicted in models such as the Nambu-Jona-Lasinio model \cite{Hatsuda:1994pi} and can be easily understood by considering a 
pion gas model, for which the matrix element $\langle \pi | \overline{s}s | \pi \rangle$ is very small \cite{Hatsuda:1990uw} 
and hence the leading order contribution of Eq.\,(\ref{eq:HRG.model}) almost vanishes. 
The fact that the error in the right plot of Fig.\,\ref{fig:chiral.cond.ratio} increases with increasing $T$, is explained from 
the relatively large error of $\langle 0| \overline{s} s| 0 \rangle$ 
in Eq.\,(\ref{eq:strange.quark.cond.value}). Once this condensate is determined with better precision, it will become possible to 
considerably decrease the error for $\langle \overline{s} s \rangle_{T} / \langle 0| \overline{s} s| 0 \rangle$. 

To summarize, the chiral condensates are by now known with rather good precision and only small systematic uncertainties from lattice QCD. 
These results can now be used in QCD sum rule analyses without having to rely on the pion gas model. 

The non-scalar condensates $\langle \overline{q} \gamma_{\mu} q \rangle_{T}$ and $\langle  \overline{s} \gamma_{\mu} s \rangle_{T}$, 
which can be related to baryon densities (see Section \,\ref{conds.at.finite.den}), remain exactly zero in a heat bath with vanishing chemical potential. 

\paragraph{Condensates of dimension 4}\mbox{}\\
At dimension four, we first discuss the thermal behavior of the scalar gluon condensate 
$\langle \frac{\alpha_s}{\pi} G^a_{\mu \nu} G^{a\mu\nu} \rangle_{T}$. In vacuum, it has been difficult to 
compute this quantity on the lattice because of renormalization issues. At finite temperature, however, 
it is relatively simple to obtain the difference $\langle \frac{\alpha_s}{\pi} G^a_{\mu \nu} G^{a\mu\nu} \rangle_{T} - \langle 0| \frac{\alpha_s}{\pi} G^a_{\mu \nu} G^{a\mu\nu} |0 \rangle$ 
as it can (within certain approximations) be related to thermodynamic quantities such as energy density and pressure. 
 
First, we follow the discussions of Refs. \cite{Shifman:1978zn,Cohen:1991nk}, where 
the trace anomaly, 
\begin{align}
T^{\mu}_{\mu} = \frac{\beta(g)}{2g} G^a_{\mu \nu} G^{a\mu\nu} + \sum_q m_q \overline{q}q.
\label{eq:trace.anomaly.1} 
\end{align}
was used. 
Here, $T^{\mu \nu}$ and $\beta(g)$ are the QCD energy momentum tensor and $\beta$-function, respectively. 
The one-loop perturbative $\beta$-function is given as 
\begin{align}
\beta(g) = -\frac{1}{(4\pi)^2}\Bigl( 11 - \frac{2}{3} N_f \Bigr) g^3 + \mathcal{O}(g^5), 
\label{eq:trace.anomaly.2} 
\end{align}
$N_f$ denoting the number of flavors. The contributions of $c$, $b$ and $t$ 
quarks to the sum in the second term on the right side of Eq.\,(\ref{eq:trace.anomaly.1}) 
can be evaluated using the heavy quark expansion, which gives, 
\begin{align}
\overline{q}q = -\frac{1}{12 m_q} \frac{\alpha_s}{\pi} G^a_{\mu \nu} G^{a\mu\nu} + \mathcal{O}(m_q^{-3}). 
\label{eq:trace.anomaly.3} 
\end{align}
The heavy quark expansion is only valid for quarks with masses larger than typical QCD scales and 
is hence not applicable to $u$, $d$ and $s$ quarks. 
Substituting the above result into Eq.\,(\ref{eq:trace.anomaly.1}), it is found that the heavy quark terms 
$\displaystyle \sum_{q=c,\,b,\,t} m_q \overline{q}q$ 
cancel exactly in the limit $m_q \to \infty$ 
with their respective contributions from the first $G^a_{\mu \nu} G^{a\mu\nu}$ term  
(the term proportional to $N_f$ in the $\beta$-function). We therefore just need to keep the light 
quark contributions in Eq.\,(\ref{eq:trace.anomaly.1}) and can set $N_f$ to 3. We thus have 
\begin{align}
T^{\mu}_{\mu} = -\frac{9}{8} \frac{\alpha_s}{\pi} G^a_{\mu \nu} G^{a\mu\nu} + m_u \overline{u}u + m_d \overline{d}d + m_s \overline{s}s + \mathcal{O}(\alpha_s^2,m_c^{-3},m_b^{-3},m_t^{-3}). 
\label{eq:trace.anomaly.4} 
\end{align}

\begin{figure}[tb]
\begin{center}
\begin{minipage}[t]{8 cm}
\vspace{0.5 cm}
\hspace{-2.0 cm}
\includegraphics[width=12cm,bb=0 0 360 252]{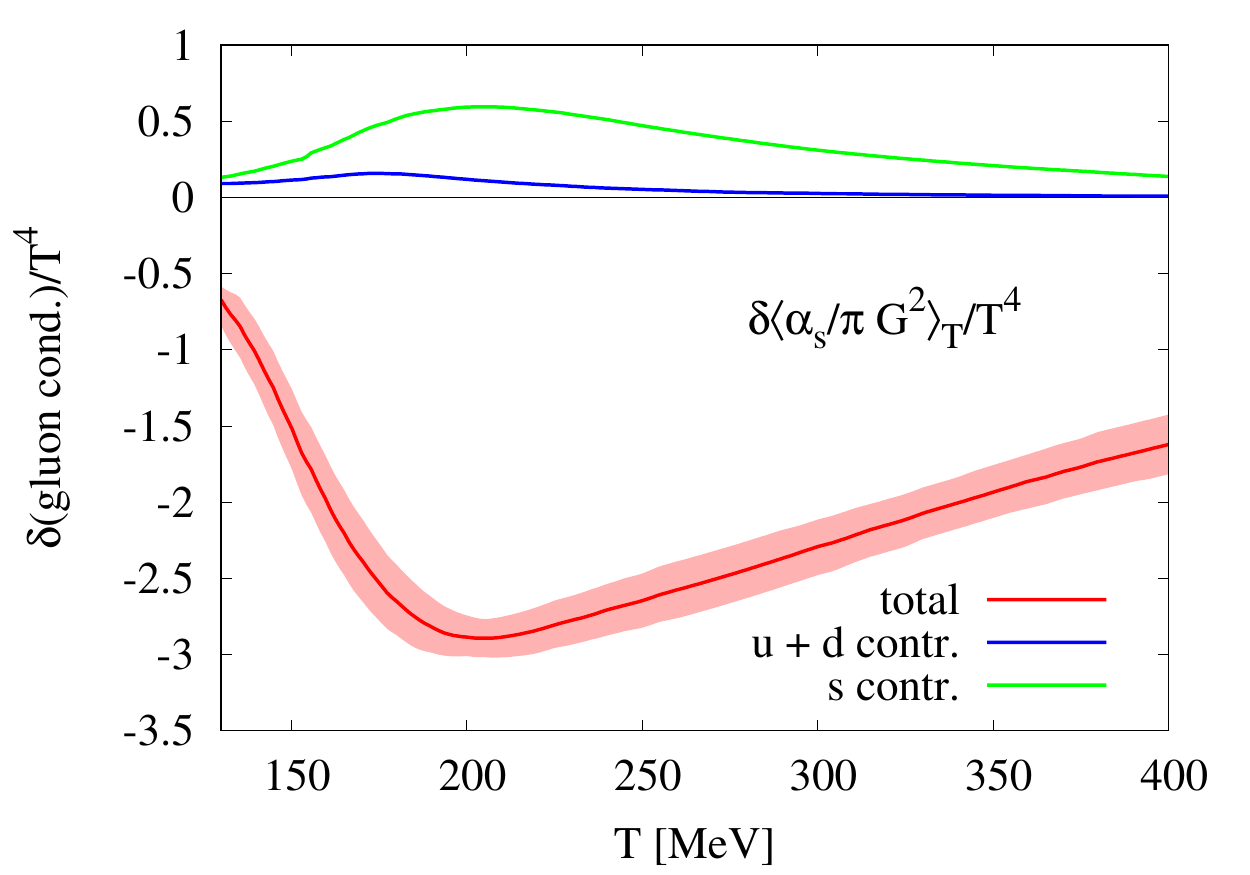}
%\vspace{1.0 cm}
\end{minipage}
\begin{minipage}[t]{16.5 cm}
\caption{The quantity 
$\delta  \langle \frac{\alpha_s}{\pi} G^a_{\mu \nu} G^{a\mu\nu} \rangle_{T}/T^4$ as a function of $T$, extracted from Eq.\,(\ref{eq:trace.anomaly.5}) and the lattice QCD 
data given in Ref.\,\cite{Bazavov:2014pvz}. The red curve shows the total right-hand side of Eq.\,(\ref{eq:trace.anomaly.5}), the red shaded area its 
uncertainty. The blue and green curves show the contributions from the $u+d$ quark and $s$ quark condensates to the same equation. See text for more details.  
\label{fig:delta.gluon.cond}}
\end{minipage}
\end{center}
\end{figure}

Based on the above trace anomaly equation, one can compute the thermal behavior of the gluon condensate. 
For simplicity of notation, we define $\delta f(T)$ as the vacuum subtracted value of the quantity $f(T)$: 
$\delta f(T) \equiv f(T) - f(0)$. 
From Eq.\,(\ref{eq:trace.anomaly.4}), we therefore obtain 
\begin{align}
\delta  \langle \frac{\alpha_s}{\pi} G^a_{\mu \nu} G^{a\mu\nu} \rangle_{T} = -\frac{8}{9} \Bigl[ \delta T^{\mu}_{\mu}(T)  - m_u  \delta  \langle \overline{u}u \rangle_{T} 
- m_d \delta  \langle \overline{d}d  \rangle_{T} - m_s \delta  \langle \overline{s}s \rangle_{T} \Bigr].  
\label{eq:trace.anomaly.5} 
\end{align}
Note that 
\begin{align}
\delta T^{\mu}_{\mu}(T) = \epsilon(T) - 3p(T),  
\label{eq:trace.anomaly.5.2} 
\end{align}   
where $\epsilon(T)$ is the energy density and $p(T)$ the pressure. Both of them are known with good precision from present day 
lattice calculations \cite{Bazavov:2014pvz,Borsanyi:2013bia}. The behavior of the quark condensates as a function of temperature is known as well, as we have seen in the 
previous Section. Applying these results to Eq.\,(\ref{eq:trace.anomaly.5}), the temperature dependence of the gluon condensate can be extracted. 
The respective results are shown in Fig.\,\ref{fig:delta.gluon.cond}, 
%%%%%%%%%%%%%%%%%%%%%%%%%%%%%%%%%%%%%%%%%%%%%%%%%%%%%%%%%%%%%%%%%%%%%%%%%%%%%%%%%%%%%%%%%%%%%%%%%%%%%%%%%%%%%%%%%%%%
%%%%%%%%%%%%%%%%%%%%%%%%%%%%%%%%%%%%%%%%%%%%%%%%%%%%%%%%%%%%%%%%%%%%%%%%%%%%%%%%%%%%%%%%%%%%%%%%%%%%%%%%%%%%%%%%%%%%
for which we have used the lattice data provided in Ref.\,\cite{Bazavov:2014pvz}. For $\epsilon(T)$ and $p(T)$ the continuum extrapolated 
results are employed. For the quark condensate terms we use the $N_{\tau} = 8$ data, which are already close to the continuum limit and for which 
a relatively large number of data points are available. It is clear from Fig.\,\ref{fig:delta.gluon.cond} that the $\epsilon(T) - 3p(T)$ term dominates the 
thermal behavior of the gluon condensate. The $u$ and $d$ condensate terms are suppressed due to their small quark masses, while the $s$ quark 
condensate term gives a non-negligible correction. Note that $\delta  \langle \frac{\alpha_s}{\pi} G^a_{\mu \nu} G^{a\mu\nu} \rangle_{T}/T^4$ 
approaches zero for large $T$ only because of the $1/T^4$ factor, whereas $\delta  \langle \frac{\alpha_s}{\pi} G^a_{\mu \nu} G^{a\mu\nu} \rangle_{T}$ is 
a negative and monotonously decreasing function of $T$. This means that the non-vacuum subtracted gluon condensate 
$\langle T |\frac{\alpha_s}{\pi} G^a_{\mu \nu} G^{a\mu\nu} |T \rangle$ will switch its sign from positive to negative and 
further continue to decrease with increasing temperature. Using Eq.\,(\ref{eq:gluon.cond.value}) for the vacuum gluon condensate, the 
transition from positive to negative sign occurs at about $T \simeq 260$ MeV. 
The thermal behavior of the gluon condensate can also be estimated based on 
the pion gas model \cite{Hatsuda:1992bv}, 
\begin{align}
\delta  \langle \frac{\alpha_s}{\pi} G^a_{\mu \nu} G^{a\mu\nu} \rangle_{T} = -\frac{m_{\pi}^2 T^2}{9} B_1 \Bigl(\frac{m_{\pi}}{T} \Bigr). 
\end{align}
The absolute value of this expression is however much too small compared to the lattice QCD result of Fig.\,\ref{fig:delta.gluon.cond}, which 
can be understood from the suppressive factor $m_{\pi}^2$, which is absent in the chiral condensate formula of Eq.\,(\ref{eq:chiral.cond.pion.gas}) 
and points to the fact that contributions of higher mass hadrons will be significant and hence need to be taken into account 
to get a better description at small temperatures. 

\begin{figure}[tb]
\begin{center}
\begin{minipage}[t]{8 cm}
\vspace{0.5 cm}
\hspace{-2.0 cm}
\includegraphics[width=12cm,bb=0 0 360 252]{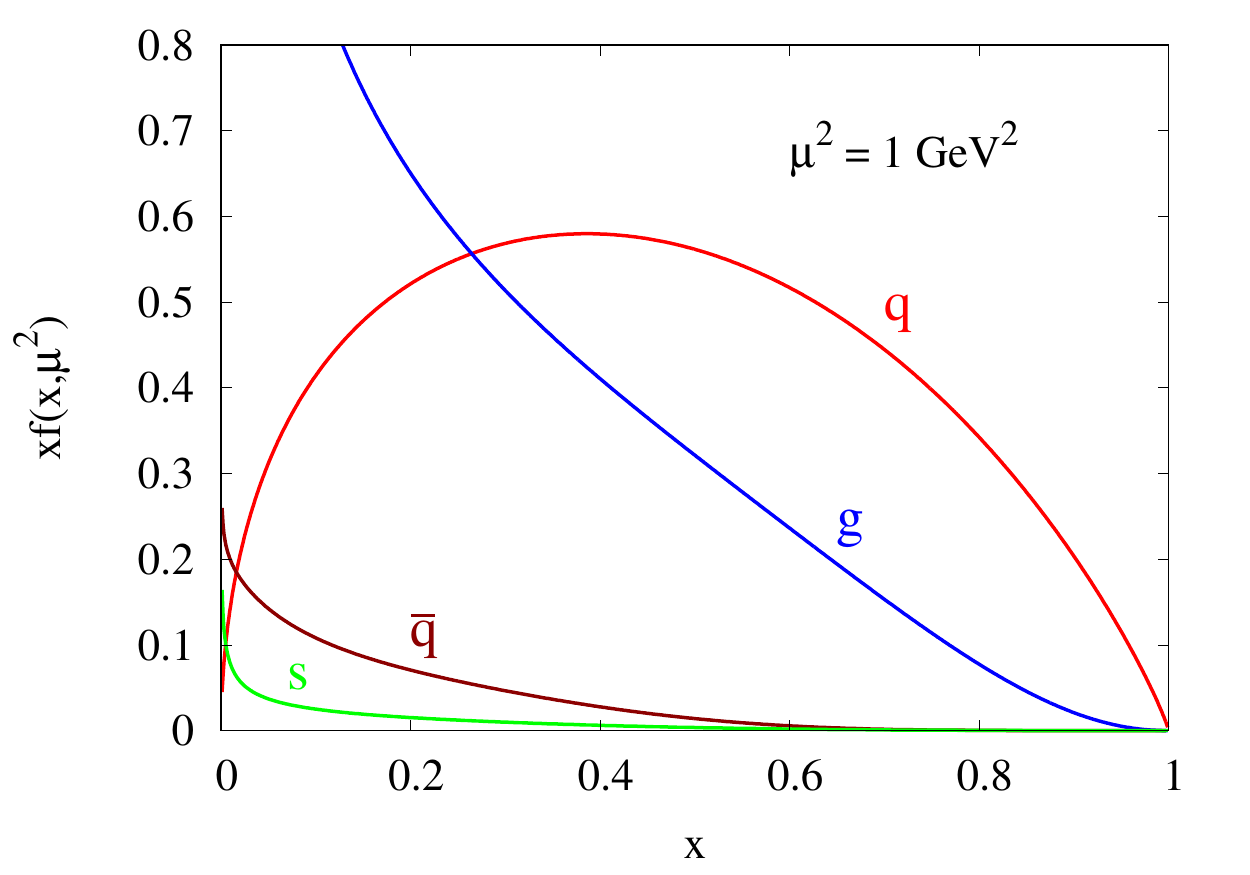}
%\vspace{1 cm}
\end{minipage}
\begin{minipage}[t]{16.5 cm}
\caption{NLO parton distributions of the pion 
as a function of Bjorken $x$, which is the fraction of total hadron momentum carried by each parton. 
The red curve ($q$) stands for the summed valence quarks, the brown curve ($\overline{q}$) for the averaged $u$ and $d$ (including their antiquarks) sea 
quarks, the green curve ($s$) for the strange sea quarks and antiquarks and the blue curve ($g$) for the gluons, respectively. All curves were 
extracted from the formulas of 
Ref.\,\cite{Gluck:1999xe} at a renormalization scale of $\mu^2 = 1\,\mathrm{GeV}^2$. 
\label{fig:parton.distributions.pion}}
\end{minipage}
\end{center}
\end{figure}

Let us next discuss the non-scalar condensates of dimension 4. The quark condensate $\langle \mathcal{ST} \bar{q} \gamma^{\mu} iD^{\nu} q \rangle_{T}$ 
represents the quark contribution to the (trace subtracted) energy-momentum tensor. To our knowledge, no lattice QCD data are presently 
available for this condensate. It is, however, possible to compute its low-temperature behavior from the pion gas model. 
In this context, it is convenient to generalize the discussion to a larger class of condensates by defining 
\begin{align}
\langle \pi^{a} (\bm{p}) | \mathcal{ST} \bar{q} \gamma_{\mu_1} D_{\mu_2} \cdots D_{\mu_n} q | \pi^{a} (\bm{p}) \rangle  
\equiv (-i)^{n - 1}  A^{\pi(q)}_n(\mu^2) \mathcal{ST}( p_{\mu_1} \cdots p_{\mu_n} ).  
\label{eq:nonlocal.dim4.temp.1}
\end{align}
The superscript $a$, which represents the three pion states, is not meant to be summed, but should be 
understood as an expectation value of a single pion state. 
For $\mathcal{ST}(p_{\mu_1} \cdots p_{\mu_n})$, the specific expressions for practically relevant cases are
\begin{align}
\mathcal{ST}(p_{\mu_1} p_{\mu_2}) &= p_{\mu_1}p_{\mu_2} - \frac{1}{4} p^2 g_{\mu_1 \mu_2}, \\
\mathcal{ST}(p_{\mu_1} p_{\mu_2} p_{\mu_3}) &= p_{\mu_1} p_{\mu_2} p_{\mu_3} 
- \frac{1}{6} p^2 (p_{\mu_1} g_{\mu_2 \mu_3} + p_{\mu_2} g_{\mu_1 \mu_3} + p_{\mu_3} g_{\mu_1 \mu_2}), \\
\mathcal{ST}(p_{\mu_1} p_{\mu_2} p_{\mu_3} p_{\mu_4}) &= p_{\mu_1} p_{\mu_2} p_{\mu_3} p_{\mu_4} 
- \frac{1}{8} p^2 (p_{\mu_1} p_{\mu_2} g_{\mu_3 \mu_4} + p_{\mu_1} p_{\mu_3} g_{\mu_2 \mu_4} + p_{\mu_1} p_{\mu_4} g_{\mu_2 \mu_3} \nonumber \\
& \,\,\,\,\,\, + p_{\mu_2} p_{\mu_3} g_{\mu_1 \mu_4} + p_{\mu_2} p_{\mu_4} g_{\mu_1 \mu_3} + p_{\mu_3} p_{\mu_4} g_{\mu_1 \mu_2}) \nonumber \\
& \,\,\,\,\,\, + \frac{1}{48} p^4 (g_{\mu_1 \mu_2} g_{\mu_3 \mu_4} + g_{\mu_1 \mu_3} g_{\mu_2 \mu_4} + g_{\mu_1 \mu_4} g_{\mu_2 \mu_3}). 
\label{eq:nonlocal.dim4.7.2}
\end{align}
These are consistent with the general expressions of Eqs.\,(\ref{eq:symmetric.traceless.1}-\ref{eq:symmetric.traceless.3}). 
Considering the theory of (fictious) deep inelastic scattering (DIS) off a pion target, the 
coefficients $A^{\pi(q)}_n(\mu^2)$ can be related to moments of pion quark distribution functions, 
\begin{align}
A^{\pi(q)}_n(\mu^2) & = 2 \int_0^1 dx x^{n-1} \bigl[q(x, \mu^2) + (-1)^n \overline{q}(x, \mu^2) \bigr]. 
\label{eq:nonlocal.dim4.temp.2}
\end{align} 
The variable $x$ is usually referred to as ``Bjorken $x$'' and in this context 
specifies the fraction of total hadron momentum carried by the considered parton (here quark $q$ or anti-quark $\overline{q}$). 
The pion quark distribution functions are not known as well as those of the nucleon (see Section \ref{conds.at.finite.den}) because 
there are no direct DIS data with a pion target. It is however possible to constrain them from Drell-Yan dilepton production and direct 
photon production in $\pi N$ reactions \cite{Gluck:1991ey,Gluck:1999xe}. 
Together with QCD evolution equations, one can thus extract the quark and gluon distributions as a function of the energy 
scale $\mu^2$. Estimates for $A^{\pi(u + d)}_2(\mu^2)$ and $A^{\pi(u + d)}_4(\mu^2)$ were given in Ref.\,\cite{Hatsuda:1992bv} based 
of the parton distribution functions provided in Ref.\,\cite{Gluck:1991ey}. We will update and slightly generalize this 
discussion here. For this purpose we use the parton distributions of Ref.\,\cite{Gluck:1999xe}, which is an update of Ref.\,\cite{Gluck:1991ey} 
and especially discriminates between $u + d$ and $s$, quarks, which is essential for obtaining an accurate estimate for 
the condensate with strange quarks. 
The NLO version of these parton distributions are shown in Fig.\,\ref{fig:parton.distributions.pion} for a scale of $\mu^2 = 1\,\mathrm{GeV}^2$. 
%%%%%%%%%%%%%%%%%%%%%%%%%%%%%%%%%%%%%%%%%%%%%%%%%%%%%%%%%%%%%%%%%%%%%%%%%%%%%%%%%%%%%%%%%%%%%%%%%%%%%%%%%%%%%%%%%%%%
%%%%%%%%%%%%%%%%%%%%%%%%%%%%%%%%%%%%%%%%%%%%%%%%%%%%%%%%%%%%%%%%%%%%%%%%%%%%%%%%%%%%%%%%%%%%%%%%%%%%%%%%%%%%%%%%%%%%
For the valence quarks (denoted as $q$ in the figure), we have $q^{\pi} = u^{\pi^+}_v + \overline{d}^{\pi^+}_v$ with 
$u^{\pi^+}_v = \overline{d}^{\pi^+}_v = \overline{u}^{\pi^-}_v = d^{\pi^-}_v$. 
Note that this is different from the treatment in Refs.\,\cite{Hatsuda:1992bv,Gluck:1991ey}, where the definition 
$q^{\pi} = u^{\pi^+}_v = \overline{d}^{\pi^+}_v$ was used. 
For the $u$ and $d$ sea quarks (denoted as $\overline{q}$) 
$\overline{q}^{\pi} = u^{\pi^{+}}_s = \overline{u}^{\pi^{+}}_s = d^{\pi^{+}}_s = \overline{d}^{\pi^{+}}_s = u^{\pi^{-}}_s = \overline{u}^{\pi^{-}}_s = d^{\pi^{-}}_s = \overline{d}^{\pi^{-}}_s$, 
therefore assuming exact isospin symmetry. For the strange quark distributions (denoted as $s$), we have 
$s^{\pi} = s^{\pi^{+}} = \overline{s}^{\pi^{+}} = s^{\pi^{-}} = \overline{s}^{\pi^{-}}$. The distributions for $\pi^0$ can be obtained as 
$f^{\pi^0} = (f^{\pi^+} + f^{\pi^-})/2$. 

To compute $A^{\pi(q)}_n(\mu^2)$, let us first define the following integrals 
\begin{align}
q_n(\mu^2) &= \int_0^1 dx x^{n-1} q^{\pi}(x, \mu^2), \\
\overline{q}_n(\mu^2) &= \int_0^1 dx x^{n-1} \overline{q}^{\pi}(x, \mu^2), \\
s_n(\mu^2) &= \int_0^1 dx x^{n-1} s^{\pi}(x, \mu^2). 
\label{eq:nonlocal.dim4.temp.3}
\end{align}
Using these, $A^{\pi(q)}_n(\mu^2) = \frac{1}{2}[A^{\pi(u)}_n(\mu^2) + A^{\pi(d)}_n(\mu^2)]$ for all three pion states ($\pi^+$, $\pi^{-}$ and $\pi^0$) can be given as 
\begin{align}
A^{\pi(q)}_n(\mu^2) = \frac{1 + (-1)^n}{2} \Bigl[ 
q_n(\mu^2) + 4\overline{q}_n(\mu^2) \Bigr]. 
\label{eq:nonlocal.dim4.temp.4}
\end{align}
For the strange quark case, one obtains 
\begin{align}
A^{\pi(s)}_n(\mu^2) = 4 \frac{1 + (-1)^n}{2} s_n(\mu^2). 
\label{eq:nonlocal.dim4.temp.5}
\end{align}
For the convenience of the reader, we tabulate $A^{\pi(q)}_n(\mu^2)$ and $A^{\pi(s)}_n(\mu^2)$ for scales $\sqrt{\mu^2} = 1\,\mathrm{GeV}$ and 
$\sqrt{\mu^2} = 2\,\mathrm{GeV}$ for both LO and NLO fits of Ref.\,\cite{Gluck:1999xe} in Table \ref{tab:A.pion.values}. 
\begin{table}
\begin{center}
\caption{$A^{\pi(q)}$ and $A^{\pi(g)}$ values as defined in Eq.\,(\ref{eq:nonlocal.dim4.temp.2}) obtained by numerically 
integrating the parton distributions of the pion provided in Ref.\,\cite{Gluck:1999xe}. Only non-zero values are shown in this table.} 
\label{tab:A.pion.values}
\begin{tabular}{ccccc}  
\toprule
 & \multicolumn{2}{c|}{LO} & \multicolumn{2}{c}{NLO} \\ \midrule
$\sqrt{\mu^2}$ & 1 GeV & \multicolumn{1}{c|}{2 GeV} & 1 GeV & 2 GeV \\ \midrule
$A^{\pi(q)}_2$ & 0.598 & 0.537 & 0.614 & 0.544 \\
$A^{\pi(s)}_2$ & 0.0255 & 0.0431 & 0.0257 & 0.0474 \\ 
$A^{\pi(g)}_2$ & 0.393 & 0.441 & 0.380 & 0.433 \\ \midrule
$A^{\pi(q)}_4$ & 0.136 & 0.103 & 0.142 & 0.104 \\
$A^{\pi(s)}_4$ & 0.00238 & 0.00274 & 0.00154 & 0.00207 \\ 
$A^{\pi(g)}_4$ &  0.0446 & 0.0282 & 0.0593 & 0.0367 \\ \midrule
$A^{\pi(q)}_6$ & 0.0645 & 0.0447 & 0.0676 & 0.0450 \\
$A^{\pi(s)}_6$ & 0.000666 & 0.000665 & 0.000351 & 0.000409 \\ 
$A^{\pi(g)}_6$ & 0.0149 & 0.00749 & 0.0222 & 0.0108 \\ 
\bottomrule
\end{tabular}
\end{center}
\end{table}
Unfortunately, no error estimates are given for these parton distributions, which is why we can only quote absolute values in 
Table \ref{tab:A.pion.values}. This situation is likely to improve in the future, due to new global fits to experimental 
data \cite{Barry:2018ort} and direct lattice QCD calculations of parton distributions \cite{Chen:2018fwa} and their moments \cite{Detmold:2003tm}. 
The latter would make it possible to compute the partonic content not only of pions, but also of other hadrons, 
for which experimental measurements are not feasible. A consistent determination of valence quark, sea quark (including strangeness) 
and gluonic parton distributions from lattice QCD remains, however, challenging. 

To estimate the corrections due to mesons with larger masses (such as kaons and $\eta$ mesons), it is useful to have at hand some 
information about their partonic components. Especially for the strange quark condensate, effects due to pions are suppressed while mesons containing 
strange valence quarks can be expected to give significant contributions. 
Even though there are some efforts to compute the parton distributions of the kaon (see for instance Refs.\,\cite{Gluck:1997ww} or \cite{Nam:2012vm} for a 
recent model based calculation), the related uncertainties are still large due to lack of experimental data. 
Here, we follow Ref.\,\cite{Hatsuda:1992bv} and, partly, Ref.\,\cite{Gluck:1997ww} and simply assume that the valence parton distributions are flavor independent, 
while the sea and gluon distributions are the same for all pseudoscalar mesons. Based on these assumptions, we get, after averaging over the 
different kaon states, 
\begin{align}
A^{K(q)}_n(\mu^2) = \frac{1 + (-1)^n}{2} \Bigl[ 
\frac{1}{2} q_n(\mu^2) + 4\overline{q}_n(\mu^2) \Bigr],  
\label{eq:nonlocal.dim4.temp.6}
\end{align}
and 
\begin{align}
A^{K(s)}_n(\mu^2) = \frac{1 + (-1)^n}{2} \Bigl[ 
q_n(\mu^2) + 4s_n(\mu^2) \Bigr].   
\label{eq:nonlocal.dim4.temp.7}
\end{align}
Equally, we obtain for the $\eta$-meson (assuming that it is a pure flavor octet state) 
\begin{align}
A^{\eta(q)}_n(\mu^2) = \frac{1 + (-1)^n}{2} \Biggl[ 
\frac{1}{3} q_n(\mu^2) + 4\overline{q}_n(\mu^2) \Biggr],  
\label{eq:nonlocal.dim4.temp.8}
\end{align}
and 
\begin{align}
A^{\eta(s)}_n(\mu^2) = \frac{1 + (-1)^n}{2} \Biggl[ 
\frac{4}{3} q_n(\mu^2) + 4s_n(\mu^2) \Biggr].   
\label{eq:nonlocal.dim4.temp.9}
\end{align}
The tabulated values corresponding to the above results are given in Tables\,\ref{tab:A.kaon.values} and \ref{tab:A.eta.values}. 
\begin{table}
\begin{center}
\caption{Same as Tab.\,\ref{tab:A.pion.values}, but for the kaon.} 
\label{tab:A.kaon.values}
\begin{tabular}{ccccc}  
\toprule
 & \multicolumn{2}{c|}{LO} & \multicolumn{2}{c}{NLO} \\ \midrule
$\sqrt{\mu^2}$ & 1 GeV & \multicolumn{1}{c|}{2 GeV} & 1 GeV & 2 GeV \\ \midrule
$A^{K(q)}_2$ & 0.371 & 0.341 & 0.379 & 0.346 \\
$A^{K(s)}_2$ & 0.295 & 0.275 & 0.301 & 0.280 \\ 
$A^{K(g)}_2$ & 0.393 & 0.441 & 0.380 & 0.433 \\ \midrule
$A^{K(q)}_4$ & 0.0719 & 0.0548 & 0.0753 & 0.0555 \\
$A^{K(s)}_4$ & 0.0506 & 0.0387 & 0.0530 & 0.0393 \\ 
$A^{K(g)}_4$ & 0.0446 & 0.0282 & 0.0593 & 0.0367 \\ \midrule
$A^{K(q)}_6$ & 0.0329 & 0.0229 & 0.0345 & 0.0231 \\
$A^{K(s)}_6$ & 0.0224 & 0.0157 & 0.0235 & 0.0158 \\ 
$A^{K(g)}_6$ & 0.0149 & 0.00749 & 0.0222 & 0.0108 \\ 
\bottomrule
\end{tabular}
\end{center}
\end{table}
\begin{table}
\begin{center}
\caption{Same as Tab.\,\ref{tab:A.pion.values}, but for the $\eta$ meson.} 
\label{tab:A.eta.values}
\begin{tabular}{ccccc}  
\toprule
 & \multicolumn{2}{c|}{LO} & \multicolumn{2}{c}{NLO} \\ \midrule
$\sqrt{\mu^2}$ & 1 GeV & \multicolumn{1}{c|}{2 GeV} & 1 GeV & 2 GeV \\ \midrule
$A^{\eta(q)}_2$ & 0.480 & 0.435 & 0.495 & 0.444 \\
$A^{\eta(s)}_2$ & 0.632 & 0.566 & 0.651 & 0.576 \\ 
$A^{\eta(g)}_2$ & 0.393 & 0.441 & 0.380 & 0.433 \\ \midrule
$A^{\eta(q)}_4$ & 0.131 & 0.0990 & 0.136 & 0.0992 \\
$A^{\eta(s)}_4$ & 0.174 & 0.131 & 0.180 & 0.132 \\ 
$A^{\eta(g)}_4$ & 0.0446 & 0.0282 & 0.0593 & 0.0367 \\ \midrule
$A^{\eta(q)}_6$ & 0.0637 & 0.0442 & 0.0664 & 0.0442 \\
$A^{\eta(s)}_6$ & 0.0847 & 0.0587 & 0.0885 & 0.0588 \\ 
$A^{\eta(g)}_6$ & 0.0149 & 0.00749 & 0.0222 & 0.0108 \\ 
\bottomrule
\end{tabular}
\end{center}
\end{table}

\begin{figure}[tb]
\begin{center}
\begin{minipage}[t]{16.5 cm}
\vspace{0.5 cm}
%\hspace{0.5 cm}
%\epsfig{file=dim4_non_scalar_q_2,scale=0.7}
%\epsfig{file=dim4_non_scalar_s_2,scale=0.7}
\includegraphics[width=8.5cm,bb=0 0 360 252]{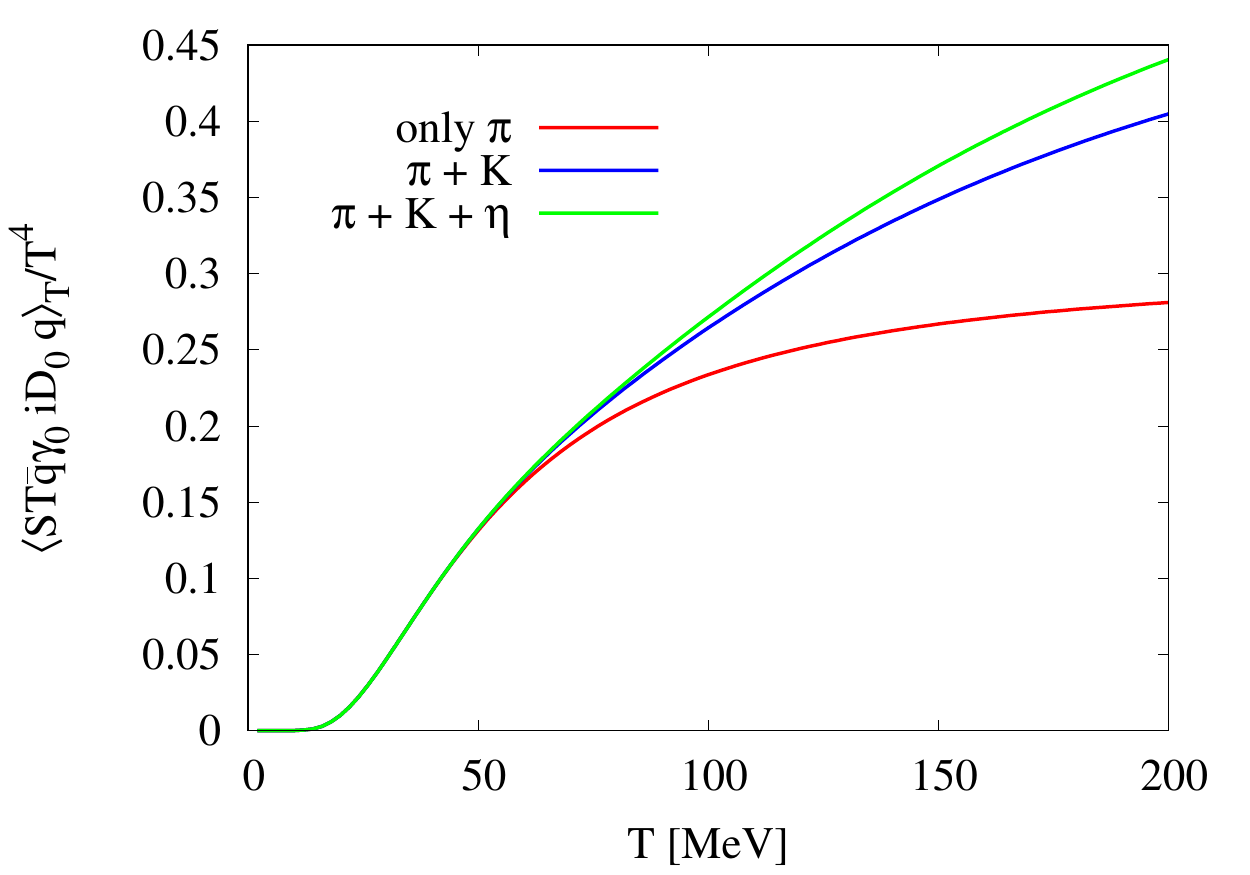}
\includegraphics[width=8.5cm,bb=0 0 360 252]{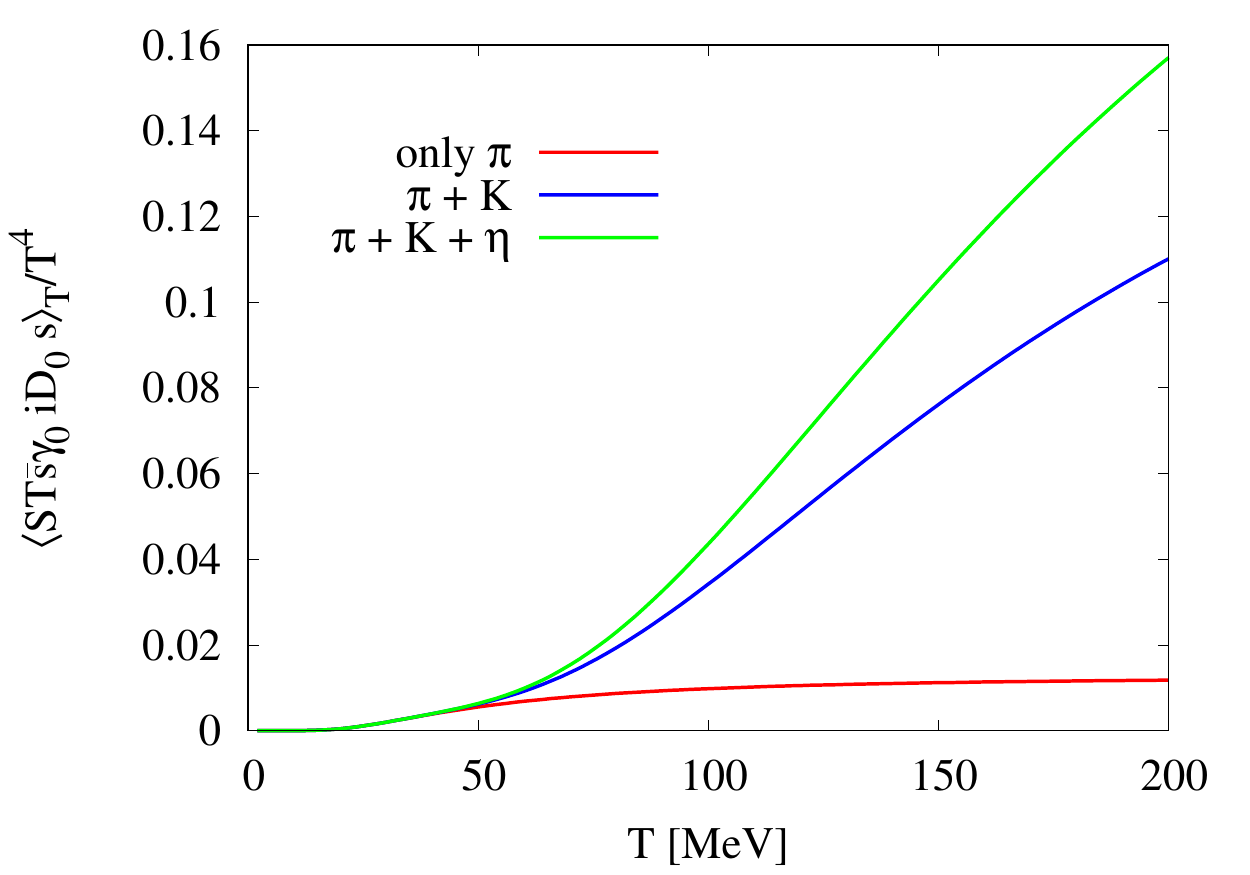}
%\vspace{1 cm}
\end{minipage}
\begin{minipage}[t]{16.5 cm}
\caption{The dimension 4, spin 2 quark condensate at finite temperature in the free hadron gas model. 
The red curve corresponds to only pion contributions, 
to which kaons are added in the blue curve and furthermore the $\eta$ contribution in the green curve. 
In the left plot the value of $\langle \mathcal{ST} \bar{q} \gamma^{0} iD^{0} q  \rangle_{T}/T^4 = 
\frac{1}{2} \bigl(\langle \mathcal{ST} \bar{u} \gamma^{0} iD^{0} u \rangle_{T} + \langle \mathcal{ST} \bar{d} \gamma^{0} iD^{0} d \rangle_{T} \bigr)/T^4$ is shown, based on Eq.\,(\ref{eq:nonlocal.dim4.temp.11})
and using the NLO values at 1 GeV in Tables \ref{tab:A.pion.values}, \ref{tab:A.kaon.values} and \ref{tab:A.eta.values}. The right plot shows the same quantity, but for 
$\langle \mathcal{ST} \bar{s} \gamma^{0} iD^{0} s \rangle_{T}/T^4$. 
\label{fig:non.scalar.dim4}}
\end{minipage}
\end{center}
\end{figure}

With the above $A$ parameter values, we can now estimate the 
\begin{align}
\langle \mathcal{ST} \bar{q} \gamma^{\mu} iD^{\nu} q \rangle_{T} = 
\frac{1}{2} \bigl(\langle \mathcal{ST} \bar{u} \gamma^{\mu} iD^{\nu} u \rangle_{T} + \langle \mathcal{ST} \bar{d} \gamma^{\mu} iD^{\nu} d \rangle_{T} \bigr)
\label{eq:nonlocal.dim4.temp.10}
\end{align}
condensate at low temperatures. 
Using Eqs.\,(\ref{eq:HRG.model}) and (\ref{eq:nonlocal.dim4.temp.1}) and performing the momentum integral, the result reads 
\begin{align}
\langle \mathcal{ST} \bar{q} \gamma^{\mu} iD^{\nu} q \rangle_{T} = \frac{d_{\pi} A^{\pi(q)}}{360} \Biggl[ 8 \pi^2 T^4 B_2\Bigl(\frac{m_{\pi}}{T}\Bigr) - 5m_{\pi}^2 T^2 B_1\Bigl(\frac{m_{\pi}}{T}\Bigr) \Biggr] \mathcal{ST}(u^{\mu}u^{\nu}), 
\label{eq:nonlocal.dim4.temp.11}
\end{align}
with $u^{\mu} = (1, 0, 0, 0)$. 
The $B_n(x)$ functions are defined in Eq.\,(\ref{eq:chiral.cond.pion.gas.2}) and $d_{\pi}$ stands for the number of degrees of freedom of pions, $d_{\pi} = 3$. 
For the strange quark case, we have, similarly, 
\begin{align}
\langle \mathcal{ST} \bar{s} \gamma^{\mu} iD^{\nu} s \rangle_{T} = \frac{d_{\pi} A^{\pi(s)}}{360} \Biggl[ 8 \pi^2 T^4 B_2\Bigl(\frac{m_{\pi}}{T}\Bigr) - 5m_{\pi}^2 T^2 B_1\Bigl(\frac{m_{\pi}}{T}\Bigr) \Biggr] \mathcal{ST}(u^{\mu}u^{\nu}).  
\label{eq:nonlocal.dim4.temp.12}
\end{align}
It is straightforward to extend the above results to include contributions of more meson states. One simply adds the same terms, replacing $d_{\pi}$, $A^{\pi(q)}$ and $m_{\pi}$ with the 
corresponding values of the kaon and $\eta$ mesons, specifically $d_{K} = 4$ and $d_{\eta} = 1$. The results of such a calculation are shown in 
Fig.\,\ref{fig:non.scalar.dim4}, for which the NLO values at 1 GeV of Tables\,\ref{tab:A.pion.values}, \ref{tab:A.kaon.values} and \ref{tab:A.eta.values} were used. 
%%%%%%%%%%%%%%%%%%%%%%%%%%%%%%%%%%%%%%%%%%%%%%%%%%%%%%%%%%%%%%%%%%%%%%%%%%%%%%%%%%%%%%%%%%%%%%%%%%%%%%%%%%%%%%%%%%%%
%%%%%%%%%%%%%%%%%%%%%%%%%%%%%%%%%%%%%%%%%%%%%%%%%%%%%%%%%%%%%%%%%%%%%%%%%%%%%%%%%%%%%%%%%%%%%%%%%%%%%%%%%%%%%%%%%%%%
The plots show that the pions dominate the thermal behavior of the condensates at temperatures below $T = 50$ MeV, above which 
the kaon and $\eta$ meson contributions start to become non-negligible. This is particularly true for the strange quark condensate, for which 
the pion contributions are strongly suppressed because of the small strange parton content of the pion. Not surprisingly, the kaons therefore 
play the dominant role for this condensate already around $T = 100$ MeV. It is expected that more hadron states come into play 
as the temperature increases above 100 MeV and approaches $T_c$. The curves shown in Fig.\,\ref{fig:non.scalar.dim4} should hence 
be understood as lower limits. 

The quark condensates $\langle \bar{q} iD^{\mu} q \rangle_{T}$ and $\langle \bar{s} iD^{\mu} s \rangle_{T}$ can be shown to 
scale with the light quark and strange baryon densities, as will be demonstrated in the discussion following Eq.\,(\ref{eq:nonlocal.dim4.1}). 
They therefore vanish exactly for the finite temperature and zero density case considered here.    

\begin{figure}[tb]
\begin{center}
\begin{minipage}[t]{8 cm}
\vspace{0.5 cm}
\hspace{-2.0 cm}
\includegraphics[width=12cm,bb=0 0 360 252]{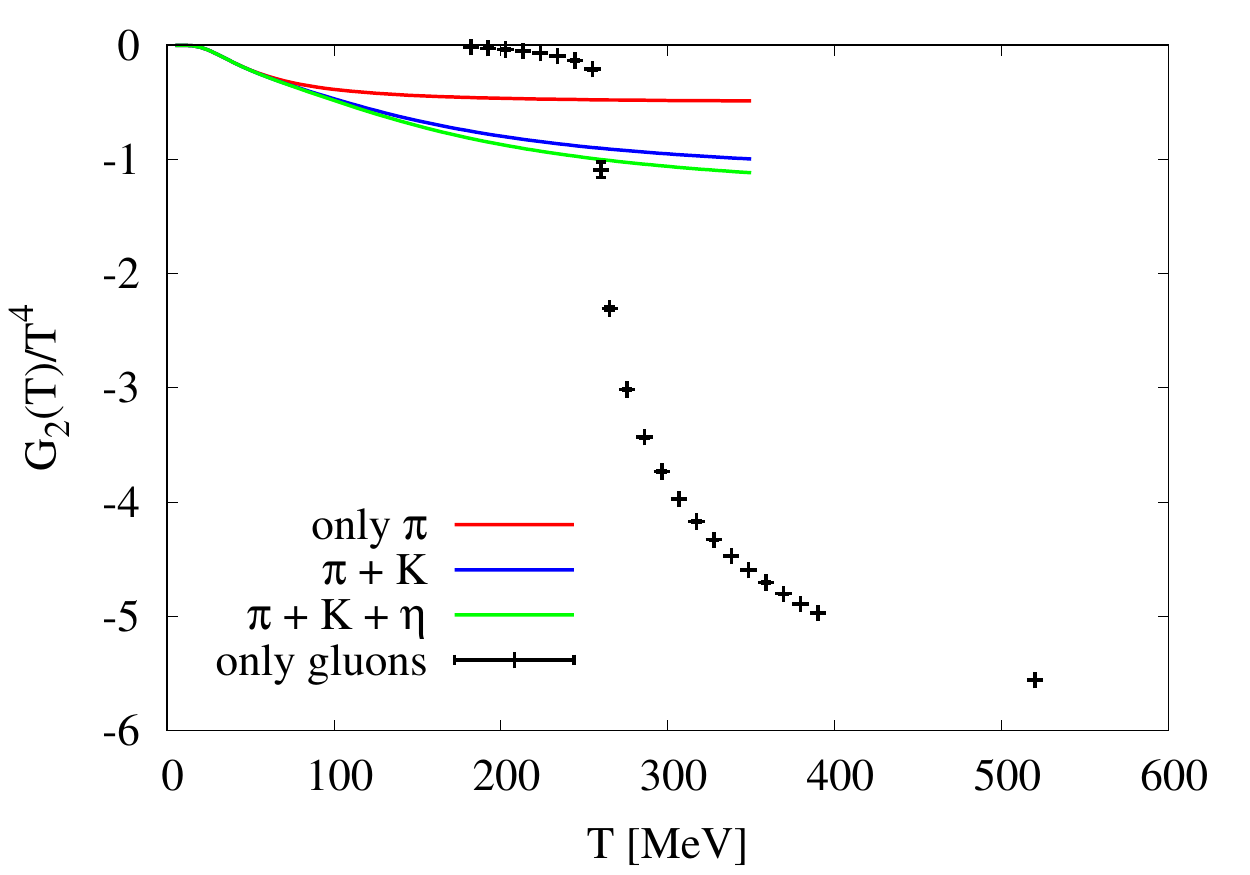}
%\vspace{1 cm}
\end{minipage}
\begin{minipage}[t]{16.5 cm}
\caption{
The dimension 4 and spin 2 gluon condensate value at finite temperature in the free hadron gas model (solid lines) and from quenched 
lattice QCD (black data points). The red, blue and green hadron gas model curves are obtained from Eq.\,(\ref{eq:nonlocal.dim4.temp.19}) 
with $A^{\pi(g)}_2$ ($A^{K(g)}_2$, $A^{\eta(g)}_2$) NLO values at 1 GeV from Tables\,\ref{tab:A.pion.values} (\ref{tab:A.kaon.values}, \ref{tab:A.eta.values}). 
The lattice QCD data points are obtained from Eq.\,(\ref{eq:nonlocal.dim4.temp.15}) and the data of Ref.\,\cite{Borsanyi:2012ve} with 
$T_c = 260$ MeV \cite{Boyd:1996bx}. 
\label{fig:dim4.non.scalar.gluon}}
\end{minipage}
\end{center}
\end{figure}

The last condensate to be discussed in this section is the spin 2 gluon condensate, $\langle \mathcal{ST} G_{\alpha}^{a\mu} G^{a\nu\alpha} \rangle_{T}$. 
Its thermal behavior is not known well, as lattice QCD calculations of this quantity including dynamical quarks have not yet been performed. 
There is, however, some information that can be extracted from quenched lattice data as well as the free hadron gas model. 
Let us start with a discussion based on quenched lattice QCD results, following the method proposed in Refs.\,\cite{Morita:2007pt,Morita:2007hv}. 
The idea is to recognize that the gluonic operator $\mathcal{ST} G_{\alpha}^{a\mu} G^{a\alpha\nu} = G_{\alpha}^{a\mu} G^{a\alpha\nu} - \frac{1}{4} g^{\mu\nu} G_{\alpha}^{a\beta} G^{a\alpha\beta}$ 
is nothing but the energy-momentum tensor of QCD without quarks [times $(-1)$], 
\begin{align}
T^{\mu \nu} = -G_{\alpha}^{a\mu} G^{a\alpha\nu} +  \frac{1}{4} g^{\mu\nu} G_{\alpha}^{a\beta} G^{a\alpha\beta}. 
\label{eq:nonlocal.dim4.temp.13}
\end{align}
The same energy-momentum tensor can be expressed using the thermodynamic quantities of energy density $\epsilon(T)$ and pressure $p(T)$, 
\begin{align}
T^{\mu \nu} = [\epsilon(T) + p(T)] \Bigl( u^{\mu}u^{\nu} - \frac{1}{4}g^{\mu \nu} \Bigr) + \frac{1}{4}[\epsilon(T) - 3p(T)] g^{\mu \nu},   
\label{eq:nonlocal.dim4.temp.14}
\end{align}
where $u^{\mu}$ is the four-velocity of the heat bath. 
Therefore, comparing the trace subtracted parts of the above equations, one obtains
\begin{align}
\langle \mathcal{ST} G_{\alpha}^{a\mu} G^{a\alpha\nu} \rangle_{T} &= G_2(T) \mathcal{ST}(u^{\mu} u^{\nu}), \label{eq:nonlocal.dim4.temp.15.2} \\
G_2(T) &= -[\epsilon(T) + p(T)]. 
\label{eq:nonlocal.dim4.temp.15}
\end{align}
Note that in Refs.\,\cite{Morita:2007pt,Morita:2007hv} $G_2(T)$ was defined with an additional factor of $\alpha_s(T)/\pi$. To avoid the uncertainties 
related to the determination of the temperature dependence of $\alpha_s(T)$, we here define the condensate without this factor. 
As mentioned earlier, the quantities $\epsilon(T)$ and $p(T)$ can nowadays be determined from lattice QCD with good precision. 
Because we are here working in the quenched approximation, quenched lattice QCD data have to be employed for consistency. 
We for this purpose use the data provided in Ref.\,\cite{Borsanyi:2012ve}, which lead to the result shown as black data points in Fig.\,\ref{fig:dim4.non.scalar.gluon}. 

To consider the same quantity in the free hadron gas model, it is useful to define the following matrix element, 
\begin{equation}
\langle \pi^{a}(\bm{p}) | \mathcal{ST} G_{\alpha \mu_1}^a D_{\mu_2} \cdots D_{\mu_{n-1}} G_{\mu_n}^{a\alpha} | \pi^{a}(\bm{p}) \rangle  \equiv
 (-i)^{n - 2} 2 A^{\pi(g)}_n(\mu^2) \mathcal{ST} ( p_{\mu_1} \cdots p_{\mu_n} ),   
\label{eq:nonlocal.dim4.temp.16}
\end{equation}
where, as before, the superscript $a$ is not summed. The theory of DIS relates this matrix element to an integral of the gluonic 
parton distributions functions of the pion, 
\begin{equation}
A^{\pi(g)}_n(\mu^2) = \frac{1 + (-1)^n}{2} \int_0^1 dx x^{n-1} g(x, \mu^2).  
\label{eq:nonlocal.dim4.temp.17}
\end{equation}
The gluonic parton distribution of the pion, given in Ref.\,\cite{Gluck:1999xe}, is shown as a blue curve in Fig.\,\ref{fig:parton.distributions.pion} 
for $\mu^2 = 1\,\mathrm{GeV}^2$ in an NLO scheme. The values of the integrals for $n=2$, $4$ and $6$ are given in Table\,\ref{tab:A.pion.values}. 
In the approximations used here, the respective values for kaons and the $\eta$ meson (given in Tables\,\ref{tab:A.kaon.values} and \ref{tab:A.eta.values}) are 
identical to those of the pion. Computing the momentum integral, we get, in analogy to Eqs.\,(\ref{eq:nonlocal.dim4.temp.11}) and (\ref{eq:nonlocal.dim4.temp.12}), 
\begin{align}
\langle \mathcal{ST} G_{\alpha}^{a\mu} G^{a\alpha\nu} \rangle_{T} = -\frac{d_{\pi} A^{\pi(g)}_2}{180} \Biggl[ 8 \pi^2 T^4 B_2\Bigl(\frac{m_{\pi}}{T}\Bigr) - 5m_{\pi}^2 T^2 B_1\Bigl(\frac{m_{\pi}}{T}\Bigr) \Biggr] \mathcal{ST}(u^{\mu}u^{\nu}). 
\label{eq:nonlocal.dim4.temp.18}
\end{align}
The minus sign in the above equation is a result of interchanging the Lorentz indices of the second gluon operator which is antisymmetric. This immediately leads to
\begin{align}
G_2(T)  = -\frac{d_{\pi} A^{\pi(g)}_2}{180} \Biggl[ 8 \pi^2 T^4 B_2\Bigl(\frac{m_{\pi}}{T}\Bigr) - 5m_{\pi}^2 T^2 B_1\Bigl(\frac{m_{\pi}}{T}\Bigr) \Biggr]. 
\label{eq:nonlocal.dim4.temp.19}
\end{align}
It is again straightforward to generalize this result to include more pseudoscalar mesons. One simply has to add further terms 
in which the $m_{\pi}$, $d_{\pi}$ and $A^{\pi(g)}_2$ are replaced by those of kaons and $\eta$ mesons. 

%%%%%%%%%%%%%%%%%%%%%%%%%%%%%%%%%%%%%%%%%%%%%%%%%%%%%%%%%%%%%%%%%%%%%%%%%%%%%%%%%%%%%%%%%%%%%%%%%%%%%%%%%%%%%%%%%%%%
%%%%%%%%%%%%%%%%%%%%%%%%%%%%%%%%%%%%%%%%%%%%%%%%%%%%%%%%%%%%%%%%%%%%%%%%%%%%%%%%%%%%%%%%%%%%%%%%%%%%%%%%%%%%%%%%%%%%
In Fig.\,\ref{fig:dim4.non.scalar.gluon}, we compare Eqs.\,(\ref{eq:nonlocal.dim4.temp.15}) and (\ref{eq:nonlocal.dim4.temp.19}), for the latter showing the curves including only pion 
contributions (red curve), pion + kaon contributions (blue curve) and pion + kaon + $\eta$ meson contributions (green curve). The quenched lattice 
QCD points do not deviate much from zero until temperatures close to $T_c$, where a sudden drop is observed, reflective of the first order phase 
transition occurring in quenched QCD. The small temperature dependence of $G_2(T)$ in quenched lattice QCD at low temperatures can be 
understood from the lowest energy excitations of the theory. These are glueballs, whose lowest mass has been estimated to be larger than 
1.5 GeV \cite{Morningstar:1999rf} and are therefore strongly suppressed at temperatures below $T_c$. The free hadron gas, which can be trusted to 
give an accurate result for $T \lesssim 100$ MeV, on the other hand gives a stronger temperature dependence for low $T$. One can expect this 
temperature dependence to become even stronger as the effects of more hadrons are taken into account. To accurately determine the 
behavior of $G_2(T)$ around $T_c$ a lattice QCD computation that includes dynamical quarks will however be needed.

\paragraph{Condensates of dimension 5}\mbox{}\\
Our knowledge of the dimension 5 condensate temperature dependences is presently still rather limited. 
Nevertheless, some pieces of information are available, which we summarize here. 
We start with the dimension 5 scalar condensate, $\langle \bar{q} \sigma_{\mu\nu} t^a G^{a\mu\nu} q  \rangle_{T}$ about which up to today 
only four works have been published in the literature: one lattice QCD study \cite{Doi:2004jp}, one based on the global color symmetry model \cite{Zhang:2004xg}, one on 
the liquid instanton model at finite T \cite{Nam:2013gza} and one on Dyson-Schwinger equations \cite{Li-Juan:2014qua}. 
As it is customary in vacuum [see Eq.\,(\ref{eq:mixed.cond.para})], this condensate is usually parametrized relative to the dimension 3 chiral condensate, 
\begin{equation}
m_0^2(T) \equiv \frac{\langle \bar{q} g \sigma G q  \rangle_{T}}{\langle \bar{q} q \rangle_{T}}.  
\label{eq:nonscalar.dim5.8}
\end{equation}
In the lattice QCD calculation of Ref.\,\cite{Doi:2004jp}, which was done using the quenched approximation and Kogut-Susskind (or staggered) fermions, it was found that 
$m_0^2(T)$ does not show any temperature dependence within errors for the probed temperature range (from zero to slightly above $T_c$). 
This means that $\langle \bar{q} g \sigma G q \rangle_{T}$, which like $\langle \bar{q} q \rangle_{T}$ is an order parameter of chiral symmetry, 
quickly (but smoothly) approaches 0 around $T_c$. As Ref.\,\cite{Doi:2004jp} is already somewhat old, it would be interesting 
to repeat it with dynamical quarks and a lattice fermion prescription with better chiral properties. Furthermore, the problem of 
potential mixing with condensates of lower dimension, which can happen on the lattice, deserves a careful investigation. The result 
nevertheless is suggestive and in essence consistent with the findings of models described in Refs.\,\cite{Zhang:2004xg,Nam:2013gza,Li-Juan:2014qua}. 

\begin{figure}[tb]
\begin{center}
\begin{minipage}[t]{16.5 cm}
\vspace{0.5 cm}
%\hspace{0.5 cm}
%\epsfig{file=dim5_non_scalar_q,scale=0.7}
%\epsfig{file=dim5_non_scalar_s,scale=0.7}
\includegraphics[width=8.5cm,bb=0 0 360 252]{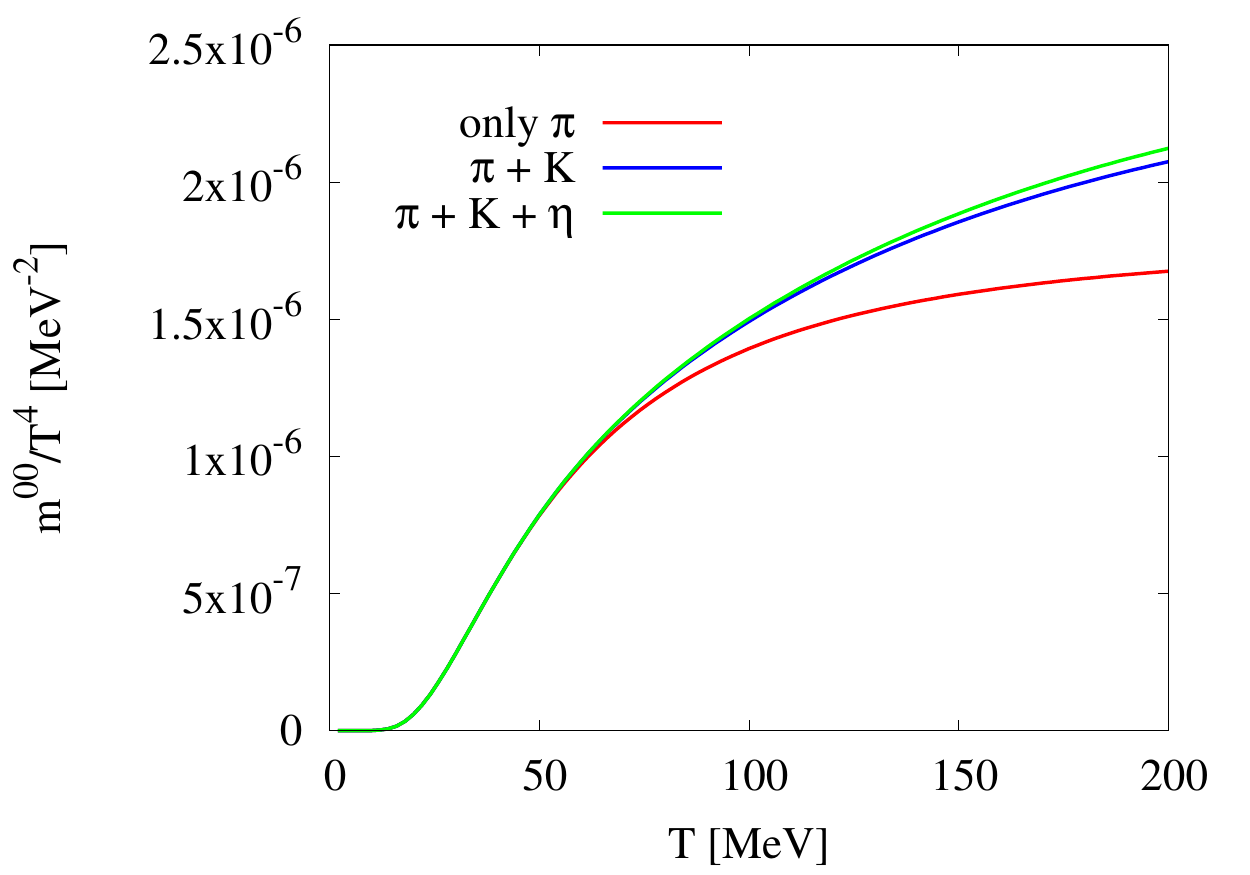}
\includegraphics[width=8.5cm,bb=0 0 360 252]{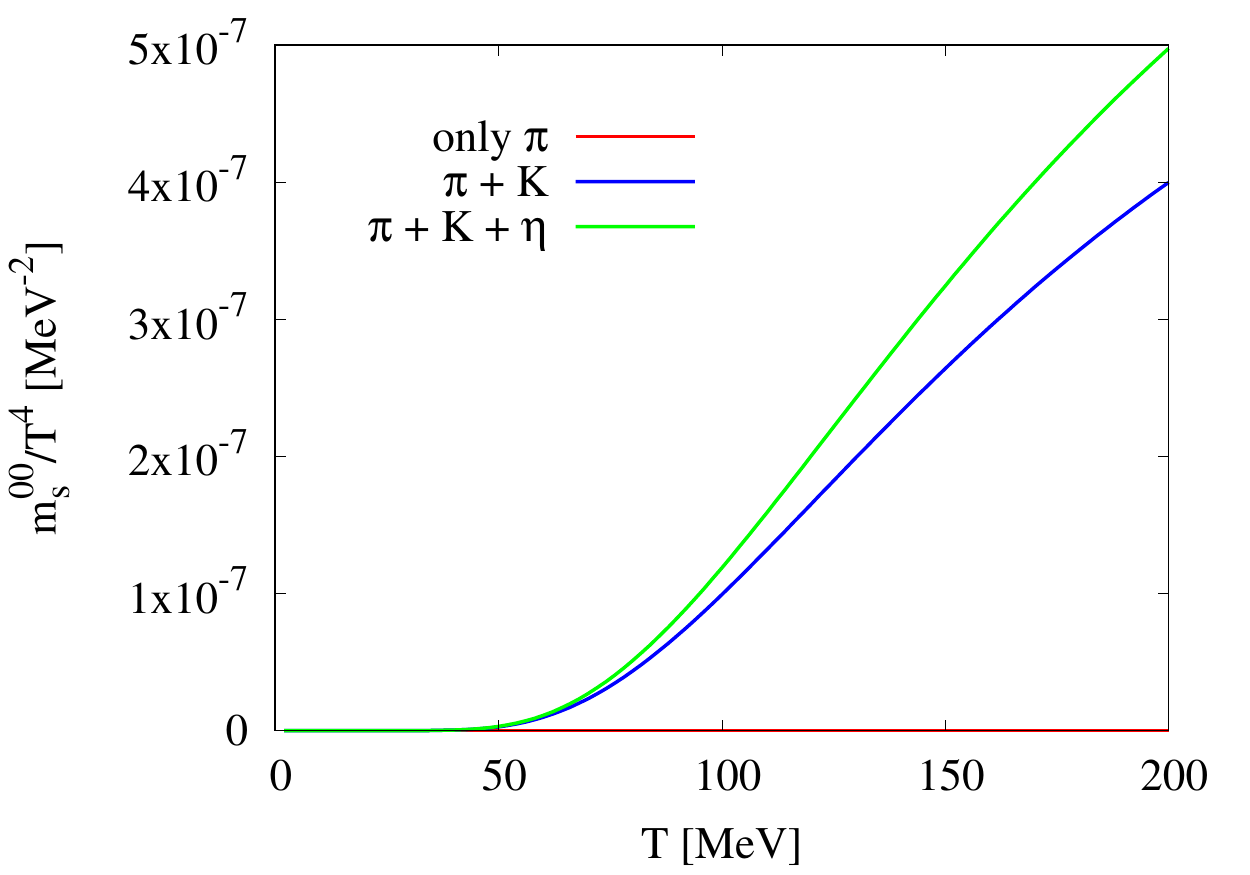}
%\vspace{1 cm}
\end{minipage}
\begin{minipage}[t]{16.5 cm}
\caption{Estimates of the dimension 5, spin 2 quark condensate value at finite temperature based on Eqs.\,(\ref{eq:nonscalar.dim5.2}-\ref{eq:nonscalar.dim5.7}). 
As in Fig.\,\ref{fig:non.scalar.dim4}, the red curve corresponds to only pion contributions, 
to which kaons are added in the blue curve and furthermore the $\eta$ contribution in the green curve. 
In the left (right) plot $m^{00} \equiv \langle \mathcal{ST} \bar{q} D^{0} D^{0} q \rangle_{T}/\langle 0 | \bar{q} q | 0 \rangle$ 
($m^{00}_s \equiv \langle \mathcal{ST} \bar{s} D^{0} D^{0} s \rangle_{T}/\langle 0 | \bar{s} s | 0 \rangle$) is shown. Note that for $m^{00}_s$ 
the pion contribution vanishes in the approximation used here.  
\label{fig:non.scalar.dim5}}
\end{minipage}
\end{center}
\end{figure}

We next turn to the non-scalar condensates [listed in the third line of Eq.\,(\ref{eq:condensates.medium})], about which unfortunately not much is known. 
Let us use Eq.\,(\ref{eq:HRG.model}) to provide a simple estimate. 
About the finite density counterpart of $\langle \mathcal{ST} \bar{q} iD^{\mu} iD^{\nu} q  \rangle_{T}$, some information was 
recently obtained from the twist-3 parton distribution function of the nucleon, $e(x)$ in Ref.\,\cite{Gubler:2015uza} (see the discussion about dimension 5 condensates 
in Section \ref{conds.at.finite.den}). 
At finite temperature, one presumably could do the same by considering the corresponding distribution function of the pion, which however presently is not known. 
We will hence have to resort to a cruder estimate. For this purpose, we follow Ref.\,\cite{Lee:1993ww} to get
\begin{align}
\langle \pi^a(\bm{p}) | \bar{q} D^{\mu} D^{\nu} q | \pi^a(\bm{p}) \rangle &\simeq - P^{q(\pi)}_{\mu} P^{q(\pi)}_{\nu} \langle \pi^a(\bm{p}) | \bar{q} q | \pi^a(\bm{p}) \rangle \nonumber \\
&\simeq - \frac{1}{16} p_{\mu} p_{\nu} \langle \pi^a(\bm{p}) | \bar{q} q | \pi^a(\bm{p}) \rangle, 
\label{eq:nonscalar.dim5.1}
\end{align}
where $P^{q(\pi)}_{\mu}$ is the average four-momentum of the quark $q$ in the pion state $|\pi^a(\bm{p}) \rangle$. Going to the second line, we assume that 
half of the momentum of the pion is carried by gluons and the rest is evenly distributed among the two valence quarks. After making the above expression traceless, using Eq.\,(\ref{eq:HRG.model}), 
carrying out the momentum integral and treating the scalar quark condensate as described in Ref.\,\cite{Hatsuda:1992bv}, one obtains
\begin{equation}
\langle \mathcal{ST} \bar{q} D^{\mu} D^{\nu} q \rangle_{T,\,\pi} \simeq \frac{d_{\pi} \langle 0 | \bar{q} q | 0 \rangle}{11520 f_{\pi}^2} 
\Biggl[ 8 \pi^2 T^4 B_2\Bigl(\frac{m_{\pi}}{T}\Bigr) - 5m_{\pi}^2 T^2 B_1\Bigl(\frac{m_{\pi}}{T}\Bigr) \Biggr] \mathcal{ST}(u^{\mu}u^{\nu}). 
\label{eq:nonscalar.dim5.2}
\end{equation}
Similarly, the contributions from kaons and the $\eta$ meson read 
\begin{align}
\langle \mathcal{ST} \bar{q} D^{\mu} D^{\nu} q  \rangle_{T,\,K} &\simeq \frac{d_{K} \langle 0 | \bar{q} q | 0 \rangle}{23040 f_{K}^2} 
\Biggl[ 8 \pi^2 T^4 B_2\Bigl(\frac{m_{K}}{T}\Bigr) - 5m_{K}^2 T^2 B_1\Bigl(\frac{m_{K}}{T}\Bigr) \Biggr] \mathcal{ST}(u^{\mu}u^{\nu}), \label{eq:nonscalar.dim5.3} \\
\langle \mathcal{ST} \bar{q} D^{\mu} D^{\nu} q \rangle_{T,\,\eta} &\simeq \frac{d_{\eta} \langle 0 | \bar{q} q | 0 \rangle}{34560 f_{\eta}^2} 
\Biggl[ 8 \pi^2 T^4 B_2\Bigl(\frac{m_{\eta}}{T}\Bigr) - 5m_{\eta}^2 T^2 B_1\Bigl(\frac{m_{\eta}}{T}\Bigr) \Biggr] \mathcal{ST}(u^{\mu}u^{\nu}).
\label{eq:nonscalar.dim5.4}
\end{align}
Here, we have assumed the momentum to scale with the number of valence quarks. Applying the same method to the respective strange 
quark condensate, one gets 
\begin{align}
\langle \mathcal{ST} \bar{s} D^{\mu} D^{\nu} s \rangle_{T,\,\pi} &\simeq 0, \label{eq:nonscalar.dim5.5} \\
\langle \mathcal{ST} \bar{s} D^{\mu} D^{\nu} s \rangle_{T,\,K} &\simeq \frac{d_{K} \langle 0 | \bar{s} s | 0 \rangle}{23040 f_{K}^2} 
\Biggl[ 8 \pi^2 T^4 B_2\Bigl(\frac{m_{K}}{T}\Bigr) - 5m_{K}^2 T^2 B_1\Bigl(\frac{m_{K}}{T}\Bigr) \Biggr] \mathcal{ST}(u^{\mu}u^{\nu}), \label{eq:nonscalar.dim5.6} \\
\langle \mathcal{ST} \bar{s} D^{\mu} D^{\nu} s \rangle_{T,\,\eta} &\simeq \frac{d_{\eta} \langle 0 | \bar{s} s | 0 \rangle}{17280 f_{\eta}^2} 
\Biggl[ 8 \pi^2 T^4 B_2\Bigl(\frac{m_{\eta}}{T}\Bigr) - 5m_{\eta}^2 T^2 B_1\Bigl(\frac{m_{\eta}}{T}\Bigr) \Biggr] \mathcal{ST}(u^{\mu}u^{\nu}). 
\label{eq:nonscalar.dim5.7}
\end{align}
It is possible to extend this approach by including further hadrons. However, doing so would not be very meaningful, as 
already Eq.\,(\ref{eq:nonscalar.dim5.1}) is not much more than a crude order of magnitude estimate. Indeed, it was shown in Ref.\,\cite{Gubler:2015uza} 
that the nucleon matrix element of the same operator estimated based on the above method turns out to be about 5 - 10 times larger than 
what is extracted from experimental information about $e(x)$ of the nucleon. 
Using Eqs.\,(\ref{eq:nonscalar.dim5.2}-\ref{eq:nonscalar.dim5.7}), the behavior of 
$m^{00} \equiv \langle \mathcal{ST} \bar{q} D^{0} D^{0} q \rangle_{T}/\langle 0 | \bar{q} q | 0 \rangle$ and 
$m^{00}_s \equiv \langle \mathcal{ST} \bar{s} D^{0} D^{0} s \rangle_{T}/\langle 0 | \bar{s} s | 0 \rangle$ are 
shown in the left and right plots of Fig.\,\ref{fig:non.scalar.dim5}, respectively, for illustration. 
%%%%%%%%%%%%%%%%%%%%%%%%%%%%%%%%%%%%%%%%%%%%%%%%%%%%%%%%%%%%%%%%%%%%%%%%%%%%%%%%%%%%%%%%%%%%%%%%%%%%%%%%%%%%%%%%%%%%
%%%%%%%%%%%%%%%%%%%%%%%%%%%%%%%%%%%%%%%%%%%%%%%%%%%%%%%%%%%%%%%%%%%%%%%%%%%%%%%%%%%%%%%%%%%%%%%%%%%%%%%%%%%%%%%%%%%%

The condensates $\langle \mathcal{ST} \bar{q} \gamma^{\mu} D^{\nu} D^{\omega} q \rangle_{T}$ and 
$\langle \mathcal{ST} \bar{s} \gamma^{\mu} D^{\nu} D^{\omega} s \rangle_{T}$ vanish in the free hadron gas model, 
as can be understood from the prefactor $1 + (-1)^2$ in Eqs.\,(\ref{eq:nonlocal.dim4.temp.4})-(\ref{eq:nonlocal.dim4.temp.9}) and 
remembering that $n$ is 3 here [see Eq.\,(\ref{eq:nonlocal.dim4.temp.1})]. A somewhat more intuitive explanation for this result can be obtained 
from an argument similar to that given in Eq.\,(\ref{eq:nonscalar.dim5.1}), where covariant derivatives are interpreted as average momenta 
of the quarks they operate on. In this picture the above two condensates become proportional to 
$\langle \mathcal{ST} \bar{q} \gamma^{\mu} q \rangle_{T}$ and $\langle \mathcal{ST} \bar{s} \gamma^{\mu} s \rangle_{T}$, which scale 
linearly with the respective quark densities and thus vanish in the zero baryon chemical potential case. 
We hence do not consider these condensates any further. 

The final condensate to be discussed at dimension 5 is $\langle \bar{q} \gamma^{\mu} \sigma_{\alpha \beta} G^{a \alpha \beta} t^a q \rangle_{T}$, 
which (in contrast to its finite density counterpart, which will be considered later), to our knowledge has so far never been studied. 
One simple estimate can be obtained by assuming that a relation similar to Eq.\,(\ref{eq:mixed.cond.para}) or Eq.\,(\ref{eq:nonscalar.dim5.8}) 
holds for this case as well. Specifically, 
\begin{align}
\langle \bar{q} \gamma^{\mu} \sigma_{\alpha \beta} G^{a \alpha \beta} t^a q \rangle_{T} & \simeq m_0^2 \langle \bar{q} \gamma^{\mu} q \rangle_{T} \\
& \simeq 0. 
\end{align}
This would suggest that $\langle \bar{q} \gamma^{\mu} \sigma_{\alpha \beta} G^{a \alpha \beta} t^a q \rangle_{T}$ is small and can be ignored for 
all practical purposes. An independent evaluation or a lattice QCD computation are however certainly needed to confirm the above rough estimate.  

\paragraph{Condensates of dimension 6}\mbox{}\\
The number of independent condensates grows considerably at dimension 6. We will not attempt to discuss all of them in full detail, 
but will give an overview over the literature and some recent progress that has been made in computing some of these condensates at finite 
temperature. 

The finite temperature behavior of the specific four-quark condensates appearing in sum rules of the vector and axial-vector channels are discussed in some 
detail in Ref.\,\cite{Hatsuda:1992bv} based on the hadron resonance gas model of Eq.\,(\ref{eq:HRG.model}). 
Besides Eq.\,(\ref{eq:HRG.model}), one uses the soft pion theorem [which can in fact be used to derive Eq.\,(\ref{eq:chiral.cond.pion.gas})], 
giving  
\begin{equation}
\lim_{\bm{p} \to 0,\,\bm{p}' \to 0} \langle \pi^a(\bm{p})| \mathcal{O} | \pi^b(\bm{p}') \rangle = - \frac{1}{f^2_{\pi}} \langle 0| [\mathcal{F}^a_5, [\mathcal{F}^b_5, \mathcal{O}]] | 0 \rangle, 
\label{eq:dim6.temp.1}
\end{equation}
with
\begin{equation}
\mathcal{F}^a = \int d^3 \bm{x} \bar{q}(x) \gamma_0 \gamma_5 \frac{\tau^a}{2} q(x).  
\label{eq:dim6.temp.2}
\end{equation}
Here, $q = (u,\,d,\,s)$ and $\tau^a$ is a $SU(3)$ matrix living in flavor space. If one only considers pions, it is enough to take into account $a =$ 1 - 3. 
The next step is to make use of current algebra to compute the double commutator of Eq.\,(\ref{eq:dim6.temp.1}). The details of this calculation can be found 
in Appendix A of Ref.\,\cite{Hatsuda:1992bv} and will not be repeated here. We here just mention the basic formulas 
 \begin{align}
[F^a_5,\, \mathcal{V}^{b,\alpha}_{\mu}] &= if^{abc} \mathcal{A}^{c,\alpha}_{\mu}, \label{eq:dim6.temp.3} \\
[F^a_5,\, \mathcal{A}^{b,\alpha}_{\mu}] &= if^{abc} \mathcal{V}^{c,\alpha}_{\mu}, \label{eq:dim6.temp.4}
\end{align}
with
 \begin{align}
\mathcal{V}^{a,\alpha}_{\mu} &= \bar{q} \gamma_{\mu} \tau^a \lambda^a q, \label{eq:dim6.temp.5} \\
\mathcal{A}^{a,\alpha}_{\mu} &= \bar{q} \gamma_{\mu} \gamma_{5} \tau^a \lambda^a q, \label{eq:dim6.temp.6}
\end{align}
where again $q = (u,\,d,\,s)$, $\tau^a$ are the $U(3)$ flavor matrices ($\tau^0 = \sqrt{1/N_f}$) and 
$\lambda^{\alpha}$ are the $SU(3)$ color matrices. 
Furthermore, the convention for which 
$f^{ab0}$ is understood to be zero, was used. 
After computing the commutators, one moreover needs to apply the factorization hypothesis of 
Eq.\,(\ref{eq:four.quark.condensate.factorization}) to obtain the final results, which can be found in Ref.\,\cite{Hatsuda:1992bv} and 
which can in principle be generalized to other four-quark condensates if necessary. 
It however has to be emphasized here that the above method only provides an order of magnitude estimate, as it relies both 
on factorization (which has systematic uncertainties that are difficult to quantify) and the hadron resonance gas model (which 
is reliable only at temperatures below $T_c$). Any QCDSR analysis that strongly depends on the behavior of the four-quark 
condensates hence has to be taken with a grain of salt. Naturally, a reliable finite temperature lattice QCD computation of these 
condensates would be very helpful. 

Next, we discuss some recent progress made in the study of the thermal behavior of dimension 6 gluonic condensates. The number of operators that can generally be 
constructed from gluonic operators and covariant derivatives is quite large. However, with the help of the equations 
of motion, symmetry properties of operator indices and the Bianchi identity, they can be reduced to just a few independent ones, which was done some time ago in Ref.\,\cite{Kim:2000kj}. 
One possible set of independent operators is the following: 
\begin{align}
\mathrm{spin}\, 0:\,\, & f^{abc} G^a_{\mu \nu} G^b_{\nu \alpha} G^c_{\alpha \mu}, \, G^a_{\mu \alpha} G^a_{\nu \alpha; \mu \nu},  \label{eq:dim6.temp.7} \\
\mathrm{spin}\, 2:\,\, & \mathcal{ST} G^a_{\kappa \lambda} G^a_{\kappa \lambda; \mu \nu}, \, \mathcal{ST} G^a_{\mu \kappa} G^a_{\nu \lambda; \lambda \kappa}, 
\, \mathcal{ST} G^a_{\mu \kappa} G^a_{\kappa \lambda; \lambda \nu}, \label{eq:dim6.temp.8} \\
\mathrm{spin}\, 4:\,\, & \mathcal{ST} G^a_{\rho \kappa} G^a_{\sigma \kappa; \mu \nu}. \label{eq:dim6.temp.9}
\end{align}
Here, the notations $G^a_{\alpha \beta; \mu} \equiv D^{ab}_{\mu} G^b_{\alpha \beta}$ and $G^a_{\alpha \beta; \mu \nu} \equiv D^{ab}_{\nu} D^{bc}_{\mu} G^c_{\alpha \beta}$ are used. 
In this paragraph, we furthermore temporarily take all Lorentz indices as lower indices to keep the notation simple. 
Making use of the equation of motion 
\begin{align}
G^a_{\alpha \beta; \beta} = g \sum_q \bar{q} \gamma_{\alpha} \frac{\lambda^a}{2}q,  
\label{eq:dim6.temp.10}
\end{align} 
the second operator of Eq.\,(\ref{eq:dim6.temp.7}) and the second and third operators of Eq.\,(\ref{eq:dim6.temp.8}) can be rewritten in terms of quark 
fields and hence vanish for pure gauge theory. The anomalous dimensions of the operators of Eq.\,(\ref{eq:dim6.temp.8}) were calculated 
only relatively recently in Ref.\,\cite{Kim:2015ywa}. Furthermore, estimates for the three operators that remain non-zero in 
pure gauge theory were given in Ref.\,\cite{Kim:2015xna}. In this work, the basic strategy was to first express the two gluonic dimension 4 condensates 
in terms of chromo-electric and chromo-magnetic fields and to translate our knowledge about the finite temperature behavior of these 
condensates into temperature dependences of chromo-electric and chromo-magnetic fields. Next, the dimension 6, spin 0 and spin 2 condensates 
are expressed using the same chromo-electromagnetic fields. Assuming that the fields are isotropic and angular correlations can be neglected, 
this then gives temperature dependences of the dimension 6 condensates. For more details, we refer interested readers to Ref.\,\cite{Kim:2015xna}. 

\subsubsection{Condensates at finite density \label{conds.at.finite.den}}
Let us start with a general discussion on our treatment of the condensates at finite density. 
We will here only consider the behavior of the condensates at densities of the order of normal nuclear matter 
density $\rho_0$. 
For such densities one can hope that the linear density approximation still gives a qualitatively correct description.  
The expectation value of a general (but for simplicity scalar) operator $\mathcal{O}$ with respect to the ground state of dense matter 
at temperature $T = 0$ and baryon density $\rho$, which we will denote as  
$\langle \mathcal{O} \rangle_{\rho}$ throughout this review, is expressed in this approximation as 
\begin{align}
\langle \mathcal{O} \rangle_{\rho} & \simeq \langle 0 | \mathcal{O}| 0 \rangle + 4 \int_{|\bm{k}| < k_F} 
\frac{d^3 \bm{k}}{(2\pi)^3} \langle N(\bm{k}) | \mathcal{O}| N(\bm{k}) \rangle \nonumber \\
& \simeq \langle 0 | \mathcal{O}| 0 \rangle + 4 \langle N(0) | \mathcal{O}| N(0) \rangle  \int_{|\bm{k}| < k_F} \frac{d^3 \bm{k}}{(2\pi)^3} \label{eq:linear.density.approximation} \\
& \simeq \langle 0 | \mathcal{O}| 0 \rangle + \rho \langle N(0) | \mathcal{O}| N(0) \rangle \nonumber, 
\end{align}
with 
\begin{equation}
k_F = \Bigl( \frac{3 \pi^2 \rho}{2} \Bigr)^{1/3}.
\label{eq:Fermi.momentum.density}
\end{equation}
Here, 
$|N(\bm{k}) \rangle$ stands for a one nucleon state with momentum $\bm{k}$. Its normalization is defined as 
\begin{equation}
\langle N(\bm{k}) | N(\bm{k}') \rangle = (2\pi)^3 \delta^{(3)}(\bm{k} - \bm{k}').
\end{equation}
In going from the first to the second line in Eq.\,(\ref{eq:linear.density.approximation}), we have ignored the dependence of $|N(\bm{k}) \rangle$ on the 
momentum $\bm{k}$. Taking this dependence explicitly into account would lead to terms on higher order in $\rho$. A $\bm{k}^2$ term in the Taylor expansion of 
$|N(\bm{k}) \rangle$ would for instance lead to a term proportional to $\rho^{5/3}$. In the above linear density approximation, the Fermi motion of 
nucleons is hence ignored completely and one in essence is working in the non-interacting Fermi gas limit. 
It is not a trivial question up to what densities this approximation can be trusted and at which densities higher order density terms 
become significant. We will discuss this issue below for the case of the chiral condensate of $u$ and $d$ quarks for which higher order terms can be 
studied systematically using chiral perturbation theory.   

\paragraph{Condensates of dimension 3}\mbox{}\\
At this dimension, we begin by studying the chiral condensates 
$\langle \overline{q} q \rangle_{\rho} = \frac{1}{2}(\langle \overline{u} u \rangle_{\rho} + \langle \overline{d} d \rangle_{\rho})$ 
and $\langle \overline{s} s \rangle_{\rho}$. 
In the linear order density approximation, discussed above, we have 
\begin{align}
\langle \overline{q} q \rangle_{\rho} & \simeq \langle 0 | \overline{q} q|0 \rangle + \rho \langle N | \overline{q} q | N \rangle = 
\langle 0 | \overline{q} q|0 \rangle + \rho \frac{\sigma_{\pi N}}{2m_q}, \label{eq:qbarq.linear} \\
\langle \overline{s} s \rangle_{\rho} & \simeq \langle 0 | \overline{s} s|0 \rangle + \rho \langle N | \overline{s} s | N \rangle =
\langle 0 | \overline{s} s|0 \rangle + \rho \frac{\sigma_{s N}}{m_s}. 
\label{eq:sbars.linear}
\end{align}
Here, we have introduced the $\pi N$ sigma term $\sigma_{\pi N} \equiv 2 m_q \langle N | \overline{q} q | N \rangle$ and the 
strange quark sigma term $\sigma_{sN} \equiv m_s \langle N | \overline{s} s | N \rangle$, which are useful because they are renormalization 
group invariant and can in principle be related to $\pi N$ \cite{Alarcon:2011zs,Hoferichter:2015dsa} or $K N$ \cite{Gasser:2000wv} scattering observables. 
The values of $\langle N | \overline{q} q | N \rangle$ and $\langle N | \overline{s} s | N \rangle$ (as well as the respective sigma 
terms) can be computed directly on the lattice. 

\begin{figure}[tb]
\begin{center}
\begin{minipage}[t]{8 cm}
\vspace{0.5 cm}
\hspace{-1.5 cm}
\includegraphics[width=11cm,bb=0 0 670 493]{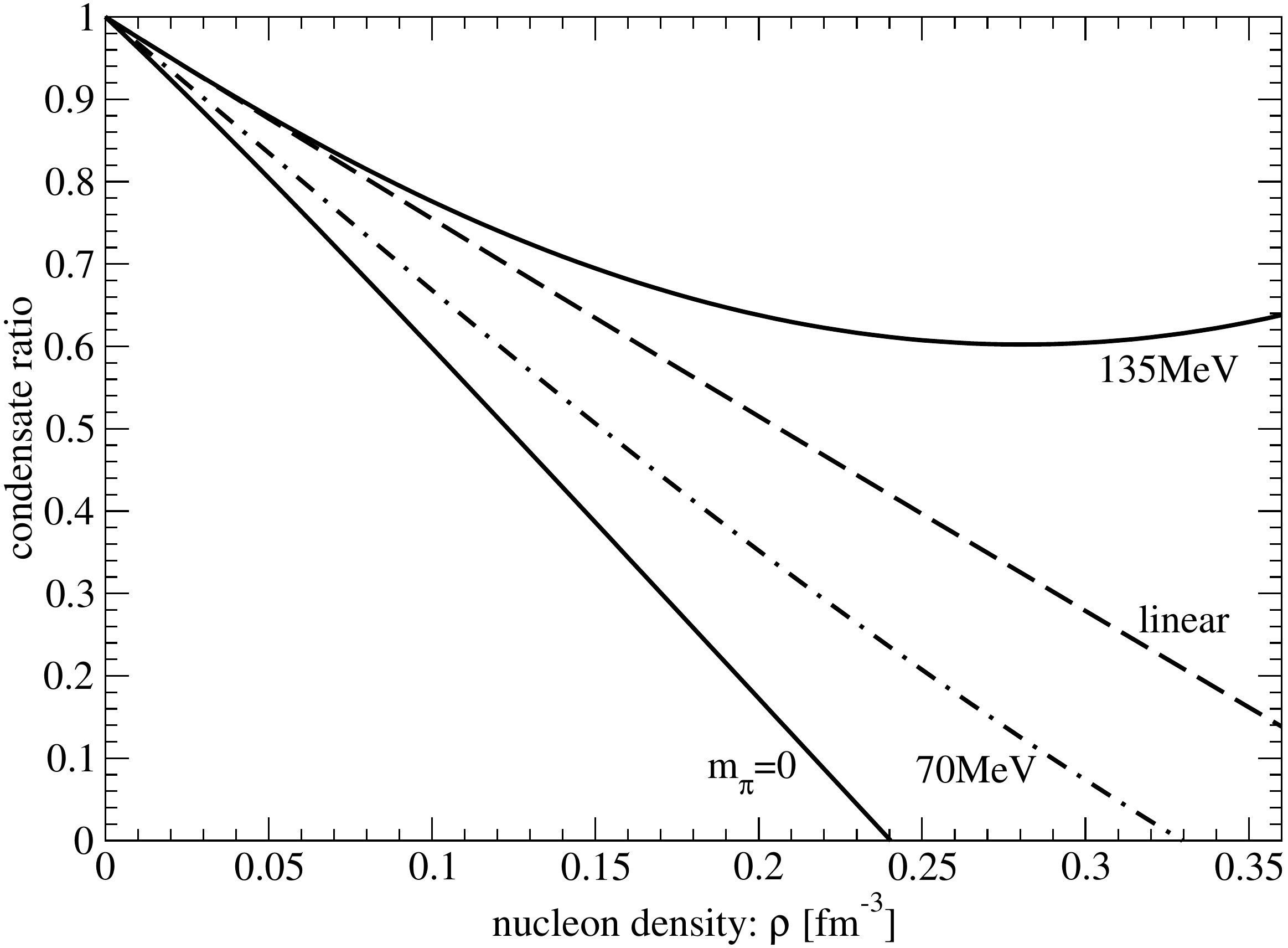}
%\vspace{-0.8cm}
\end{minipage}
\begin{minipage}[t]{16.5 cm}
\caption{The ratio $\langle \overline{q} q \rangle_{\rho}/\langle 0 | \overline{q} q|0 \rangle$ as a function of baryon density $\rho$ including 
non-linear terms computed by chiral perturbation theory, 
for three different pion mass values, $m_{\pi} = 0$, $70\,\mathrm{MeV}$ and $135\,\mathrm{MeV}$. 
The dashed curve 
corresponds to the linear density approximation using the sigma term value $\sigma_{\pi N} = 45\,\mathrm{MeV}$. 
Taken from Fig.\,5 of Ref.\,\cite{Kaiser:2007nv}.  
\label{fig:chiral.cond.nonlinear}}
\end{minipage}
\end{center}
\end{figure}

Before discussing the sigma term values in detail, let us first examine the reliability of the linear density approximation for 
$\langle \overline{q} q \rangle_{\rho}$. This is the only quantity for which terms beyond linear order in density are available 
and thus the deviation from the linear behavior can be systematically studied and the density range for which the linear approximation breaks down 
can be estimated. 
Calculations of $\langle \overline{q} q \rangle_{\rho}$ based on chiral perturbation theory that go beyond linear order in $\rho$ were performed 
in Refs.\,\cite{Kaiser:2007nv,Goda:2013bka}. Following here Ref.\,\cite{Kaiser:2007nv}, one can express the ratio of $\langle \overline{q} q \rangle_{\rho}$ 
and $\langle 0 | \overline{q} q|0 \rangle$ as 
\begin{equation}
\frac{\langle \overline{q} q \rangle_{\rho}}{\langle 0 | \overline{q} q|0 \rangle} \simeq 1 - \frac{\rho}{f_{\pi}^2} 
\Bigl[
\frac{\sigma_{\pi N}}{m_{\pi}^2}\bigl(1 - \frac{3k_F^2}{10M_N^2} + \frac{9k_F^4}{56M_N^4} \bigr) 
+ D(k_F)
\Bigr], 
\end{equation}
for which the relation between the Fermi momentum and the density is given in Eq.\,(\ref{eq:Fermi.momentum.density}). 
Keeping only the term of leading order in density and using the Gell-Mann-Oakes-Renner relation of Eq.\,(\ref{eq:GellMann.Oakes}), 
a result equivalent to Eq.\,(\ref{eq:qbarq.linear}) is recovered. 
The function $D(k_F)$ is related to the derivative of the interaction energy per particle $\bar{E}(k_F)$ with respect to the 
pion mass $m_{\pi}$, 
\begin{equation}
D(k_F) = \frac{1}{2 m_{\pi}} \frac{\partial \bar{E}(k_F)}{\partial m_{\pi}}. 
\end{equation}
For more details, see Ref.\,\cite{Kaiser:2007nv}. Here, we simply show the final result in 
Fig.\,\ref{fig:chiral.cond.nonlinear}.  
%%%%%%%%%%%%%%%%%%%%%%%%%%%%%%%%%%%%%%%%%%%%%%%%%%%%%%%%%%%%%%%%%%%%%%%%%%%%%%%%%%%%%%%%%%%%%%%%%%%%%%%%%%%%%%%%%%%%
%%%%%%%%%%%%%%%%%%%%%%%%%%%%%%%%%%%%%%%%%%%%%%%%%%%%%%%%%%%%%%%%%%%%%%%%%%%%%%%%%%%%%%%%%%%%%%%%%%%%%%%%%%%%%%%%%%%%
It is seen in this figure that for physical pion masses, 
the non-linear terms weaken 
the reduction of the chiral condensate by about 20\,\% 
at normal nuclear matter density $\rho_0 = 0.17\,\mathrm{fm}^{-3}$. 
At higher densities, 
the linear behavior is modified significantly. At the same time, however, the chiral expansion becomes less reliable at 
high densities, meaning that terms of even higher orders in $\rho$ might further change this behavior (if the expansion is 
convergent at all). 
For further developments concerning the ``stabilization'' of the chiral condensate at high baryon density, see Refs.\,\cite{Drews:2013hha,Drews:2014spa}. 
The authors of Ref.\,\cite{Goda:2013bka}, which treat the chiral expansion differently and make use of the chiral Ward identity, 
obtain qualitatively compatible results with a reduced chiral restoration due to the non-linear terms. 
Ref.\,\cite{Goda:2013bka}, however, gives reduced non-linear corrections, which are smaller than 10\,\% at normal nuclear matter density. 
The difference between the two approaches gives an approximate estimate of the systematic uncertainties related to the non-linear 
terms in chiral perturbation theory. 
In this context, it is worth mentioning past \cite{Itahashi:1999qb,Suzuki:2002ae} and ongoing \cite{Nishi:2017chb} experimental efforts to measure 
deeply bound pionic atom spectra, which, if precise enough, can strongly constrain the chiral condensate value at finite density. 
For related theoretical work, see also Refs.\,\cite{Kolomeitsev:2002gc,Jido:2008bk,Yamazaki:2012zza}. 

For $\langle \overline{s} s \rangle_{\rho}$ no systematic computation of non-linear terms has yet been performed, 
even though a similar approach based on chiral perturbation theory would in principal be possible. 
For all other condensates to be discussed in later sections, it is presently not known how to systematically treat terms beyond linear order. 
We will therefore focus on the linear terms in the following. 

Let us consider the $\pi N$ sigma term, appearing in Eq.\,(\ref{eq:qbarq.linear}). The traditionally 
quoted and still widely used value for this parameter is 
\begin{equation}
\sigma_{\pi N} = 45\,\mathrm{MeV} \hspace{0.5cm} \cite{Gasser:1990ce}, 
\end{equation}
which was based on chiral perturbation theory and $\pi N$ scattering data. 
In the more than 25 years after this estimate was given, progress has been made both in lattice QCD and the analysis of 
$\pi N$ scattering data, which led to a number of novel and more precise determinations of $\sigma_{\pi N}$. 
It should be emphasized here that $\langle N | \overline{q} q|N \rangle$ is not a finite density object, but the expectation value of a 
one-nucleon-state, which can hence be computed on the lattice. 
Furthermore, making use of the fact that the Feynman-Hellmann theorem relates the $\pi N$ sigma term to the 
quark mass dependence of the nucleon mass $m_N$, 
\begin{equation}
\sigma_{\pi N} = m_q \frac{\partial m_N}{\partial m_q},  
\label{eq:Feynman.Hellmann.q}
\end{equation}
many studies have been conducted that combine lattice data of nucleon masses at several quark masses with 
chiral perturbation theory fits to extrapolate the nucleon mass derivative of the quark mass to the physical point. 
What has emerged from all this is that direct computations of $\sigma_{\pi N}$ from lattice QCD and analyses based on 
experimental information about the $\pi N$ interaction do not agree, the former 
getting values around 30 to 40 MeV, while the latter obtain values close to 60 MeV. The corresponding results 
are summarized in Table\,\ref{tab:sigma.term.values}, in which 
we only show works published after 2011. Furthermore, we only quote the most recent result for each collaboration.   
\begin{table}
\begin{center}
\caption{Recent $\sigma_{\pi N}$ values from direct lattice QCD calculations, chiral fits to lattice QCD data and works analyzing experimental information about 
pionic atoms and the low energy $\pi N$ interaction.} 
\label{tab:sigma.term.values}
\begin{tabular}{cccc}  
\toprule
Method & Collaboration, Year & $\sigma_{\pi N}$ [MeV] & Reference \\ \midrule
Lattice QCD & BMW, 2016 & 38(3)(3) & \cite{Durr:2015dna} \\
Lattice QCD & $\chi$QCD, 2016 & 45.9(7.4)(2.8) & \cite{Yang:2015uis} \\
Lattice QCD & ETM, 2016 & 37.2(2.6)($^{4.7}_{2.9}$) & \cite{Abdel-Rehim:2016won} \\
Lattice QCD & RQCD, 2016 & 35(6) & \cite{Bali:2016lvx} \\
Lattice QCD & JLQCD, 2018 & 26(3)(5)(2) & \cite{Yamanaka:2018uud} \\ \midrule
Lattice QCD data + ChPT & 2012 & 32(2) & \cite{Semke:2012gs} \\ 
Lattice QCD data + ChPT & 2013 & 52(7)/45(6)\footnotemark & \cite{Chen:2012nx} \\ 
Lattice QCD data + ChPT & 2013 & 45(6) & \cite{Shanahan:2012wh} \\ 
Lattice QCD data + ChPT & 2013 & 41(5)(4) & \cite{Alvarez-Ruso:2013fza} \\ 
Lattice QCD data + ChPT & 2015 & 55(1)(4) & \cite{Ren:2014vea} \\ 
Lattice QCD data + ChPT & 2017 & 64.9(1.5)/51.7(4.3)\footnotemark & \cite{Alexandrou:2017xwd} \\ 
Lattice QCD data + ChPT & 2017 & 50.3(1.2)(3.4) & \cite{Ling:2017jyz} \\ \midrule
ChPT & 2012 & 59(7) & \cite{Alarcon:2011zs} \\
Roy-Steiner Eqs. (pionic atoms) & 2015 & 59.1(3.5) & \cite{Hoferichter:2015dsa} \\
Roy-Steiner Eqs. ($\pi N$ scat. data) & 2018 & 58(5) & \cite{RuizdeElvira:2017stg} \\
\bottomrule
\end{tabular}
\end{center}
\end{table} 
%\addtocounter{footnote}{-2}
%\stepcounter{footnote}
%\footnotetext{In this work, the authors give values for a fit without and with an explicit $\Delta(1232)$ contribution. 
%The former gives 52(7) MeV, while the latter leads to a value of 45(6) MeV.}
%\stepcounter{footnote}
%\footnotetext{The values quoted in this reference correspond to two separate fits to the same lattice data, using $\mathcal{O}(p^3)$ and 
%$\mathcal{O}(p^4)$ chiral perturbation theory approaches.} 
Notably, works of both lattice QCD and $\pi N$ scattering analyses appear to be roughly consistent with each other, while there is a clear 
tension between the two. What the origin of this disagreement is, remains presently unknown. 
Moreover, calculations using a combination of chiral perturbation theory and lattice QCD data (with a few notable exceptions) lie roughly 
between the other two approaches. 
A potential solution to the above discrepancy was recently proposed in Ref.\,\cite{Dmitrasinovic:2016hup}, in which the 
nucleon was described as a superposition of two distinct chiral multiplets and the $\pi N$ sigma term was computed making use of 
chiral algebra considerations. We relegate detailed explanations to Ref.\,\cite{Dmitrasinovic:2016hup}, 
but here just mention that the admixture of the second (non-standard) chiral multiplet is key for the present discussion as it enhances 
$\sigma_{\pi N}$ to a value close to 60 MeV, consistent with those obtained from $\pi N$ experimental data (Refs.\,\cite{Alarcon:2011zs,Hoferichter:2015dsa,RuizdeElvira:2017stg}). 
It would therefore be interesting to study what sort of chiral multiplets are taken into account in the current lattice 
QCD studies of the $\pi N$ sigma term. 

\addtocounter{footnote}{-2}
\stepcounter{footnote}
\footnotetext{In this work, the authors give values for a fit without and with an explicit $\Delta(1232)$ contribution. 
The former gives 52(7) MeV, while the latter leads to a value of 45(6) MeV.}
\stepcounter{footnote}
\footnotetext{The values quoted in this reference correspond to two separate fits to the same lattice data, using $\mathcal{O}(p^3)$ and 
$\mathcal{O}(p^4)$ chiral perturbation theory approaches.} 
Next, we discuss what is known about the strange quark sigma term $\sigma_{s N}$ showing up in Eq.\,(\ref{eq:sbars.linear}). 
Similar to $\sigma_{\pi N}$, this quantity can be directly computed in lattice QCD and can at the same time be related to some 
combination of $\pi N$ and $K N$ scattering processes and/or pionic and kaonic atoms. 
Analyses relating $\sigma_{s N}$ to experimental observables however are 
less developed compared to the $\sigma_{\pi N}$ discussion of the last paragraph. 
Lattice QCD also had (and remains to have) its problems (mainly because of the difficulty in 
computing disconnected diagrams), but has in recent years shown considerable progress in estimating 
$\sigma_{s N}$ at physical pion masses. 
Besides calculating $\sigma_{s N}$ directly on the lattice, some groups have also used 
the Feynman-Hellmann theorem, which, analogous to Eq.\,(\ref{eq:Feynman.Hellmann.q}), gives 
\begin{equation}
\sigma_{s N} = m_s \frac{\partial m_N}{\partial m_s}.  
\label{eq:Feynman.Hellmann.s}
\end{equation}
We here focus on recent direct lattice QCD computations and results based on chiral fits 
to lattice QCD data. 
At the end, we will briefly discuss the possibility of determining $\sigma_{s N}$ based on experimental information. 

Other than $\sigma_{s N}$, there are quite a large number of parameters used 
to quantify the ``strangeness content of the nucleon", $\langle N | \overline{s} s|N \rangle$. 
Another frequently employed variable is 
\begin{equation}
y = \frac{\langle N | \overline{s} s|N \rangle}{\langle N | \overline{q} q|N \rangle} = \frac{2 m_q}{m_s} \frac{\sigma_{s N}}{\sigma_{\pi N}}. 
\end{equation}
Other parametrizations are
\begin{equation}
\sigma_{0} = (1 - y) \sigma_{\pi N}, 
\end{equation}
or
\begin{equation}
f_{T_s} = \frac{\sigma_{sN}}{M_N}, 
\end{equation}
where $M_N$ is the average of proton and neutron masses. 
We here focus only on $\sigma_{s N}$, firstly because it is renormalization group invariant and secondly does not 
depend on $\sigma_{\pi N}$, which has its own uncertainties as we have seen in the preceding discussion. 
If needed, the quantities $y$, $\sigma_0$ and $f_{T_s}$ can easily be obtained from the above formulas. 

Recent results for $\sigma_{s N}$ are summarized in Table\,\ref{tab:strange.sigma.term.values}. 
Here, we again only show results published after 2011 and quote only the most recent result 
of each collaboration. 
\begin{table}
\begin{center}
\caption{Recent $\sigma_{s N}$ values from lattice QCD and ChPT fits to lattice QCD data.} 
\label{tab:strange.sigma.term.values}
\begin{tabular}{cccc}  
\toprule
Method & Collaboration, Year & $\sigma_{s N}$ [MeV] & Reference \\ \midrule
Lattice QCD (Feynman-Hellmann) & BMW, 2016 & 105(41)(37) & \cite{Durr:2015dna} \\
Lattice QCD (direct) & $\chi$QCD, 2016 & 40.2(11.7)(3.5) & \cite{Yang:2015uis} \\
Lattice QCD (direct) & ETM, 2016 & 41.1(8.2)($^{7.8}_{5.8}$) & \cite{Abdel-Rehim:2016won} \\
Lattice QCD (direct) & RQCD, 2016 & 35(12) & \cite{Bali:2016lvx} \\
Lattice QCD (direct) & JLQCD, 2018 & 17(18)(9) & \cite{Yamanaka:2018uud} \\ \midrule
Lattice QCD data + ChPT & 2012 & 22(20) & \cite{Semke:2012gs} \\
Lattice QCD data + ChPT & 2013 & 21(6) & \cite{Shanahan:2012wh} \\ 
Lattice QCD data + ChPT & 2015 & 27(27)(4) & \cite{Ren:2014vea} \\ 
\bottomrule
\end{tabular}
\end{center}
\end{table}
Among the values shown in the table, the first four are pure lattice QCD calculations that do not rely on 
chiral perturbation theory fits, while the latter three use a combination of the Feynman-Hellmann theorem, lattice 
data of the nucleon at several quark masses and chiral perturbation theory to obtain their result. 
One observes that the latter works all have the tendency to give relatively small values for $\sigma_{s N}$. 
Overall, the numerical errors are still rather large in comparison to the $\sigma_{\pi N}$ results 
of Table\,\ref{tab:sigma.term.values}. For the direct lattice QCD calculations this is due to the large numerical 
cost and noisiness of the disconnected diagrams, that are the sole contribution to $\sigma_{s N}$. 
For works that rely on the Feynman-Hellmann theorem of Eq.\,(\ref{eq:Feynman.Hellmann.s}), the lack of 
precision is related to the fact that the nucleon mass $m_N$ only depends very weakly on the strange 
quark mass $m_s$, which means that $m_N$ needs to be calculated with extremely high precision to 
reliably compute the derivative of Eq.\,(\ref{eq:Feynman.Hellmann.s}). 
Because of these issues, the results of Table\,\ref{tab:strange.sigma.term.values} are still spread over a wide range 
and more precise calculations will be needed to pin down the exact value of $\sigma_{s N}$. 

As a further point, let us mention the possibility of determining the strange quark sigma term $\sigma_{s N}$ from 
experimental data. Given the recent and precise measurement of the kaonic hydrogen by the SIDDHARTA experiment \cite{Bazzi:2011zj}, 
and the planned hadronic deuterium measurement by the SIDDHARTA-2 collaboration at LNF \cite{Curceanu:2013bxa,Iliescu:2016zgx} 
and the E57 experiment at J-PARC \cite{Zmeskal:2015efj}, it should, in principle, be possible to go through the same program as 
in Ref.\,\cite{Hoferichter:2015dsa}, which was already described schematically in Ref.\,\cite{Gasser:2000wv}. 
To our knowledge this task has not yet been carried out and it remains to be seen whether its outcome would 
in terms of precision be able to compete with the lattice QCD approaches discussed above. 

There is one more condensate at dimension 3, namely $\langle \bar{q} \gamma^{\mu} q \rangle_{\rho}$, the 
lowest dimensional Lorentz violating condensate as shown in Eq.\,(\ref{eq:condensates.medium}). This condensate must 
be proportional to the four-velocity of nucleon matter $u^{\mu}$, hence we have 
\begin{equation}
\langle \bar{q} \gamma^{\mu} q \rangle_{\rho} =  \langle \bar{q} \slash{u} q  \rangle_{\rho} u^{\mu}.  
\label{eq:nonlocal.dim3}
\end{equation}
Going to the (most natural) frame in which the medium is at rest, $u^{\mu} = (1, 0, 0, 0)$, we obtain 
\begin{align}
\langle \bar{q} \gamma^{\mu} q \rangle_{\rho} & =  \langle q^{\dagger} q  \rangle_{\rho} \delta^{\mu 0} \nonumber \\
& = \frac{3}{2} \rho  \delta^{\mu 0}, 
\label{eq:nonlocal.dim3.2}
\end{align}
where $\rho$ is the nucleon density. This expression is exact. For the strange quark case, we have 
from an analogous discussion  
\begin{equation}
\langle \bar{s} \gamma^{\mu} s \rangle_{\rho} =  0,   
\end{equation}
which is also exact. 

\paragraph{Condensates of dimension 4}\mbox{}\\
We start with the finite density behavior of the dimension 4 gluon condensate 
$\langle \frac{\alpha_s}{\pi} G^a_{\mu \nu} G^{a\mu\nu} \rangle_{\rho} = 
\langle \frac{\alpha_s}{\pi} G^2 \rangle_{\rho}$. Here we use the conventional 
definition which includes the factor $\frac{\alpha_s}{\pi}$, hence eliminating the scale dependence of this operator. 
Not much is known about the behavior of the gluon condensate going beyond linear order in density, as 
it is presently not known how to compute 
higher order terms in a systematic way. 
There are nevertheless a few relatively old model calculations, which suggest that the linear behavior is 
accurate to a good degree at normal nuclear matter density and non-linear terms start to become 
significant only at larger densities \cite{Mishra:1993bs,Saito:1997bt}. 
We will here focus on the linear density term, about which model independent statements can be made. 
Making, as before, use of Eq.\,(\ref{eq:linear.density.approximation}), we can write 
\begin{equation}
\langle \frac{\alpha_s}{\pi} G^2 \rangle_{\rho} \simeq 
\langle 0 | \frac{\alpha_s}{\pi} G^2 |0 \rangle + \rho \langle N | \frac{\alpha_s}{\pi} G^2 | N \rangle. 
\label{eq:gluon.linear}
\end{equation}
To compute the quantity $\langle N | \frac{\alpha_s}{\pi} G^2 | N \rangle$, Eq.\,(\ref{eq:trace.anomaly.4}) 
can be used. This equation is based 
on the trace anomaly, where higher order $\alpha_s$ terms are neglected and contributions due to heavy quarks $c$, $b$ and $t$ are 
converted into the squared gluon field term via the heavy quark expansion. If one does not wish to rely on 
the heavy quark expansion, the same discussion can be straightforwardly repeated keeping the explicit heavy quark terms 
$m_c \overline{c} c$, $m_b \overline{b} b$ and $m_t \overline{t} t$. 
Following Ref.\,\cite{Cohen:1991nk}, we write 
\begin{align}
\langle T^{\mu}_{\mu} \rangle_{\rho} & = \langle 0 | T^{\mu}_{\mu} | 0 \rangle + e(\rho) \nonumber \\
& \simeq \langle 0 | T^{\mu}_{\mu} | 0 \rangle + \rho M_N. 
\label{eq:energy.momentum.tensor.expect}
\end{align}
in the first line, which is exact, $e(\rho)$ is the energy density of matter with baryon 
density $\rho$, which in the linear density approximation used in the second line, becomes $\rho M_N$. 
In the first line pressure contributions vanish because we are considering 
matter in equilibrium. 
One can then obtain the linear density term by computing 
$(\langle T^{\mu}_{\mu} \rangle_{\rho} - \langle 0 | T^{\mu}_{\mu} | 0 \rangle)/\rho$, both using Eqs.\,(\ref{eq:trace.anomaly.4}) 
and (\ref{eq:energy.momentum.tensor.expect}). 
We thus get 
\begin{equation}
M_N = -\frac{9}{8} \langle N | \frac{\alpha_s}{\pi} G^2 | N \rangle + \sigma_{\pi N} + \sigma_{s N}, 
\label{eq:gluon.cond.density.1}
\end{equation}
and hence 
\begin{equation}
\langle N | \frac{\alpha_s}{\pi} G^2 | N \rangle = -\frac{8}{9}(M_N -  \sigma_{\pi N} - \sigma_{s N}). 
\label{eq:gluon.cond.density.2}
\end{equation}
Looking at the values of $\sigma_{\pi N}$ and $\sigma_{s N}$ in Tables\,\ref{tab:sigma.term.values} and \ref{tab:strange.sigma.term.values}, 
which are of the order of 50 MeV, it is clear that $\langle N | \frac{\alpha_s}{\pi} G^2 | N \rangle$ is to a large degree determined 
by the nucleon mass value $M_N$. The fact that $\sigma_{\pi N}$ and $\sigma_{s N}$ are not yet determined with good precision however 
leads to some uncertainty for $\langle N | \frac{\alpha_s}{\pi} G^2 | N \rangle$. 

Next, we consider the non-scalar condensates of dimension 4, of which there are three.  
We begin with $\langle \bar{q} iD^{\mu} q \rangle_{\rho}$, which is most straightforward. 
This condensate must be proportional to $u^{\mu}$, hence 
\begin{equation}
\langle \bar{q} iD^{\mu} q \rangle_{\rho} = \langle \bar{q} u_{\alpha} iD^{\alpha} q  \rangle_{\rho} u^{\mu}.
\label{eq:nonlocal.dim4.1}
\end{equation}
As described in Ref.\,\cite{Jin:1992id}, we can furthermore use 
\begin{equation}
D^{\alpha} = \frac{1}{2}(\gamma^{\alpha} \slashb{D} + \slashb{D} \gamma^{\alpha})
\label{eq:nonlocal.dim4.2}
\end{equation}
and
\begin{align}
\overline{q} \slashb{D} \Gamma q & = - \overline{q} \cev{\slashb{D}} \Gamma q, \\
i \slashb{D} q & = m_q q, \\
\overline{q} i \cev{\slashb{D}} & = -m_q \overline{q}, 
\label{eq:nonlocal.dim4.3}
\end{align}
where in the first line $\Gamma$ is an arbitrary gamma matrix. Using the above relations and equations of motion, we 
obtain 
\begin{align}
\langle \bar{q} u_{\alpha} iD^{\alpha} q \rangle_{\rho} & = m_q \langle \bar{q} \slash{u} q  \rangle_{\rho} \nonumber \\
& = m_q \langle q^{\dagger}  q  \rangle_{\rho} \\
& = \frac{3}{2} m_q \rho,  \nonumber
\label{eq:nonlocal.dim4.4}
\end{align}
where in the second and third line we have used again $u^{\mu} = (1, 0, 0, 0)$ 
and have proceeded in the same way as in Eq.\,(\ref{eq:nonlocal.dim3.2}). 
We therefore have 
\begin{equation}
\langle \bar{q} iD^{\mu} q \rangle_{\rho} =  \frac{3}{2} m_q \rho \delta^{\mu 0}.
\label{eq:nonlocal.dim4.5}
\end{equation}
No approximations were used in this 
derivation. 
From similar considerations, we also obtain the exact result  
\begin{equation}
\langle \bar{s} iD^{\mu} s \rangle_{\rho} =  0.
\label{eq:nonlocal.dim4.6}
\end{equation}

\begin{figure}[tb]
\begin{center}
\begin{minipage}[t]{16.5 cm}
\vspace{0.5 cm}
%\hspace{0.5 cm}
%\epsfig{file=fx_quarks_mu,scale=0.7}
%\epsfig{file=fx_gluons_mu,scale=0.7}
\includegraphics[width=8.5cm,bb=0 0 360 252]{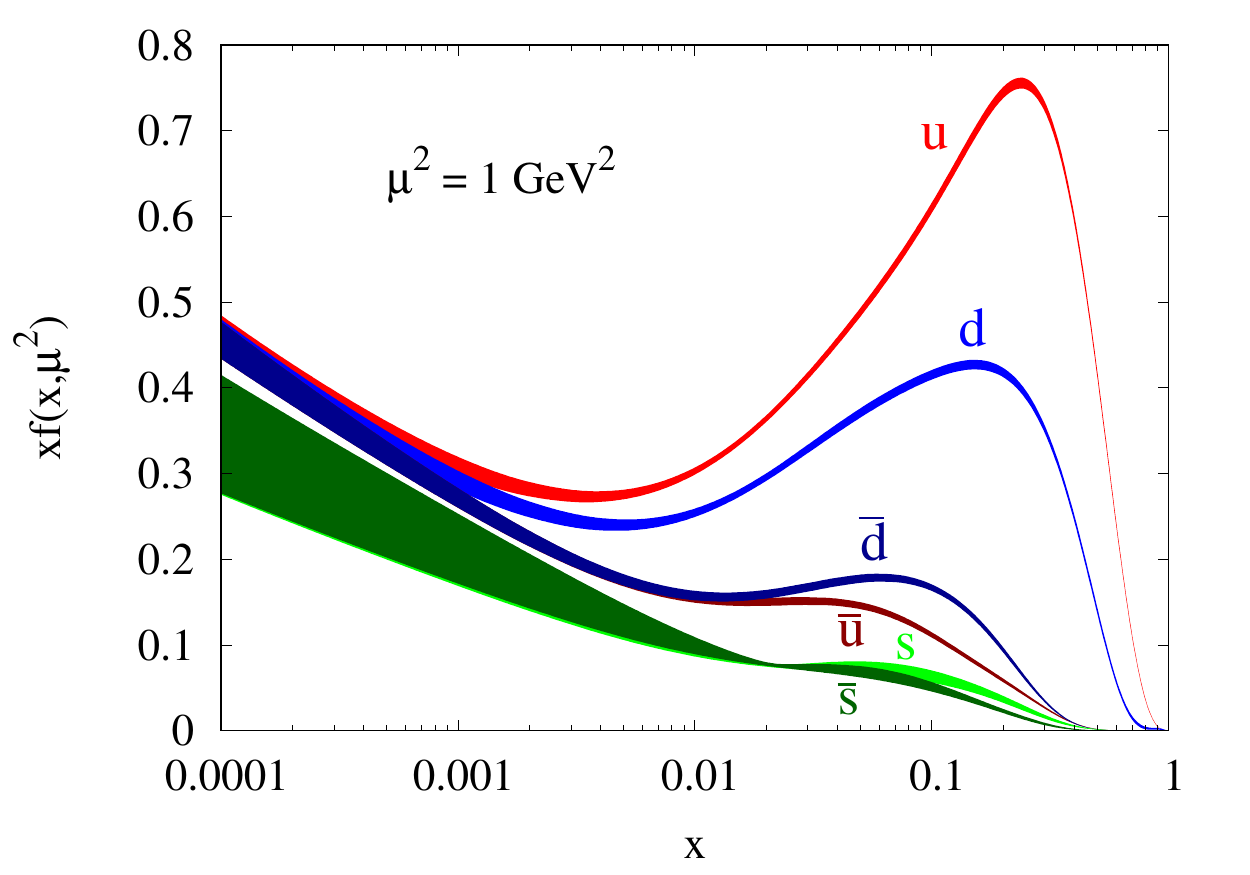}
\includegraphics[width=8.5cm,bb=0 0 360 252]{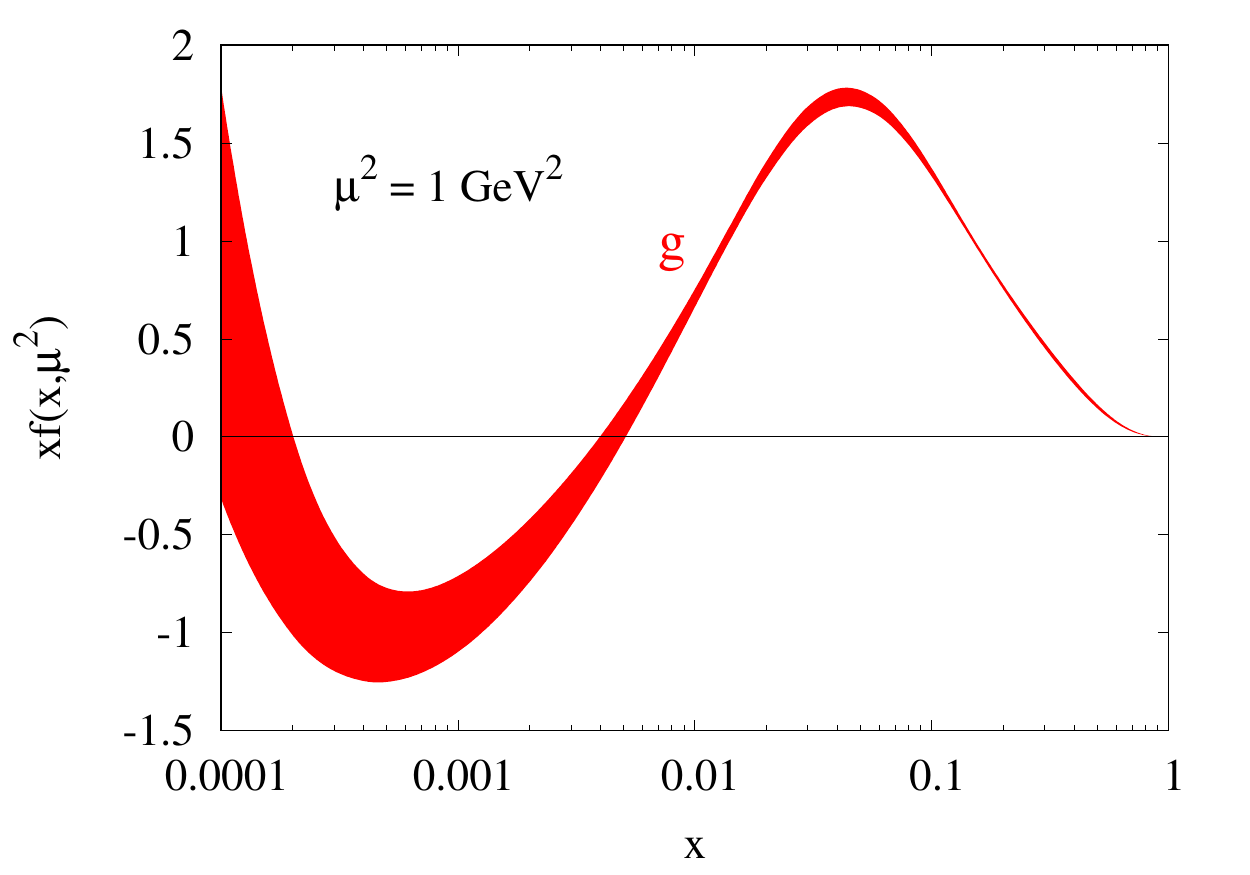}
%\vspace{1 cm}
\end{minipage}
\begin{minipage}[t]{16.5 cm}
\caption{Parton distributions of various quark species (left plot) and gluon (right plot) as a function of Bjorken $x$, extracted from the fit of 
Ref.\,\cite{Harland-Lang:2014zoa} and the corresponding codes of Ref.\,\cite{Harland-Lang.page} for NLO results 
at $\mu^2 = 1\,\mathrm{GeV}^2$. 
\label{fig:parton.distributions}}
\end{minipage}
\end{center}
\end{figure}

We next look at the first and third condensates in Eq.\,(\ref{eq:condensates.medium}). 
Studying these, it is convenient to discuss a more general class of operators, with arbitrary 
numbers of covariant derivatives, which can be related to moments of specific nucleonic parton distribution functions. 
From DIS theory, we have \cite{Jin:1992id,Efremov:1976ih,Efremov:1980ub,Collins:1981uw}
\begin{align}
\langle N | \mathcal{ST} \bar{q} \gamma_{\mu_1} D_{\mu_2} \cdots D_{\mu_n} q | N \rangle & 
\equiv (-i)^{n - 1} \frac{1}{2M_N} A^q_n(\mu^2) \mathcal{ST}( p_{\mu_1} \cdots p_{\mu_n} ), \label{eq:nonlocal.dim4.7.1} \\
\langle N | \mathcal{ST} G_{\alpha \mu_1}^a D_{\mu_2} \cdots D_{\mu_{n-1}} G_{\mu_n}^{a\alpha} | N \rangle & \equiv 
 (-i)^{n - 2} \frac{1}{M_N} A^g_n(\mu^2) \mathcal{ST} ( p_{\mu_1} \cdots p_{\mu_n} ).
\label{eq:nonlocal.dim4.7}
\end{align}
Here, $p_{\mu}$ is the four-momentum of the nucleon state $|N\rangle$ and 
$q$ can stand for all three quark species or their averages. 
The coefficients $A^q_n(\mu^2)$ and $A^g_n(\mu^2)$ are each related to momenta of quark and gluon parton 
distributions at renormalization scale $\mu^2$: 
\begin{align}
A^q_n(\mu^2) & = 2 \int_0^1 dx x^{n-1} \bigl[q(x, \mu^2) + (-1)^n \bar{q}(x, \mu^2) \bigr], \label{eq:nonlocal.dim4.8.1} \\
A^g_n(\mu^2) & = \frac{1 + (-1)^n}{2} \int_0^1 dx x^{n-1} g(x, \mu^2).  
\label{eq:nonlocal.dim4.8}
\end{align}
In practice, mostly operators with $n$ values ranging from 2 to 4 (corresponding to operators with dimensions 4 to 6) will be of 
importance. 
Their expectation values can be obtained by numerically integrating the parton distributions fitted in Ref.\,\cite{Harland-Lang:2014zoa} 
to a vast amount of experimental data using LO, NLO and NNLO QCD results. The parton distributions can be extracted 
for wide $\mu^2$ ranges from the codes provided in Ref.\,\cite{Harland-Lang.page}. 
As an example and for illustration, the NLO distributions (times $x$) are shown in Fig.\,\ref{fig:parton.distributions} 
at $\mu^2 = 1\,\mathrm{GeV}^2$. 
%%%%%%%%%%%%%%%%%%%%%%%%%%%%%%%%%%%%%%%%%%%%%%%%%%%%%%%%%%%%%%%%%%%%%%%%%%%%%%%%%%%%%%%%%%%%%%%%%%%%%%%%%%%%%%%%%%%%
%%%%%%%%%%%%%%%%%%%%%%%%%%%%%%%%%%%%%%%%%%%%%%%%%%%%%%%%%%%%%%%%%%%%%%%%%%%%%%%%%%%%%%%%%%%%%%%%%%%%%%%%%%%%%%%%%%%%
For the convenience of the reader, we give the $A^q$ and $A^g$ values for data fits employing LO, NLO and 
NNLO expressions and scales $\sqrt{\mu^2} = 1\,\mathrm{GeV}$ and $\sqrt{\mu^2} = 2\,\mathrm{GeV}$ in Table\,\ref{tab:A.values}. 
\begin{table}
\begin{center}
\caption{$A^q$ and $A^g$ values as defined in Eqs.\,(\ref{eq:nonlocal.dim4.8.1}) and (\ref{eq:nonlocal.dim4.8}) obtained by numerically 
integrating the (proton) parton distributions provided in Refs.\,\cite{Harland-Lang:2014zoa} and \cite{Harland-Lang.page}. The errors 
are computed by integrating the respective distribution errors. For $A^q_n = \frac{1}{2}(A^u_n + A^d_n)$, 
the errors of $A^u_n$ and $A^d_n$ are added in quadrature.} 
\label{tab:A.values}
\begin{tabular}{ccccccc}  
\toprule
 & \multicolumn{2}{c|}{LO} & \multicolumn{2}{c|}{NLO} & \multicolumn{2}{c}{NNLO} \\ \midrule
$\sqrt{\mu^2}$ & 1 GeV & \multicolumn{1}{c|}{2 GeV} & 1 GeV & \multicolumn{1}{c|}{2 GeV} & 1 GeV & 2 GeV \\ \midrule
$A^u_2$ & 0.735(19) & 0.640(15) & 0.784(17) & 0.679(14) & 0.819(18) & 0.696(14) \\
$A^d_2$ & 0.424(21) & 0.380(17) & 0.430(18) & 0.385(14) & 0.448(18) & 0.394(14) \\
$A^q_2$ & 0.580(14) & 0.510(11) & 0.607(12) & 0.532(10) & 0.634(13) & 0.545(10) \\
$A^s_2$ & 0.0378(94) & 0.0585(79) & 0.053(13) & 0.072(11) & 0.050(16) & 0.071(13) \\ 
$A^g_2$ & 0.401(35) & 0.454(21) & 0.367(23) & 0.425(16) & 0.341(23) & 0.411(16) \\ \midrule
$A^u_3$ & 0.2171(54) & 0.1633(39) & 0.2178(48) & 0.1640(35) & 0.2278(51) & 0.1663(36) \\
$A^d_3$ & 0.0812(61) & 0.0611(44) & 0.0782(52) & 0.0589(37) & 0.0836(56) & 0.0610(38) \\
$A^q_3$ & 0.1492(41) & 0.1122(29) & 0.1480(36) & 0.1114(25) & 0.1557(38) & 0.1136(26) \\
$A^s_3$ & 0.00110(92) & 0.00082(81) & 0.0016(18) & 0.0012(14) & 0.0017(23) & 0.0012(17) \\ 
$A^g_3$ & 0 & 0 & 0 & 0 & 0 & 0 \\ \midrule
$A^u_4$ & 0.0991(23) & 0.0701(15) & 0.0945(21) & 0.0668(14) & 0.0984(22) & 0.0670(14) \\
$A^d_4$ & 0.0357(27) & 0.0257(18) & 0.0327(25) & 0.0233(16) & 0.0348(28) & 0.0240(17) \\
$A^q_4$ & 0.0674(17) & 0.0479(12) & 0.0636(16) & 0.0450(11) & 0.0666(18) & 0.0455(11) \\
$A^s_4$ & 0.00040(18) & 0.00105(19) & 0.00121(44) & 0.00122(31) & 0.00099(57) & 0.00110(39) \\ 
$A^g_4$ & 0.0338(48) & 0.0177(21) & 0.0208(23) & 0.0125(11) & 0.0283(36) & 0.0158(16) \\ \midrule
$A^u_5$ & 0.0494(12) & 0.03265(75) & 0.0449(11) & 0.02990(70) & 0.0464(11) & 0.02960(69) \\
$A^d_5$ & 0.0141(14) & 0.00934(89) & 0.0123(15) & 0.00818(92) & 0.0139(17) & 0.0089(10) \\
$A^q_5$ & 0.03179(90) & 0.02100(58) & 0.02860(92) & 0.01904(58) & 0.0301(10) & 0.01923(61) \\
$A^s_5$ & 0.000066(46) & 0.000043(61) & 0.00016(14) & 0.000105(91) & 0.00020(19) & 0.00012(12) \\ 
$A^g_5$ & 0 & 0 & 0 & 0 & 0 & 0 \\ \midrule
$A^u_6$ & 0.02819(67) & 0.01786(41) & 0.02472(64) & 0.01578(40) & 0.02531(63) & 0.01544(38) \\
$A^d_6$ & 0.00741(80) & 0.00477(50) & 0.0062(10) & 0.00396(60) & 0.0073(12) & 0.00447(68) \\
$A^q_6$ & 0.01780(52) & 0.01132(32) & 0.01544(59) & 0.00987(36) & 0.01629(67) & 0.00995(39) \\
$A^s_6$ & 0.000021(13) & 0.000113(25) & 0.000092(51) & 0.000091(32) & 0.000069(74) & 0.000092(47) \\ 
$A^g_6$ & 0.0074(16) & 0.00288(55) & 0.00402(67) & 0.00184(26) & 0.0089(17) & 0.00352(61) \\ 
\bottomrule
\end{tabular}
\end{center}
\end{table} 
These values are all obtained for the proton. Therefore, to study symmetric nuclear matter, 
the average of $A^u$ and $A^d$, $A^q_n = \frac{1}{2}(A^u_n + A^d_n)$, is needed, which 
can also be found in Table\,\ref{tab:A.values}. 

Making use of Eqs.\,(\ref{eq:linear.density.approximation}) and (\ref{eq:nonlocal.dim4.7}), we have
\begin{align}
\langle \mathcal{ST} \bar{q} \gamma^{\mu} iD^{\nu} q \rangle_{\rho} & \simeq \rho \langle N | \mathcal{ST} \bar{q} \gamma^{\mu} iD^{\nu} q | N \rangle \nonumber \\
&  = \frac{\rho}{2M_N} A^q_2 \Bigl(p^{\mu} p^{\nu} - \frac{p^2}{4} g^{\mu \nu} \Bigr) \nonumber \\
&  = \frac{\rho M_N}{2} A^q_2 \Bigl(\delta^{\mu 0} \delta^{\nu 0 } - \frac{1}{4} g^{\mu \nu} \Bigr), \\
\langle \mathcal{ST} \bar{s} \gamma^{\mu} iD^{\nu} s \rangle_{\rho} & \simeq \frac{\rho M_N}{2} A^s_2 \Bigl(\delta^{\mu 0} \delta^{\nu 0 } - \frac{1}{4} g^{\mu \nu} \Bigr), \\
\langle \mathcal{ST} G_{\alpha}^{a\mu} G^{a\nu\alpha} \rangle_{\rho} & \simeq  \rho M_N A^g_2 \Bigl(\delta^{\mu 0} \delta^{\nu 0 } - \frac{1}{4} g^{\mu \nu} \Bigr), 
\end{align}
where we have in the third, fourth and fifth line employed $p^{\mu} = M_N u^{\mu}$, which is valid only at leading order in $\rho$. 

\paragraph{Condensates of dimension 5}\mbox{}\\
We begin again with the density dependence of the only scalar condensate of this dimension, $\langle \bar{q} g \sigma G q  \rangle_{\rho}$. 
Worse than 
its vacuum counterpart, not much first hand information is available for this quantity even at leading order in 
density. The only estimate 
given in the literature is from Ref.\,\cite{Jin:1992id}, which we will update here. 
It was assumed in Ref.\,\cite{Jin:1992id} 
that the parameter 
$m_0^2$, introduced earlier in Eq.\,(\ref{eq:mixed.cond.para}), is independent of density. Therefore, one has 
\begin{align}
\langle \bar{q} g \sigma G q  \rangle_{\rho} \simeq \langle 0 | \bar{q} g \sigma G q | 0 \rangle + \rho \langle N | \bar{q} g \sigma G q | N \rangle
\end{align}
with
\begin{align}
\langle N | \bar{q} g \sigma G q | N \rangle & = m_0^2 \langle N | \bar{q}q | N \rangle \nonumber \\
& = m_0^2 \frac{\sigma_{\pi N}}{2 m_q} \nonumber \\
& = 3.8 \pm 1.7\,\mathrm{GeV}^2. 
\label{eq:mixed.condensate.nucleon}
\end{align}
Here, we have used Eq.\,(\ref{eq:mixed.cond.value}), $\sigma_{\pi N} = 45 \pm 15$ MeV, which encompasses most 
of the values in Table\,\ref{tab:sigma.term.values}, and $m_q = 4.7 \pm 0.7$ MeV \cite{Patrignani:2016xqp}, which 
is the averaged $u$ and $d$ quark mass in the $\overline{\mathrm{MS}}$ scheme at a renormalization scale of 1 GeV. 
The value given here is larger than that quoted in Ref.\,\cite{Jin:1992id} because of the 
smaller quark mass used here, but consistent within errors. Furthermore, the error is larger 
than in Ref.\,\cite{Jin:1992id} because we have assumed a larger 
uncertainty for $\sigma_{\pi N}$. It should in any case be kept in mind that the above value is not more 
than a rough estimate, as the validity of the assumption of $m_0^2$ not to depend on $\rho$ is not clear. 
Again, a direct lattice QCD calculation of $\langle N | \bar{q} g \sigma G q | N \rangle$ would be very 
helpful.  

The strange counterpart $\langle N | \bar{s} g \sigma G s | N \rangle$ can be estimated in a similar way. 
We get 
\begin{align}
\langle N | \bar{s} g \sigma G s | N \rangle & = m_1^2 \langle N | \bar{s}s | N \rangle \nonumber \\
& = m_1^2 \frac{\sigma_{s N}}{m_s} \nonumber \\
& = 0.37 \pm 0.28\,\mathrm{GeV}^2,  
\label{eq:strange.mixed.condensate.nucleon}
\end{align}
where we have used Eq.\,(\ref{eq:strange.mixed.cond.value.2}), 
$\sigma_{s N} = 60 \pm 40\,\mathrm{MeV}$, which approximately represents the values given in Table\,\ref{tab:strange.sigma.term.values} 
and $m_s = 130 \pm 8 \,\mathrm{MeV}$ \cite{Patrignani:2016xqp}, which is again the most recent 
PDG value of the $s$ quark mass in the $\overline{\mathrm{MS}}$ scheme at a renormalization scale 1 GeV. 
The large error of this estimate obviously originates from the large uncertainty of $\sigma_{s N}$. 
It moreover relies on the somewhat arbitrary assumption that 
$m_1^2$ does not depend on the density.  

We next consider 
the non-scalar condensate $\langle N | \bar{q} \gamma^{\mu} g \sigma G q | N \rangle$, 
which can be treated as in Eq.\,(\ref{eq:nonlocal.dim3}, 
\begin{align}
\langle N | \bar{q} \gamma^{\mu} g \sigma G q | N \rangle = \langle N | \bar{q} \slash{u} g \sigma G q | N \rangle u^{\mu}
\end{align}
and, going to the nuclear rest frame, 
\begin{align}
\langle N | \bar{q} \gamma^{\mu} g \sigma G q | N \rangle = \langle N | q^{\dagger} g \sigma G q | N \rangle \delta^{\mu 0}.
\end{align}
About $\langle N | q^{\dagger} g \sigma G q | N \rangle$ 
some older works are available, providing an idea about its order of magnitude. 
In Refs.\,\cite{Shuryak:1981dg,Shuryak:1981kj,Braun:1986ty} the operator 
\begin{align}
\mathcal{O}^{\mathrm{S}}_{\mu} = \bar{u} g \lambda^a \tilde{G}_{\mu \alpha}^a \gamma^{\alpha} \gamma_5 u + \bar{d} g \lambda^a \tilde{G}_{\mu \alpha}^a \gamma^{\alpha} \gamma_5 d 
\end{align}
was studied. Here, $\tilde{G}_{\mu \nu}^{a} = \frac{1}{2} \epsilon_{\mu \nu \alpha \beta} G^{a \alpha \beta}$. 
Its nucleon expectation value can be related to the above condensate as 
\begin{align}
\langle N |\mathcal{O}^{\mathrm{S}}_{\mu}| N \rangle = 2 \langle N | \bar{q} \gamma^{\mu} g \sigma G q | N \rangle
\end{align}
using the convention $\epsilon_{0123} = 1$ employed in these works (also note that in Refs.\,\cite{Shuryak:1981dg,Shuryak:1981kj} 
$t^a$ stands for $\lambda^a$ in our notation). 
Furthermore, adjusting their normalization convention to ours, the 
results of Refs.\,\cite{Shuryak:1981dg,Shuryak:1981kj}, 
using the Gross-Llewellyn Smith sum rule and experimental data, become 
\begin{align}
\langle N | q^{\dagger} g \sigma G q | N \rangle = -0.5\,\mathrm{GeV}^2. 
\end{align}
The sign of this value is different from that quoted in 
Ref.\,\cite{Jin:1992id}. 
On the other hand, Ref.\,\cite{Braun:1986ty} gets 
\begin{align}
\langle N | q^{\dagger} g \sigma G q | N \rangle = 0.33\,\mathrm{GeV}^2 
\end{align}
from a QCD sum rule analysis, while also obtaining 
\begin{align}
\langle N | q^{\dagger} g \sigma G q | N \rangle = 0.22\,\mathrm{GeV}^2  
\end{align}
from vector dominance and 
\begin{align}
\langle N | q^{\dagger} g \sigma G q | N \rangle = 0.2\,\mathrm{GeV}^2  
\end{align}
from a non-relativistic quark model. 
We here give a new estimate \cite{MorathLeeWeise}, that so far has 
not been discussed in published works. The idea is to simply 
assume that a relation analogous to Eq.\,(\ref{eq:mixed.condensate.nucleon}) holds 
with $m_0^2$ of equal order of magnitude. 
We hence have
\begin{align}
\langle N | q^{\dagger} g \sigma G q | N \rangle & \simeq m_0^2 \langle N | q^{\dagger} q | N \rangle \nonumber \\
& = \frac{3}{2} m_0^2 \nonumber \\
& \simeq 1.2\,\mathrm{GeV}^2. 
\label{eq:mixed.nonscalar.condensate.nucleon}
\end{align}
All the above results are summarized in Table\,\ref{tab:non.scalar.mixed.cond.values}. 
\begin{table}
\begin{center}
\caption{Values of $\langle N | q^{\dagger} g \sigma G q | N \rangle$ obtained from different 
approaches. For details, see the text and the references cited here.} 
\label{tab:non.scalar.mixed.cond.values}
\begin{tabular}{ccc}  
\toprule
$\langle N | q^{\dagger} g \sigma G q | N \rangle$ $[\mathrm{GeV}^2]$ & method  & reference \\ \midrule
$-0.5$ & Gross-Llewellyn Smith SR & \cite{Shuryak:1981dg,Shuryak:1981kj} \\
$0.33$ & QCD sum rule & \cite{Braun:1986ty} \\
$0.22$ & vector dominance & \cite{Braun:1986ty} \\
$0.2$ & non-rel. quark model & \cite{Braun:1986ty} \\
$1.2$ & $m_0^2$ & this work \\
\bottomrule
\end{tabular}
\end{center}
\end{table} 
We see that the results largely differ depending on the employed method, 
even its sign is uncertain. It can, however, 
be conjectured that its absolute value is of the order of $\sim 1\,\mathrm{GeV}^2$ or smaller. 

As for the strange condensate $\langle N | s^{\dagger} g \sigma G s | N \rangle$, to our knowledge 
no estimate is currently available. The simplest way of estimating this matrix element 
is to use the same strategy as in Eq.\,(\ref{eq:mixed.nonscalar.condensate.nucleon}). We then have 
\begin{align}
\langle N | s^{\dagger} g \sigma G s | N \rangle & \simeq m_1^2 \langle N | s^{\dagger} s | N \rangle \nonumber \\
& = 0.  
\label{eq:strange.mixed.nonscalar.condensate.nucleon}
\end{align}
We hence see that this condensate will likely be small. Another way of estimating $\langle N | s^{\dagger} g \sigma G s | N \rangle$ 
is to assume 
\begin{align}
\frac{\langle N | s^{\dagger} g \sigma G s | N \rangle}{\langle N | q^{\dagger} g \sigma G q | N \rangle}  
& \simeq  \frac{\langle N | \bar{s} g \sigma G s | N \rangle}{\langle N | \bar{q} g \sigma G q | N \rangle} \nonumber \\
& \simeq 0.1.  
\label{eq:strange.mixed.nonscalar.condensate.nucleon.ratio}
\end{align}
Therefore, 
\begin{align}
|\langle N | s^{\dagger} g \sigma G s | N \rangle|   
& \simeq 0.1\times |\langle N | q^{\dagger} g \sigma G q | N \rangle| \nonumber \\
& \lesssim 0.1\,\mathrm{GeV}^2.
\label{eq:strange.mixed.nonscalar.condensate.nucleon.2}
\end{align}

\begin{figure}[tb]
\begin{center}
\begin{minipage}[t]{8 cm}
\vspace{0.5 cm}
\hspace{-2.0 cm}
\includegraphics[width=12cm,bb=0 0 360 252]{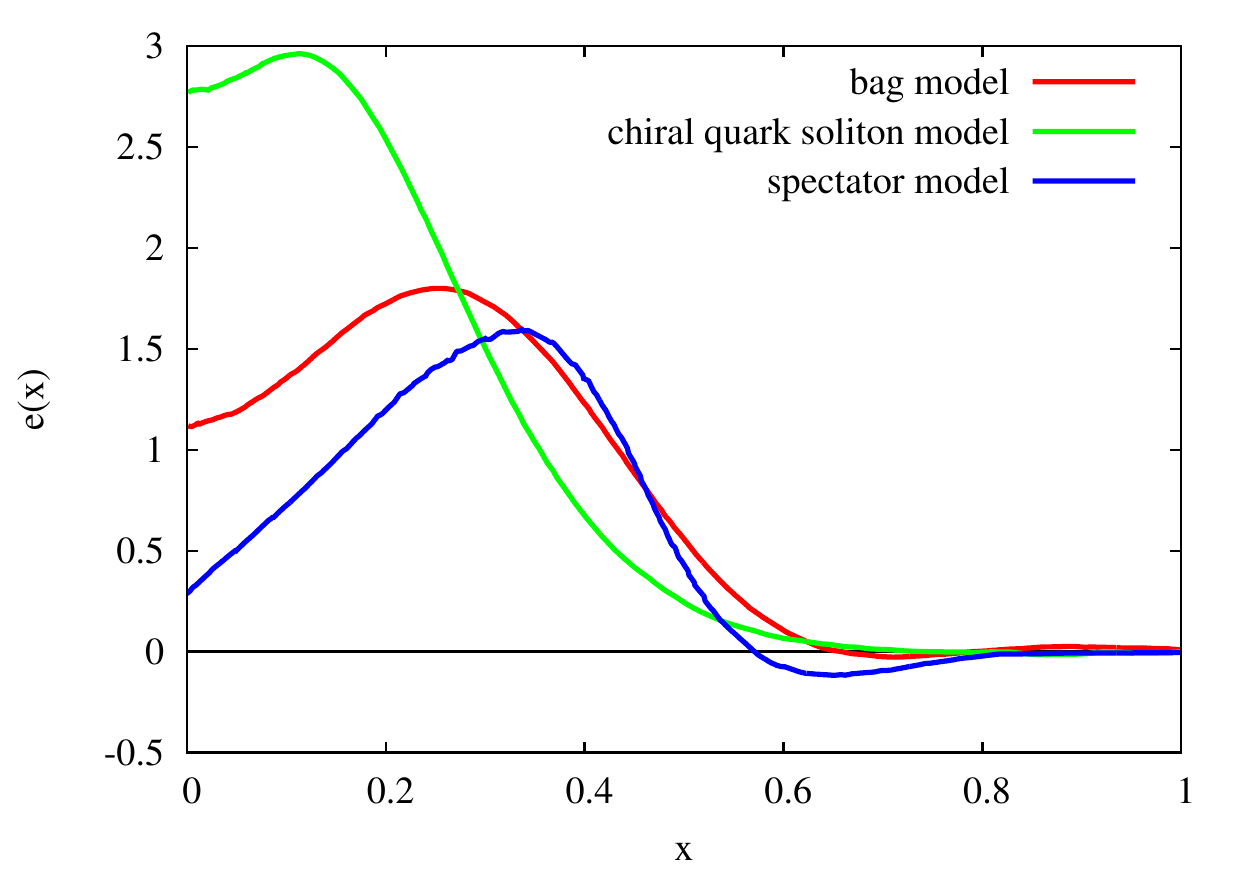}
%\vspace{1 cm}
\end{minipage}
\begin{minipage}[t]{16.5 cm}
\caption{The function $e_{n}(x, \mu^2)$ as a function of Bjorken $x$, computed in the 
bag model (red curve), the chiral quark soliton model (green curve) and the spectator model (blue curve)
with $n =$ odd and $\mu^2$ taken at a typical hadronic scale (see text). 
The numerical
data needed for this plot have been extracted from Fig. 9 of Ref.\,\cite{Cebulla:2007ej}. 
Taken from Fig.\,2 Ref.\,\cite{Gubler:2015uza}.
\label{fig:e.function.model}}
\end{minipage}
\end{center}
\end{figure}

Another non-scalar condensate appearing at dimension 5 is $\langle \mathcal{ST} \bar{q} iD^{\mu} iD^{\nu} q  \rangle_{\rho}$, which 
is presently constrained only up to leading order in density, 
\begin{align}
\langle \mathcal{ST} \bar{q} iD^{\mu} iD^{\nu} q  \rangle_{\rho} \simeq \rho \langle N |\mathcal{ST} \bar{q} iD^{\mu} iD^{\nu} q | N \rangle.
\label{eq:two.covariant.derivatives}
\end{align}
When discussing $\langle N |\mathcal{ST} \bar{q} iD^{\mu} iD^{\nu} q | N \rangle$, it is again 
useful to define a more general matrix element with an arbitrary number of covariant derivatives, which can be related to 
specific moments of the twist-3 distribution function $e(x, \mu^2)$ \cite{Jin:1992id,Jaffe:1991ra}, 
\begin{align}
\langle N |\mathcal{ST} \bar{q} iD_{\mu_1} iD_{\mu_2} \cdots iD_{\mu_n} q | N \rangle \equiv e_n(\mu^2) \mathcal{ST}(p_{\mu_1} \cdots p_{\mu_n}), 
\label{eq:two.covariant.derivatives.2}
\end{align}
with 
\begin{align}
e_n(\mu^2) = \int_0^1 dx x^n e_n(x, \mu^2). 
\label{eq:two.covariant.derivatives.3}
\end{align}
As usual, we consider the matrix element of Eq.\,(\ref{eq:two.covariant.derivatives.2}) as the average over $u$ and $d$ quarks. In that case 
$e(x,\mu^2)$ can be decomposed as 
\begin{align}
e_n(x, \mu^2) = \frac{1}{2} \bigl[ e^u(x, \mu^2) + e^d(x, \mu^2) + (-1)^n e^{\bar{u}}(x, \mu^2) + (-1)^n e^{\bar{d}}(x, \mu^2) \bigr], 
\label{eq:two.covariant.derivatives.4}
\end{align}
where the contributions from the individual quarks are given as \cite{Jaffe:1991ra}
\begin{align}
e^q(x, \mu^2) = \frac{1}{2 M_N} \int \frac{d \lambda}{2\pi} e^{i \lambda x}  
\langle N | \bar{q}(0) [0, \lambda n] q(\lambda n)   | N \rangle,  
\label{eq:two.covariant.derivatives.5}
\end{align}
where $[0, \lambda n]$ is the gauge link to make the above quantity gauge invariant. 
The symbol $n$ here stands for 
a null vector with mass dimension $-1$. 
It should hence not be confused with the $n$ of Eqs.\,(\ref{eq:two.covariant.derivatives.2}-\ref{eq:two.covariant.derivatives.4}). 
Eqs.\,(\ref{eq:two.covariant.derivatives.3}), (\ref{eq:two.covariant.derivatives.4}) and (\ref{eq:two.covariant.derivatives.5}) 
have for a long time not been of much practical use, as $e_n(x, \mu^2)$ and/or $e^q(x, \mu^2)$ was essentially unknown, 
and had only been obtained from models, such as the 
bag model \cite{Jaffe:1991ra}, the chiral quark soliton model \cite{Ohnishi:2003mf} and the 
spectator model \cite{Jakob:1997wg}.  
For illustration, we show in Fig.\,\ref{fig:e.function.model} the function $e_{n}(x, \mu^2)$ computed in these models.  
%%%%%%%%%%%%%%%%%%%%%%%%%%%%%%%%%%%%%%%%%%%%%%%%%%%%%%%%%%%%%%%%%%%%%%%%%%%%%%%%%%%%%%%%%%%%%%%%%%%%%%%%%%%%%%%%%%%%
%%%%%%%%%%%%%%%%%%%%%%%%%%%%%%%%%%%%%%%%%%%%%%%%%%%%%%%%%%%%%%%%%%%%%%%%%%%%%%%%%%%%%%%%%%%%%%%%%%%%%%%%%%%%%%%%%%%%
In this figure $n$ is taken to be an even number, $e_n(x, \mu^2)$ is simply denoted as $e(x)$ and the renormalization scale $\sqrt{\mu^2}$ can be assumed to be close 
to a typical hadronic scale of $\sqrt{\mu^2} \simeq 0.5 \sim 1.0$ GeV. 
At this preliminary stage of the analysis, the renormalization scale is usually not seriously considered. We will therefore ignore it in the following discussion. 

\begin{figure}[tb]
\begin{center}
\begin{minipage}[t]{8 cm}
\vspace{0.5 cm}
\hspace{-2.0 cm}
\includegraphics[width=12cm,bb=0 0 360 252]{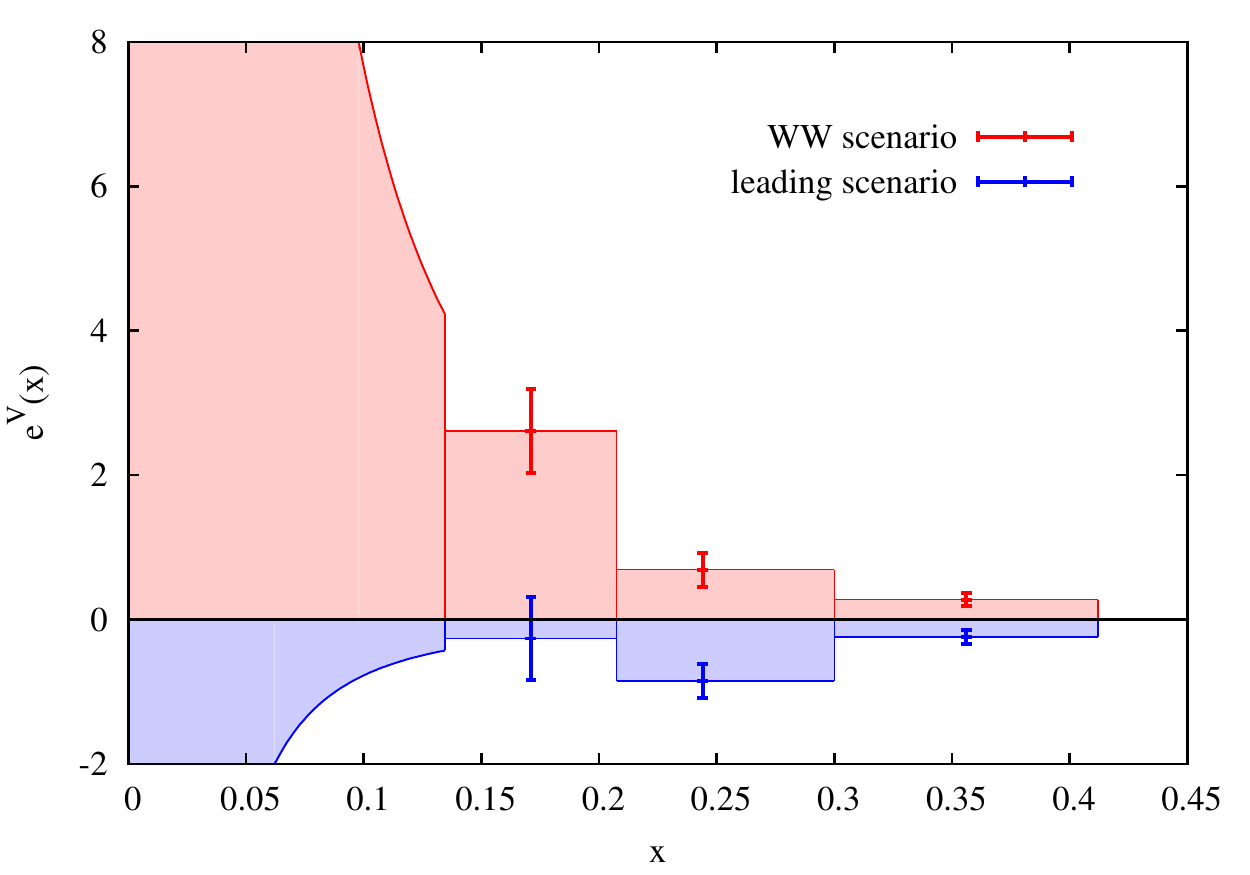}
%\vspace{1 cm}
\end{minipage}
\begin{minipage}[t]{16.5 cm}
\caption{The three data points of $e^{\mathrm{V}}(x)$ obtained from the beam-spin asymmetry of di-hadron semi-inclusive
DIS at CLAS. 
In addition to the data points, the extrapolation used for $e^{\mathrm{V}}(x)$ at small $x$, 
assuming a $1/x^2$ form, is illustrated below $x \simeq 0.13$. 
Taken from Fig.\,1 of Ref.\,\cite{Gubler:2015uza}.
\label{fig:e.function.experiment}}
\end{minipage}
\end{center}
\end{figure}

In past works, only rather crude estimates for $e_2$ have been provided, such as 
$e_2 = 0.36$ \cite{Lee:1993ww} or $e_2 = 1.95$ \cite{Jin:1992id} (see also Ref.\,\cite{Gubler:2015uza}). 
Recently, the situation has however improved, as some experimental information about $e^q(x)$ has become 
available. To be precise, a few data points of the function 
\begin{align}
e^{\mathrm{V}}(x) = \frac{4}{9} \Bigl[ e^u(x) - e^{\bar{u}}(x) \Bigr] - \frac{1}{9} \Bigl[ e^d(x) - e^{\bar{d}}(x) \Bigr]
\label{eq:two.covariant.derivatives.6}
\end{align}
were measured by analyzing 
experimental data on the beam-spin asymmetry of di-hadron semi-inclusive
DIS obtained by the CLAS experiment at Jefferson Lab \cite{Courtoy:2014ixa}. 
In Ref.\,\cite{Courtoy:2014ixa} two schemes were used to extract $e^{\mathrm{V}}(x)$, giving results that even have 
different signs. 
The obtained data points are shown in Fig.\,\ref{fig:e.function.experiment}. 
%%%%%%%%%%%%%%%%%%%%%%%%%%%%%%%%%%%%%%%%%%%%%%%%%%%%%%%%%%%%%%%%%%%%%%%%%%%%%%%%%%%%%%%%%%%%%%%%%%%%%%%%%%%%%%%%%%%%
%%%%%%%%%%%%%%%%%%%%%%%%%%%%%%%%%%%%%%%%%%%%%%%%%%%%%%%%%%%%%%%%%%%%%%%%%%%%%%%%%%%%%%%%%%%%%%%%%%%%%%%%%%%%%%%%%%%%
It is thus clear that at present no precise estimate on any $e_n$ value can be given. It is, however, possible to 
get order of magnitude estimates by making reasonable assumptions about the relative strengths of the $u$, $d$, $\bar{u}$ and $\bar{d}$ 
contributions. We refer the interested reader to Ref.\,\cite{Gubler:2015uza} and here just quote the final results, 
\begin{align}
e_2 & = (1.7 \pm 4.7) \times 10^{-2}, \label{eq:e2.value} \\
e_3 & = (1.4 \pm 7.5) \times 10^{-3}. 
\label{eq:e3.value}
\end{align}
The extraction of Eq.\,(\ref{eq:e2.value}) is explained in detail in Ref.\,\cite{Gubler:2015uza}, while Eq.\,(\ref{eq:e3.value}) was obtained 
using the same method. We refrain from giving $e_n$ values for higher $n$ because we have no direct information about the behavior of $e(x)$ 
at large $x$, which leads to even larger uncertainties. More detailed experimental information about $e^{\mathrm{V}}(x)$ will become available 
soon \cite{Pisano} through the analysis of the CLAS12 data, which hopefully will make it possible to get more precise $e_n$ values and to 
go to higher $n$. 

About the strange condensate $\langle \mathcal{ST} \bar{s} iD^{\mu} iD^{\nu} s \rangle_{\rho}$ 
(as well as $\langle N |\mathcal{ST} \bar{s} iD^{\mu} iD^{\nu} s | N \rangle$, 
its linear order density coefficient) no direct information 
is presently available. It is, however, possible to get an estimate by considering strange - non-strange ratios of 
the similar and better known condensates of Eq.\,(\ref{eq:nonlocal.dim4.7.1}) \cite{Gubler:2015uza}. 
For $e^s_2$ [defined as in Eq.\,(\ref{eq:two.covariant.derivatives.2}), but with 
strange quarks], we have
\begin{align}
e^s_2 & \simeq e_2 \times \frac{A^s_2}{A^q_2} \nonumber \\
& = (1.5 \pm 4.1) \times 10^{-3},    
\end{align}
and for $e^s_3$ 
\begin{align}
e^s_3 & \simeq e_3 \times \frac{A^s_3}{A^q_3} \nonumber \\
& = (1.5 \pm 8.1) \times 10^{-5},     
\end{align}
where we have used the NLO values of Table\,\ref {tab:A.values} at 
a renormalization scale of 1 GeV and have ignored their respective uncertainties. 

Finally, the linear density terms of the condensates $\langle \mathcal{ST} \bar{q} \gamma^{\alpha} iD^{\mu} iD^{\nu} q \rangle_{\rho}$ 
and $\langle \mathcal{ST} \bar{s} \gamma^{\alpha} iD^{\mu} iD^{\nu} s \rangle_{\rho}$ were already discussed 
around and after Eqs.\,(\ref{eq:nonlocal.dim4.7.1}) and (\ref{eq:nonlocal.dim4.7}). 

\paragraph{Condensates of dimension 6}\mbox{}\\
Raising the mass dimension to 6, the number of independent condensates suddenly increases to a fairly large number. 
Therefore, a consistent and systematic discussion of the density dependences of all these condensates has never been attempted, 
even though some specific classes of condensates have been 
studied, as will be seen below. 
One can reasonably argue that a general discussion of density dependences of such condensates is somewhat premature at the current stage 
at which their values are not well known even in vacuum. 
We will hence not attempt such a discussion here, but only mention some general features and refer the 
reader to the works, in which some of these condensates have been studied. 
To study the dimension 6 condensates, it would as a first step be useful 
to determine the respective independent operators. 
For instance, four-quark operators can be related to each other by 
applying Fierz-transformations, as it was discussed in detail in Ref.\,\cite{Thomas:2007gx}. 
Furthermore, equations of motion and Bianchi identities can be used to relate different operators, 
as shown in Refs.\,\cite{Kim:2015ywa,Kim:2000kj} for scalar and non-scalar purely gluonic 
dimension 6 operators. 

Let us review some basic strategies used to study these condensates. 
For the four quark condensates, the usual method is to employ a generalized vacuum saturation approximation 
similar to Eq.\,(\ref{eq:four.quark.condensate.factorization}) (see Appendix A of Ref.\,\cite{Jin:1992id} for a 
detailed discussion)  
\begin{align}
\langle \overline{q}^{i}_{\alpha} \overline{q}^{k}_{\beta} q^{l}_{\gamma} q^{m}_{\delta} \rangle_{\rho} & \simeq 
\langle \overline{q}^{i}_{\alpha} q^{m}_{\delta} \rangle_{\rho} \langle \overline{q}^{k}_{\beta} q^{l}_{\gamma} \rangle_{\rho}
- \langle \overline{q}^{i}_{\alpha} q^{l}_{\gamma} \rangle_{\rho}  \langle \overline{q}^{k}_{\beta} q^{m}_{\delta} \rangle_{\rho}, \label{eq:four.quark.condensate.factorization.density.1} \\
\langle \overline{q}^{i}_{\alpha} \overline{q}'^{k}_{\beta} q'^{l}_{\gamma} q^{m}_{\delta} \rangle_{\rho} & \simeq 
\langle \overline{q}^{i}_{\alpha} q^{m}_{\delta} \rangle_{\rho} \langle \overline{q}'^{k}_{\beta} q'^{l}_{\gamma} \rangle_{\rho}, 
\label{eq:four.quark.condensate.factorization.density}
\end{align}
where in the first line, all four quark operators have the same flavor, 
while in the second line the operators $q$ and $q'$ represent different flavors. 
The various two-quark expectation values are further expanded into color singlet and 
Lorentz scalar and vector pieces. One then substitutes for instance Eqs.\,(\ref{eq:qbarq.linear}) 
or (\ref{eq:sbars.linear}) and expands the result to linear order in $\rho$. This gives a crude order of magnitude estimate, 
but as it was the case for the same vacuum saturation approximation, it is not clear to what degree 
this approximation is realized in nature. 
Indeed, a study of the nucleon QCD sum rules at finite density suggests that the density dependence of the scalar-scalar four-quark 
condensate should be considerably weaker than the estimate obtained from Eq.\,(\ref{eq:four.quark.condensate.factorization.density}) \cite{Furnstahl:1992pi}. 
A first principle lattice QCD calculation 
will therefore be needed in the future 
to have better control over the systematic uncertainties. 
Besides the method explained above, certain operators that appear in the OPE of the electromagnetic current correlator 
$j^{\mathrm{em}}_{\mu} = \sum_{q} e_q \overline{q} \gamma_{\mu} q$ can be constrained from lepton-hadron deep 
inelastic scattering data \cite{Choi:1993cu}. 

As a reference for the interested reader, we in the following give a brief guide to the literature dealing with the density dependence of dimension 6 condensates. 
Note however that the list of works mentioned here is not necessarily complete. 
The factorization hypothesis of the four-quark condensates applied to finite density, was discussed in Ref.\,\cite{Jin:1992id}, while 
Ref.\,\cite{Thomas:2007gx} studied the algebraic relations between different four-quark condensates and their evaluation using factorization 
and the perturbative chiral quark model \cite{Drukarev:2003xd}. Further discussions on the role of four-quark condensates and their values 
are given in Ref.\,\cite{Buchheim:2014rpa}. Experimental constraints of dimension 6 condensates appearing the vector channel OPE are 
studied in Ref.\,\cite{Choi:1993cu}, while the same condensates containing strange quarks were considered in Ref.\,\cite{Gubler:2015uza}. 
Estimates of the nucleon expectation values of gluonic dimension 6 operators are provided in Ref.\,\cite{Kim:2000kj}.

\subsubsection{Condensates in a homogenous and constant magnetic field \label{sec:conds.in.magn.field}}
In recent years, the effects of a strong magnetic field have attracted the interest of the hadron physics community because 
of the potential existence of such strong fields in heavy-ion collisions and magnetars \cite{Deng:2012pc,Harding:2006qn}. 
In this context, interesting phenomena such as the chiral magnetic effect \cite{Kharzeev:2007jp,Fukushima:2008xe} or 
the magnetic catalysis \cite{Gusynin:1994re} have been widely discussed (see for instance Refs.\,\cite{Kharzeev:2013jha,Hattori:2016emy} for 
recent reviews and further references). 
This has motivated practitioners of both lattice QCD and QCD sum rules to study the behavior of hadrons under a strong 
(of the order of a typical QCD scale) and constant magnetic field. 
For QCD sum rule studies, this means that condensates must be determined as a function of the magnetic field to be used as 
input. The present status of what is known about the magnetic behavior of the condensates is briefly reviewed here. 
In what follows, we use the notation $B \equiv |\bm{B}|$ and will, if not stated otherwise assume the magnetic field to 
point into the direction of the z-axis: $\bm{B} = (0,\,0,\,B)$. 

Besides the scalar condensates given in Eq.\,(\ref{eq:condensates.vacuum}), which will get modified as the magnetic field is 
increased, there are also novel condensates that appear once the magnetic field is switched on. These condensates are 
different from those in Eq.\,(\ref{eq:condensates.medium}), which 
emerge at finite temperature or density. This can be understood as follows. 
Non-scalar condensates in hot or dense matter are constructed by considering all positive parity, gauge invariant and independent  
combinations of quark fields, gluon field strengths and covariant derivatives that do not vanish when contracted with 
$u^{\mu}$, the four-velocity of the heat bath or the dense medium. In the magnetic field case, $u^{\mu}$ is replaced by 
$F_{\mu \nu}$ the electromagnetic field strength tensor (or combinations thereof). The non-scalar condensates obtained in this way are
\begin{align}
\mathrm{dimension}\, 3:\,\, & \langle \bar{q} \sigma_{\mu \nu} q \rangle_{B}, \nonumber \\
\mathrm{dimension}\, 4:\,\, & \langle \mathcal{ST} \bar{q} \gamma_{\mu} iD_{\nu} q \rangle_{B}, \, \langle \mathcal{ST} G_{\alpha}^{a\mu} G^{a\nu\alpha} \rangle_{B}, \label{eq:condensates.magn.field} \\
\mathrm{dimension}\, 5:\,\, & \langle \bar{q} t^a G^a_{\mu \nu} q \rangle_{B}, \, \langle \bar{q} \gamma_5 t^a \tilde{G}^a_{\mu \nu} q \rangle_{B}, \, \dots  \nonumber \\
\dots & \nonumber
\end{align}
Here, $\langle \mathcal{O} \rangle_{B}$ stands for the expectation value of the operator $\mathcal{O}$ with respect to the QCD ground state with zero temperature and 
zero baryon density, but with a constant and homogenous magnetic background field. 
These condensates can be further categorized according to their C-parity. Those with negative C-parity will be proportional to odd numbers of $F_{\mu \nu}$. 
For small magnetic fields they will generally be proportional to $B$. Those with positive C-parity have to be proportional to even numbers of $F_{\mu \nu}$ and at small magnetic 
field will be proportional to $B^2$. 
In the list shown above, the dimension 3 and 5 condensates have negative C-parity, while the dimension 4 ones have positive C-parity. It is possible 
to construct positive C-parity operators at dimension 5, which are not shown here. 
For a more extended discussion of operators with higher dimension and negative C-parity (whose properties are, however, at present practically unknown), see Ref.\,\cite{Ioffe:1983ju}. 

\paragraph{Condensates of dimension 3}\mbox{}\\
We start with the magnetic behavior of the chiral condensate $\langle \bar{q} q \rangle_{B}$, about which we currently have the most 
detailed information. This is mainly thanks to chiral perturbation theory and recent lattice QCD calculations, where it is relatively straightforward 
to introduce constant magnetic fields. First, let us give the chiral perturbation theory result based on Refs.\,\cite{Shushpanov:1997sf,Cohen:2007bt}, which 
can be cast in a simple and analytic form, 
\begin{equation}
\frac{\langle \bar{q} q \rangle_{B}}{\langle 0 | \bar{q} q | 0 \rangle} = 1 + \frac{\log(2) eB}{16 \pi^2 f_{\pi}^2} I_{H} \Biggl( \frac{m_{\pi}^2}{eB} \Biggr),   
\end{equation}
with
\begin{equation}
I_{H}(y) = - \frac{1}{\log(2)} \int_0^{\infty} \frac{dz}{z^2} e^{-yz} \Biggl[\frac{z}{\sinh(z)} - 1 \Biggr]. 
\end{equation}
According to the lattice QCD calculations to be shown below, this expression is accurate up to magnetic field values of about $eB \simeq 0.1\,\mathrm{GeV}^2$. 

\begin{figure}[tb]
\begin{center}
\begin{minipage}[t]{16.5 cm}
\vspace{0.5 cm}
%\hspace{1.0 cm}
%\epsfig{file=cond_magn,scale=0.3}
%\includegraphics[width=8.5cm,natwidth=610,natheight=642]{contfitT0.pdf}
\includegraphics[width=8.5cm,bb=0 0 574 436]{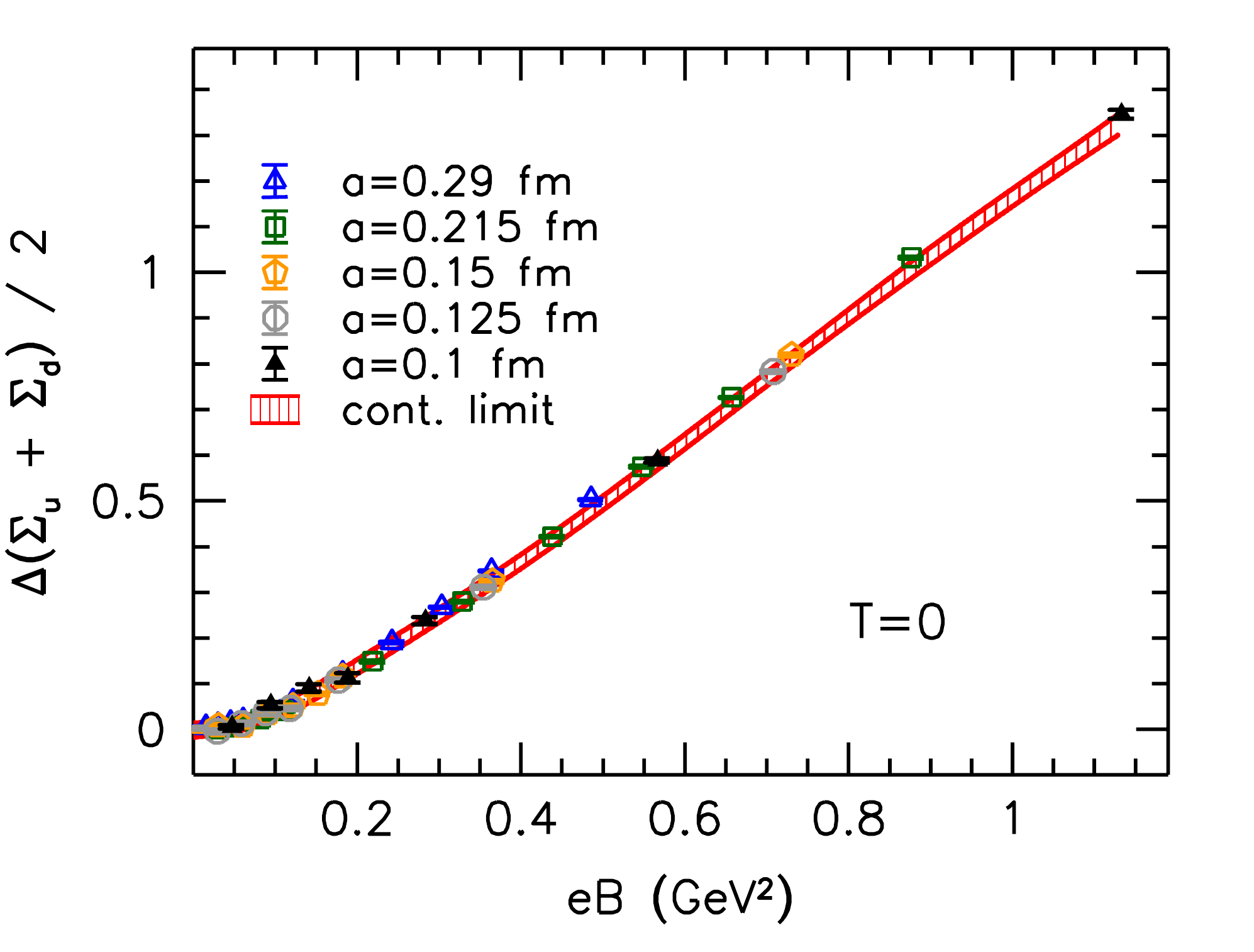}
%\hspace{-1.5 cm}
%\epsfig{file=cond_magn_T,scale=0.3}
%\includegraphics[width=8.5cm,natwidth=610,natheight=642]{pbp_B_T.pdf}
\includegraphics[width=8.5cm,bb=0 0 574 436]{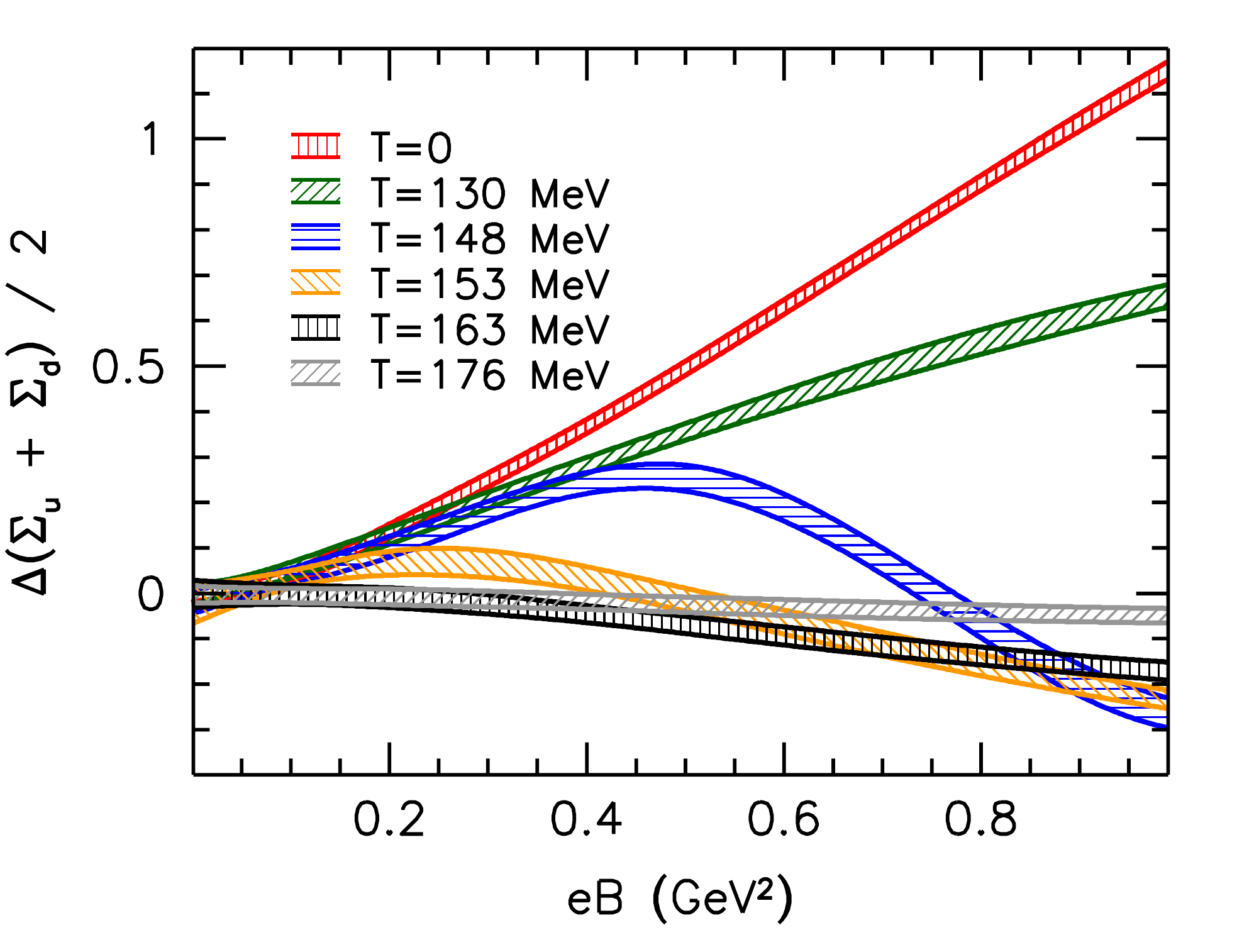}
%\vspace{1 cm}
\end{minipage}
\begin{minipage}[t]{16.5 cm}
\caption{Left plot: The change of the renormalized chiral condensate in a magnetic field at $T=0$. 
For the precise definition, see Eqs.\,(\ref{eq:dim3.magn.field.1}) and (\ref{eq:dim3.magn.field.1.2}). 
The various data points show results for different lattice spacings $a$, while the red band corresponds to 
the continuum limit. 
Right plot: The same as for the left plot, but showing only the continuum extrapolated results at finite temperature. 
Taken from Figs.\,1 and 2 of Ref.\,\cite{Bali:2012zg}.  
\label{fig:cond.magnetic.field}}
\end{minipage}
\end{center}
\end{figure}

A relatively recent high precision lattice QCD calculation of $\langle \bar{q} q  \rangle_{B}$ can be found in Ref.\,\cite{Bali:2012zg}, where 
staggered fermions were used to simulate $1+1+1$ dynamical quarks at the physical point. The results were furthermore extrapolated to the 
continuum limit. For an earlier result based on the quenched approximation, see Ref.\,\cite{Braguta:2010ej}. 
The behavior of the chiral condensate as a function of $eB$ is shown on the left plot of Fig.\,\ref{fig:cond.magnetic.field}. 
%%%%%%%%%%%%%%%%%%%%%%%%%%%%%%%%%%%%%%%%%%%%%%%%%%%%%%%%%%%%%%%%%%%%%%%%%%%%%%%%%%%%%%%%%%%%%%%%%%%%%%%%%%%%%%%%%%%%
%%%%%%%%%%%%%%%%%%%%%%%%%%%%%%%%%%%%%%%%%%%%%%%%%%%%%%%%%%%%%%%%%%%%%%%%%%%%%%%%%%%%%%%%%%%%%%%%%%%%%%%%%%%%%%%%%%%%
The definition of $\Delta \Sigma_q \equiv \Delta (\Sigma_u + \Sigma_d)/2$ depicted in this figure is
\begin{align}
\Sigma_{u}(B,T) &= \frac{2m_{ud}}{m_{\pi}^2 f_{\pi}^2} \Bigl(\langle B,T | \bar{u} u | B,T \rangle  - \langle 0,0 | \bar{u} u | 0,0 \rangle \Bigr) + 1, \label{eq:dim3.magn.field.1} \\
\Delta \Sigma_{u}(B,T) &= \Sigma_{u}(B,T)  - \Sigma_{u}(0,T),   
\label{eq:dim3.magn.field.1.2}
\end{align}
and analogously for the $d$ quark. $m_{ud}$ is the mass of the degenerate $u$ and $d$ quarks, $m_{ud} = m_{u} = m_{d}$. 
We here have kept the notation and convention of Ref.\,\cite{Bali:2012zg}, where $\langle 0,0 | \bar{u} u | 0,0 \rangle$ is a positive number. 
The above definitions are used to eliminate additive and multiplicative divergences that appear in the lattice computations of the condensates. 
With the help of the Gell-Mann-Oakes-Renner relation of Eq.\,(\ref{eq:GellMann.Oakes}) and keeping in mind the changed sign convention, Eq.\,(\ref{eq:dim3.magn.field.1})
can be rewritten as 
\begin{equation}
\Sigma_{u}(B,T) = \frac{\langle B,T | \bar{u} u | B,T \rangle}{\langle 0,0 | \bar{u} u | 0,0 \rangle}. 
\end{equation}
The left plot of Fig.\,\ref{fig:cond.magnetic.field} shows that the average $u$ and $d$ condensate increases with and increasing magnetic field. 
This phenomenon is commonly referred to as ``magnetic catalysis''. 
On the lattice, it is possible to study how the magnetic field dependence changes with increasing temperature. 
The corresponding results are shown in the right plot of Fig.\,\ref{fig:cond.magnetic.field}. It is interesting to 
see that the magnetic field dependence almost completely vanishes at temperatures around $T_c$.  
The presence of a magnetic field breaks isospin symmetry, hence 
causing the $u$ and $d$ quarks to behave differently. Therefore, it is not sufficient to only consider the average $\Delta \Sigma_q$, but 
also the difference between $\Delta \Sigma_u$ and $\Delta \Sigma_d$ which grows with increasing $B$. For more details about the 
quark flavor dependence and 
a comparison of the lattice results with chiral perturbation theory and models, see Ref.\,\cite{Bali:2012zg}. 
The behavior of the strange quark condensate has so far not been studied in lattice QCD. 

Next, we discuss the non-scalar condensate given in the first line of Eq.\,(\ref{eq:condensates.magn.field}). 
This quantity is not only important as an input in QCD sum rule studies, but also 
for determining the response of the QCD free energy density to magnetic fields. 
As it is common, we assume 
the magnetic field to be parallel to the z-axis, which means that only the component $\langle \bar{q} \sigma_{12} q \rangle_{B}$ will 
be of relevance here. This condensate was studied on the lattice for the first time in Ref.\,\cite{Braguta:2010ej} in the quenched approximation and 
later in Ref.\,\cite{Bali:2012jv} in full QCD, by the same group and under the same conditions as 
the chiral condensate discussed above and in Ref.\,\cite{Bali:2012zg}. At relatively small magnetic fields, two parametrizations have been 
used to quantify $\langle \bar{q} \sigma_{12} q  \rangle_{B}$:
\begin{align}
\langle \bar{q} \sigma_{12} q \rangle_{B} & = q_f B \langle 0 | \bar{q} q | 0 \rangle \chi_f, \label{eq:dim3.magn.field.2} \\
\langle \bar{q} \sigma_{12} q \rangle_{B} & = q_f B \tau_f. 
\label{eq:dim3.magn.field.3}
\end{align}
Here, the quark field $q$ represents any of the quark flavors $u$, $d$ and $s$, while $q_f$ is the respective electric charge. 
$\chi_f$ is commonly referred to as the ``magnetic susceptibility of the condensate'', while $\tau_f$ is called 
``tensor coefficient'' in Ref.\,\cite{Bali:2012jv}. The full QCD lattice results obtained in this reference are 
\begin{align}
\chi_u & = -(2.08 \pm 0.08) \, \mathrm{GeV}^{-2}, \nonumber \\
\chi_d & = -(2.02 \pm 0.09) \, \mathrm{GeV}^{-2}, \label{eq:dim3.magn.field.31} \\
\chi_s & = -(3.4 \pm 1.4) \, \mathrm{GeV}^{-2},  \nonumber
\end{align}
and
\begin{align}
\tau_u & = 40.7 \pm 1.3 \, \mathrm{MeV}, \nonumber \\
\tau_d & = 39.4 \pm 1.4 \, \mathrm{MeV}, \\
\tau_s & = 53.0 \pm 7.2 \, \mathrm{MeV},  \nonumber
\end{align}
at a renormalization scale of 2 GeV in the $\overline{\mathrm{MS}}$ scheme\footnote{In Ref.\,\cite{Bali:2012jv}, the conventions of 
positive $\langle 0 | \bar{q} q | 0 \rangle$ and $\sigma_{\mu \nu} = \frac{1}{2i}[\gamma_{\mu}, \gamma_{\nu}]$ were used. To adjust to our 
conventions with negative $\langle 0 | \bar{q} q | 0 \rangle$ and $\sigma_{\mu \nu} = \frac{i}{2}[\gamma_{\mu}, \gamma_{\nu}]$, we 
changed the sign of the $\tau_{f}$ values given in Ref.\,\cite{Bali:2012jv}.}. 
It was furthermore shown that the linear behavior of Eqs.\,(\ref{eq:dim3.magn.field.2}) and (\ref{eq:dim3.magn.field.3}) is 
valid up to magnetic fields of about $eB = 0.3\,\mathrm{GeV}^2$, above which the effects of higher order terms [$\mathcal{O}(B^3)$] become 
visible. 
For practical applications, the $\tau_f$ values are the preferred choice as their evaluation does not depend on a separate 
calculation of $\langle 0 | \bar{q} q | 0 \rangle$ and furthermore have only a mild renormalization scale dependence \cite{Bali:2012jv}. 
Indeed, the anomalous dimension of the operator $\bar{q} \sigma_{\mu \nu} q$ is only $-\frac{4}{27}$ for three active flavors \cite{Ioffe:1983ju,Shifman:1980dk}. 

\paragraph{Condensates of dimension 4}\mbox{}\\
At this dimension, we start with the scalar gluon condensate $\langle \frac{\alpha_s}{\pi} G^2 \rangle_{B}$, which was recently studied using  
lattice QCD in Ref.\,\cite{Bali:2013esa}. This was done via measuring the interaction measure (in other words, the trace of the energy momentum tensor), 
of which the gluonic part is proportional to the scalar gluon condensate. In this calculation it was found that, analogously to the quark condensate, 
the gluon condensate is enhanced with an increasing magnetic field (see the left plot of Fig.\,1 in Ref.\,\cite{Bali:2013esa}). Quantitatively, the gluon 
condensate value is roughly increased by about 30\,\% at $eB = 0.8\,\mathrm{GeV}^2$ compared to the vacuum, assuming the vacuum value 
of Eq.\,(\ref{eq:gluon.cond.value}). This behavior does not agree with the earlier study of Ref.\,\cite{Agasian:1999sx} (which got a decreasing gluon condensate value 
with increasing $B$), but agrees with the more recent works of Refs.\,\cite{Ozaki:2013sfa,Agasian:2016hcc}. 
The change of the magnetic field behavior of the gluon condensate with increasing temperature was also studied in Ref.\,\cite{Bali:2013esa} and again 
a behavior similar to the one found for the quark condensate was obtained: the dependence on the magnetic field weakens as the temperature 
approaches $T_c$ and switches its sign for even larger temperatures, giving rise to a decreasing gluon condensate with an increasing magnetic field. 

We next discuss $\langle \mathcal{ST} \bar{q} \gamma_{\mu} iD_{\nu} q \rangle_{B}$, which has positive C-parity and hence is 
expected to behave as $\mathcal{O}(B^2)$ for small $B$. Unfortunately, there are presently no lattice QCD calculations available for this condensate. 
Moreover, to our knowledge only one simple quark model estimate has been reported so far. 
This estimate is given in Appendix E of Ref.\,\cite{Gubler:2015qok}, which should be consulted for more details. 
Schematically, the method employed in Ref.\,\cite{Gubler:2015qok} can be summarized as 
\begin{equation}
\langle \mathcal{O} \rangle_{B} - \langle 0 | \mathcal{O} | 0 \rangle = - \int^{\Lambda} \frac{d^4p}{(2\pi)^4} \mathrm{Tr}_{C,\,D}[\mathcal{O} S(p)_B], 
\end{equation}
where $\mathcal{O}$ represents a general operator, that can contain gamma matrices or covariant derivatives, 
$S(p)_B$ stands for the quark propagator with one or more magnetic field insertions and $\mathrm{Tr}_{C,\,D}$ for the color and Dirac trace. 
In this model it is possible to reproduce the magnetic field dependence with rather good accuracy when setting the (constituent) quark mass to 
$m_q = 300$ MeV and the (Euclidean) cutoff to $\Lambda = 1$ GeV. 
For $\mathcal{O} = \gamma_{\mu} iD_{\nu}$, the final result reads 
\begin{equation}
\langle \mathcal{ST} \bar{q} \gamma^{\mu} iD^{\nu} q \rangle_{B} = \frac{1}{8\pi^2} q_f^2 B^2 (g_{\parallel}^{\mu \nu} - g_{\perp}^{\mu \nu}) A, 
\label{eq:dim3.magn.field.4}
\end{equation}
with
\begin{equation}
A = \log \Biggl(\frac{\Lambda^2}{m^2_q}\Biggr) - \frac{4}{3} + \log \Biggl(1 + \frac{m^2_q}{\Lambda^2}\Biggr) 
      + \frac{3}{2} \frac{m^2_q}{\Lambda^2(1 + m^2_q/\Lambda^2)} - \frac{1}{6} \frac{m^6_q}{\Lambda^6(1 + m^2_q/\Lambda^2)^3}. 
\end{equation}
The tensors $g_{\parallel}^{\mu \nu}$ and $g_{\perp}^{\mu \nu}$ appearing in Eq.\,(\ref{eq:dim3.magn.field.4}) are defined as 
$g_{\parallel}^{\mu \nu} = \mathrm{diag}(1,0,0,-1)$ and $g_{\perp}^{\mu \nu} = \mathrm{diag}(0,-1,-1,0)$. 
The form $g_{\parallel}^{\mu \nu} - g_{\perp}^{\mu \nu}$ can be understood as part of the electromagnetic counterpart of 
the gluonic operator $\mathcal{ST} G_{\alpha}^{a\mu} G^{a\nu\alpha}$. Indeed, 
\begin{equation}
F^{\mu \alpha} F^{\nu}_{\alpha} - \frac{1}{4} g^{\mu \nu} F^{\alpha \beta}F_{\alpha \beta} = - \frac{1}{2} B^2 (g_{\parallel}^{\mu \nu} - g_{\perp}^{\mu \nu})  
\label{eq:dim3.magn.field.5}
\end{equation}
can be derived for $F^{\mu \nu}$ containing only a magnetic field pointing in the direction of the z-axis. 
Naturally, the above result is only valid for B values, for which higher order $\mathcal{O}(B^4)$ terms can be neglected. 

For the second condensate at dimension 4, $\langle \mathcal{ST} G_{\alpha}^{a\mu} G^{a\nu\alpha} \rangle_{B}$ which can also be 
expected to behave as $\mathcal{O}(B^2)$ for small $B$, we presently do not have much information. 
As gluons do not couple directly to the magnetic field, this condensate vanishes exactly in the quenched approximation. 
In full QCD with dynamical quarks, it does not necessarily vanish, but can be 
expected to be strongly suppressed. To obtain a quantitative estimate, a lattice QCD or model calculation will be needed in the future. 

\paragraph{Condensates of dimension 5}\mbox{}\\
At dimension 5, we will discuss only the two condensates given in the third line of Eq.\,(\ref{eq:condensates.magn.field}), 
as the behavior of possible other scalar and non-scalar operator expectation values are presently not known. 
These two have been considered already a long time ago in Ref.\,\cite{Ioffe:1983ju} 
(and also partly in Ref.\,\cite{Balitsky:1983xk}), based on a QCD sum rule calculation of  
the nucleon magnetic moments. 
They are traditionally parametrized as 
\begin{align}
\langle \bar{q} t^a G^a_{\mu \nu} q \rangle_{B} &= q_f \kappa F_{\mu \nu} \langle 0 | \bar{q} q | 0 \rangle, \label{eq:dim5.magn.field.1} \\
\langle \bar{q} \gamma_5 t^a \tilde{G}^a_{\mu \nu} q  \rangle_{B} &= \frac{i}{2} q_f \xi F_{\mu \nu} \langle 0 | \bar{q} q | 0 \rangle. 
\label{eq:dim5.magn.field.2}
\end{align}
This parametrization is only valid for small electromagnetic field values. Higher order terms in $F_{\mu \nu}$ have so far 
not been studied. The two operators $\bar{q} t^a G^a_{\mu \nu} q$ and $\bar{q} \gamma_5 t^a \tilde{G}^a_{\mu \nu} q$ generally 
mix when changing the renormalization scale. Respective eigenvalues and eigenvectors of the corresponding anomalous dimension 
matrix are given in Refs.\cite{Balitsky:1989ry,Ball:2002ps}. The parameters $\kappa$ and $\xi$ have been discussed 
in many QCD sum rule studies over the years. In particular Chiu, Pasupathy and Wilson have studied them in series of papers 
in the eighties \cite{Chiu:1985ey,Chiu:1986cf,Pasupathy:1986pw,Chiu:1987jm}, where they have used QCD sum rules of various 
channels to determine $\kappa$ and $\xi$. In Ref.\,\cite{Chiu:1985ey}, the vector channel sum rules in combination with the 
vector-dominance model was used in a one pole and two pole approximation, respectively. 
The obtained results were
\begin{align}
\kappa &= 0.22, & \xi &= -0.44 & (\text{one pole}), \\
\kappa &= 0.4, & \xi &= -0.8 & (\text{two poles}).
\end{align}
In the same paper, they further carried out two different fits of baryonic magnetic moment sum rules to experimental data to 
obtain $\kappa - 2\xi = 5.73$ and $\kappa - 2\xi = 8.93$. In subsequent papers (Refs.\,\cite{Pasupathy:1986pw,Chiu:1987jm}), 
they took further baryons into account for their fit, which led to 
\begin{align}
\kappa &= 0.75, \label{eq:dim5.magn.field.3} \\
\xi &= -1.5. 
\label{eq:dim5.magn.field.4}
\end{align}
These values remain rather popular and are widely used even today. 
In the same work, a simple parametrization was also given for the strange counterparts of $\kappa$ and $\xi$ 
[and of $\chi_q$ defined in Eq.\,(\ref{eq:dim3.magn.field.2})]: 
\begin{align}
\phi &= \frac{\kappa_s}{\kappa} =  \frac{\xi_s}{\xi} = \frac{\chi_s}{\chi_q}, \\
& = 0.6.
\end{align}
It however has to be noted here that the newest lattice QCD results for $\chi_s$ and $\chi_q$ [see Eq.\,(\ref{eq:dim3.magn.field.31})] 
give $\chi_s/\chi_q \simeq 1.66$ and do not agree with the above value, which therefore needs to be taken with a grain of salt. 
Moreover, besides the most often used values of Eqs.\,(\ref{eq:dim5.magn.field.3}) and (\ref{eq:dim5.magn.field.4}), 
\begin{align}
\kappa &= 0.2, \label{eq:dim5.magn.field.5} \\
\xi &= -0.4,  
\label{eq:dim5.magn.field.6}
\end{align}
given in Ref.\,\cite{Ball:2002ps} and partly based on Ref.\cite{Balitsky:1989ry}, are also sometimes quoted in the literature (see, for instance, Ref.\,\cite{Wang:2017dce}). 
In all, it can be said that $\kappa$ and $\xi$ likely have orders of magnitude as given in Eqs.\,(\ref{eq:dim5.magn.field.3}) and (\ref{eq:dim5.magn.field.4})
or Eqs.\,(\ref{eq:dim5.magn.field.5}) and (\ref{eq:dim5.magn.field.6}). Their precise values remain, however, presently unknown. 

\section{Analysis strategies \label{sec:analysis.strategies}}
Once the condensates that appear in the OPE of a specific channel are identified,  the calculation of the corresponding Wilson coefficients is completed and 
the condensate values are determined with sufficient precision, the remaining task is to extract information about the spectral function 
from the sum rules given for instance in Eqs.\,(\ref{eq:derive.correlator.3}) and (\ref{eq:derive.correlator.4}). Obviously, this is not a simple task 
as $\Pi(q^2)$ or $\widetilde{\Pi}(q^2)$ are not known exactly, but only as an expansion in $1/q^2$, with coefficients that by themselves have 
uncertainties due to incomplete knowledge about the condensates and higher order perturbative $\alpha_s$ corrections in the Wilson 
coefficients. 
At most, what one can hope for is to extract some basic features of the spectral function, but not its detailed 
form. How to extract these features will be discussed in this section.  

\subsection{Derivation of sum rules for practical numerical analysis}
In most present day QCDSR studies, the sum rules of Eqs.\,(\ref{eq:derive.correlator.3}) and (\ref{eq:derive.correlator.4}) are usually not 
analyzed in the form shown in these equations, but are further modified, partly to improve the OPE convergence and/or to enhance the contribution 
of the low energy region of the spectral function to the sum rules. There are multiple ways of doing this, the most popular one being the 
use of the Borel transform, which was already introduced in the very first QCD sum rules papers by Shifman et al. \cite{Shifman:1978bx,Shifman:1978by}. 
We will discuss here this Borel transform method in some detail, but will later also introduce alternatives, which for certain purposes can be 
more effective in practice. 

The Borel transform is defined as 
\begin{equation}
\Pi(M^2) \equiv \widehat{L}_M \Pi(q^2) \equiv  \lim_{\genfrac{}{}{0pt}{}{-q^2,n \to \infty,}{
-q^2/n=M^2}}
\frac{(-q^2)^{n}}{(n-1)!} \Bigg(\frac{d}{dq^2}\Bigg)^n 
\Pi(q^2),  
\label{eq:Boreltrans1}
\end{equation}
where the newly introduced parameter $M$ is referred to as the ``Borel mass'' because it has units of mass. Note, however, that 
$M$ is just an artificial parameter, which has nothing to do with the mass of any physical object. 
Some typical and often used examples of the Borel transform are shown below, 
\begin{align}
\widehat{L}_M (q^2)^k &= 0,  
\label{eq:Boreltrans3} \\
\widehat{L}_M (q^2)^k \ln (-q^2) &= - k! (M^2)^k, 
\label{eq:Boreltrans6} \\
\widehat{L}_M \Bigl(\frac{1}{q^2}\Bigr)^k &= \frac{(-1)^k}{(k-1)!}\Bigl(\frac{1}{M^2}\Bigr)^k, 
\label{eq:Boreltrans4} \\
\widehat{L}_M \Bigl(\frac{1}{s-q^2}\Bigr)^k &= \frac{1}{(k-1)!}\Bigl(\frac{1}{M^2}\Bigr)^k e^{-s/M^2}. 
\label{eq:Boreltrans5}
\end{align}
Here, $k$ is a positive integer. 
For more related formulas, see Refs.\,\cite{Narison:2007spa,Langwallner:2005}. 
After applying Eq.\,(\ref{eq:Boreltrans5}) to the dispersion relation of Eq.\,(\ref{eq:derive.correlator.3}) [or Eq.\,(\ref{eq:derive.correlator.4})], one obtains 
\begin{equation}
\Pi_{\mathrm{OPE}}(M^2) = \frac{1}{\pi M^2} \displaystyle \int^{\infty}_{0}ds
e^{-s/M^2} \rho(s). 
\label{eq:Boreltrans2}
\end{equation}
As seen in Eq.(\ref{eq:Boreltrans1}), the Borel transform contains an infinite number of derivatives. 
All subtraction terms thus automatically vanish. Moreover, it causes the high energy part of the dispersion integral to be 
exponentially suppressed, meaning that the integral converges to a finite value, as long as the spectral function 
itself does not grow exponentially, which does not happen for QCD. 
It is furthermore observed in Eq.\,(\ref{eq:Boreltrans4}) that higher dimensional terms of the OPE, 
which are proportional to $(1/q^2)^k$, 
are suppressed by a factor of $1/(k-1)!$, hence improving the convergence of the OPE. 
The Borel transformed sum rules of Eq.\,(\ref{eq:Boreltrans2}) are presently most commonly employed in 
practical QCDSR analyses.  

Eq.\,(\ref{eq:Boreltrans2}), however, is not the unique QCDSR form. Indeed, sum rules with a Gaussian kernel were 
derived in Ref.\,\cite{Bertlmann:1984ih}, commonly referred to as ``Gaussian sum rules''. We will not repeat the somewhat lengthy derivation here, 
but just give the final form, which reads 
\begin{equation}
\Pi(t,\tau) = \frac{1}{\pi \sqrt{4\pi\tau}} \displaystyle \int_0^{\infty} ds e^{-\frac{(s-t)^2}{4\tau}} \rho(s).  
\label{eq:Gaussian}
\end{equation} 
Here, $t$ and $\tau$ are free parameters that roughly correspond to the Borel mass $M$ in Eq.\,(\ref{eq:Boreltrans2}). 
The advantage of the Gaussian kernel is that two parameters can be varied, which makes it possible to extract more detailed information 
about the spectral function from the sum rules. Furthermore, the Gaussian kernel has 
a distinct peak at $t=s$, which means that any structure that might be present in the spectral 
function is more likely to be preserved in $\Pi(s,\tau)$, rather than smeared out as it is usually the case for 
the Borel sum rule. 
This situation is similar to what has occurred in nuclear structure
studies, where the Lorentz kernel has proven to be useful \cite{Efros:1994iq,Efros:2007nq}. 
The Gaussian sum rule was successfully applied in Refs.\,\cite{Ohtani:2011zz,Suzuki:2015est} to the 
nucleon and D meson sum rules, in combination with a numerical maximum entropy method (MEM) analysis to be discussed below. 

Another way to increase the amount of information that can be extracted from the sum rules 
is to promote the parameters appearing in the kernels (such as $M$ or $t$), which are usually treated as real valued, 
to complex numbers. This causes the kernels to become oscillating functions with varying frequencies, which can be useful 
for constraining spectral fits or for MEM analyses. This idea has in recent years been applied to multiple MEM 
analyses of sum rules in various channels: the parity projected Gaussian sum rules for the nucleon \cite{Ohtani:2012ps}, 
the Borel sum rules of the $\phi$ meson \cite{Araki:2014qya} and the finite temperature Borel sum rules of S-wave 
charmonia \cite{Araki:2017ebb}.

The analyticity of the correlator can also be used to derive sum rules with an analytic, however not 
explicitly specified kernel. In the past, this has been done mainly to derive sum rules in a hot or dense medium, 
see for instance Refs.\,\cite{Furnstahl:1992pi,Ohtani:2012ps,Bochkarev:1985ex,Jido:1996ia}. 
The in-medium sum rules are usually formulated using the energy variable $\omega$ instead of $s$. 
Using for instance the retarded correlator $\Pi^{\mathrm{R}}(\omega, \bm{p})$ at finite temperature, 
which is analytic in the upper half of the complex $\omega$ plane in combination with a 
function $W(\omega)$, which is analytic in the same region, one can derive 
\begin{equation}
\int_{-\infty}^{\infty} d\omega W(\omega) \rho(\omega, \bm{p}) = \int_{-\infty}^{\infty} d\omega W(\omega) \mathrm{Im}\Pi^{\mathrm{R}}_{\mathrm{OPE}}(\omega, \bm{p}), 
\label{eq:general.sum.rule}
\end{equation} 
where $\Pi^{\mathrm{R}}_{\mathrm{OPE}}(\omega, \bm{p})$ is the retarded correlator calculated from the operator product expansion. 
For more details, see Ref.\,\cite{Bochkarev:1985ex}, for a similar derivation for the finite density case, see Refs.\,\cite{Furnstahl:1992pi,Ohtani:2012ps,Jido:1996ia} and 
for an application in the context of the unitary fermi gas, see Ref.\,\cite{Gubler:2015iva}. 
The most important feature of Eq.\,(\ref{eq:general.sum.rule}) is that $W(\omega)$ is arbitrary as long as it is analytic and can hence be 
chosen depending on what region of the spectral function one wants to study. Some care, however is needed when making this choice as the 
convergence of the OPE will depend on $W(\omega)$. For instance if one chooses a kernel analogous to the one used in Eq.\,(\ref{eq:Gaussian}), 
it at first sight would seem advantageous to choose a small value for $\tau$, such that the spectral function can be 
extracted with a good resolution. It however turns out that higher order OPE terms are proportional to increasingly high powers of $1/\sqrt{\tau}$, therefore 
destroying the OPE convergence for too small $\tau$ values. The choice of $W(\omega)$ thus always has to be a compromise between 
the resolution of the extracted spectral function and the OPE convergence. 

\subsection{Conventional analysis strategy \label{sec:conventional.analysis}}
The method employed most often in QCD sum rule studies will be described here. 
As this method has already been discussed many times in previous reviews, we keep this part brief and refer the interested  
reader to older works (see e.g. Ref.\,\cite{Colangelo:2000dp}) for more details. 

We first consider the right hand side of the dispersion relation of Eq.\,(\ref{eq:derive.correlator.3}). Using the optical theorem 
and inserting a complete set of intermediate hadronic states, one gets 
\begin{equation}
\rho(q^2) = \frac{1}{2}\displaystyle \sum_n
\langle 0|J|n(p_n)\rangle
\langle n(p_n)|\overline{J}|0\rangle
d\tau_n(2\pi)^4 \delta^{(4)}(q-p_n), 
\label{eq:spectral}
\end{equation}
where $n$ is summed over all hadronic states which couple to the interpolating field $J$, including sums over polarizations and $d\tau_n$ symbolically denotes the 
phase space integration of the states $|n\rangle$. 
The sum rules discussed in the previous section generally only provide information on an integral of the spectral function $\rho(s)$. 
One hence can only hope to extract some bulk properties of the spectrum, but not all its detailed features. 
It therefore has traditionally been the custom in practical sum rule analyses to make a deliberated guess about the form  
of the spectral function, parametrize it with a small number of parameters and then fit these parameters with the help of the sum rules. 
The most frequently used ansatz in present-day QCDSR studies is referred to as the ``pole + continuum'' ansatz and reads 
\begin{equation}
\rho(s) = \pi |\lambda|^2 \delta(s - m^2) + \theta(s - s_{th}) \mathrm{Im} \Pi_{\mathrm{OPE}}(s). 
\label{eq:pole.plus.continuum}
\end{equation}
Here, $m$ is the mass of the ground state, which is assumed to be manifested as a narrow peak, and $|\lambda|^2$ is the coupling 
strength of this ground state to the operator $J$. 
The variable 
$s_{th}$ is referred to as the threshold parameter. 
While usually not much attention is payed to its physical meaning, its modification at finite temperature or 
density has been discussed in the context of the finite energy sum rules as a probe of deconfinement \cite{Carlomagno:2016bpu} 
or chiral symmetry restoration \cite{Marco:1999xz,Kwon:2008vq}. Note that the 
above ansatz completely ignores the width of the ground state and 
potential excited states (including a continuum) below $s_{th}$. $\Pi_{\mathrm{OPE}}(s)$ stands 
for the correlator calculated at high energy using the OPE. Due to asymptotic freedom, it is known that the spectral 
function will approach this limit at $s \to \infty$. Based on the quark-hadron duality \cite{Shifman:2000jv,Poggio:1975af} (see also Section \ref{sec:duality}), 
the second term approximately parametrizes all excited states that couple to $J$. 

In the ``pole + continuum'' ansatz, 
the parameters $|\lambda|^2$, $m$ and $s_{th}$ need to be determined from the sum rules. Usually, one is most 
interested in $m$, which can be obtained as follows. First, one substitutes Eq.\,(\ref{eq:pole.plus.continuum}) into 
Eq.\,(\ref{eq:Boreltrans2}), which leads to 
\begin{align}
|\lambda|^2 e^{-m^2/M^2} &= M^2\Pi_{\mathrm{OPE}}(M^2) - \frac{1}{\pi} \int_{s_{th}}^{\infty} ds e^{-s/M^2}  \mathrm{Im} \Pi_{\mathrm{OPE}}(s) \nonumber \\
&\equiv f(M^2, s_{th}).  
\label{eq:pole.plus.continuum.mass.1}
\end{align}
From this equation, $m^2$ can be derived as 
\begin{equation}
m^2  = \frac{1}{ f(M^2, s_{th})} \frac{\partial f(M^2, s_{th})}{\partial (-1/M^2)}.  
\label{eq:pole.plus.continuum.mass.2}
\end{equation}
After obtaining $m^2$, $|\lambda|^2$ can be easily extracted from Eq.\,(\ref{eq:pole.plus.continuum.mass.1}). 
As a result, $m^2$ and $|\lambda|^2$ become functions of $M^2$ and $s_{th}$, which are parameters that physical observables should not depend on. 
The Borel mass $M$ in particular is an artificially introduced unphysical parameter, which has nothing to do with the 
ground state mass $m$. It is therefore customary to show in QCDSR papers the so-called Borel mass curve, which is 
nothing but a plot of Eq.\,(\ref{eq:pole.plus.continuum.mass.2}) with changing $M$ values. The degree of the (non-)dependence of 
$m$ on the Borel mass M provides a criterion for determining the quality of the sum rules. 

However, just checking the Borel mass dependence of $m$ is not enough. There are at least two 
more criteria that always need to be checked to ensure the reliability and validity of the sum rules. The first 
one is the confirmation of an existent Borel window, which 
corresponds to a region of $M$ where some of the approximations used to derive Eq.\,(\ref{eq:pole.plus.continuum.mass.2}) 
can be considered to be reliable. The lower bound of the Borel window is determined by the convergence of 
the OPE. As the OPE is (likely) an asymptotic series, which is hard to compute up to high orders due to our 
lack of knowledge about the high-dimensional condensates, it is not possible to define a rigorous convergence 
criterion. Instead, an often advocated choice (see, for instance, Ref.\,\cite{Colangelo:2000dp}) is to demand that the contribution of 
the highest dimensional term is less than 10\,\% of all the OPE terms, 
\begin{equation}
\frac{\Pi^{\text{terms of highest dim}}_{\mathrm{OPE}}(M^2)}{\Pi^{\text{all terms}}_{\mathrm{OPE}}(M^2)} < 0.1.
\label{eq:pole.plus.continuum.window.1}
\end{equation}
As the OPE after the Borel transform becomes an expansion in
$1/M^2$, this gives a lower bound for the Borel mass $M$. The upper bound of the Borel window is determined 
from the relative contribution of the ground state to the whole sum rules. As can be understood from 
Eq.\,(\ref{eq:Boreltrans2}), a larger Borel mass $M$ means a smaller suppression of high energy contributions 
to the sum rule. 
The most frequently employed condition is  
\begin{equation}
\frac{\int_0^{s_{th}}ds e^{-s/M^2} \rho(s)}{\int_0^{\infty}ds e^{-s/M^2} \rho(s)} > 0.5, 
\label{eq:pole.plus.continuum.window.2}
\end{equation}
which can be rewritten using Eqs.\,(\ref{eq:Boreltrans2}) and (\ref{eq:pole.plus.continuum.mass.1}) as
\begin{equation}
\frac{|\lambda|^2 e^{-m^2/M^2}}{M^2 \Pi_{\mathrm{OPE}}(M^2)} > 0.5.  
\label{eq:pole.plus.continuum.window.3}
\end{equation}
This gives an upper bound for $M$. If there is a region in $M$, which satisfies both Eq.\,(\ref{eq:pole.plus.continuum.window.1}) 
and Eq.\,(\ref{eq:pole.plus.continuum.window.3}), this is referred to as ``Borel window''. The numbers 
on the right hand sides of Eqs.\,(\ref{eq:pole.plus.continuum.window.1}) and (\ref{eq:pole.plus.continuum.window.3}) are somewhat arbitrary,  
indeed other values (or even other kinds of conditions) sometimes are used in the literature. 
To allow the reader to make a 
reasonable judgement about the accuracy of the QCDSR approach in each studied case, it is important that the used conditions and the resulting Borel window 
are explicitly stated. A simple Borel curve plot [showing the left hand side of Eq.\,(\ref{eq:pole.plus.continuum.mass.2}) as a function of the Borel mass $M$] 
in this sense contains not sufficient information. 

Besides the above conditions related to the Borel window, which have been considered as standard for QCDSR 
studies, we will advocate here one more criterion that should be checked to ensure the reliability of the method. 
This is related to the $s_{th}$ dependence of $m^2$ in Eq.\,(\ref{eq:pole.plus.continuum.mass.2}). If the ``pole + continuum'' 
ansatz of Eq.\,(\ref{eq:pole.plus.continuum}) is a reasonably good approximation of the real spectral function, this dependence should 
be small. If, however, the spectrum for $s < s_{th}$ is dominated by a continuum or several broad peaks instead of a single sharp 
peak, an increase in $s_{th}$ should lead to an increasing $m^2$ value. 
Especially for exotic channels with more than three quarks, the contribution of the continuum is potentially large 
because leading order perturbation theory and dimensional analysis dictate it to increase with a larger power of 
$s$ compared three-quark baryon or two quark meson channels. Indeed, this issue 
was pointed out some time ago in the context 
of pentaquark sum rules in Refs.\,\cite{Kojo:2006bh,Gubler:2009ay,Gubler:2009iq}. 

\subsection{Alternative analysis strategies}
The method discussed in the previous section is the most popular approach used in 
current QCDSR studies. Nevertheless, this does not mean that it is unconditionally the ideal choice. 
First of all, the ``pole + continuum'' ansatz is certainly not for all channels an appropriate assumption. 
The most straightforward and natural way of improvement is to introduce a non-zero width to 
the ground state in Eq.\,(\ref{eq:pole.plus.continuum}) by replacing the delta-function with a Gaussian or 
a Breit-Wigner peak and to treat the peak width as a free parameter that should, ideally, be determined 
by the sum rules. Such a strategy was followed for instance in Ref.\,\cite{Leupold:1997dg} to study the $\rho$ meson at 
finite density or in Refs.\,\cite{Morita:2007pt,Morita:2007hv} to study charmonia at finite temperature. 
It is usually found in such fits that it is not possible to uniquely determine 
both the peak mass and its width. Instead, one obtains multiple combinations of mass shifts and widths, which 
all reproduce the sum rule well within its uncertainties.  

Another method to analyze QCD sum rules was introduced in Ref.\,\cite{Leinweber:1995fn}. 
The essential idea of this approach is to carry out a proper Monte-Carlo error analysis by generating 
Gaussian distributions for the different condensate input values and fitting the sum rules to 
each generated condensate value configuration. This then gives distributions for 
$|\lambda|^2$, $m^2$ and $s_{th}$, which allows one to perform an uncertainty analysis 
for these parameters. Furthermore, this also makes it possible to investigate correlations between, say, 
$m^2$ and the chiral condensate, which can be useful when considering the relation between 
hadron masses and the spontaneous breaking of chiral symmetry. After this method was proposed, 
it has been applied by several groups. See for instance Refs.\,\cite{Lee:1997ix,Lee:1997ne,Lee:1997jk,Erkol:2007sj,Erkol:2008gp} 
for a few representative papers. 

A few years ago, still another alternative analysis strategy was proposed in Refs.\,\cite{Lucha:2009uy,Lucha:2009jk,Lucha:2009et,Lucha:2010wu} and 
subsequently further developed and applied by the same group to various channels \cite{Lucha:2010ea,Lucha:2011zp,Lucha:2013gta,Lucha:2014xla,Lucha:2015xua,Lucha:2016nzv}. 
The essential idea of this approach is to promote the threshold parameter $s_{th}$, which conventionally is considered as a constant, to become a function which depends on the 
Borel mass $M$. As an ansatz, $s_{th}$ was proposed to be a power series of $\tau = 1/M^2$. The respective coefficients are obtained by demanding that the computed hadron mass 
value [Eq.\,(\ref{eq:pole.plus.continuum.mass.2})] is as close as possible to the experimental value over the whole range of the Borel window. 
In this approach, the hadron mass therefore is regraded as an input. Instead, it is possible to compute the residue $|\lambda|^2$ (often referred to as 
``decay constant'') with improved precision compared to the conventional approach. 

\subsubsection{The maximum entropy method \label{MEM.section}}
Recently, a novel prescription to analyze QCDSRs, based on Bayesian inference theory, was proposed in 
Ref.\,\cite{Gubler:2010cf}. The advantage of this approach, which is commonly referred to as the maximum 
entropy method (MEM), is that it does not require any explicit assumption about the form of the spectral function 
such as the ``pole + continuum'' ansatz of Eq.\,(\ref{eq:pole.plus.continuum}). As this approach is still relatively new 
and differs from the previously mentioned methods in many respects, we will briefly recapitulate it here. 
For more details, see Refs.\,\cite{Gubler:2013moa,Gubler:2010cf} and the references cited 
therein. 

The basic problem to be solved by MEM can be written down as 
\begin{equation}
G(x) = \int_{0}^{\infty} d\omega K(x,\omega) \rho(\omega), 
\label{eq:MEM.1}
\end{equation}
where $G(x)$ is given for a limited range of $x$ (for Borel-type QCDSRs, $x = M$) or for a finite number 
of data points (this happens in the imaginary time formalism of Monte-Carlo approaches, such as lattice QCD, 
where $x$ stands for imaginary time), with an attached error. $K(x,\omega)$ is the kernel, which for the sum rules 
of Eq.\,(\ref{eq:Boreltrans2}) becomes $2 \omega e^{-\omega^2/M^2}/M^2$ with $s = \omega^2$. Solving Eq.\,(\ref{eq:MEM.1}) for $\rho(\omega)$ is generally an 
ill-posed problem. The strategy often adopted is hence to make an educated guess about the form of 
$\rho(\omega)$, parametrize it with a small number of parameters and then to fit these parameters such that Eq.\,(\ref{eq:MEM.1}) 
is satisfied as accurately as possible. This is what is done in the conventional QCDSR analysis 
described in Section,\ref{sec:conventional.analysis}. 
In cases where the ``pole + continuum'' description is qualitatively accurate, it will likely be useful 
and produce approximately correct findings. However, if, say, the spectral function at low energy is dominated 
by a flat continuum instead of a narrow peak, the ``pole + continuum'' can potentially lead to misleading results. 

In contrast, MEM does not need any strong assumption about $\rho(\omega)$, but instead aims at 
providing its most probable form, given all the available information, such as asymptotic values 
of $\rho(\omega)$ and the positive definiteness of this function. For this purpose, one makes use of Bayes' theorem, 
which in the present context reads
\begin{equation}
P[\rho|G I] = \frac{P[G| \rho I] P[\rho|I]}{P[G |I]}.  
\label{eq:bayes.theorem}
\end{equation}
Here, $P[A|B]$ is a conditional probability, for the event $A$ to be realized under the condition of event $B$. In Eq.\,(\ref{eq:bayes.theorem}),  
``$\rho$'' is a specific form of the spectral function, 
``$G$'' the information from the OPE [the left hand side of Eq.\,(\ref{eq:Boreltrans2})] and ``I'' prior information 
about the spectral function such as positive definiteness and asymptotic values. 
Finding the maximum of $P[\rho|G I]$ 
will give the most probable form of $\rho(x)$. $P[G |\rho I]$ is usually referred to as ``likelihood 
function" and $P[\rho|I]$ as the ``prior probability". Ignoring the prior probability and 
maximizing only the likelihood function corresponds to ordinary $\chi^2$-fitting. 

Let us discuss the forms usually used for the likelihood function and the prior probability and 
in particular how they can be formulated for QCDSRs. For simplicity, we will here only consider the 
application to the Borel sum rules of Eq.\,(\ref{eq:Boreltrans2}). The MEM treatment for Gaussian sum rules 
is discussed in Ref.\,\cite{Ohtani:2011zz}. 
To determine the likelihood function, 
we assume that the values of the function $G(x)$ are distributed according to uncorrelated Gaussian distributions. 
For the QCDSR analysis discussed here, we will numerically generate uncorrelated values for each used 
data point of $G(x)$ that follow a Gaussian distribution and hence satisfy this assumption. 
One then has   
\begin{equation}
\begin{split}
P[G|\rho I] &= e^{-L[\rho]}, \\
L[\rho] &= \frac{1}{2(x_{\mathrm{max}} - x_{\mathrm{min}})}  
\displaystyle \int_{x_{\mathrm{min}}}^{x_{\mathrm{max}}} 
dx \frac{ \bigl[G(x) - G_{\rho}(x) \bigr]^2}{\sigma^2(x)}.   
\end{split}
\label{eq:likelihood}
\end{equation}
If $G(x)$ is obtained using Monte-Carlo methods such as in lattice QCD, the correlation between 
the values of $G(x)$ at different $x$ have to be taken into account by the use of the covariance 
matrix \cite{Jarrell:1996rrw,Asakawa:2000tr}. 
$\sigma(x)$ stands for the uncertainty of $G(x)$ at $x$ and 
$G_{\rho}(x)$ is defined as the integral on the left hand side of Eq.\,(\ref{eq:MEM.1}). 

The prior probability should quantify the prior knowledge of $\rho(\omega)$ such as positivity 
and asymptotic values. While several parametrizations have been proposed in the literature 
(see for instance Ref.\,\cite{Burnier:2013nla}), the one used most frequently makes use of 
the Shannon-Jaynes entropy $S[\rho]$, giving 
\begin{equation}
\begin{split}
P[\rho|I] &= e^{\alpha S[\rho]}, \\
S[\rho] &= \displaystyle \int_0^{\infty} d\omega \Bigr[ \rho(\omega) - m(\omega) - \rho(\omega)\log \Bigl( \frac{\rho(\omega)}{m(\omega)} \Bigr) \Bigl].
\end{split}
\label{eq:prior}
\end{equation}
Here, the function $m(\omega)$, which an input in the MEM analysis,  is referred to as ``default model". 
In the case of no available data $G(x)$, MEM just gives $m(\omega)$ for $\rho(\omega)$ because this function maximizes $P[\rho|I]$.
The default model is often used to fix asymptotic values of the spectral function to analytically known results. 
In MEM studies of both QCD sum rules and lattice QCD, the default model is often set to the asymptotic high energy 
limit of the spectral function, which is known from perturbation theory. 
The scaling factor $\alpha$, introduced in Eq.\,(\ref{eq:prior}), will be integrated out in a later step 
of the MEM procedure. 
The Shannon-Jaynes entropy of Eq.\,(\ref{eq:prior}) can be derived from the law of large numbers or 
axiomatically constructed from requirements such as locality, coordinate invariance, system independence and scaling \cite{Gubler:2013moa,Asakawa:2000tr}. 
For the actual calculations, the integrals of Eqs.\,(\ref{eq:MEM.1}), 
(\ref{eq:likelihood}) and (\ref{eq:prior}) will be approximated as sums over discrete points using the 
trapezoidal rule. 

From the above results, the needed probability $P[\rho|G I]$ can be obtained as 
\begin{align}
P[\rho|G I] & \propto P[G|\rho I] P[\rho|I] = e^{Q[\rho]}, \\
Q[\rho] & \equiv  \alpha S[\rho] - L[\rho].
\label{eq:finalprob}
\end{align}
Determining the form of $\rho(\omega)$ which maximizes $Q[\rho]$ and, therefore, is the 
most probable $\rho(\omega)$ given $G(x)$ and $I$, is now merely a numerical problem, for which the 
so-called Bryan algorithm is frequently used \cite{Bryan1990}. This algorithm, 
which uses the singular-value decomposition to reduce the dimension of the configuration space of $\rho(\omega)$ and 
therefore largely reduces the calculation time, has indeed been employed in all 
studies applying MEM to QCDSRs so far.  
It can moreover be proven that the maximum of $Q[\rho]$ is unique if it exists and, therefore, the problem of 
local minima does not occur \cite{Asakawa:2000tr}. 
Once $\rho_{\alpha}(x)$, which maximizes $Q[\rho]$ at a specific $\alpha$ value is found, this parameter is 
eliminated by  
averaging $\rho(\omega)$ over a range of $\alpha$ and assuming that $P[\rho|G I]$ is sharply peaked around 
its maximum $P[\rho_{\alpha}|G I]$. 
The details of this step, which we will not discuss here, are for instance explained in Ref.\,\cite{Gubler:2013moa}. 

\begin{figure}[tb]
\begin{center}
\begin{minipage}[t]{8 cm}
\vspace{0.5 cm}
\hspace{0.5 cm}
\includegraphics[width=9cm,bb=0 0 477 401]{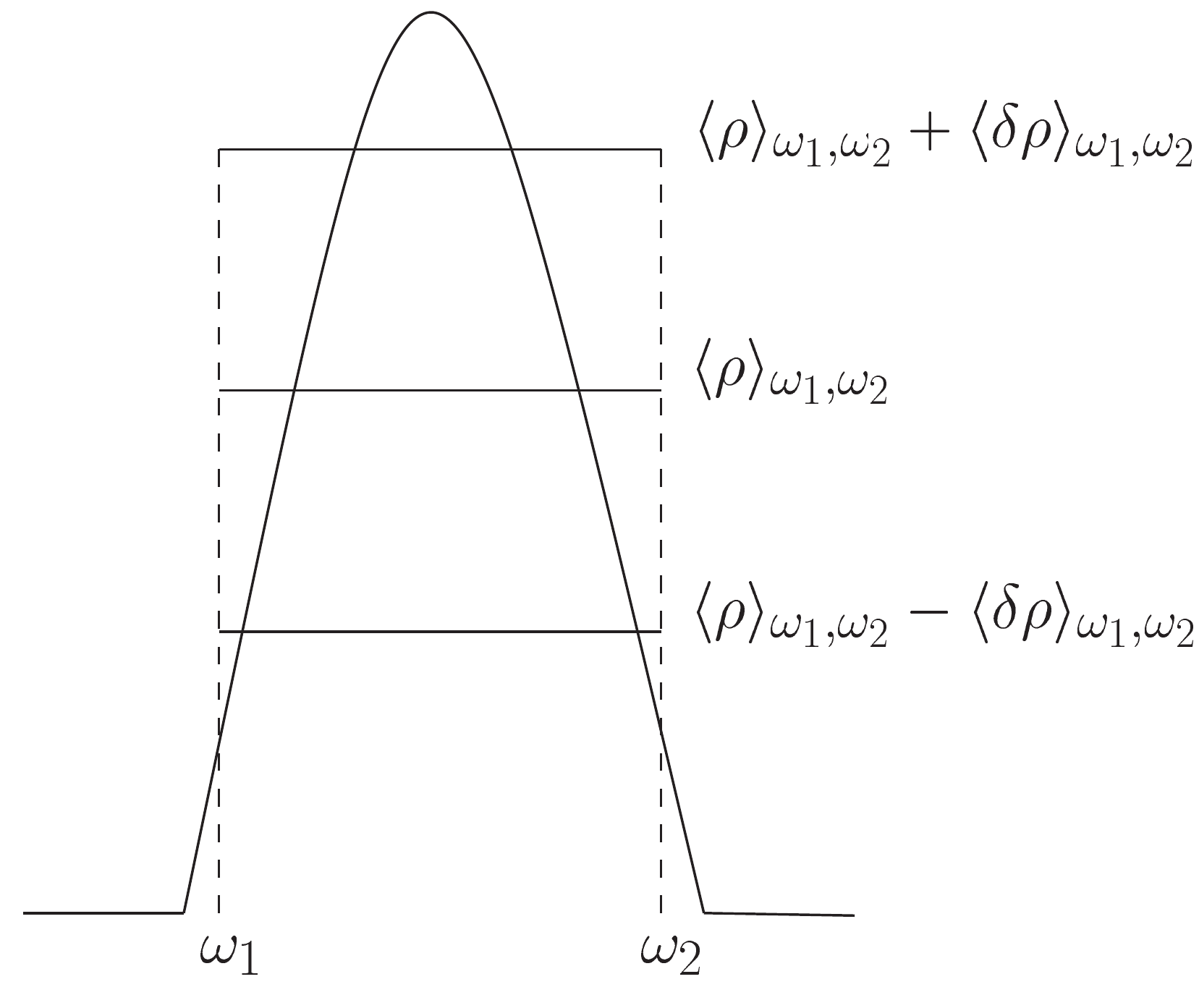}
\end{minipage}
\begin{minipage}[t]{16.5 cm}
\caption{
Illustration of a typical error bar adapted in MEM studies. The solid 
line depicting a peak stands for the spectral function $\rho(\omega)$. 
The three horizontal bars indicate the height of the mean value 
of the spectral function over the 
interval ($\omega_1$, $\omega_2$), $\langle \rho \rangle_{\omega_1,\omega_2}$, and of the corresponding errors added to and subtracted from it. 
Taken from Fig.\,4.2 of Ref.\,\cite{Gubler:2013moa}. 
\label{fig:error.MEM}}
\end{minipage}
\end{center}
\end{figure}

One important and useful feature of MEM is its ability to provide 
error estimates for averages of the spectral function over some range of $\omega$. 
Defining the variance of $\rho(\omega)$ from its most probable form for fixed $\alpha$ as  
$\delta \rho(\omega)$, its squared average over the interval ($\omega_1$, $\omega_2$)
can be given as 
\begin{align}
\langle (\delta \rho)^2 \rangle_{\omega_1,\,\omega_2} & \equiv \frac{1}{(\omega_2 - \omega_1)^2}
\displaystyle \int [d \rho] \displaystyle \int_{\omega_1}^{\omega_2} d\omega d\omega^{\prime} 
\delta \rho(\omega) \delta \rho(\omega^{\prime}) P[\rho|G I] \nonumber \\
& = - \frac{1}{(\omega_2 - \omega_1)^2} \displaystyle \int_{\omega_1}^{\omega_2} d\omega d\omega^{\prime} 
\biggl( \frac{\delta^2 Q}{\delta \rho(\omega) \delta \rho(\omega^{\prime})} \biggr)^{-1} \bigg|_{\rho=\rho_{\alpha}},
\label{eq:error} 
\end{align}
where the definition 
\begin{equation}
[d \rho] \equiv \prod_i \frac{d \rho_i}{\sqrt{\rho_i}},  
\label{eq:measure.rho}
\end{equation}
was used. The 
$\rho_i$ here stands for the value of $\rho(\omega_i)$ at the discretized position $\omega_i$. 
In going from the first to the second line in Eq.\,(\ref{eq:error}), the Gaussian approximation 
for the probability $P[\rho|G I]$ was employed. The final error $\langle \delta \rho \rangle_{\omega_1,\,\omega_2}$ can then 
be obtained by taking the average of $\sqrt{\langle (\delta \rho)^2 \rangle_{\omega_1,\,\omega_2}}$ over $\alpha$. 
Usually, the interval ($\omega_1$, $\omega_2$) is taken to cover a peak or some other structure of interest, as 
illustrated in Fig.\,\ref{fig:error.MEM}. 
%%%%%%%%%%%%%%%%%%%%%%%%%%%%%%%%%%%%%%%%%%%%%%%%%%%%%%%%%%%%%%%%%%%%%%%%%%%%%%%%%%%%%%%%%%%%%%%%%%%%%%%%%%%%%%%%%%%%
%%%%%%%%%%%%%%%%%%%%%%%%%%%%%%%%%%%%%%%%%%%%%%%%%%%%%%%%%%%%%%%%%%%%%%%%%%%%%%%%%%%%%%%%%%%%%%%%%%%%%%%%%%%%%%%%%%%%
Information about the error of the spectral function is valuable, for instance, to make an informed judgement about the statistical 
significance of an extracted peak. 

\begin{figure}[tb]
\begin{center}
\begin{minipage}[t]{8 cm}
\vspace{0.5 cm}
\hspace{-1.5 cm}
\includegraphics[width=11cm,bb=0 0 360 252]{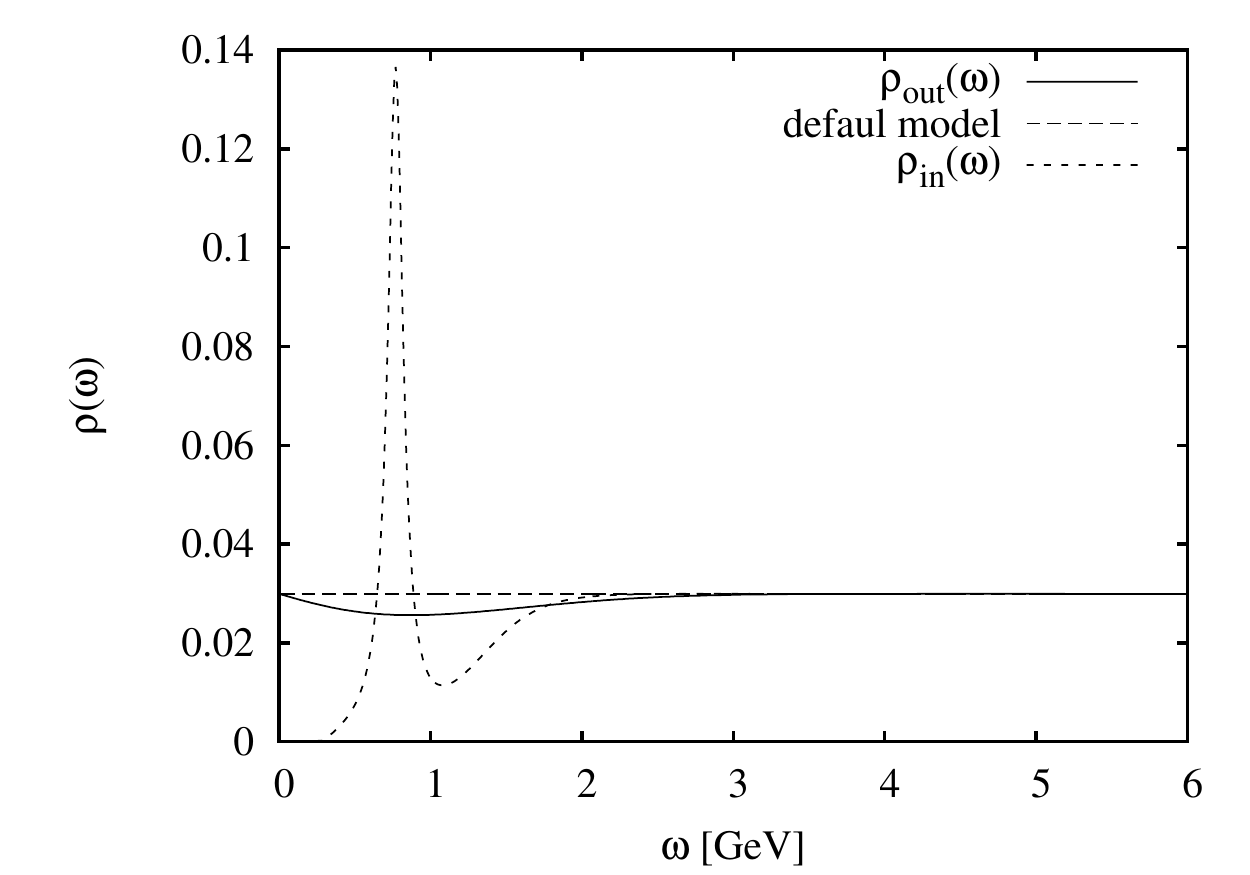}
\end{minipage}
\begin{minipage}[t]{16.5 cm}
\caption{Result of an MEM analysis using a constant default model with its value fixed to the
perturbative QCD limit. $\rho_{\mathrm{in}}(\omega)$ is the function that was used to produce the mock data, and 
$\rho_{\mathrm{out}}(\omega)$ shows the spectral function extracted by MEM. Taken from Fig.\,3 of Ref.\,\cite{Gubler:2010cf}. 
\label{fig:MEM.rho.1}}
\end{minipage}
\end{center}
\end{figure}

\begin{figure}[tb]
\begin{center}
\begin{minipage}[t]{8 cm}
%\vspace{1 cm}
\vspace{0.5 cm}
\hspace{-4.0 cm}
\includegraphics[width=16cm,bb=0 0 360 252]{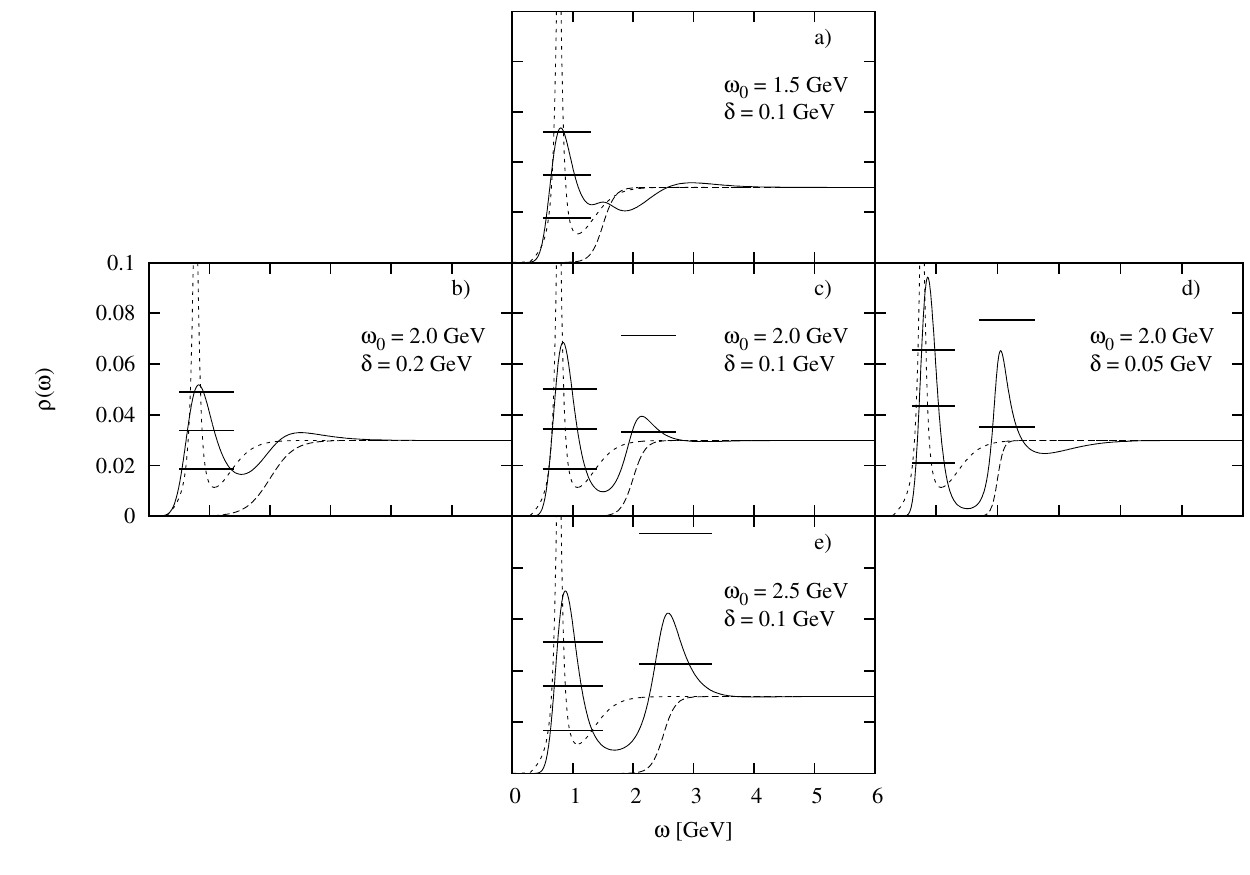}
\vspace{-0.8 cm}
\end{minipage}
\begin{minipage}[t]{16.5 cm}
\caption{Results of the MEM analyses of mock data with various default models. As in Fig.\,\ref{fig:MEM.rho.1},
the solid lines stand for the output of the analysis $\rho_{\mathrm{out}}(\omega)$, the long-dashed lines for the default
model with the parameters shown in the figure, and the short-dashed lines for the input spectral function, $\rho_{\mathrm{in}}(\omega)$. 
The horizontal bars show the values of the spectral function, averaged over
the peaks and the corresponding ranges as illustrated in Fig.\,\ref{fig:error.MEM}. 
For figures c), d) and e), the lower error bars of the second
peak are not shown because they lie below the $\omega$ axis. Taken from Fig.\,4 of Ref.\,\cite{Gubler:2010cf}. 
\label{fig:MEM.rho.2}}
\end{minipage}
\end{center}
\end{figure}

Finally, we review some representative findings of QCDSR MEM analyses. The first one 
was carried out for the $\rho$ meson channel in Ref.\,\cite{Gubler:2010cf}. It was found in this work that it is indeed 
possible to apply MEM to QCDSRs, but only with a default model that has the correct behavior in the low and high energy 
limits. In the $\rho$ meson channel, the spectral function is known to vanish in the low energy limit ($\omega \to 0$), as there 
are no massless states with $\rho$ meson quantum numbers. 
At high energy ($\omega \to \infty$) the spectral function has to approach 
the perturbative QCD limit. 
This finding is illustrated in Figs.\,\ref{fig:MEM.rho.1} and \ref{fig:MEM.rho.2}, where results of test MEM analyses of mock data are shown. 
%%%%%%%%%%%%%%%%%%%%%%%%%%%%%%%%%%%%%%%%%%%%%%%%%%%%%%%%%%%%%%%%%%%%%%%%%%%%%%%%%%%%%%%%%%%%%%%%%%%%%%%%%%%%%%%%%%%%
%%%%%%%%%%%%%%%%%%%%%%%%%%%%%%%%%%%%%%%%%%%%%%%%%%%%%%%%%%%%%%%%%%%%%%%%%%%%%%%%%%%%%%%%%%%%%%%%%%%%%%%%%%%%%%%%%%%%
A Borel kernel was used with a range of the Borel mass equivalent to the actual Borel window in the $\rho$ 
meson sum rule to obtain these results. Furthermore, the error used in this analysis was generated from the uncertainties of the condensates via 
the OPE expression of the $\rho$ channel.  
The correct spectral function, denoted as $\rho_{\mathrm{in}}(\omega)$ (which should be reproduced if MEM works perfectly) 
is depicted as a short-dashed line. 
In Fig.\,\ref{fig:MEM.rho.1}, an analysis with a constant default model (long-dashed line) matched to the perturbative high energy limit 
is shown. As this model has the wrong low energy limit, the MEM analysis does not work 
well and does not reproduce any significant $\rho$ meson peak. 
%%%%%%%%%%%%%%%%%%%%%%%%%%%%%%%%%%%%%%%%%%%%%%%%%%%%%%%%%%%%%%%%%%%%%%%%%%%%%%%%%%%%%%%%%%%%%%%%%%%%%%%%%%%%%%%%%%%%
%%%%%%%%%%%%%%%%%%%%%%%%%%%%%%%%%%%%%%%%%%%%%%%%%%%%%%%%%%%%%%%%%%%%%%%%%%%%%%%%%%%%%%%%%%%%%%%%%%%%%%%%%%%%%%%%%%%%
In Fig.\,\ref{fig:MEM.rho.2}, default models with correct low and 
high energy limits are used, leading to approximate reproductions of the $\rho$ meson peak. 
Specifically, 
\begin{equation}
m(\omega) = \frac{1}{4 \pi^2} \Bigl(1 + \frac{\alpha_s}{\pi} \Bigr) \frac{1}{1 + e^{(\omega_0 - \omega)/\delta}}
\label{eq:default.model.rho}
\end{equation}
was employed as the default model with various values for $\omega_0$ and $\delta$. 
These are shown in Fig.\,\ref{fig:MEM.rho.2} as long-dashed lines. 
As can be seen in Fig.\,\ref{fig:MEM.rho.2}, the details of the spectral functions extracted by MEM (solid lines) clearly 
depend on the chosen default model. The position of the lowest peak, however, approximately stays at the same position. 
Furthermore, MEM is not able to reproduce the $\rho$ meson width correctly, but instead produces a somewhat broader peak. 
This is a general feature of MEM. If the input data are precise enough, the position of the lowest peak can usually be  
reproduced quite well. The spectral function is however often smeared out, such that narrow peak widths are difficult 
to extract. 
This in some sense corresponds to the use of the pole term in the ``pole + continuum'' ansatz of Eq.\,(\ref{eq:pole.plus.continuum}), where the 
ground state is approximated by a delta function and one does not attempt to extract the peak width from the sum rules. 

Let us review one concrete example of an application of MEM to QCDSRs. 
The main advantage of the MEM approach compared to conventional methods is that one does not have to assume 
any specific functional form for the spectral function. MEM is therefore especially useful when one does not have 
any prior knowledge about the spectral function or when one wants to study the (unknown) modification of some spectral 
function in an extreme environment such as a hot or dense medium. 
As an example, we here summarize a study of charmonium at finite temperature \cite{Gubler:2011ua}. 
The melting of charmonium has long been considered to be a signal of the quark gluon plasma 
formation in heavy-ion collisions \cite{Matsui:1986dk,Hashimoto:1986nn} and has thus attracted much interest 
from both theoreticians and experimentalists. 
However, directly computing the charmonium spectral function at finite temperature from first principles of 
QCD is challenging even today. This is partly due to the fact that even though lattice QCD is by now able to perform 
precise calculations at finite temperature, it is only directly applicable to static quantities and not dynamical ones such as spectral functions. 
Lattice QCD can so far only compute correlators at imaginary time, which are related to certain integrals of the respective spectral function. 

The OPE side of QCD sum rules for charmonium (and similarly, bottomonium) of any channel $J$ can, after applying the Borel transform, be cast 
in the following form: 
\begin{align}
\mathcal{M}^{\mathrm{J}}(\nu) =  e^{-\nu}A^{\mathrm{J}}(\nu)[1 + \alpha_s(\nu) a^{\mathrm{J}}(\nu) + b^{\mathrm{J}}(\nu) \phi_b(T) + c^{\mathrm{J}}(\nu) \phi_c(T) +  d^{\mathrm{J}}(\nu) \phi_d(T)].
\label{eq:OPEFT}
\end{align}
Here, $\nu \equiv 4m_c^2/M^2$, with the charm quark mass $m_c$ and the Borel mass $M$. Because the heavy quark 
condensates can all be expressed as gluonic condensates with the help of the heavy quark expansion, the OPE 
only contains gluonic condensates as non-perturbative contributions. Light quark condensates can in principle appear 
at higher orders in $\alpha_s$, but are expected to be strongly suppressed. 
The first two terms in 
Eq.\,(\ref{eq:OPEFT}) are the leading order perturbative term and its first order 
$\alpha_s$ correction. The third and fourth terms contain the scalar and 
spin-2 gluon condensates of mass dimension 4: 
\begin{align}
\phi_b(T) &= \frac{4\pi^2}{9(4m_c^2)^2} G_0, \\
\phi_c(T) &= \frac{4\pi^2}{3(4m_c^2)^2} G_2, 
\end{align}
where 
\begin{equation}
G_0 = \langle \frac{\alpha_s}{\pi} G^a_{\mu\nu} G^{a\mu\nu} \rangle_{T}, 
\end{equation}
which includes both vacuum and temperature dependent parts discussed around 
Eqs.\,(\ref{eq:gluon.cond.value}) and (\ref{eq:trace.anomaly.5}), respectively. 
$G_2$ is defined similarly to Eq.\,(\ref{eq:nonlocal.dim4.temp.15.2}), but with an additional 
factor of $\alpha_s(T)/\pi$. 
The detailed expressions of the Wilson coefficients of the first four 
terms are given in Ref.\,\cite{Morita:2009qk}. 
In Ref.\,\cite{Gubler:2011ua}, only one dimension 6 term was 
taken into account, namely, 
\begin{equation}
\phi_d(T) = \frac{1}{(4m_c^2)^3}\langle g^3 f^{abc} G^{a\nu}_{\mu} G^{b\lambda}_{\nu} G^{c\mu}_{\lambda} \rangle_{T}. 
\end{equation}
The more complete list of dimension 6 terms together with the corresponding OPE expressions is given in Ref.\,\cite{Kim:2000kj}. 
Their influence on the sum rule results is discussed in Ref.\,\cite{Kim:2015xna}. The temperature dependences of the dimension 4 
gluonic condensates can be obtained as explained in Sec.\,\ref{conds_T}. In Ref.\,\cite{Gubler:2011ua} quenched lattice QCD data were used 
for this purpose. The dimension 6 term was estimated using the dilute instanton gas model. For more details, see Refs.\,\cite{Gubler:2013moa,Gubler:2011ua}. 

\begin{figure}[tb]
\begin{center}
\begin{minipage}[t]{16.5 cm}
\vspace{0.5 cm}
\includegraphics[width=8.0cm,bb=0 0 259 290]{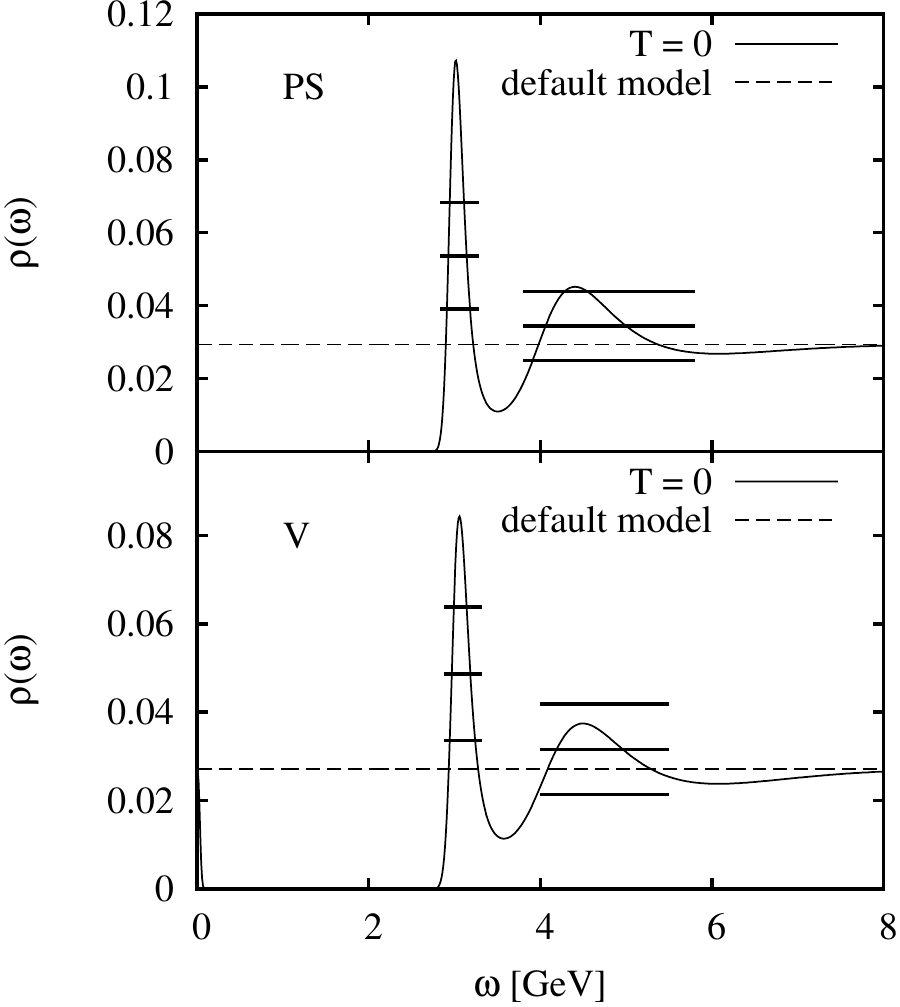}
\includegraphics[width=8.0cm,bb=0 0 259 290]{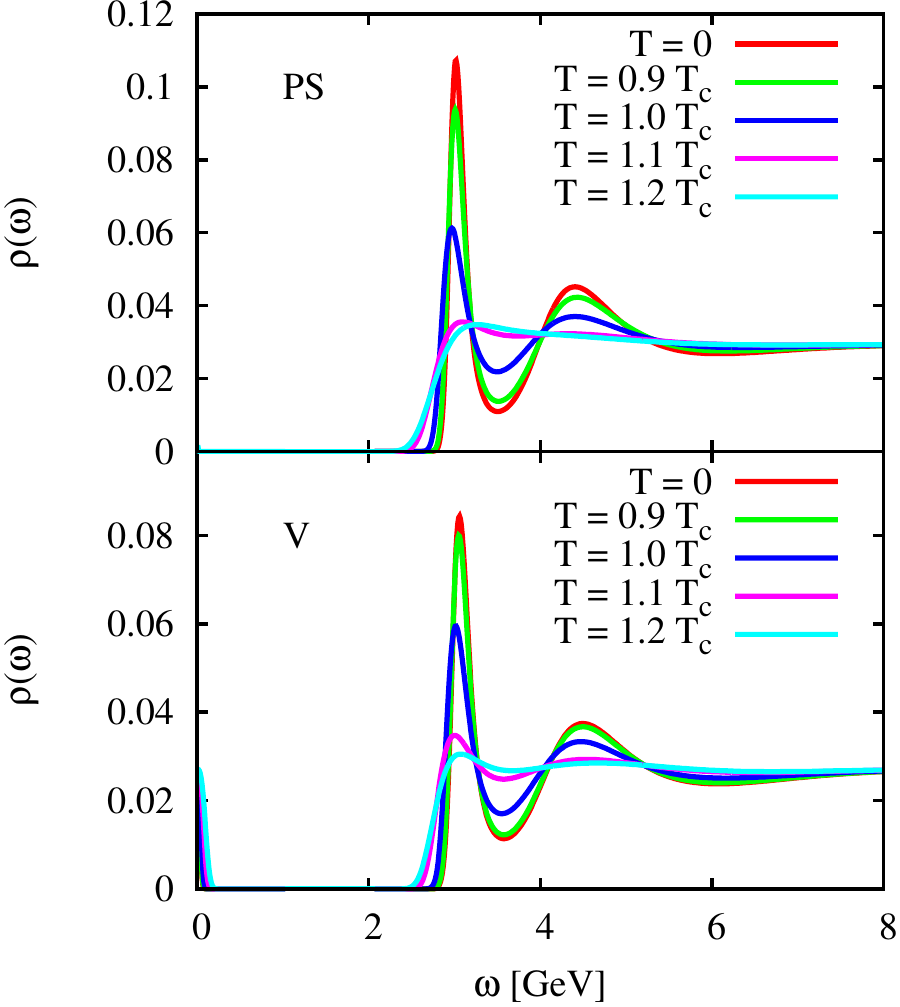}
\end{minipage}
\begin{minipage}[t]{16.5 cm}
\caption{Left plot: spectral functions in the pseudoscalar (upper plot) and vector (lower plot) channel at $T = 0$, 
with corresponding errors, as illustrated in Fig.\,\ref{fig:error.MEM}. The dashed lines show the default model used in the 
MEM analysis. Right plot: the same spectral functions at temperatures around $T_c$. Taken from Fig.\,1 of Ref.\,\cite{Gubler:2011ua}.   
\label{fig:charmonium_0_T}}
\end{minipage}
\end{center}
\end{figure}

In the notation of Eq.\,(\ref{eq:OPEFT}), the sum rule can be expressed as 
\begin{equation}
\mathcal{M}^{\mathrm{J}}(\nu) = \displaystyle \int_0^{\infty}dx^2  e^{-x^2 \nu} \rho^{\mathrm{J}}(2m_h x).
\label{eq:dispersionFT}
\end{equation} 
With Eq.\,(\ref{eq:OPEFT}) at hand and gluon condensates determined, 
the remaining task is to use MEM to extract the spectral function from Eq.\,(\ref{eq:dispersionFT}). 
The results of such an analysis are shown in Fig.\,\ref{fig:charmonium_0_T}. 
%%%%%%%%%%%%%%%%%%%%%%%%%%%%%%%%%%%%%%%%%%%%%%%%%%%%%%%%%%%%%%%%%%%%%%%%%%%%%%%%%%%%%%%%%%%%%%%%%%%%%%%%%%%%%%%%%%%%
%%%%%%%%%%%%%%%%%%%%%%%%%%%%%%%%%%%%%%%%%%%%%%%%%%%%%%%%%%%%%%%%%%%%%%%%%%%%%%%%%%%%%%%%%%%%%%%%%%%%%%%%%%%%%%%%%%%%
Let us make a few comments about the obtained spectral functions, focusing especially on the vector 
channel (lower plots in Fig.\,\ref{fig:charmonium_0_T}), which is most relevant for experiment. 
For the vacuum spectrum shown on the left side, a clear peak is 
observed slightly above 3 GeV. This peak in essence corresponds to the $J/\Psi$ state, 
but also contains some contributions from its first and second excited states, $\Psi'$ and $\Psi(3770)$. 
This is related to the large and artificial width that is generated by MEM due to its limited resolution. 
For a more detailed discussion about this point, including MEM analyses of mock data, see Ref.\,\cite{Gubler:2013moa}. 
The finite temperature results shown on the right plot of Fig.\,\ref{fig:charmonium_0_T} show a sudden disappearance 
(melting) of the lowest peak right above $T_c$. This sudden change of the spectrum is caused by 
the strong temperature dependence of the gluon condensates around $T_c$. For a similar calculation 
for bottomonium, see Ref.\,\cite{Suzuki:2012ze}. For more recent work with an improved kernel and hence an MEM analysis with 
better resolution, consult Ref.\,\cite{Araki:2017ebb}. 

%%%%%%%%%%%%%%%%%%%%
\section{Hadrons at finite density \label{sec:HadronsDensity}} %meson, nucleon
%Philipp will write.
\subsection{Physics motivation}
Understanding the behavior of hadrons in a dense environment such as nuclear matter has been 
the motivation not only for theoretical 
studies, but also for dedicated experimental projects for several decades (see Refs.\,\cite{Hayano:2008vn,Leupold:2009kz,Metag:2017yuh} for recent reviews). 
Worldwide, there are at present multiple 
experimental facilities that plan to investigate the properties of dense matter and its 
influence on hadrons. These include the J-PARC \cite{Sato:2009zze} facility in Japan, 
CBM \cite{Friman:2011zz,Ablyazimov:2017guv} and 
PANDA \cite{Lutz:2009ff} experiments at FAIR in Germany, HIAF \cite{Yang:2013yeb} in China as well as NICA \cite{Senger:2016wfb} in Russia. 

One of the goals of these experimental efforts is to detect signatures of the (partial) 
restoration of chiral symmetry at finite density. Defining such signals that are 
sufficiently simple and experimentally measurable in practice is, however, not a trivial task. 
One proposal to relate the restoration of chiral symmetry with physical observables was made in 
the early nineties by Brown and Rho in Ref.\,\cite{Brown:1991kk}, where hadron masses 
were conjectured to scale according to the behavior of the chiral condensate at finite density 
(the so-called Brown-Rho scaling). Not much later, more evidence for this scaling behavior 
was found in Ref.\,\cite{Hatsuda:1991ez} for the $\rho$ meson and other vector mesons, 
based on a QCDSR calculation at finite density. Later, QCDSR studies however found 
that they do \textit{not} necessarily imply a decreasing $\rho$ meson mass with finite 
density, but are also consistent with a scenario in which it is primarily 
broadened \cite{Leupold:1997dg,Klingl:1997kf}. 

The above history illustrates the basic motivation for studying the behavior of hadrons at finite density 
within QCDSRs, but also shows the limitations of the method. As QCDSRs provide a 
relation between integrals of the spectral function and various QCD condensates, it also 
relates the modification of the spectral function with the behavior of the condensates as a 
function of density. Therefore, the effects of the reduction of the chiral condensate in 
dense matter (and hence of the restoration of chiral symmetry) on the hadronic spectrum 
can in principle be studied. However, sum rules only provide information about certain integrals 
of the spectral function, not about detailed features of the structures that appear in them. Conclusions 
about the behavior of specific particles at finite density are thus not necessarily unique, 
as it was found in studies of the $\rho$ meson mentioned above. 
Furthermore, QCDSRs do not only involve the chiral condensate, but also other condensates 
such as the gluon condensate, four-quark condensates, the mixed condensate (involving both quark 
and gluon fields) and more condensates with higher mass dimensions, 
which might or might not be directly related to chiral symmetry and its restoration and hence can 
obscure the relation between the modification of spectral function and the restoration of 
chiral symmetry. 
In this context, especially the four-quark condensates have been considered frequently in recent 
studies \cite{Thomas:2007gx,Buchheim:2014rpa,Leupold:2005eq,Thomas:2005dc,Hilger:2010cn,Buchheim:2015yyc}. 

Keeping the above issues in mind, we will 
in the following discuss the behavior of various hadrons at finite density from a 
QCDSR perspective and review the progress that has been made during recent years. 
Wherever possible, we will furthermore try to assess what information about the QCD vacuum structure and 
its modification at finite density can be extracted from such QCDSR analyses for each specific channel. 

\subsection{Light hadrons}
We define light hadrons as hadrons containing $u$, $d$ or $s$ quarks or anti-quarks as valence quarks. 
The behavior of these hadrons at finite density has been studied intensively during the years, as they 
are relatively easy to produce in comparison with hadrons containing one or more heavy quarks. 
Among them, the vector mesons have attracted the most attention because they decay into di-leptons which 
do not feel the strong interaction and hence are not strongly distorted due to the surrounding 
nuclear medium. 
Therefore, their properties in a dense environment are one of the most suitable targets for experimental study. 
We will in this section thus focus on the light vector mesons, but also discuss other light hadron 
species in later Sections.  

\subsubsection{The $\rho$ meson}
Among the various light mesons, 
the modification of the $\rho$ meson spectral function has been investigated most extensively both in theory and 
experiment because its mass shift at finite density was originally regarded as the most promising candidate to detect 
the partial restoration of chiral symmetry in nuclear matter \cite{Hatsuda:1991ez}. Studies based on 
hadronic effective theory later however indicated that the $\rho$ meson (which is already 
rather broad in vacuum with a width of about $\Gamma_{\rho} \simeq 148\,\mathrm{MeV}$ \cite{Patrignani:2016xqp}), 
is more likely to be modified in a more complicated manner, that cannot be described by a simple mass shift and/or broadening. 
Typically, these calculations find an enhancement of the spectral strength in the low energy region below the original 
$\rho$ meson peak \cite{Klingl:1997kf,Rapp:1997fs,Peters:1997va,Post:2000qi,Post:2003hu}. The detailed form of the spectrum depends however 
quite strongly on the channel (longitudinal or transverse), the value of the spatial momentum and, most importantly, on the 
details of the employed model. 
Furthermore, as already mentioned earlier, it was demonstrated that QCDSRs are consistent not only with a negative 
mass shift of the $\rho$ meson at finite density, but also with a scenario in which the $\rho$ is primarily broadened and 
receives only a very small mass shift \cite{Leupold:1997dg,Klingl:1997kf}.  

We will here not go into the details of these past calculations, but mention some more recent studies that have been 
conducted based on the QCDSR approach. In Ref.\,\cite{Kwon:2008vq}, the usefulness and potential importance 
of spectral moments was emphasized in a study that made use of finite energy sum rules (see also the earlier work of Ref.\,\cite{Marco:1999xz}). 
It was moreover checked in the same work to what degree the sum rules are satisfied by phenomenologically obtained spectral functions. 
More about the spectral moments will be discussed later in Section \ref{sec:phi.meson.density} about the $\phi$ meson. 
It will for the moment suffice to 
mention that moments are directly related to QCD condensates of a specific dimension. This is different 
form the more widely used Borel sum rules, which involve expansions in inverse powers of the Borel mass and contain an 
infinite series of condensates with arbitrary dimension. Computing moments of experimentally measurable spectra therefore in principle 
allows one to ``measure" condensates of a specific dimension. For this to become a realistic possibility, a precise measurement of 
the spectral function in a wide energy range is however necessary, which is not an easy task. 
The authors of Ref.\,\cite{Hilger:2010cn} focused on the role of the four-quark condensates in the $\rho$ meson sum 
rules when the chiral symmetry gets restored. Specifically, they distinguished between chiral even (invariant) and odd (variant) 
four-quark condensates and studied the scenario in which only the chiral odd condensates vanish as chiral 
symmetry gets restored while the chiral even ones remain at their vacuum values. 
In Ref.\,\cite{Mishra:2014rha} the behavior of the $\rho$ (together with the $\omega$ and the $\phi$) was studied 
not in normal nuclear matter, but in hadronic matter containing strangeness, using a chiral SU(3) model to describe the behavior 
of the condensates for this case. 

\subsubsection{The $\omega$ meson}
Not much theoretical work based QCDSRs has been devoted to the $\omega$ meson in recent years. Even though its width 
of $\Gamma_{\omega} \simeq 8.5\,\mathrm{MeV}$ \cite{Patrignani:2016xqp} is more than an order of magnitude smaller 
than that of the $\rho$, the corresponding OPE expression is in fact almost the same as that of the $\rho$, the only 
difference coming from four-quark condensate terms, vanishing completely once factorization is assumed. 
This exemplifies the fact that QCDSRs generally only have a limited 
sensitivity to the decay widths of resonances. 
It also means that many conclusions obtained for the $\rho$ from QCDSR studies also apply for the $\omega$. 

On the experimental side, however, valuable new information about the behavior of the $\omega$ in 
nuclear matter has been obtained during the last few years. Namely, the mass shift and width 
of the $\omega$ at normal nuclear matter density $\rho_0$ have been measured with high precision \cite{Metag:2011ji,Thiel:2013cea,Friedrich:2014lba,Metag:2015lza,Kotulla:2008aa}. 
Recently, even results about the momentum dependence of its width at $\rho_0$ have become available \cite{Friedrich:2016cms}. 
It would therefore be meaningful to revisit the earlier sum rule calculations and to study how the new experimental findings could constrain the behavior 
of the condensates at finite density. 

Another interesting topic related to the $\omega$ is the study of its chiral partner (or partners) and how their spectra will eventually approach 
each other as chiral symmetry gets restored. Generally, it is known that the chiral partner of the $\omega$ will be an axial vector meson containing 
both $u$, $d$ and $s$ components, that is presumably a mixed state of the $f_1(1285)$ and the $f_1(1420)$. 
The $f_1(1285)$ [$f_1(1420)$] is widely believed to be 
dominated by $u$ and $d$ ($s$) quark components. In the recent work of Ref.\,\cite{Gubler:2016djf}, it was argued that if disconnected diagrams 
can be neglected, the chiral partner of the $\omega$ is the $f_1(1285)$ and that they therefore should approach each other with 
increasing density (see Section \ref{sec:f1.density}). 

\subsubsection{The $\phi$ meson \label{sec:phi.meson.density}}
The behavior of the $\phi$ meson in nuclear matter has recently attracted renewed theoretical interest, in part because of the various experimental 
studies that have been performed in the past few years \cite{Ishikawa:2004id,Muto:2005za,Sakuma:2006xc,Wood:2010ei,Polyanskiy:2010tj,Hartmann:2012ia} 
or that are planed for the future \cite{Aoki:2015qla}. 
Recent theoretical studies include works based on QCDSRs \cite{Gubler:2014pta}, 
various effective field theories \cite{Gubler:2015yna,Gubler:2016itj,Cabrera:2016rnc,Cobos-Martinez:2017vtr,Cabrera:2017agk} 
and a work examining 
the possibility of $\phi$-nucleus bound states \cite{Cobos-Martinez:2017woo}. 
Here, we will focus on theoretical investigations related to QCD sum rules and review them in some detail. 

We start from the correlator 
\begin{equation}
\Pi_{\mu\nu}(q) = i\displaystyle \int d^{4}x \,\,e^{iqx} \langle \mathrm{T} [j_{\mu}(x) j_{\nu}(0)] \rangle_{\rho}  
\label{eq:veccorr1}
\end{equation}
for the operator $j_{\mu}(x) = \overline{s}(x) \gamma_{\mu} s(x)$, which predominantly couples to the $\phi$ meson in the 
vicinity of its pole. 
We consider the correlator in contracted form,   
\begin{equation}
\Pi(q^2) = \frac{1}{3q^2} \Pi^{\mu}_{\mu}(q), 
\label{eq:veccorr2}
\end{equation}
which is sufficient when studying the $\phi$ meson at rest with respect to the nuclear medium. 
After computing the OPE, $\Pi(q^2)$ can generally be 
expressed as 
\begin{equation}
\Pi(q^2=-Q^2) =  -c_0 \log \Big(\frac{Q^2}{\mu^2}\Big) + \frac{c_2}{Q^2} + \frac{c_4}{Q^4} + \frac{c_6}{Q^6} +  \dots.   
\label{eq:ope1}
\end{equation}
We first consider the $\phi$ meson in vacuum ($\rho = 0$), 
where the first few coefficients $c_n$ are obtained as 
\begin{align} 
c_0 & =  \frac{1}{4 \pi^2}\Big(1 + \frac{\alpha_s}{\pi} \Big), \hspace{1cm}
c_2 = -\frac{3 m_s^2}{2 \pi^2}, \label{eq:operesult1} \\
c_4 & = \frac{1}{12} \langle 0| \frac{\alpha_s}{\pi} G^2 | 0 \rangle + 2m_s \langle 0| \overline{s} s | 0 \rangle, \label{eq:c4} \\
c_6 & = -2 \pi \alpha_s \Big[ \langle 0| (\overline{s}\,\gamma_{\mu} \gamma_5 \,\lambda^a\,s)^2 | 0 \rangle  + \frac{2}{9} 
         \langle 0 | (\overline{s}\,\gamma_{\mu} \,\lambda^a \,s) \sum_{q=u,d,s} (\overline{q}\,\gamma_{\mu} \,\lambda^a \,q) | 0 \rangle \Big]. 
\label{eq:operesult3}
\end{align}
Here, we have kept only the most important terms. Higher order corrections due to the strange quark mass $m_s$ or the strong coupling constant $\alpha_s$ 
have been considered for instance in Ref.\,\cite{Gubler:2014pta} and shown not to change the qualitative behavior of the result. 
Also, numerical analyses show that the above expression is consistent with the $\phi$ meson dominating the spectral function at low 
energy and with a vacuum mass close to its experimental value. 
Especially the $m_s^2$ term is crucial in generating a $\phi$ mass that is heavier compared to the $\rho$ or $\omega$. 

Next, we turn to the finite density case, where the condensates already present in the vacuum get 
modified. Furthermore, new condensates appear due to the breaking of Lorentz symmetry related to 
the presence of nuclear matter. The details of these condensate modifications (within the linear density approximation) 
are discussed in Section \ref{conds.at.finite.den}. As a result, the above coefficients $c_n$ are modified as follows 
at linear order in density,  
\begin{align} 
\delta c_0 & = 0, \hspace{1cm}
\delta c_2 = 0, \label{eq:operesult4} \\
\delta c_4 & = \Big(- \frac{2}{27} M_N + \frac{56}{27} \sigma_{sN} + \frac{2}{27} \sigma_{\pi N} + A^s_2 \,M_N \Big) \rho , \label{eq:dc4} \\
\delta c_6 & = - \pi \alpha_s  \frac{448}{81} \kappa_N(\rho) \frac{\sigma_{sN}}{m_s} \, \langle 0 | \overline{s} s | 0 \rangle \, \rho , 
\label{eq:operesult5}
\end{align}
where again only the most essential terms have been taken into account. A more complete 
compilation can be found in Refs.\,\cite{Kim:2017nyg,Gubler:2014pta}. Especially, Ref.\,\cite{Kim:2017nyg} 
compiles the complete list of all possible operators and their Wilson coefficients at leading order in $\alpha_s$ up to 
dimension 6. 
Eqs.\,(\ref{eq:operesult4}-\ref{eq:operesult5}) are 
true only at leading order in $\rho$ and should hence not be 
trusted for densities much larger than normal nuclear matter density. 

\begin{figure}[tb]
\begin{center}
\begin{minipage}[t]{8 cm}
\vspace{0.5 cm}
\hspace{-2.0 cm}
\includegraphics[width=12cm,bb=0 0 360 252]{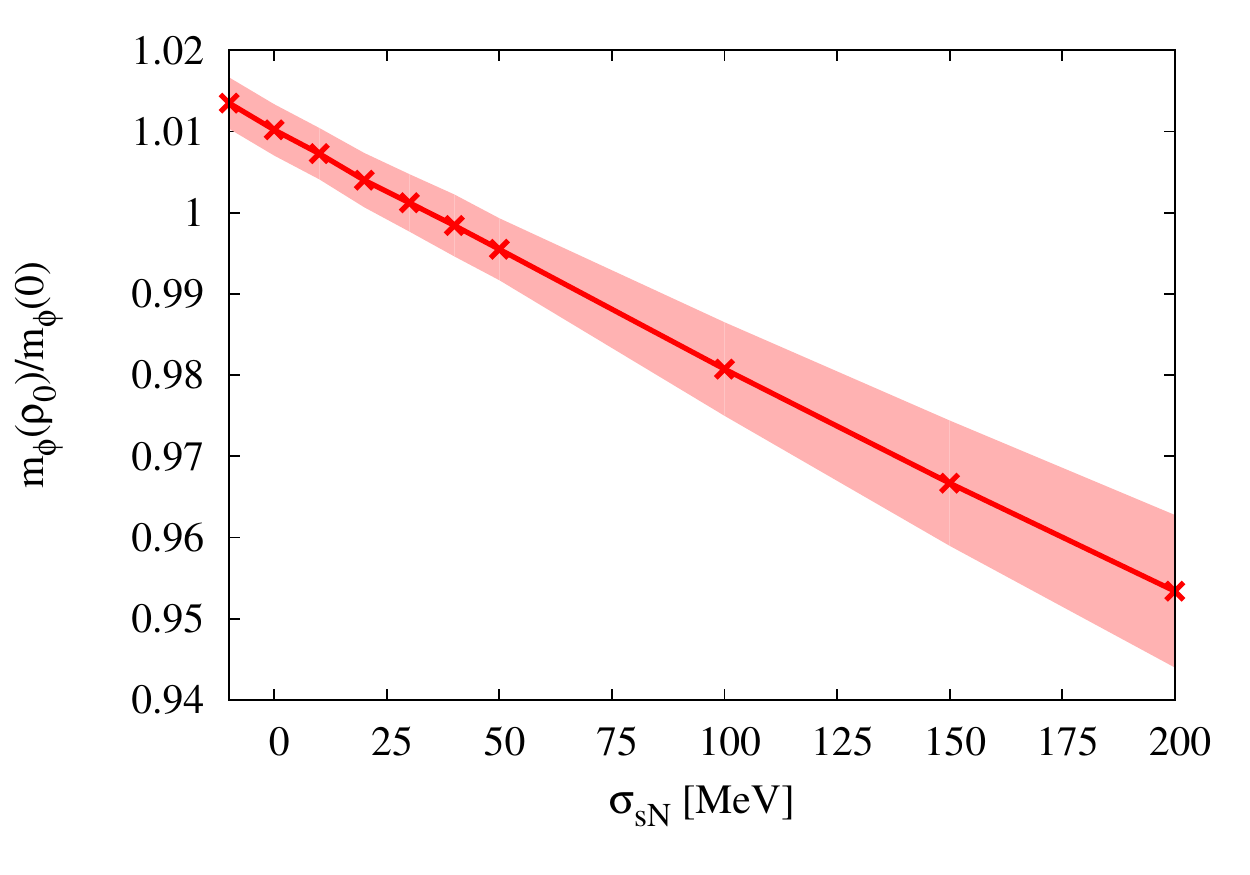} 
\vspace{-0.8 cm}
\end{minipage}
\begin{minipage}[t]{16.5 cm}
\caption{The $\phi$ meson mass at normal nuclear matter density, normalized by its vacuum value, as a function of the strange quark sigma 
term $\sigma_{sN}$. Taken from the lower plot in Fig.\,4 of Ref.\,\cite{Gubler:2014pta}. 
\label{fig:phi.sigmaNs}}
\end{minipage}
\end{center}
\end{figure}

With the above input and the numerical values of the parameters discussed in Section \ref{conds.at.finite.den}, one can now study the 
sum rules of the $\phi$ meson channel both in vacuum and nuclear matter. 
The most important quantity to be studied in such an analysis will be 
the mass shift of the 
$\phi$ peak. Such work was carried out in Ref.\,\cite{Gubler:2014pta}, where MEM was used for the analysis of the sum rules. 
The central result is reproduced in Fig.\,\ref{fig:phi.sigmaNs}, which shows the $\phi$ meson mass (normalized by its vacuum value) at normal nuclear 
matter density as a function of the strange quark sigma term $\sigma_{sN}$. 
%%%%%%%%%%%%%%%%%%%%%%%%%%%%%%%%%%%%%%%%%%%%%%%%%%%%%%%%%%%%%%%%%%%%%%%%%%%%%%%%%%%%%%%%%%%%%%%%%%%%%%%%%%%%%%%%%%%%
%%%%%%%%%%%%%%%%%%%%%%%%%%%%%%%%%%%%%%%%%%%%%%%%%%%%%%%%%%%%%%%%%%%%%%%%%%%%%%%%%%%%%%%%%%%%%%%%%%%%%%%%%%%%%%%%%%%%
It is observed in this figure that the $\phi$ meson mass shift is rather sensitive to the value of $\sigma_{sN}$, which we 
have discussed in Section \ref{conds.at.finite.den}. Even the sign of the mass shift depends crucially on $\sigma_{sN}$. This means that 
a measurement of the $\phi$ meson mass shift could help constraining the value of $\sigma_{sN}$, which still has large uncertainties 
even in state-of-the-art lattice QCD calculations, as can be seen in Table\,\ref{tab:strange.sigma.term.values}. 

An alternative point of view, which was already emphasized in Ref.\,\cite{Kwon:2008vq} for the $\rho$ meson, was 
discussed for the $\phi$ meson spectral function in Refs.\,\cite{Gubler:2015yna,Gubler:2016itj}. In these works, the importance and 
usefulness of spectral moments was stressed, which have been discussed in the QCDSR literature under the name of 
finite energy sum rules. For the $\phi$ meson spectral function discussed above, they can be written down as 
\begin{align} 
\int_0^{s_0} ds \rho(s) & = c_0 s_0 + c_2, \label{eq:moment0} \\
\int_0^{s_0} ds s \rho(s) & = \frac{c_0}{2} s_0^2 - c_4, \label{eq:moment1} \\
\int_0^{s_0} ds s^2 \rho(s) & = \frac{c_0}{3} s_0^3 + c_6. 
\label{eq:moment2}
\end{align}
Here, $s_0$ represents a scale that divides the low- and high-energy part of the spectrum. It needs to 
be determined from the (finite energy) sum rules themselves. 
The advantage of Eqs.\,(\ref{eq:moment0}-\ref{eq:moment2}) is that they relate spectral moments 
only to condensates of specific dimensions. Terms with condensates of higher dimension such as in the Borel transformed sum rules 
do not appear. Hence, in cases where the spectral function is a priori known, Eqs.\,(\ref{eq:moment0}-\ref{eq:moment2}) 
can in principle be used to determine certain combinations of condensates of some specific dimension. 
Conversely, they can also be used to check whether a spectral function computed by some phenomenological model is 
consistent with basic requirements of QCD. How this can be done, was demonstrated in detail in Refs.\,\cite{Gubler:2015yna,Gubler:2016itj}. 

\subsubsection{The $f_1(1285)$ \label{sec:f1.density}}
The behavior of the axial vector, isospin zero meson $f_1(1285)$ in nuclear matter has so far not been much studied 
in QCDSRs or any other method. Motivated by a measurement of this particle in photoproduction from a proton target, which resulted in a 
relatively small width of $18.4 \pm 1.4$ MeV by the CLAS collaboration \cite{Dickson:2016gwc}, its modification 
in a nuclear medium was recently studied in Ref.\,\cite{Gubler:2016djf} in a QCDSR approach. 
The main focus of this work was to regard the $\omega$ and $f_1(1285)$ as chiral partners, to determine how 
the partial restoration of chiral symmetry manifests itself for these particles and to what degree they can 
play the role as experimental probes for this restoration. 
This is an especially pressing issue now, as the behavior of $\omega$ in nuclear matter 
has been studied in detail in experiments \cite{Thiel:2013cea,Friedrich:2014lba,Kotulla:2008aa,Friedrich:2016cms} 
and analogous studies on the $f_1(1285)$ might become possible by replacing a proton with a nucleon target at the CLAS experiment. 

To be precise, $\omega$ and $f_1(1285)$ can only be regarded as chiral 
partners when chiral symmetry is extended to three flavors. 
In such a scenario $\phi$ and $f_1(1420)$ [the latter being the (mostly) strange counterpart of the $f_1(1285)$] 
have to be included in the chiral partner structure. In Ref.\,\cite{Gubler:2016djf}, it was however argued that 
even if taking into account only flavor $SU(2)$, $\omega$ and $f_1(1285)$ can be regarded as chiral partners 
in the limit where disconnected diagrams are neglected. In this limit, the difference between the $\omega$ and 
$f_1(1285)$ current correlators indeed vanishes when chiral symmetry is completely restored. Based on this 
approximation, one can expect that the $\omega$ and $f_1(1285)$ spectra should approximately approach each other in nuclear 
matter where chiral symmetry is at least partially restored. 

\begin{figure}[tb]
\begin{center}
\begin{minipage}[t]{8 cm}
\vspace{0.5 cm}
\hspace{-2.0 cm}
\includegraphics[width=12cm,bb=0 0 360 252]{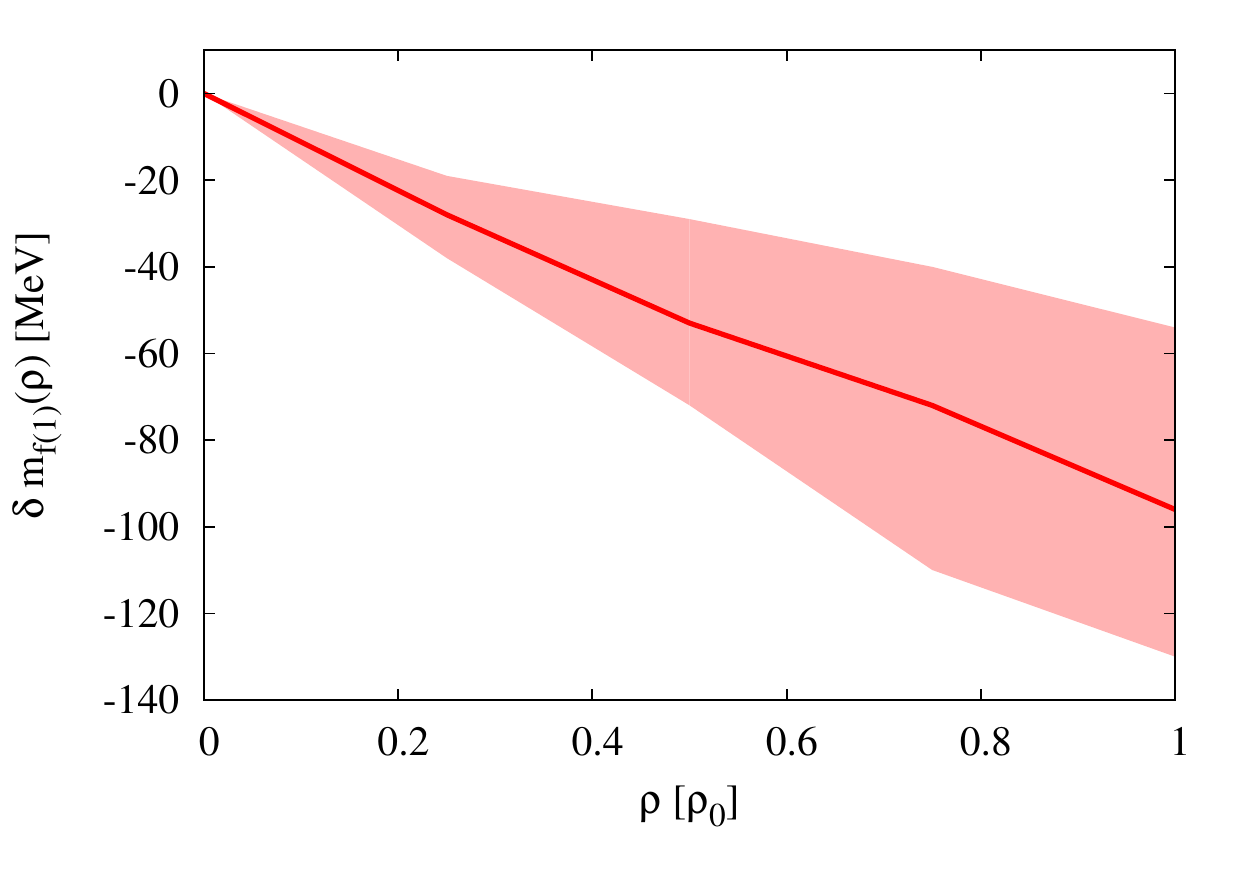}
\vspace{-0.8 cm}
\end{minipage}
\begin{minipage}[t]{16.5 cm}
\caption{The $f_1(1285)$ meson mass shift as a function of density, in units of $\rho_0$, the normal nuclear matter density. 
The red shaded area is obtained from the uncertainty of the $\pi N$ sigma term,  $\sigma_{\pi N} = 45 \pm 15$ MeV. 
Taken from Fig.\,3 of Ref.\,\cite{Gubler:2016djf}. 
\label{fig:f1.rho}}
\end{minipage}
\end{center}
\end{figure}

The mass of the $f_1(1285)$ as function of density was then studied in Ref.\,\cite{Gubler:2016djf} using a 
conventional QCDSR analysis relying on the ``pole + continuum'' assumption of Eq.\,(\ref{eq:pole.plus.continuum}). 
The corresponding result is shown in Fig.\,\ref{fig:f1.rho},  
%%%%%%%%%%%%%%%%%%%%%%%%%%%%%%%%%%%%%%%%%%%%%%%%%%%%%%%%%%%%%%%%%%%%%%%%%%%%%%%%%%%%%%%%%%%%%%%%%%%%%%%%%%%%%%%%%%%%
%%%%%%%%%%%%%%%%%%%%%%%%%%%%%%%%%%%%%%%%%%%%%%%%%%%%%%%%%%%%%%%%%%%%%%%%%%%%%%%%%%%%%%%%%%%%%%%%%%%%%%%%%%%%%%%%%%%%
where it is observed that the $f_1(1285)$ potentially receives a negative mass shift of about 100 MeV at normal 
nuclear matter density. It however has to be kept in mind that this result is obtained by assuming
a delta function in the ``pole + continuum'' ansatz even in nuclear medium. 
As it was shown in Ref.\,\cite{Leupold:1997dg} for the $\rho$ meson, changes of the OPE at finite density 
can also be satisfied with a smaller change in the mass and a simultaneous increase of the width.  
A similar effect likely applies to the $f_1(1285)$. The above result should hence be understood 
as the maximum mass shift that can expected in nuclear matter.

\subsubsection{The nucleon}
The study of nucleon properties at finite density in a QCDSR approach already has quite a long history. 
See, for instance, the early nineties works of Refs.\,\cite{Cohen:1994wm,Jin:1992id,Furnstahl:1992pi,Drukarev:1988kd,Cohen:1991js,Jin:1993up}. 
We will not review these older studies here, but focus on recent progress made during about the last ten years. 

\begin{figure}[tb]
\begin{center}
\begin{minipage}[t]{8 cm}
\vspace{0.5 cm}
\hspace{-2.3 cm}
\includegraphics[width=12cm,bb=0 0 360 252]{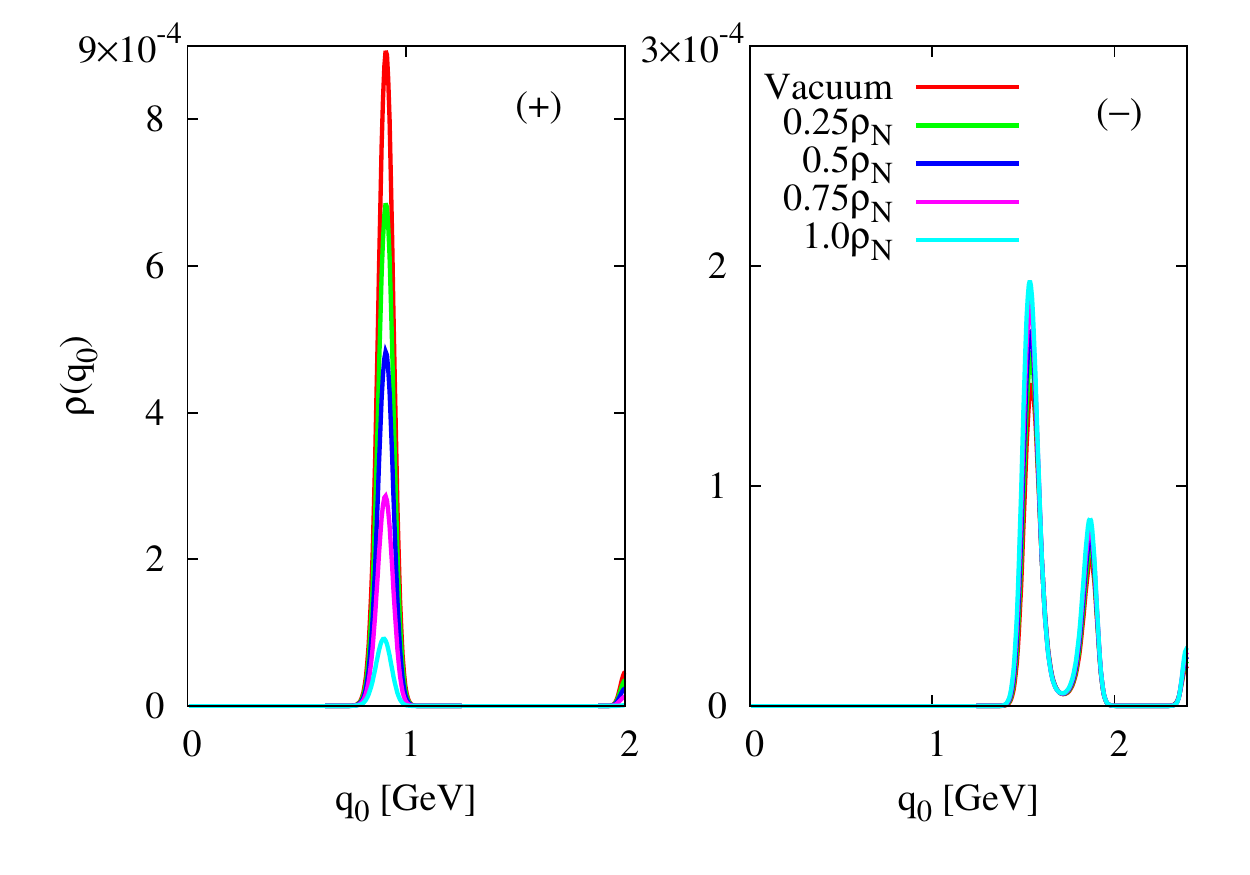}
\vspace{-1.0 cm}
\end{minipage}
\begin{minipage}[t]{16.5 cm}
\caption{The positive (left) and negative (right) parity nucleon spectral functions extracted from parity-projected QCD sum rules and MEM.
The red, green, blue, magenta and light blue lines correspond to spectral functions at densities $0.0$, $0.25$, $0.5$, 
$0.75$ and $1.0$ units of normal nuclear matter density, denoted here as $\rho_N$. 
Taken from Fig.\,2 of Ref.\,\cite{Ohtani:2016pyk}, where the variable $q_0$ was adopted instead of $\omega$ used in this review.   
\label{fig:nucleon.finite.density}}
\end{minipage}
\end{center}
\end{figure}

In the first QCDSR studies of baryons \cite{Krasnikov:1982ea,Ioffe:1981kw,Chung:1981cc}, a proper parity 
projection was not included in the formalism, but thanks to 
Ref.\,\cite{Jido:1996ia}, it is now possible to construct parity projected baryonic sum rules, and hence to study not only the 
positive parity ground state, but also its lowest negative parity excited state (see also Refs.\,\cite{Ohtani:2012ps,Kondo:2005ur} for related discussions). 
Generalizing this technique to finite density, one can study the behavior of the lowest positive and negative parity nuclear excitations 
and can especially examine to what degree the positive and negative parity spectra approach each other as chiral symmetry  
gets partially restored. Similar questions were recently studied in lattice QCD in the finite temperature regime \cite{Aarts:2015mma}. 
A related QCDSR study at finite density was carried out in Ref.\,\cite{Ohtani:2016pyk}. 
In this work, parity-projected in-medium nucleon QCD sum rules were constructed and subsequently 
analyzed with MEM. The positions of the lowest peaks in the obtained vacuum spectral functions are consistent 
with the ground state N(939) and its lowest negative parity excitation N(1535). See Fig.\,\ref{fig:nucleon.finite.density}. 
%%%%%%%%%%%%%%%%%%%%%%%%%%%%%%%%%%%%%%%%%%%%%%%%%%%%%%%%%%%%%%%%%%%%%%%%%%%%%%%%%%%%%%%%%%%%%%%%%%%%%%%%%%%%%%%%%%%%
%%%%%%%%%%%%%%%%%%%%%%%%%%%%%%%%%%%%%%%%%%%%%%%%%%%%%%%%%%%%%%%%%%%%%%%%%%%%%%%%%%%%%%%%%%%%%%%%%%%%%%%%%%%%%%%%%%%%
Increasing the density, these peaks exhibit a somewhat surprising behavior. 
Their positions namely turn out to be almost density independent, meaning that the total energies of both
the positive and negative-parity states are not much modified by nuclear matter effects up to normal nuclear matter
density. The residue of the positive parity nucleon ground state on the other hand 
decreases while that of the negative parity first excited state remains almost unchanged with increasing density. 
It is shown in detail in Ref.\,\cite{Ohtani:2016pyk} that this behavior is closely related to the modifications of the 
condensates $\langle \bar{q} q \rangle_{\rho}$ and $\langle q^{\dagger} q \rangle_{\rho}$ at finite density, which demonstrates 
that these condensates are important for the description of the in-medium properties of the nucleon and its negative parity 
excited state. An intuitive picture for the behavior shown in Fig.\,\ref{fig:nucleon.finite.density}, however, has so far not been found 
and requires further investigations. 

Besides the abovementioned work, recent studies of the nucleon at finite density based on QCDSRs include 
the ongoing series of papers by Drukarev and collaborators \cite{Drukarev:2004zg,Drukarev:2004fn,Drukarev:2009ac,Drukarev:2016ius,Drukarev:2017kil,Drukarev:2018xjb} 
and other groups \cite{Mallik:2009zz,Azizi:2014yea}, including generalizations 
to decuplet baryons and hyperons \cite{Marques:2018mic,Azizi:2016hbr,Jeong:2016qlk,Azizi:2015ica}. 
The nuclear symmetry energy is another interesting quantity that was studied using QCDSRs during the last few years. 
Details can be found in Refs.\,\cite{Jeong:2012pa,Jeong:2014lsa,Jeong:2015ima}. 

\subsection{Heavy hadrons}
Heavy hadrons are defined in this work as hadrons with at least one $c$ or $b$ valence quark or anti-quark. 
The finite density behavior of such hadrons will be discussed in this section, starting first with mesons and finishing with baryons.  

\subsubsection{Charmonium}
With the exception of Refs.\,\cite{Morita:2010pd,Kumar:2010hs}, the behavior of charmonium states in nuclear matter has 
not been much studied within QCDSRs in recent years. 
The earlier works of Refs.\,\cite{Kim:2000kj,Klingl:1998sr} therefore still remain the state-of-the-art today. 
Generally, charmonium states are not expected to be much affected by nuclear matter as they are tightly bound systems with no 
$u$ or $d$ valence quarks which are expected to be most strongly perturbed by surrounding nuclei. In QCDSRs, finite density 
effects enter the calculation through the density dependence of gluonic condensates. Light quark condensates appear 
in charmonium (and bottomonium) sum rules only at second order in $\alpha_s$ and are therefore suppressed. 
In Ref.\,\cite{Kim:2000kj}, where gluonic condensates up to dimension 6 were taken into account, 
the $J/\Psi$ was found to receive a negative mass shift of
\begin{equation}
\Delta m_{J/\Psi} = - 4\,\mathrm{MeV}
\end{equation}  
at normal nuclear matter density. It remains to be seen whether such a small mass shift can be observed in 
future experiments. With such a measurement, it would be possible to constrain the finite density dependence of 
a certain combination of gluonic condensates. 

In this context, we mention the subject of charmonium in a magnetic field, which has recently attracted much 
attention, especially because of the large magnetic field which is generated at the initial stage of non-central 
heavy-ion collisions \cite{Deng:2012pc}. In QCDSRs, this was studied for the first time in Refs.\,\cite{Cho:2014exa,Cho:2014loa}. 
In these works, a special emphasis was laid on the mixing effects between $\eta_c$ and $J/\Psi$, which occur because of the existence 
of a homogenous and constant magnetic field. 
According to the findings of Refs.\,\cite{Cho:2014exa,Cho:2014loa}, the modifications of the correlators due to the magnetic field 
are saturated to a large degree by these mixing effects. 
Another related direction of work is to study the combined effect of a magnetic field and finite density 
which was partly done in Ref.\,\cite{Kumar:2018ujk}. 

\subsubsection{$D$ and $B$ mesons}
During the last decade, the finite density behavior of $D$ (and $B$) mesons have been studied quite intensively and controversially in QCDSRs 
and various other approaches. 
The reason for this interest lies in the possibility of probing the modification of such mesons produced in nuclei or 
high density matter at FAIR, J-PARC or other similar facilities. For such an experimental study to be meaningful, it is important to produce 
$D$ mesons in nuclei with sufficiently small momentum such that they remain in the region of high density long enough. Only 
then can potential spectral modifications have a large enough effect to be experimentally measurable. This currently still 
appears to be a challenge for experiments and new ideas might be needed \cite{Noumi:2017hbc}. 

We will here concentrate on theoretical works based on QCDSRs. The discussion given in this section should 
hence not be understood as a complete summary of all works about the D and B mesons at finite density. For more general 
discussions and more references, see Refs.\,\cite{Metag:2017yuh,Hosaka:2016ypm}. 
We furthermore focus on $D$ mesons, which are presently far more relevant for experiments than $B$ mesons, 
because they are lighter and therefore easier to produce with high statistics. While this might not be an essential issue 
for high-energy heavy ion experiments at the LHC, where a large amount of bottom quarks can be created and 
where the behavior of matter at high temperature can be studied, 
lower energy collisions with much fewer bottom quarks are needed to create matter at high densities \cite{Nara:1999dz,Akamatsu:2018olk}. 
We will therefore mention $B$ mesons only briefly 
at the end of this section. 

The study of $D$ mesons in nuclear matter with QCDSRs started with the paper of Hayashigaki \cite{Hayashigaki:2000es}, 
which found a (negative) mass shift of $-48$ MeV for the $D$ at normal nuclear matter density. In this work, OPE terms up to 
dimension 4 were taken into account. Later, dimension 5 OPE terms and further terms that break charge 
symmetry were included in the analysis of Ref.\,\cite{Hilger:2008jg}, which however led to the opposite conclusion 
of $+45$ MeV, albeit with large uncertainties related to the determination of the threshold parameter in the 
``pole + continuum'' ansatz. The more recent works of Refs.\,\cite{Wang:2015uya,Azizi:2014bba} are qualitatively 
consistent with the earlier results of Ref.\,\cite{Hayashigaki:2000es}, obtaining negative mass shifts of $-46$ MeV and $-72$ MeV, respectively. 
Furthermore, Refs.\,\cite{Chhabra:2016vhp,Chhabra:2017rxz} employ a chiral SU(3) model to compute 
the dimension 3 quark condensate and the dimension 4 gluon condensate at finite density, which are then used 
as input in the QCD sum rule analysis. As a result, they obtain negative mass shifts for both $D$ and $D_s$ mesons 
(as well as  $B$ and $B_s$) of the same order as Ref.\,\cite{Hayashigaki:2000es}. 

\begin{figure}[tb]
\begin{center}
\begin{minipage}[t]{8 cm}
\vspace{0.5 cm}
\hspace{-2.3 cm}
\includegraphics[width=13cm,bb=0 0 557 287]{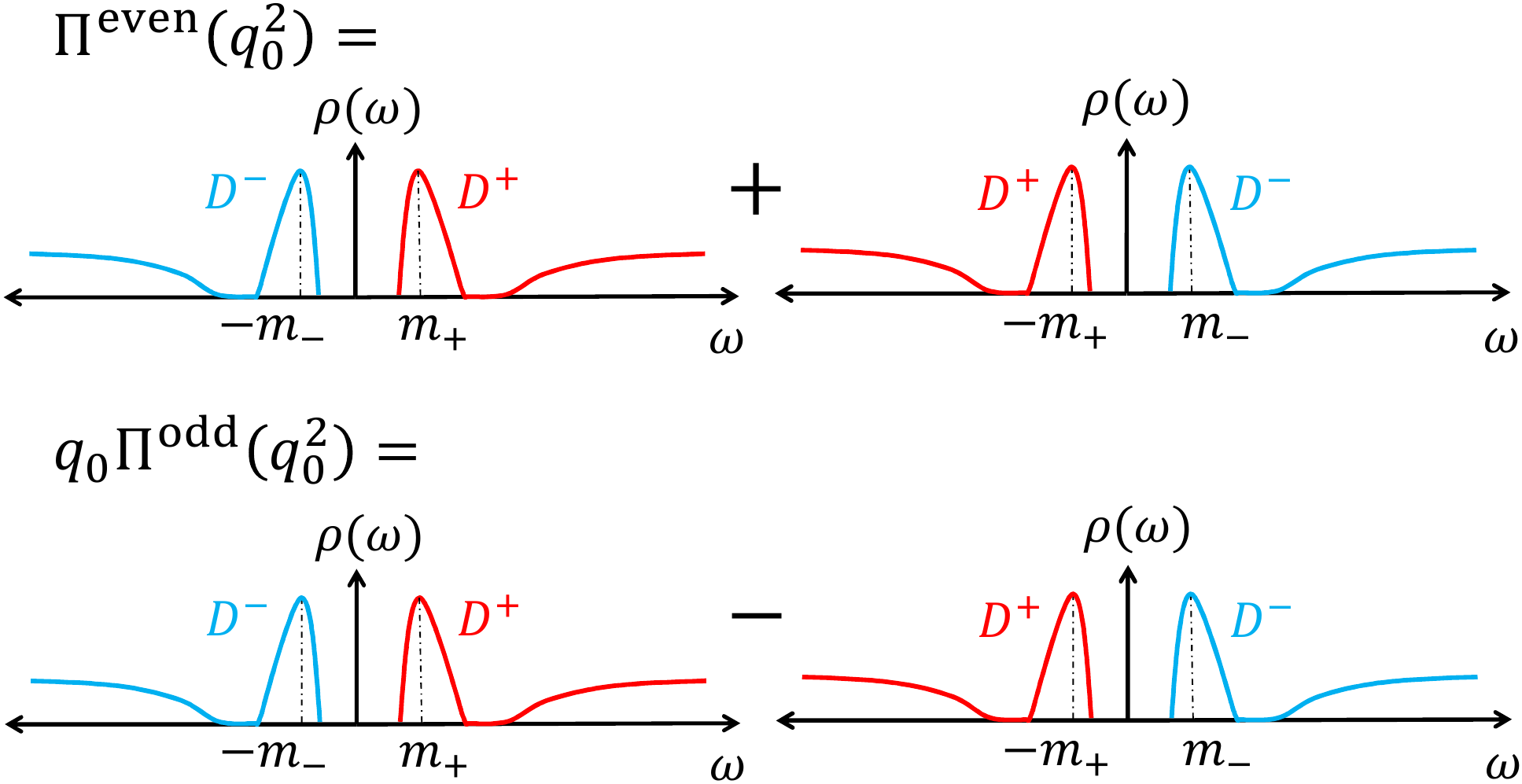}
\vspace{-0.3 cm}
\end{minipage}
\begin{minipage}[t]{16.5 cm}
\caption{Schematic illustration of spectral contributions to $\Pi^{\mathrm{even}}(\omega^2)$ and $\Pi^{\mathrm{odd}}(\omega^2)$ for 
the $D^+$ correlator. $D^+$ and $D^-$ contributions are simply interchanged for the $D^-$ correlator. 
Spectral functions of the old-fashioned correlator of Eq.\,(\ref{eq:old.fashioned.correlator}) include only spectra at positive $\omega$. 
Taken from Fig.\,1 of Ref.\,\cite{Suzuki:2015est}, where $q_0$ was used instead of $\omega$ in this review. 
\label{fig:D.illust}}
\end{minipage}
\end{center}
\end{figure}

Finally, we will here summarize the findings of Ref.\,\cite{Suzuki:2015est}, in which MEM was used to study 
charge-conjugate-projected Gaussian sum rules [see Eq.\,(\ref{eq:Gaussian}) for the specific form of the Gaussian kernel]. 
The charge conjugate projection, proposed in Ref.\,\cite{Suzuki:2015est} for the first time, makes it possible to disentangle the $D^{+}$ and $D^{-}$ 
spectra and hence to study the respective states independently. To discuss this method, let us consider the correlator of 
Eq.\,(\ref{eq:correlator}) with the current $J(x)$ coupling to the $D$ meson of interest, for instance $J^{D^+}(x) = i \bar{d}(x) \gamma_5 c(x)$ or 
$J^{D^-}(x) = i \bar{c}(x) \gamma_5 d(x)$. In vacuum, the correlators of $J^{D^+}(x)$ and $J^{D^-}(x)$ are, of course, identical and will 
depend only on $q^2$ because of Lorentz invariance. 
Replacing the vacuum $| 0 \rangle$ expectation value of Eq.\,(\ref{eq:correlator}) with that of finite baryon density matter $\rangle_{\rho}$, 
the two correlators will be different and furthermore depend on $\omega$ [we here us the notation $q = (\omega, \bm{p})$ and set the momentum $\bm{p}$ to zero 
for simplicity], 
\begin{equation}
\Pi^{J}(\omega) = \Pi^{\mathrm{even}}(\omega^2) + \omega \Pi^{\mathrm{odd}}(\omega^2).
\end{equation} 
Here, $\Pi^{\mathrm{odd}}(\omega^2)$ contains only non-scalar condensates, which vanish in the zero density limit, 
such as $\langle \bar{q} \gamma^{\mu} q \rangle_{\rho}$, 
$\langle \mathcal{ST} \bar{q} \gamma^{\alpha} iD^{\mu} iD^{\nu} q \rangle_{\rho}$ or $\langle \bar{q} \gamma^{\mu} \sigma_{\alpha \beta} G^{a \alpha \beta} t^a q \rangle_{\rho}$. 
Note that in Ref.\,\cite{Suzuki:2015est} the variable $q_0$ was used instead of $\omega$ here. 
$\Pi^{\mathrm{even}}(\omega^2)$ and $\Pi^{\mathrm{odd}}(\omega^2)$ for $\Pi^{D^+}(\omega)$ can be related to $D^{+}$ and $D^{-}$ as follows, 
\begin{align}
\Pi^{\mathrm{even}}(\omega^2) & = \frac{1}{2} \Bigl[ \Pi^+(\omega) + \Pi^-(\omega) \Bigr], \\
\omega \Pi^{\mathrm{odd}}(\omega^2) & = \frac{1}{2} \Bigl[ \Pi^+(\omega) - \Pi^-(\omega) \Bigr],  
\end{align} 
where $\Pi^+(\omega)$ carries the $D^+$ spectrum at positive $\omega$ and the $D^-$ spectrum 
at negative $\omega$ and vice versa for $\Pi^-(\omega)$. See Fig.\,\ref{fig:D.illust} for a schematic 
illustration. 
%%%%%%%%%%%%%%%%%%%%%%%%%%%%%%%%%%%%%%%%%%%%%%%%%%%%%%%%%%%%%%%%%%%%%%%%%%%%%%%%%%%%%%%%%%%%%%%%%%%%%%%%%%%%%%%%%%%%
%%%%%%%%%%%%%%%%%%%%%%%%%%%%%%%%%%%%%%%%%%%%%%%%%%%%%%%%%%%%%%%%%%%%%%%%%%%%%%%%%%%%%%%%%%%%%%%%%%%%%%%%%%%%%%%%%%%%
For the $D^-$ correlator $\Pi^{D^-}(\omega)$, $D^+$ and $D^-$ contributions are simply interchanged. 

\begin{figure}[tb]
\begin{center}
\begin{minipage}[t]{8 cm}
\vspace{0.5 cm}
\hspace{-2.0 cm}
\includegraphics[width=12cm,bb=0 0 360 252]{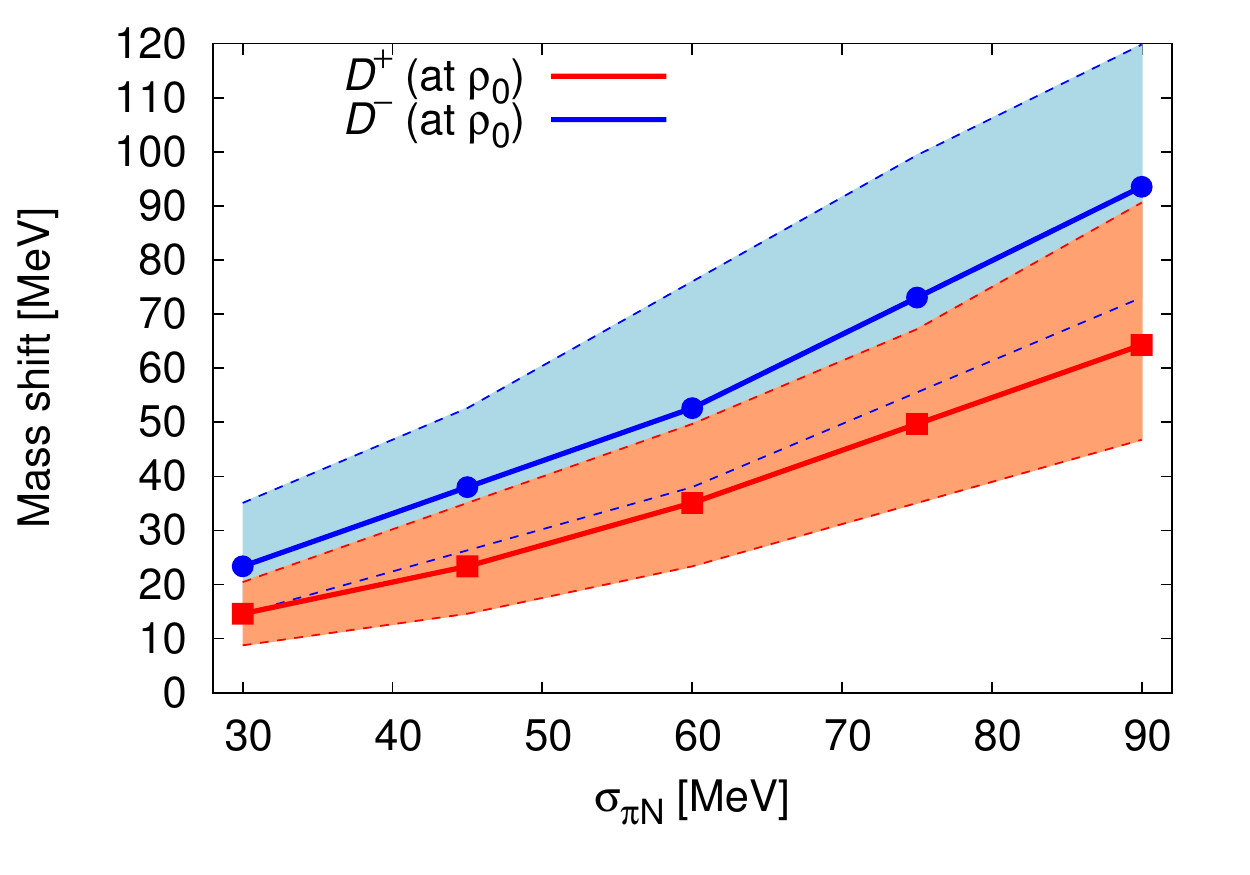}
\vspace{-0.8 cm}
\end{minipage}
\begin{minipage}[t]{16.5 cm}
\caption{The $\pi N$ sigma term dependence of $D^+$ and $D^-$ meson mass shifts at 
normal nuclear matter density $\rho_0$. Dashed lines and shaded areas
correspond to errors related to uncertainties of in-medium condensates
excluding the error of $\sigma_{\pi N}$. 
Taken from Fig.\,6 of Ref.\,\cite{Suzuki:2015est}. 
\label{fig:D.mass.sigma.term}}
\end{minipage}
\end{center}
\end{figure}

To disentangle the $D^+$ and $D^-$ spectra, the charge-conjugate-projected sum rule is constructed by a method analogous to 
parity-projection for baryonic sum rules \cite{Ohtani:2012ps,Jido:1996ia}. The idea is to introduce the so-called old-fashioned correlator, 
which for zero-momentum is defined as 
\begin{equation}
\Pi^{\mathrm{old}}(\omega) = i \int d^4 x e^{i \omega x} \theta(x_0) \langle T[J(x) J^{\dagger}(0)]  \rangle_{\rho}. 
\label{eq:old.fashioned.correlator}
\end{equation}
Here, $\theta(x_0)$ represents the Heaviside step function, which is introduced to remove the negative energy contribution from the correlator. 
Using $\Pi^{\mathrm{old}}(\omega)$, the correlators that have only $D^{+}$ or $D^{-}$ contributions can be constructed as 
\begin{equation}
\Pi^{D^{\pm}}(\omega) = \Pi^{\mathrm{even,}\,\mathrm{old}}(\omega^2) + \omega \Pi^{\mathrm{odd,}\,\mathrm{old}}(\omega^2). 
\label{eq:old.fashioned.correlator.2}
\end{equation}
Making use of the 
analyticity of this function, one can formulate sum rules for the $D^{+}$ and $D^{-}$ spectra, 
as explained in Ref.\,\cite{Suzuki:2015est} for the Gaussian sum rule case. 
The resulting sum rules were analyzed using MEM, as discussed in Section\,\ref{MEM.section}. 
We refer the interested reader to Ref.\,\cite{Suzuki:2015est} for detailed discussions about adopted input 
condensate parameters and error analyses and here only show the most important result about 
the $D$ meson masses at normal nuclear matter density 
as a function of the $\pi N$ sigma term $\sigma_{\pi N}$ 
in Fig.\,\ref{fig:D.mass.sigma.term}.  
%%%%%%%%%%%%%%%%%%%%%%%%%%%%%%%%%%%%%%%%%%%%%%%%%%%%%%%%%%%%%%%%%%%%%%%%%%%%%%%%%%%%%%%%%%%%%%%%%%%%%%%%%%%%%%%%%%%%
%%%%%%%%%%%%%%%%%%%%%%%%%%%%%%%%%%%%%%%%%%%%%%%%%%%%%%%%%%%%%%%%%%%%%%%%%%%%%%%%%%%%%%%%%%%%%%%%%%%%%%%%%%%%%%%%%%%%
As can be seen in this figure, both $D^+$ and $D^-$ mesons receive a positive mass shift. Its magnitude ranges from 10 MeV to almost 100 MeV, depending on 
the $\sigma_{\pi N}$ value. This shows that rate of restoration of chiral symmetry, which is governed by $\sigma_{\pi N}$ [see Eq.\,(\ref{eq:qbarq.linear})], 
determines the size of the D meson mass shift. A simple quark model picture, that explains this initially 
surprising finding, was given in Ref.\,\cite{Park:2016xrw}. 
Irrespective of the $\sigma_{\pi N}$ value, the $D^{-}$ mass shift is always larger than that of the $D^{+}$. In the sum rules, this difference is generated 
due to the chiral odd terms in the OPE, particularly $\langle \bar{q} \gamma^{\mu} q \rangle_{\rho}$, which is proportional to baryon density. 
It is rather straightforward to think of an intuitive quark based picture to understand why the 
$D^{-}$ receives more repulsion than the $D^{+}$ 
at finite density. 
The $D^{-}$ meson has a $d$ valence quark, which can be expected to interact repulsively with the same $d$ quark existing in nuclear matter, 
due to Pauli blocking. For $D^{+}$, with a $\bar{d}$ valence quark, such a Pauli blocking effect is absent and the repulsion hence becomes weaker. 

In summary QCDSR results so far do not appear to be conclusive. While Refs.\,\cite{Wang:2015uya,Hayashigaki:2000es,Azizi:2014bba} 
obtain a negative mass shift, it is positive for Refs.\,\cite{Suzuki:2015est,Hilger:2008jg}. It is however not difficult to identify the reason for this discrepancy. Namely, 
Refs.\,\cite{Wang:2015uya,Hayashigaki:2000es,Azizi:2014bba} employ a QCDSR approach proposed in Refs.\,\cite{Koike:1993mx,Kondo:1993py}, which 
extracts the $D$-$N$ scattering amplitude in the zero-momentum limit. Refs.\,\cite{Suzuki:2015est,Hilger:2008jg} on the other hand use the more conventional 
method, partly explained above, which directly analyses the spectral function and the modification of the $D$ meson peak at finite density. 
The application of the former method to light vector mesons was criticized in Ref.\,\cite{Hatsuda:1995dy} and also later in Ref.\,\cite{Suzuki:2015est} for 
issues related to the Borel window and specifically for the apparent lack of the pole (or ground state) contribution in this approach. 
This criticism has so far not been refuted. 

Let us briefly discuss the $B$ meson, which was studied in Refs.\,\cite{Wang:2015uya,Suzuki:2015est,Hilger:2008jg,Azizi:2014bba,Chhabra:2016vhp}. 
The general trends are the same as for the $D$ meson, namely Refs.\,\cite{Wang:2015uya,Azizi:2014bba,Chhabra:2016vhp} obtain a negative mass shift for the averaged 
$B^+$ and $B^-$ masses, while Refs.\,\cite{Suzuki:2015est,Hilger:2008jg} get a positive one. Numerically, the negative mass shifts of Refs.\,\cite{Wang:2015uya,Azizi:2014bba,Chhabra:2016vhp} 
are of the order of several hundreds of MeV, while the positive ones in Refs.\,\cite{Suzuki:2015est,Hilger:2008jg} are below 100 MeV. 
An interesting finding 
about the masses of the individual $B^+$ and $B^-$ states 
was furthermore reported in Refs.\,\cite{Suzuki:2015est,Hilger:2008jg}. In both works, 
the mass splitting between the two states rapidly increases with increasing heavy quark mass, leading to a larger positive mass shift for $B^-$ and a small negative 
mass shift for $B^+$. This effect is related to the $\omega$-odd terms, which for $B^-$ have the same sign as the density dependent $\omega$-even terms. 
For $B^+$, the two contributions almost completely cancel, leaving only a small negative mass shift. 

For further results about other $D$ and $B$ meson channels, such as $D_s$, $D^{\ast}$, $D_0$, $D_1$, $B_s$, $B^{\ast}$, $B_0$ and $B_1$, which 
we will not discuss here, see Refs.\,\cite{Wang:2015uya,Hilger:2008jg,Chhabra:2017rxz}. 

As a last point, it is worth mentioning related works studying $D$ mesons in a constant magnetic field. Such a study was first performed in 
Ref.\,\cite{Machado:2013yaa} for the $B$ meson and later in Ref.\,\cite{Gubler:2015qok} for the $D$ meson where more condensates were taken 
into account and some trivial mistakes in the calculation of Ref.\,\cite{Machado:2013yaa} were pointed out. As a result, it was shown that, similar to the 
charmonium case discussed at the end of the previous subsection, mixing effects between pseudoscalar and vector channels are important to obtain 
spectral functions that are consistent with the sum rules. For charged $D$ mesons, Landau level effects furthermore need to be taken into account. 
It was found in Ref.\,\cite{Gubler:2015qok}, that the above two effects saturate the sum rules for the charged $D$ mesons, while for neutral 
ones a further positive mass shift is needed to be consistent with the OPE.  

\subsubsection{Heavy baryons}
Studies about the finite density behavior of heavy hadrons, that is, hadrons with at least one $c$ or $b$ valence quark, have only begun recently. 
The $\Lambda_c$ and $\Lambda_b$ state properties in nuclear matter were studied first in Ref.\,\cite{Wang:2011hta} and 
subsequently in Refs.\,\cite{Azizi:2016dmr,Ohtani:2017wdc}. 
The first two works, Refs.\,\cite{Wang:2011hta,Azizi:2016dmr} obtained an increasing $\Lambda_c$ ($\Lambda_b$) 
mass at finite density, leading 
to 85 MeV (92 MeV) repulsion (sum of scalar and vector self-energies) in Ref.\,\cite{Wang:2011hta} and a considerably larger 
432 MeV (1089 MeV) repulsion in Ref.\,\cite{Azizi:2016dmr} at normal nuclear matter density. 
In Ref.\,\cite{Ohtani:2017wdc} the sum rules were improved by taking into account $\alpha_s$ corrections to the Wilson coefficients and  
employing the parity projected sum rules with a Gaussian kernel. In the same work, the treatment of the density 
dependence of four-quark condensates was studied in detail. Specifically, two treatments of the four-quark condensates were considered: 
the traditional factorization ansatz of Eqs.\,(\ref{eq:four.quark.condensate.factorization.density.1}) and (\ref{eq:four.quark.condensate.factorization.density}) 
and a parametrization based on the perturbative chiral quark model (PCQM) \cite{Thomas:2007gx,Drukarev:2003xd}. 
Applying these two four-quark condensate specifications first to the finite density QCDSRs for the $\Lambda$ (with an $s$ quark instead of a $c$ quark), 
it was found that only the latter PCQM prescription gives a small and negative mass shift for the $\Lambda$ at normal nuclear 
matter density that is consistent with our knowledge from $\Lambda$ hypernucleon spectroscopy \cite{Hashimoto:2006aw}. It was therefore concluded in 
Ref.\,\cite{Ohtani:2017wdc} that only the PCQM prescription is suitable for this specific sum rule and hence also for the one of the $\Lambda_c$. 
This then leads to an about 20 MeV attraction of the $\Lambda_c$ at normal nuclear matter density. For $\Lambda_b$ the attraction turns out 
to be practically zero. 
The studies performed up to now are far from being consistent and more work will be needed 
to clarify the origin of the various discrepancies. 

Similarly, the finite density behavior of $\Sigma_c$ and $\Sigma_b$ has been studied in Refs.\,\cite{Azizi:2016dmr,Wang:2011yj}. 
For $\Sigma_c$ ($\Sigma_b$), a strong repulsion of 323 MeV (401 MeV) was found in Ref.\,\cite{Wang:2011yj}, while 
an equally strong attraction of -450 MeV (-232 MeV) was obtained in Ref.\,\cite{Azizi:2016dmr} for the sum of scalar and vector self energies 
at normal nuclear matter density. Again, the results are in complete disagreement. Further studies are warranted for reaching a final conclusion 
on this issue. 

As for the behavior of $\Xi_c$ and $\Xi_b$ in nuclear matter, only the results of Ref.\,\cite{Azizi:2016dmr} are presently available. 
In this work, only a very weak attraction of -4 MeV (-2 MeV) was obtained for $\Xi_c$ ($\Xi_b$) at normal nuclear matter density. 
Furthermore, spin-$\frac{3}{2}$ $\Sigma_{Q}^{\ast}$, $\Xi_{Q}^{\ast}$ and $\Omega_{Q}^{\ast}$ ($Q$ here stands for a $c$ or $b$ quark) 
baryons in nuclear matter were studied in Ref.\,\cite{Azizi:2018dtb}. While for the scalar self-energies of $\Sigma_{c}^{\ast}$, $\Sigma_{b}^{\ast}$ 
and $\Xi_{b}^{\ast}$ some attraction was obtained, the total of scalar and vector self-energies turned out to be repulsive for all studied states.  
Independent calculations will be needed in the future to check and confirm these findings. 

Finally, doubly heavy spin-$\frac{1}{2}$ baryons, specifically $\Xi_{QQ}$ and $\Omega_{QQ}$ (where again $Q = c$ or $b$), in nuclear matter 
were studied in Ref.\,\cite{Wang:2012xk}. In this paper, the scalar self-energies had the tendency to be much larger than their vector 
counterparts. The sum of scalar and vector self-energy turned out to be attractive for all investigated channels. 
At normal nuclear matter density, the obtained values for this sum are $-0.97$ GeV for $\Xi_{cc}$, $-0.34$ GeV for $\Omega_{cc}$, 
$-2.86$ GeV for $\Xi_{bb}$ and $-1.04$ GeV for $\Omega_{bb}$. Here, it is especially worth noting the remarkably large attraction in 
the $\Xi_{bb}$ channel. It will be interesting to see if it can be reproduced in future works based on the same or other methods 
and if such a large mass shift could perhaps be measured in a future experiment. 
Very recently, the finite density behavior of doubly heavy spin-$\frac{3}{2}$ baryons,  $\Xi^{\ast}_{QQ}$, $\Omega^{\ast}_{QQ}$, $\Xi^{\ast}_{QQ'}$ and $\Omega^{\ast}_{QQ'}$  
(for the last two, $Q \neq Q'$) were investigated in Ref.\,\cite{Er:2019hhk}. The reported results are qualitatively different from the spin-$\frac{1}{2}$ case 
of Ref.\,\cite{Wang:2012xk}. For all channels, the absolute values of the scalar and vector self-energies are of the same order of magnitude. For 
the $\Omega^{\ast}_{QQ}$ and $\Omega^{\ast}_{QQ'}$ channels, both scalar and vector parts have the size of at most a few percent of the respective 
vacuum masses, leading for their sum to a weak repulsion in nuclear matter. For the $\Xi^{\ast}_{QQ}$ and  $\Xi^{\ast}_{QQ'}$ states, the self energies are 
larger, namely around 20\,\% of the vacuum masses at normal nuclear matter density $\rho_0$. Their sum however largely cancel, giving only a very 
small effect of at most 2\,\% at $\rho_0$. 

%%%%%%%%%%%%%%%%%%%%
\section{Exact sum rules at finite temperature \label{sec:exact-sumrule}} 

In addition to the conventional sum rules reviewed in the previous sections, it was recently attempted to derive and make use of exact sum rules. 
The novel feature of these sum rules is to use the infrared (IR) behavior of the Green function, 
which is correctly described by hydrodynamics if we consider channels with conserved currents 
at finite temperature/density, as well as the ultraviolet (UV) behavior described by the OPE. 
Originally, such sum rules were derived for the energy-momentum tensor channel \cite{Romatschke:2009ng}, 
and recently also for the vector current channel \cite{Gubler:2016hnf, Gubler:2017qbs}. 
In both cases, the sum rules were used to improve related lattice QCD analyses. 
We will review these works in the next two subsections, focusing 
on the finite temperature and zero chemical potential case unless otherwise specified. 
As will be discussed below, 
the shape of the spectral function at finite temperature becomes rather complicated compared with that at $T=0$. 
Constraints obtained from exact sum rules can therefore be very helpful. 

The physical motivation to investigate the finite temperature and zero chemical potential case is related to the 
research of quark-gluon plasma, which was realized in the early universe and is now being created terrestrially in 
heavy ion collision experiments. 
Even though much was learned over the years, there still remain some unsolved problems in this field. 
For example, at what temperature ground state and excited state charmonia melt, is still a controversial topic. 
Also, hydrodynamics has proven to be useful for describing heavy ion collision experiments. 
The determination of its parameters, transport coefficients, is a theoretically interesting and phenomenologically necessary task. 
Especially, the bulk viscosity is believed to behave in a way that is closely related to the QCD phase transition. 
The sum rules introduced in this section have the potential to contribute to the current research of these topics.

%%%%%%%%%%%%%%%%%%%%%%%%%%%%%%
\subsection{Energy-momentum tensor channel \label{sec:Tmunu}}

In this subsection, we review the derivation of sum rules and their application in the channel of the energy-momentum tensor, 
which is a conserved current. 
Two sum rules in the shear sector and one in the bulk sector will be discussed.
%%%%%%%%%%%%%%%%%%%%%%%%%%%%%%

\subsubsection{Derivation \label{deriv.of.exaxt.sr}}
\begin{figure}[tb]
\begin{center}
\begin{minipage}[t]{8 cm}
\vspace{-12.0 cm}
%\vspace{1.0 cm}
\hspace{3.0 cm}
\includegraphics[width=10cm,bb=137 462 492 698]{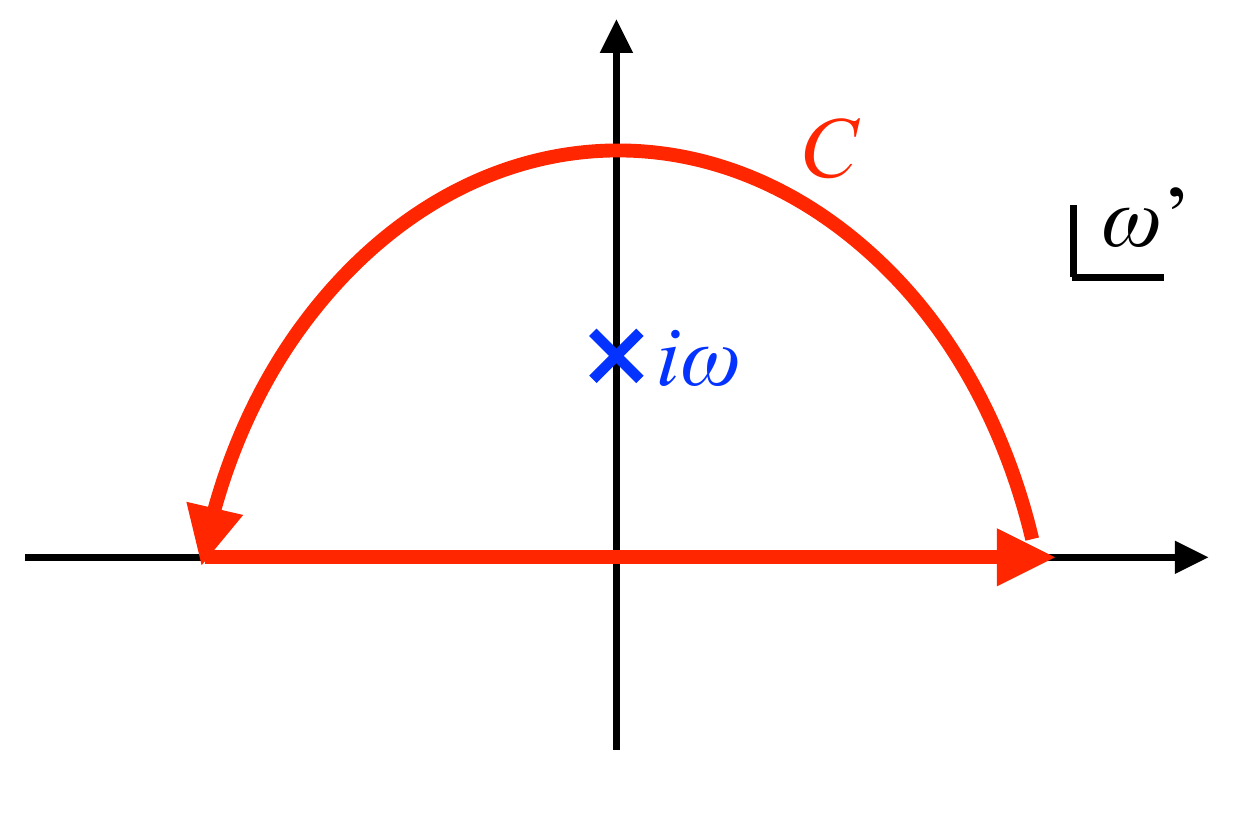}
\vspace{12.5 cm}
%\vspace{-0.8 cm}
\end{minipage}
\begin{minipage}[t]{16.5 cm}
\caption{The contour $C$ in the complex $\omega'$ energy plane, used to derive the exact sum rules. 
The contour runs infinitesimally above the real axis, so that it does not overlap with the singularities on the real axis. 
Taken from Fig.\,1 of Ref.\,\cite{Gubler:2017qbs}. 
\label{fig:contour}}
\end{minipage}
\end{center}
\end{figure}

As the derivation of the exact sum rules has so far only been outlined a few times in the literature, we recapitulate it here \cite{Romatschke:2009ng}. 
The starting point is to consider the integral on the contour $C$ drawn in Fig.~\ref{fig:contour}.
As the integrand, we consider the quantity $[\delta G^R_{\mu\nu,\alpha\beta}(\omega,\vp)-\delta G^R_{\mu\nu,\alpha\beta}(\omega\rightarrow\infty,\vp)]/(\omega-i\omega')$, 
where the $\delta$ stands for the subtraction of the $T=0$ part, $\delta G^R\equiv G^R-G^R_{T=0}$. The retarded Green function in the energy-momentum tensor 
sector\footnote{In this channel, the Green function can alternatively be defined in curved space-time~\cite{Romatschke:2009ng} instead of the flat one. The two definitions differ by 
a contact term, which is proportional to $\delta^{(4)}(x-y)$ in coordinate space.
This contact term does not affect the final form of the sum rule, because, as we will see, it is canceled by a similar term coming from the Green function in the IR limit. } 
is defined as 
\begin{equation} 
G^R_{\mu\nu,\alpha\beta}(p) \equiv i \int d^4x e^{ip\cdot x} \theta(x^0)\langle [T_{\mu\nu}(x), T_{\alpha\beta}(0)]\rangle_{T}. 
\label{eq:retarded.correlator.tensor.channel}
\end{equation}
As the retarded function is analytic in the upper half of the complex energy plane, the residue theorem gives 
\bea
\label{eq:residue-theorem}
\delta G^R_{\mu\nu,\alpha\beta}(i\omega,\vp)-\delta G^R_{\mu\nu,\alpha\beta}(\omega''\rightarrow\infty,\vp)
= \frac{1}{2\pi i}\oint_C d\omega' 
\frac{\delta G^R_{\mu\nu,\alpha\beta}(\omega',\vp)-\delta G^R_{\mu\nu,\alpha\beta}(\omega''\rightarrow\infty,\vp)}{\omega'-i\omega} ,
\eea
The subtractions of the $T=0$ part and the $\omega''\rightarrow\infty$ limit remove any potential UV divergence, such that the contribution from the half circle on 
the contour $C$ can be neglected when we take its radius to infinity. 
The above equation thus reduces to 
\bea
\delta G^R_{\mu\nu,\alpha\beta}(0,\vp)-\delta G^R_{\mu\nu,\alpha\beta}(\infty,\vp)
= \frac{1}{\pi i}{\rm P} \int^{\infty}_{-\infty} \frac{d\omega}{\omega}
\delta G^R_{\mu\nu,\alpha\beta}(\omega,\vp),
\eea
where we have used $1/(\omega'-i\omega)\rightarrow {\rm P}(1/\omega')+i\pi\delta(\omega')$.
In the following, we will only consider the Green function of two identical operators, $(\mu,\nu)=(\alpha,\beta)$.
In this case, the real part of $G^R(p)$ is even in $\omega$ while the imaginary part is odd. We thus have 
\bea
\label{eq:basiceq-sumrule}
\delta G^R_{\mu\nu,\alpha\beta}(0,\vp)-\delta G^R_{\mu\nu,\alpha\beta}(\infty,\vp)
= \frac{2}{\pi}  \int^{\infty}_{0} \frac{d\omega}{\omega}
\delta \rho_{\mu\nu,\alpha\beta}(\omega,\vp),
\eea
where we have introduced the spectral function as $\rho_{\mu\nu,\alpha\beta}(p)={\rm {Im}}G^R_{\mu\nu,\alpha\beta}(p)$. 
It is seen in Eq.\,(\ref{eq:basiceq-sumrule}) that, the integral of the spectral function is constrained by the asymptotic behavior of the Green function 
in the UV and IR limits. These are correctly described by the OPE and hydrodynamics, respectively, as long as $|\vp|$ is small enough. 
We note that the OPE expression obtained has an ambiguity in form of a contact term \cite{CaronHuot:2009ns}. However, such an ambiguity does not appear in 
the final sum rule, as it vanishes on the left-hand side of Eq.\,(\ref{eq:basiceq-sumrule}).

We next proceed to a more concrete discussion in the shear and bulk sectors. 
We set the direction of $\vp$ to the $z$-axis, in which the corresponding components reduce to the simple forms, 
$G_{\eta}\equiv G^R_{12,12}$ and $G_{\zeta}\equiv g^{\mu\nu} g^{\alpha\beta} G^{R }_{\mu \nu,\alpha\beta}$, with $g^{\mu\nu}=$diag$(1,-1,-1,-1)$.
We consider only these two components, where the above assumption [$(\mu,\nu)=(\alpha,\beta)$] is valid. 
For the general tensor decomposition of $G^R_{\mu\nu,\alpha\beta}$, see Ref.~\cite{Kovtun:2005ev}. 
We will derive the sum rule in the shear sector\footnote{This channel is sometimes referred to as the tensor channel.} first, and move to the bulk sector thereafter. 
%%%%%%%%%%%%%%%%%
%%%%%%%%%%%%%%%%%%%%%%%%%%%%%%%%%%%%%%%%%%%%%%%%
 
\paragraph{Shear sector}\mbox{}\\
In this sector, we confine our discussion to pure Yang-Mills theory, as the sum rule in full QCD has not been obtained yet.
The OPE at leading order reads \cite{Osborn:1993cr} 
\bea 
\label{eq:OPE-shear}  
G_{\eta}(p) 
&= -3p + A\langle \alpha_s G^2\rangle_{T} , 
\eea 
where $A$ is an undetermined constant, which can in principle be obtained from a higher-order calculation. 
$p$ stands for the pressure. 
The relation between p, the energy density $\epsilon$ and the traceless component of energy-momentum tensor can be given as $\langle T^{00} \rangle_{T} =3(\epsilon+p)/4$. 
The gluon condensate is also related to these thermodynamic quantities as $\epsilon-3p = -b_0\langle \alpha_s G^2\rangle_{T}/(8\pi)$ for weak coupling, 
where $b_0\equiv 11\Nc/3$. Note the difference to Eqs.\,(\ref{eq:trace.anomaly.5}) and (\ref{eq:trace.anomaly.5.2}), where we use $N_f = 3$ instead of $N_f = 0$ here. 
 
On the other hand, second order hydrodynamics provides the following expression about the IR behavior of the 
retarded correlator \cite{Baier:2007ix}, 
\bea
\label{eq:hydro-shear} 
G_{\eta}(p) 
&= -p+i\eta\omega
+\left(\eta\tau_\pi-\dfrac{1}{2} \kappa \right)\omega^2
-\dfrac{1}{2}\kappa\vp^2.  
\eea 
Here, $\eta$ is a first order transport coefficient (shear viscosity), while $\tau_\pi$ and $\kappa$ are of second order. 
The terms proportional to $\omega^2$ and $\vp^2$ are valid only in the conformal limit, and are expected to be modified in the non-conformal case. 
If the long time tail caused by the interaction among the hydro modes is taken into account, second order hydrodynamics is modified so that a 
non-analytic term ($\sim\omega^{3/2}$) enters the expression above.
Such an effect is suppressed in the large $\Nc$ limit, both in the weak and strong coupling limits \cite{Kovtun:2003vj}.

Combining these two expressions with Eq.~(\ref{eq:basiceq-sumrule}), we get the first sum rule (sum rule 1) in the shear sector, which reads~\cite{Romatschke:2009ng} 
\bea
\label{eq:sumrule-1-shear} 
 \frac{\epsilon+p}{2}
 +B(\epsilon-3p)
= \frac{2}{\pi}  \int^{\infty}_{0} \frac{d\omega}{\omega}
\delta \rho_{\eta}(\omega,\vzero),
\eea
where $B$ is an undetermined constant, which we introduced because the OPE expression has an undetermined coefficient for the gluon condensate term.
In this derivation, we used only the asymptotic behavior of the Green function in the UV and IR energy regions.
The former is given by the OPE, in which the Wilson coefficients are evaluated exactly at infinitely large energy, while the latter is given by hydrodynamics, 
which is a reliable low energy effective theory for channels of conserved quantities.
Thus, this sum rule is exact, once the undetermined constant is fixed by a higher order OPE calculation. 
This sum rule was generalized to the case with a lattice discretization later in Refs.\,\cite{Meyer:2011gj, Meyer:2010gu}.

Compared to the more conventional QCDSRs discussed in detail in previous sections, the exact sum rules do not introduce an UV cutoff, so that the leading order 
OPE result becomes exact due to asymptotic freedom. 
The UV divergence is removed by subtracting the spectral function at $T=0$, instead of a cutoff. 
Furthermore, hydrodynamics is used to describe the IR behavior, unlike in the conventional sum rules, for which IR quantities do not appear. 
It is worth mentioning here a 
similar approach, which was used in Ref.\,\cite{Kapusta:1993hq} to construct 
a sum rule from the difference between the vector and the axial vector spectral functions, in order to discuss the effect of chiral 
symmetry, its breaking and restoration. Taking this difference, the UV divergence is removed as it happens for the sum rules in this paper. 
These so-called Weinberg sum rules (proposed first in Ref.\,\cite{Weinberg:1967kj} for the vacuum case) are frequently discussed in 
the context of chiral symmetry breaking and its restoration at finite temperature or density as the vector and axial vector spectral functions 
should become identical in a situation of completely restored chiral symmetry. 
Recent studies related to this topic can be found for instance in Refs.\cite{Hohler:2013eba,Cabrera:2009qr,Hohler:2012xd,Holt:2012wr}. 

The derivation of the second sum rule is somewhat non-trivial. 
Equation\,(\ref{eq:residue-theorem}) for the shear channel is first rewritten as
\bea
\nonumber
\delta G_{\eta}(i\omega,\vp)-\delta G_{\eta}(\infty,\vp) 
&=& \dfrac{1}{\pi }\int^\infty_{0} d\omega' 
\dfrac{\omega' \delta \rho_{\eta}(\omega',\vp)+\omega[\re\delta G_{\eta}(\omega',\vp)-\delta G_{\eta}(\infty,\vp)]}{\omega'{}^2+\omega^2}\\
\label{eq:derivation}
&=& \dfrac{2}{\pi }\int^\infty_{0} d\omega' 
\dfrac{\omega' \delta \rho_{\eta}(\omega',\vp)}{\omega'{}^2+\omega^2} ,
\eea
where we have used the relation 
\bea
0 =  \int^\infty_{-\infty} d\omega' 
\frac{\omega' \delta \rho_{\eta}(\omega',\vp)-\omega[\re\delta G_{\eta}(\omega',\vp)-\delta G_{\eta}(\infty,\vp)]}{\omega'{}^2+\omega^2},
\eea
 in the last line, which is obtained by using the residual theorem for the integral $\oint_C d\omega' 
[\delta G_{\eta}(\omega',\vp)-\delta G_{\eta}(\infty,\vp)]/(\omega'+i\omega)$.
Subtracting $i\omega \delta G'_\eta = 2i\omega^2\delta G'_\eta \int^\infty_0 d\omega'/[\pi(\omega^2+\omega'{}^2)]$ from Eq.~(\ref{eq:derivation}), which is necessary to regularize the IR singularity in the integral, we get 
\bea
\delta G_{\eta}(i\omega,\vp)-\delta G_{\eta}(0,\vp) -i\omega \delta G'_\eta(0,\vp) 
= \frac{2}{\pi }\omega^2 \int^\infty_{0} d\omega' \frac{1}{\omega^2+\omega'{}^2}
\left[ \delta \rho_{\eta}(\omega',\vp)\frac{-1}{\omega'} 
+\delta \rho'_\eta{}(0,\vp)\right],
\eea
where $'$ stands for the derivative in terms of energy ($\omega$, $\omega'$). 
Taking the $\omega \to 0$ limit, this reduces to
\bea 
\frac{1}{2}\delta G''_{\eta}(0,\vp)
= \frac{2}{\pi } \int^\infty_{0} d\omega \frac{1}{\omega^3}
\left[ \delta \rho_{\eta}(\omega,\vp)
-\omega \delta \rho'_\eta{}(0,\vp)\right]. 
\eea
Here, we have changed the integration variable from $\omega'$ to $\omega$ for simplicity.
We hence obtain the following second sum rule (sum rule 2) \cite{Romatschke:2009ng} by using the expressions of the Green function in the IR limit [see Eq.\,(\ref{eq:hydro-shear})], 
\bea
\label{eq:sumrule-2-shear.1}
\frac{1}{2}\delta G''_{\eta}(0,\vp)
&=& \frac{2}{\pi } \int^\infty_{0} d\omega \frac{1}{\omega^3}
\left[ \delta \rho_{\eta}(\omega,\vp)
-\omega \delta \rho'_\eta{}(0,\vp)\right]. 
\eea
Taking furthermore the $|\vp|=0$ limit, we have 
\bea
\label{eq:sumrule-2-shear}
\eta\tau_\pi-\dfrac{1}{2} \kappa
&=&\frac{2}{\pi } \int^\infty_{0} d\omega \frac{1}{\omega^3}
\left[ \delta \rho_{\eta}(\omega,\vzero)
-\eta\omega \right]. 
\label{eq:final_sum_rule_shear}
\eea
In this sum rule, 
the $\omega^2$ term obtained from hydrodynamics in Eq.\,(\ref{eq:hydro-shear}) was used. Eq.\,(\ref{eq:final_sum_rule_shear}) is thus expected to be modified for finite $\Nc$. 
%This sum rule is using the $\omega^2$ term obtained from hydrodynamics in Eq.\,(\ref{eq:hydro-shear}) and will thus be modified for finite $\Nc$. 
%%%%%%%%%%%%%%%%%%%%%%%%%%%%%%%%%%%%%%%%%%%%%%%%

\paragraph{Bulk sector}\mbox{}\\ 
The OPE as before provides the UV behavior in the bulk sector \cite{CaronHuot:2009ns}, predicting that $G_\zeta(p)$ vanishes in the $\omega\rightarrow\infty$ limit. 
On the other hand, the IR behavior is obtained from hydrodynamics \cite{Kharzeev:2007wb, Karsch:2007jc} as 
\bea
\label{eq:hydro-bulk}
G_{\zeta}(\omega=0,\vp\rightarrow\vzero)
= -\left(T\dfrac{\partial}{\partial T}-4\right)(\epsilon - 3p)
-\left(T\dfrac{\partial}{\partial T}-2\right) \sum_f m_f \delta \langle \overline{\psi}_f \psi_f\rangle_{T} ,
\eea 
where ${\cal{O}}(m^2)$ terms are neglected. 

Making use of these asymptotic expressions, Eq.\,(\ref{eq:basiceq-sumrule}) yields the following sum rule for the bulk sector, 
\bea
\label{eq:sumrule-bulk}
-\left(T\frac{\partial}{\partial T}-4\right)(\epsilon-3p) 
-\left(T\frac{\partial}{\partial T}-2\right) \sum_f m_f \delta \langle \overline{\psi}_f \psi_f\rangle_{T} 
= \frac{2}{\pi}  \int^{\infty}_{0} \frac{d\omega}{\omega}
\delta \rho_{\zeta}(\omega,\vzero).
\eea 
Because we have omitted ${\cal{O}}(m^2)$ terms, this sum rule is only valid for light quarks, and becomes exact for the massless or pure glue case. 
This sum rule was derived for the first time in Refs.\,\cite{Kharzeev:2007wb, Karsch:2007jc} for infinitesimal $|\vp|$, and was generalized to the case of finite 
density in Ref.\,\cite{Kadam:2014cua} and to a non-zero magnetic field in Ref.\,\cite{Kadam:2014xka}. 
Later, the sum rule for the case where the $\vp=\vzero$ limit is taken first so that the sound peak does not appear in $\delta \rho_{\zeta}(\omega,\vp)$, 
was obtained in Ref.\,\cite{Romatschke:2009ng}. 

Let us furthermore mention that in addition to the sum rules in the shear and bulk components, similar sum rules were derived for other components in the 
energy-momentum tensor channel in Ref.\,\cite{Meyer:2007fc}.
%%%%%%%%%%%%%%%%%%%%%%%%%%%%%%

\subsubsection{Applications}
We here review a few possible applications of the above sum rules, starting with the shear sector.
%%%%%%%%%%%%%%%%%%%%%%%%%%%%%%%%%%%%%%%%%%%%%%%%
 
\paragraph{Shear sector}\mbox{}\\ 
\begin{enumerate}
\item In both strong coupling ${\cal N}=4$ super Yang-Mills theory \cite{Romatschke:2009ng} and the weakly coupled QCD \cite{Romatschke:2009ng,Arnold:2003zc,York:2008rr}, 
the inequality $\eta\tau_\pi>\kappa/2$ holds. 
Through sum rule 2, this 
property constrains the shear spectral function $\delta\rho_\eta(p)$ to be larger than $\eta\omega$ at least in some $\omega$ region \cite{Romatschke:2009ng}. 
Especially, the simplest ansatz, for which the spectral function is saturated by a Lorentzian peak 
[$\delta\rho_\eta(\omega,\vp=\vzero)=\Gamma^2\eta\omega/(\omega^2+\Gamma^2)$]\footnote{The overall coefficient is determined so that 
it matches with the definition of $\eta$, $\rho_\eta(\omega,\vzero)\simeq \eta\omega$, which can be read off from Eq.\,(\ref{eq:hydro-shear}).} 
at zero momentum, was shown to be inconsistent with sum rule 2: 
the integrand in this sum rule becomes $\delta\rho_\eta-\eta\omega = -\eta\omega^3/(\omega^2+\Gamma^2)$, which is negative while the left-hand side of the sum rule is positive.

\item In pure Yang-Mills theory, it was confirmed that the shear spectral function calculated at NLO accuracy satisfies sum rule 1 of Eq.\,(\ref{eq:sumrule-1-shear}) \cite{Schroder:2011ht}. 
This is one example, in which 
an exact sum rule is used as a consistency check of an explicit spectral function calculation. 
\end{enumerate}
Finally, we make a few remarks on possible future applications. 
First, the left-hand side of sum rule 1 can be calculated with lattice QCD without having to deal with the problem of analytic continuation. 
Once the constant $B$ is fixed, it will constrain the spectral function and may be used to improve spectral fits to lattice QCD data. 
Next, sum rule 2 can potentially be of help in determining $\kappa$, making use of the spectral function obtained from lattice QCD. 
Actually, some attempts in this direction have already been tried in the vector channel, as will be seen in the next subsection.
%%%%%%%%%%%%%%%%%%%%%%%%%%%%%%%%%%%%%%%%%%%%%%%% 
 
\paragraph{Bulk sector}\mbox{}\\
To obtain dynamical or real time quantities from a lattice QCD calculation, one has to overcome the well known problem of analytical continuation, 
as already mentioned earlier. 
Namely, lattice QCD cannot evaluate quantities defined in real time such as spectral functions directly, but can only compute imaginary time objects. 
The Green function $G_E$ in Euclidian time for instance can be obtained on the lattice and is related to the spectral function as 
\bea
G_E(\tau)
& = \displaystyle \int^\infty_0 \dfrac{d\omega}{2\pi} \rho(\omega)
\dfrac{\cosh[\omega(\tau-1/2T)]}{\sinh(\omega/2T)}
\eea
with Euclidian time $\tau$. 
Thus, assuming an ansatz about the form of the spectral function, or attempting to get a model-independent result 
from numerical methods such as MEM \cite{Gubler:2010cf, Jarrell:1996rrw,  Asakawa:2000tr, Tripolt:2018xeo} becomes necessary. 

\begin{enumerate}
\item There are already several studies attempting to use exact sum rules to evaluate the bulk viscosity from lattice QCD. 
Substituting the simplest ansatz for the spectral function\footnote{$\Gamma(T)$ was set to the scale at which the values 
for the running coupling evaluated by lattice QCD \cite{Kaczmarek:2004gv} and perturbation theory coincide.}, 
$\delta\rho_\zeta(\omega,\vp=\vzero)=9\zeta \Gamma^2\omega/[\pi(\omega^2+\Gamma^2)]$, to a preliminary version of 
the sum rule and matching it with thermodynamic quantities and the chiral condensate evaluated by lattice QCD, the bulk 
viscosity $\zeta$ was evaluated in Refs.\,\cite{Kharzeev:2007wb, Karsch:2007jc}. 
Later, the sum rule was corrected in Ref.\,\cite{Romatschke:2009ng}, and the abovementioned simple ansatz was criticized because at least in pure Yang-Mills theory, the 
left-hand side of the sum rule was shown to be negative in lattice QCD \cite{Boyd:1996bx}, which is inconsistent with the simple Lorentzian ansatz, 
which always yields a positive contribution to the sum rule. 
Subsequently, a similar method was attempted by using the correct version of the sum rule and a more sophisticated ansatz 
for $\delta\rho_\zeta$ \cite{Meyer:2010ii}.

\item It was shown that the spectral function calculated at LO \cite{CaronHuot:2009ns}, and later at NLO \cite{Laine:2011xm} satisfies the sum rule. 
This provides a cross-check for the perturbative result, as it was the case in the shear sector. 
\end{enumerate}
%%%%%%%%%%%%%%%%%%%%%%%%%%%%%%

\subsection{Vector current channel}
In this subsection, we review the exact sum rules and their applications for the correlator of the conserved vector current. 
%%%%%%%%%%%%%%%%%%%%%%%%%%%%%%

\subsubsection{Derivation}
The derivation of the sum rules given in Refs.\,\cite{Gubler:2016hnf,Gubler:2017qbs} is similar to that in the previous energy-momentum tensor case. 
The basic equation is still Eq.\,(\ref{eq:basiceq-sumrule}), where $G^R_{\mu\nu,\alpha\beta}$ should be replaced with $G^R_{\mu\nu}$, which is the Green function of the vector current. 
This function has two independent channels, called transverse and longitudinal. Namely
\bea
G^R_{\mu\nu}(p) = P^T_{\mu\nu}(p) G_T(p)
+P^L_{\mu\nu}(p) G_L(p), 
\eea
where $P^T_{\mu\nu}(p)\equiv g^{\mu i} g^{\nu j}\left(\delta^{ij}-\frac{p^i p^j}{\vp^2}\right)$ 
and $P^L_{\mu\nu}(p)\equiv P^{\mu\nu}_0(p) -P^{\mu\nu}_T(p)$ with $P^{\mu\nu}_0(p) \equiv -\left(g^{\mu\nu}-\frac{p^\mu p^\nu}{p^2}\right)$, are the projection tensors for transverse and longitudinal channels.
The OPE of both components can be found in Appendix \ref{sec:Appendix1}. 
We first derive the sum rules in the transverse sector, and then continue with the longitudinal sector. 
%%%%%%%%%%%%%%%%%%%%%%%%%%%%%%%%%%%%%%%%%%%%%%%%

\paragraph{Transverse sector}\mbox{}\\
The UV behavior is given by the OPE result of Eq.\,(\ref{eq:OPE-w-RG-T}) in the Appendix. The IR asymptotic behavior is described by hydrodynamics as \cite{Kovtun:2012rj} 
\bea
G_{T}(p) = i\sigma\omega-\sigma\tau_J\omega^2+\kappa_B\vp^2
+{\cal O}(\omega^3, \omega\vp^2, \vp^4).  
\eea
Here, $\sigma$ is the electrical conductivity, $\tau_J$ a second order transport 
coefficient corresponding to the $\partial_0\vE$ term in the current, and $\kappa_B$ the transport coefficient corresponding to the $\nabla\times\vB$ term, respectively. 
Combining these expressions, we get the first sum rule (sum rule 1) in the transverse sector from Eq.\,(\ref{eq:basiceq-sumrule}), 
\bea
\label{eq:sumrule-1-transverse}
\kappa_B \vp^2
+{\cal O}(\vp^4)
= \frac{2}{\pi}\int^\infty_{0} d\omega
\frac{\delta \rho_{T}(\omega,\vp)}{\omega}.
\eea
This sum rule at $|\vp|=0$ was first obtained from the current conservation law in Ref.\,\cite{Bernecker:2011gh}. 

Using $\omega^2 G_T(p)$ instead of $G_T(p)$ in Eq.\,(\ref{eq:basiceq-sumrule}), we obtain the second sum rule (sum rule 2) in the transverse channel as \cite{Gubler:2016hnf,Gubler:2017qbs}, 
\bea
\label{eq:sumrule-2-transverse}
-e^2\sum q^2_f 
\Biggl[\left\{2m_f \delta\left\langle\overline{\psi}_f \psi_f \right\rangle_{T}
+\frac{1}{12}\delta\left\langle \frac{\alpha_s}{\pi}G^2\right\rangle_{T} \right\}
+\frac{8}{3}\frac{1}{4\Cf+\Nf}\delta\left\langle T^{00} \right\rangle_{T} 
\Biggr] 
= \frac{2}{\pi}\int^\infty_{0} d\omega \omega
\delta \rho_{T}(\omega,\vp). 
\eea
A preliminary version of this sum rule at $|\vp|=0$ was in fact derived long time ago in Refs.\,\cite{Huang:1994fs, Huang:1994vd}, however with incorrect coefficients. 
We note that $T^{00}$, appearing in the above sum rule is not the energy-momentum tensor itself, but its trace subtracted version. 

Finally, making use of the same method as in the derivation of sum rule 2 in the shear channel [Eq.\,(\ref{eq:sumrule-2-shear})], 
we can obtain another sum rule (sum rule 3) in the transverse channel \cite{Gubler:2016hnf,Gubler:2017qbs}, 
\bea
\label{eq:sumrule-3-transverse.1}
\frac{1}{2}\delta G''_{T}(0,\vp) 
&=& \frac{2}{\pi } \int^\infty_{0} d\omega \frac{1}{\omega^3}
\left[ \delta \rho_{T}(\omega,\vp)
-\omega \delta \rho'_T{}(0,\vp)\right]. 
\eea
Taking again the $|\vp|=0$ limit, we derive 
\bea
\label{eq:sumrule-3-transverse}
-\sigma\tau_J
&=& \frac{2}{\pi } \int^\infty_{0} d\omega \frac{1}{\omega^3}
\left[ \delta \rho_{T}(\omega,\vzero)
-\sigma\omega \right].
\eea
As in the shear channel, this sum rule holds only in the large $\Nc$ limit.

It was shown and discussed in detail in Refs.\,\cite{Gubler:2016hnf,Gubler:2017qbs} that the spectral function calculated at leading order in the weak coupling expansion 
satisfies the above three sum rules. 
%%%%%%%%%%%%%%%%%%%%%%%%%%%%%%%%%%%%%%%%%%%%%%%%

\paragraph{Longitudinal sector}\mbox{}\\
Hydrodynamics gives the following IR behavior \cite{Kovtun:2012rj}, 
\bea
\label{eq:G00-hydro}
G^R_{00}(p) = i\sigma\vp^2\frac{1+{\cal O}(\omega,\vp^2)}{\omega+iD\vp^2+{\cal O}(\omega\vp^2,\vp^4)}, 
\eea
where $D$ is the diffusion constant.
Before discussing the sum rules, let us remember that the retarded Green function in the longitudinal channel is exactly known 
at zero momentum \cite{Ding:2010ga} from the charge conservation law, 
\begin{align}
\label{eq:rho-L-p=0-exact}
\rho_{00}(\omega,\vzero) 
&= \pi \chi_q \omega\delta(\omega) ,
\end{align}
where $\chi_q\equiv \int d^3\vx \langle j^0(\vx) j^0(\vzero) \rangle/T$ is the charge susceptibility. 
Therefore, sum rules in the longitudinal channel provide nontrivial information only when $\vp$ is finite.
We hence consider only the finite momentum case in this subsection. 
Furthermore, matching Eq.\,(\ref{eq:rho-L-p=0-exact}) with the hydro result of Eq.\,(\ref{eq:G00-hydro}), we obtain $\sigma/D=\chi_q$. 

From Eq.\,(\ref{eq:G00-hydro}), the OPE result of Eq.\,(\ref{eq:OPE-w-RG-L}), and Eq.\,(\ref{eq:basiceq-sumrule}), we derive the first sum rule (sum rule 1) in the longitudinal channel,
\bea
\label{eq:sumrule-1-longitudinal}
\frac{\sigma}{D}
+{\cal O}(\vp^2) 
= \frac{2}{\pi}\int^\infty_{0} d\omega
\frac{\delta \rho_{00}(\omega,\vp)}{\omega}. 
\eea 

Next, considering the integral in Eq.\,(\ref{eq:basiceq-sumrule}) with two more powers of $\omega$, we are led to  
\bea
\label{eq:sumrule-2-longitudinal}
0= \frac{2}{\pi}\int^\infty_{0} d\omega \omega
\delta \rho_{00}(\omega,\vp), 
\eea
which is the second sum rule (sum rule 2) in the longitudinal channel. 
Sum rule 2 was first derived using the current conservation in Ref.\,\cite{Bernecker:2011gh} 
(note that $\rho_{00}=\rho_{33}\vp^2/\omega^2$).
This implies that the sum rule is in fact exact for any momentum value, and one does not need to assume that it is small here. 

Finally, increasing the powers of $\omega$ by two in the integral of Eq.\,(\ref{eq:basiceq-sumrule}), we obtain the third sum rule (sum rule 3) in the longitudinal channel 
as
\bea
&& -e^2\sum q^2_f \vp^2 
\Bigl[\left\{2m_f \delta\left\langle\overline{\psi}_f \psi_f \right\rangle_{T}
+\frac{1}{12}\delta\left\langle \frac{\alpha_s}{\pi}G^2\right\rangle_{T} \right\} 
+\frac{8}{3}\frac{1}{4\Cf+\Nf} \delta\left\langle T^{00} \right\rangle_{T}  
\Bigr] \nonumber \\
&=& \frac{2}{\pi}\int^\infty_{0} d\omega \omega^3
\delta \rho_{00}(\omega,\vp). 
\label{eq:sumrule-3-longitudinal}
\eea

%%%%%%%%%%%%%%%%%%%%%%%%%%%%%%
\subsubsection{Applications}
Let us briefly review possible applications of the exact sum rules derived in this subsection. 
Applications currently exist only for the zero momentum case, for which the transverse and longitudinal sum rules are degenerate.
Therefore, we will not distinguish the two channels below, and refer only to the sum rules in the former channel. 

\begin{enumerate}
\item The earliest vector channel sum rules application is to our knowledge the study of spectral properties at the chiral phase transition in Refs.\,\cite{Huang:1994fs, Huang:1994vd}. 
The authors of these works discussed the structure of the spectral function at the transition temperature/density in the context of soft modes 
related to the phase transition \cite{Hatsuda:1985eb}.

\item Using perturbative QCD, it 
was shown in Refs.\,\cite{Gubler:2016hnf,Gubler:2017qbs} that the spectral function calculated at LO satisfies all three sum rules, 
therefore demonstrating again that these sum rules can be used as a consistency check for perturbative calculations.  

\item The sum rules can be applied to the analysis of the spectral function and transport coefficients in lattice QCD. 
As mentioned in Section\,\ref{sec:Tmunu}, 
because of the issue of analytic continuation 
an ansatz for the form of the spectral function often needs to be assumed in lattice QCD analyses of the spectral function. 
Earlier works such as Ref.\,\cite{Ding:2010ga} have proposed an ansatz motivated by weak coupling results, namely\footnote{Note that their conventions differ from ours by a factor of $1/6$.}
\bea
\label{eq:rho-lattice-naive}
\frac{\rho(\omega)}{\Cem}
=  A_T \rho_{\mathrm{peak}}(\omega)
+\kappa \rho_{\mathrm{cont}}(\omega), 
\eea
where $\rho(\omega)\equiv\rho_T(\omega,\vp=\vzero)$. 
The two parts, 
\bea
\rho_{\mathrm{peak}}(\omega) &\equiv& \dfrac{1}{3} \dfrac{\omega\Gamma/2}{\omega^2+(\Gamma/2)^2} , \\
\rho_{\mathrm{cont}}(\omega) &\equiv & \dfrac{\omega^2}{4\pi }\left(1-2\nf\left(\dfrac{\omega}{2}\right)\right), 
\eea
correspond to the transport peak and the continuum, which can be derived in the weak coupling limit. 
The former is a Lorentzian peak appearing at an energy scale governed by transport processes [given approximately as 
$(\text{mean free path})^{-1}\ll T$], while the latter appears at a scale of the order of $T$ 
and is caused by the process $\gamma\rightarrow q \bar{q}$.
However, Eq.\,(\ref{eq:rho-lattice-naive}) generally does not satisfy sum rule 1 of Eq.\,(\ref{eq:sumrule-1-transverse}). 
To satisfy it, the simple relation $A_T=\kappa T^2$ needs to hold. 
Furthermore, it cannot satisfy sum rules 2 [Eq.\,(\ref{eq:sumrule-2-transverse})] and 3 [Eq.\,(\ref{eq:sumrule-3-transverse})], 
because it would generate UV and IR divergences in the respective integrals. 
This happens because the transport peak and the continuum are simply summed in Eq.\,(\ref{eq:rho-lattice-naive}), 
while in principle there should be a smooth crossover between the two at least in the weak coupling case.

In a later analysis, a more sophisticated form for the spectral function was suggested in Refs.\,\cite{Brandt:2015aqk,Brandt:2012jc}. 
Here, we mention only the one given in Ref.\,\cite{Brandt:2015aqk}, which reads
\bea 
\label{eq:rho-lattice-T=0}
\frac{\rho_{T\simeq 0}(\omega) }{\Cem}
&=& \dfrac{\pi}{3}a_V \delta(\omega-m_V)
+ \kappa_0 \rho_{\mathrm{cont}}(\omega)\theta(\omega-\Omega_0),\\
\frac{\rho(\omega) }{\Cem}
&= & A_T\rho_{\text{peak}}(\omega)
+ \dfrac{\pi}{3} a_T \delta(\omega-m_T)
+ \theta(\omega-\Omega_T) \tilde{\kappa}_0\rho_{\text{cont}}(\omega) \nonumber \\
&& + \theta(\omega-\Omega_O)\kappa_O \rho'_{\text{tail}}(\omega). 
\label{eq:rho-lattice-T.not.0}
\eea
Here, $\rho'_{\text{tail}}(\omega) \equiv 1/(4\pi \omega^2)$, while the coefficient of the continuum term is modified to
\bea 
\tilde{\kappa}_0 &\equiv \kappa_0 
+\kappa_1\left[1-\tanh\left( \dfrac{\omega}{\Omega_0\eta}\right)^2\right] ,
\eea
to obtain a better fit. 
The former expression of Eq.\,(\ref{eq:rho-lattice-T=0}) is the ansatz for zero temperature (or a temperature sufficiently below $T_c$), 
while the latter is an ansatz suitable for high temperature. 
The former in essence corresponds to the ``pole + continuum'' ansatz of conventional QCDSR studies, 
while the latter is motivated by the transport 
peak and the UV tail at weak coupling, briefly mentioned in Section \ref{sec:OPE}. 
$\delta\rho$ is obtained by subtracting the former from the latter. Requiring that it satisfies sum rule 1, the authors of Ref.\,\cite{Brandt:2015aqk} derived a constraint on the parameters 
appearing in Eqs.\,(\ref{eq:rho-lattice-T=0}) and (\ref{eq:rho-lattice-T.not.0}). 
Moreover, the spectral function can be adjusted consistently with sum rule 3, as the potential IR divergence is regularized by the cutoff parameters ($\Omega_0$, $\Omega_T$, $\Omega_O$). 
However, it still violates sum rule 2 because the transport peak and the UV tail cause a UV divergence. 

Recently, it was attempted in Ref.\,\cite{Gubler:2017qbs} to improve this ansatz such that it can satisfy both sum rules 2 and 3. Specifically, 
the proposed ansatz reads 
\bea 
\label{eq:rho-improved}
\frac{\rho(\omega) }{\Cem}
&= A_T\rho_{\text{peak}}(\omega)[1-A(\omega)] 
+ \dfrac{\pi}{3} a_T \delta(\omega-m_T)
+ \tilde{\kappa}_0\rho_{\text{cont}}(\omega) A(\omega)
+ \theta(\omega-\Omega_O)\kappa_O \rho_{\text{tail}}(\omega).
\eea 
Compared with the previous version of Eq.\,(\ref{eq:rho-lattice-T.not.0}), two features are modified. 
First, the transport peak and the continuum are smoothly connected by the function $A(\omega) \equiv \tanh\left(\omega^2/\Delta^2\right)$, instead of the jump at $\omega=\Omega_T$.
Second, the UV tail term is modified to 
\bea
\rho_{\text{tail}}(\omega)
&\equiv &\frac{1}{4\pi}\frac{1}{ \omega^2 [\ln(\omega/\Lambda_{\text{QCD}})]^{1+\tilde{a}}} ,
\eea
which more closely resembles the OPE expression given in Eq.\,(\ref{eq:UVtail-T}), 
that includes a logarithmic dependence.
These two improvements help to regularize the UV divergence, such that, as a whole, the spectral 
function can be consistent with sum rule 2.
The spectrum at $T=0$ remains the same as in Eq.\,(\ref{eq:rho-lattice-T=0}). 
In Ref.\,\cite{Gubler:2017qbs}, 
sum rules 1 and 2 were furthermore employed 
to reduce the number of independent fitting parameters in Eq.\,(\ref{eq:rho-improved}). 
Specifically, thermodynamic quantities of Ref.\,\cite{Bazavov:2014pvz} were used to express the condensates and $T^{00}$ on the 
left hand side of Eq.\,(\ref{eq:sumrule-2-transverse}). 
For illustration, we show the resulting fitted spectral functions at various temperatures in Fig.\,\ref{fig:rho.lattice.fit}. 
With the spectral function fixed, 
sum rule 3 [Eq.\,(\ref{eq:sumrule-3-transverse})] was subsequently used to evaluate the second-order transport coefficient $\tau_J$. 
These results demonstrate that the sum rules are helpful for spectral fits to 
lattice QCD data. 
As shown in Ref.\,\cite{Gubler:2017qbs}, more precise and a larger number of data points will however be needed for a 
conclusive determination of the spectral function at finite temperature. 
\end{enumerate}

Let us conclude this section with a few remarks about possible future directions.
First, the spectral function at $T=0$ can in principle be extracted from the experimental cross section for $e^+ e^-\rightarrow $ hadron processes \cite{Tanabashi:2018oca}. 
Therefore, once physical point lattice data become available, such experimental data can be used instead of the simple ansatz of Eq.\,(\ref{eq:rho-lattice-T=0}). 
Also, as mentioned above, more data points should be studied in future lattice QCD analyses, such that more accurate spectral functions can be obtained with the help of the sum rules. 
Finally, we mention potential applications at finite momentum.
The transport coefficient $\kappa_B$ can be determined with lattice QCD nonperturbatively without suffering from the problem of analytic continuation (see Ref.\,\cite{Brandt:2013faa}). 
Therefore, once the 
lattice QCD data for the vector current propagator at finite momentum become available, 
we expect that sum rule 1 in the transverse channel [Eq.\,(\ref{eq:sumrule-1-transverse})] will become useful to constrain the shape of the spectral function at 
non-zero momentum. 
%%%%%%%%%%%%%%%%%
\begin{figure}[tb]
\begin{center}
\begin{minipage}[t]{8 cm}
\vspace{0.5 cm}
\hspace{-2.0 cm}
\includegraphics[width=12cm,bb=0 0 360 252]{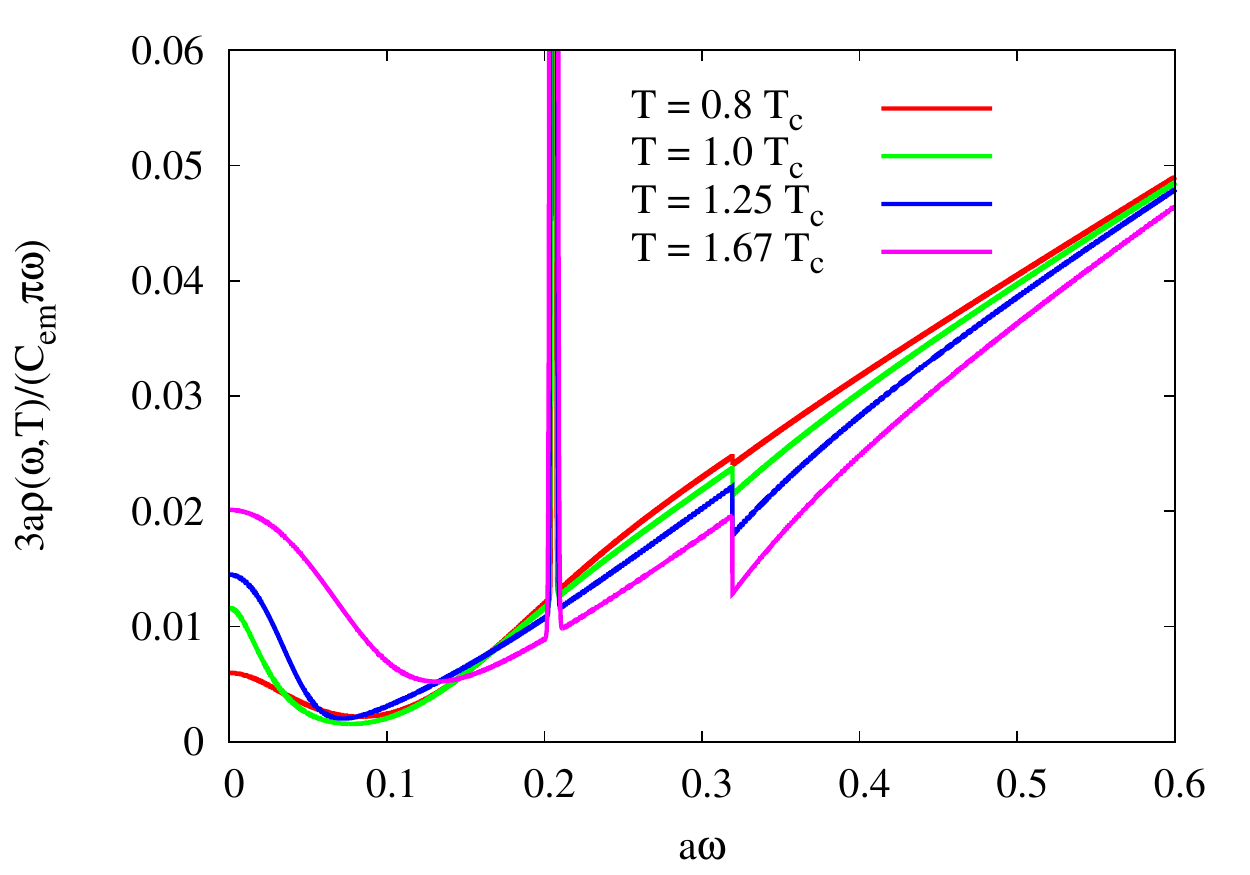}
\vspace{-0.5 cm}
\end{minipage}
\begin{minipage}[t]{16.5 cm}
\caption{The appropriately normalized spectral function of the $\rho$ meson channel, $\rho(\omega)$, obtained by fitting Eq.\,(\ref{eq:rho-improved}) to the lattice QCD data of Ref.\,\cite{Brandt:2015aqk}. 
Here $a$ is the lattice spacing and $C_{\text{em}}\equiv e^2\sum_f q^2_f$. 
Taken from Fig.\,5 of Ref.\,\cite{Gubler:2017qbs}. 
\label{fig:rho.lattice.fit}}
\end{minipage}
\end{center}
\end{figure}
%%%%%%%%%%%%%%%%%%%%

\section{Summary and Outlook \label{sec:SummandOutl}}
In this review article, we have given an overview of recent developments in QCDSR studies. 
Particular focus has been laid on reviewing determinations of QCD condensates based on methods such as lattice QCD 
or chiral perturbation theory and, where possible, on experimental data. In doing this, we have attempted to provide a 
comprehensive survey of the most recent and relevant literature. 

We have furthermore not only critically examined the traditional QCDSR analysis method which makes use 
of the Borel transform and subsequently of the so-called Borel-curves for hadron masses and residues, 
but have also looked at alternatives that are presently being used in the QCDSR community. These include, 
for instance, the use of alternative kernels different from the Laplace-type obtained using the Borel transform, 
or the application of the maximum entropy method for extracting the spectral function from the 
sum rules. 

As areas of QCDSR applications have grown and multiplied over the years, we necessarily had to limit ourselves to a 
limited range of QCDSR applications to be discussed in this review, in order not to let the article become 
inhumanly lengthy. 
We have hence focused on applications, for which QCDSRs can produce relevant results 
for experiments and theoretical practitioners of related methods. 
With this guiding principle in mind, we have summarized recent works employing QCDSRs to investigate 
properties of hadrons at finite density, particularly in nuclear matter. Even though such calculations 
have their limitations in terms of precision and lack of ability to obtain detailed features of the 
in-medium spectral functions, they are nevertheless useful as they can provide interpretations of 
observed hadron spectra in dense matter in terms of QCD condensates. In channels containing 
light (valence) quarks, this often leads to direct connections between modifications of the spectral function 
and the (partial) restoration of chiral symmetry in dense matter. Such calculations are moreover 
relevant in view of the fact that lattice QCD studies at finite density are still challenging due to 
the existence of the sign problem. 
Besides the above topic, we have furthermore given a brief overview on the derivation and 
applications of exact sum rules, a topic that that has been studied already long ago, but has 
attracted renewed interest in recent years. 
Some of these exact sum rules are presently being used in spectral fits to lattice QCD data and 
can in the future potentially be used to determine certain combinations of QCD condensates 
or hydrodynamic transport coefficients. 
Another area, where QCDSRs continue to be used frequently, but which we have not covered in 
this review, is the study of exotics, i.e. channels with four, five or even more valence quarks or hadronic molecules. 
We refer interested readers to Ref.\,\cite{Nielsen:2009uh} for an earlier review. 
Another interesting and important topic is the behavior of hadrons at finite temperature \cite{Ayala:2016vnt}, 
especially for understanding experimental measurements from heavy-ion collisions. Here, QCDSRs however have 
to compete with lattice QCD and new ideas such as those proposed in Refs.\,\cite{Araki:2014qya,Gubler:2010cf} 
are needed in order to be competitive.   

Finally, let us give an outlook about how QCDSR studies might develop in the future. 
Certainly, the fields described in the previous paragraph will remain the ones where QCDSR can 
provide the most meaningful contributions to the field of hadron physics and QCD. 
Furthermore, as we have emphasized in this article, the determination of QCD 
condensates has advanced considerably during the last decade. It is especially worth mentioning the 
very precise information now available about the dimension 3 quark (or chiral) condensate, in vacuum, 
at finite temperature and in a constant and homogenous magnetic field\footnote{There is however still rather large uncertainty about its 
behavior at finite density. 
Even the value of its linear order density coefficient, the 
$\pi N$ sigma term, is still controversially discussed.}. Such results, as well as similar ones for other condensates, 
can and are often being taken into account in modern QCDSR analyses. 
Together with the novel analysis methods that have been developed over the years, this shows that 
the field of QCDSRs is continuously evolving and will hopefully continue to do so in the future. 

In all, we hope that this article will be useful for QCD practitioners as a reference for the most 
up-to-date QCD condensate values, for researchers of adjacent fields to get an idea about the present status 
of QCDSR studies and for interested beginners as a starting point in their study of this subject.  

%%%%%%%%%%%%%%%%%%%%%%%%%%%%%%%%%%%%%%%%%%%%%%%%
\section*{Acknowledgements}
The authors thank Wolfram Weise 
for his encouragement and critical reading of the manuscript. 
They furthermore thank HyungJoo Kim for his valuable comments about the article. 
D.S. thanks the Alexander von Humboldt Foundation for supporting his research by its fellowship, and Johann Wolfgang Goethe-Universit\"at for its warm hospitality during the fellowship period.
P.G. thanks Keio University for its hospitality at the time when he started to write this review and for the support by the Mext-Supported Program for
the Strategic Foundation at Private Universities, ``Topological Science" (No. S1511006). 
P.G. is supported by KAKENHI under Contract No. 18K13542 and the Leading Initiative for Excellent Young Researchers (LEADER) of the Japan Society for the Promotion of Science (JSPS).

%%%%%%%%%%%%%%%%%%%%
\appendix
\section{Operator product expansion of correlator and UV tail in the vector channel \label{sec:Appendix1}}
In this Appendix, we provide OPE and UV tail expressions for various vector correlators. 
The OPE is obtained as 
\begin{align}
\label{eq:OPE-wo-RG-T}
\delta G_{T}(\omega,\vp) 
&=e^2\sum q^2_f \frac{1}{p^2} 
 \Bigl[\left\{2m_f \delta\left\langle\overline{\psi}_f \psi_f \right\rangle_{T}
+\frac{1}{12}\delta\left\langle \frac{\alpha_s}{\pi}G^2\right\rangle_{T} \right\}
+\frac{8}{3}\frac{\omega^2+\vp^2}{p^2}\delta\left\langle T^{00}_f \right\rangle_{T} 
\Bigr]
+{\cal O}(\omega^{-4}),\\
\label{eq:OPE-wo-RG-L}
\delta G^{R}_{00}(\omega,\vp) 
&=e^2\sum q^2_f \frac{1}{p^2}\frac{\vp^2}{p^2} 
 \Bigl[\left\{2m_f \delta\left\langle\overline{\psi}_f \psi_f \right\rangle_{T}
+\frac{1}{12}\delta\left\langle \frac{\alpha_s}{\pi}G^2\right\rangle_{T} \right\} 
+\frac{8}{3} \delta\left\langle T^{00}_f \right\rangle_{T}  
\Bigr]
+{\cal O}(\omega^{-6}).
\end{align}
We decompose the quark component of the traceless energy-momentum tensor as 
\begin{align}
\label{eq:Tf-decompose}
T^{00}_{f}=T'{}^{00}_{f}+\frac{1}{4\Cf+\Nf}\left(T^{00}+\frac{2}{\Nf}\tilde{T}^{00}\right),
\end{align} 
where 
\begin{align}
T'{}^{00}_{f}\equiv T^{00}_{f}-\frac{1}{\Nf}\sum_{f'} T^{00}_{f'},\\
T^{00}\equiv \sum_{f'} T^{00}_{f'}+T^{00}_g,\\
\tilde{T}^{00}\equiv 2C_F  \sum_{f'} T^{00}_{f'}-\frac{\Nf}{2} T^{00}_g .
\end{align} 
Here, $T^{\mu\nu}_g\equiv -G^{\mu\alpha}_{a}G^\nu{}_{\alpha a} +g^{\mu\nu}G^2/4$ is the gluon component of the traceless part of the energy-momentum tensor.
A standard renormalization group (RG) analysis yields the following scaling properties~\cite{Peskin:1995ev}: 
\begin{align}
\label{eq:scaling-T'-Ttilde}
\begin{split} 
T'{}^{00}_{f}(\kappa)&=  \left[\frac{\ln\left(\kappa^2_0/\Lambda^2_{\text{QCD}} \right)}{\ln\left(\kappa^2/\Lambda^2_{\text{QCD}} \right)}\right]^{a'} 
T'{}^{00}_{f}(\kappa_0),\\
 \tilde{T}^{00}(\kappa)
&= \left[\frac{\ln\left(\kappa^2_0/\Lambda^2_{\text{QCD}} \right)}{\ln\left(\kappa^2/\Lambda^2_{\text{QCD}} \right)}\right]^{\tilde{a}}
 \tilde{T}^{00}(\kappa_0),
\end{split} 
\end{align} 
where $\kappa$ and $\kappa_0$ are renormalization scales, $\Lambda_{\text{QCD}}$ is the QCD scale parameter, 
$a'\equiv 8\Cf/(3b_0)$, and $\tilde{a}\equiv 2(4\Cf+\Nf)/(3b_0)$, where $b_0\equiv (11\Nc-2\Nf)/3$, 
which appears in the expression 
\begin{align}
\label{eq:RG-alphas}
\alpha_s(\kappa)=\frac{4\pi}{b_0 \ln(\kappa^2/\Lambda^2_{\text{QCD}})}.
\end{align}
Note that $T^{00}$ is independent of $\kappa$. 
In the $\omega\rightarrow \infty$ limit, it is natural to choose the RG scale as $\kappa^2=\omega^2$.
We see that, except for the $T^{00}$ term, all terms in Eq.~(\ref{eq:Tf-decompose}) are suppressed logarithmically at large $\omega$. 
Thus, Eqs.~(\ref{eq:OPE-wo-RG-T}) and (\ref{eq:OPE-wo-RG-L}) become
\begin{align} 
\label{eq:OPE-w-RG-T}
\delta G_{T}(\omega,\vp) 
&=e^2\sum q^2_f \frac{1}{p^2} 
 \Bigl[\left\{2m_f \delta\left\langle\overline{\psi}_f \psi_f \right\rangle_{T}
+\frac{1}{12}\delta\left\langle \frac{\alpha_s}{\pi}G^2\right\rangle_{T} \right\}
+\frac{8}{3}\frac{1}{4\Cf+\Nf}\frac{\omega^2+\vp^2}{p^2}\delta\left\langle T^{00} \right\rangle_{T} 
\Bigr] \nonumber \\
&~~~+{\cal O}(\omega^{-4}),\\
\label{eq:OPE-w-RG-L}
\delta G^{R}_{00}(\omega,\vp) 
&=e^2\sum q^2_f \frac{1}{p^2}\frac{\vp^2}{p^2} 
 \Bigl[\left\{2m_f \delta\left\langle\overline{\psi}_f \psi_f \right\rangle_{T}
+\frac{1}{12}\delta\left\langle \frac{\alpha_s}{\pi}G^2\right\rangle_{T} \right\} 
+\frac{8}{3}\frac{1}{4\Cf+\Nf} \delta\left\langle T^{00} \right\rangle_{T}  
\Bigr]
+{\cal O}(\omega^{-6}). 
\end{align} 

Next, we briefly explain basic idea of the derivation 
of the spectral UV tail at high energy.   
The UV behavior of the retarded vector current correlator is described by the OPE expressions of Eqs.\,(\ref{eq:OPE-wo-RG-T}) and (\ref{eq:OPE-wo-RG-L}).
Among the three terms, only $\langle T^{00}_f\rangle_{T}$ is not RG invariant.
This operator yields imaginary parts of the retarded correlator, as can be understood as follows. 
The scaling relations of Eq.\,(\ref{eq:scaling-T'-Ttilde}) can be rewritten as
\begin{align}
\begin{split} 
T'{}^{00}_{f}(\kappa)
&\simeq  T'{}^{00}_{f}(\kappa_0)
+a'\ln\left(\frac{\kappa^2_0}{\kappa^2}\right)
\frac{b_0}{4\pi}\alpha_sT'{}^{00}_{f} ,\\
 \tilde{T}^{00}(\kappa)
&\simeq  \tilde{T}^{00}(\kappa_0)
+\tilde{a}\ln\left(\frac{\kappa^2_0}{\kappa^2}\right)
\frac{b_0}{4\pi}\alpha_s\tilde{T}^{00}_{f},
\end{split} 
\end{align}
when $\kappa$ is close to $\kappa_0$.
It was shown in Ref.\,\cite{CaronHuot:2009ns} 
that the factor $\ln(\kappa^2_0/\kappa^2)$ generates an imaginary contribution $i\pi$, due to the analytic continuation to real time. 
Following this prescription, the imaginary parts of the retarded correlators of Eqs.\,(\ref{eq:OPE-wo-RG-T}) and (\ref{eq:OPE-wo-RG-L}) read
\begin{align}
\label{eq:deltarho_T_UVtail}
\delta \rho_T(p) 
&=e^2\sum q^2_f \frac{8}{9}\frac{\omega^2+\vp^2}{(p^2)^2}
\alpha_s(\omega) 
  \left( 2\Cf \delta\left\langle T'{}^{00}_{f}(\omega)  \right\rangle_{T}
+\frac{1}{\Nf}  \delta\left\langle \tilde{T}^{00}(\omega) \right\rangle_{T} \right) ,\\
\delta \rho_{00}(p)
&=  e^2\sum q^2_f \frac{8}{9}\frac{\vp^2}{(p^2)^2}
\alpha_s(\omega) 
 \left( 2\Cf \delta\left\langle T'{}^{00}_{f}(\omega)  \right\rangle_{T}
+\frac{1}{\Nf}  \delta\left\langle \tilde{T}^{00}(\omega) \right\rangle_{T} \right) .
\end{align}
This expression is valid when the OPE is reliable, that is, for $\omega\gg T,\Lambda_{\text {QCD}}$. 

Especially, in the chiral and weak coupling limits, the operator expectation values at the renormalization scale $\kappa_0\sim T$ read
\begin{align}
\label{eq:T00f-free}
\langle T^{00}_f\rangle_{T} &= \Nc\frac{7\pi^2T^4}{60},\\
\label{eq:T00g-free}
\langle T^{00}_g\rangle_{T} &=2\Cf \Nc \frac{\pi^2T^4}{15},
\end{align}
which, by using the scaling relation of Eq.\,(\ref{eq:scaling-T'-Ttilde}), leads to 
\begin{align}
\label{eq:UVtail-T}
\delta \rho_T(p) 
&=\Cem \frac{1}{\omega^2} 
\left(1+3\frac{\vp^2}{\omega^2}\right)
\alpha_s(\kappa_0) \Nc\Cf\frac{4\pi^2T^4}{27} 
 \left[\frac{\ln\left(\kappa_0/\Lambda_{\text{QCD}} \right)}{\ln\left(\omega/\Lambda_{\text{QCD}} \right)}\right]^{\tilde{a}+1}  ,\\
\label{eq:UVtail-00}
\delta \rho_{00}(p)
&=\Cem \frac{\vp^2}{\omega^4} 
\left(1+2\frac{\vp^2}{\omega^2}\right)
\alpha_s(\kappa_0) \Nc\Cf \frac{4\pi^2T^4}{27}
 \left[\frac{\ln\left(\kappa_0/\Lambda_{\text{QCD}} \right)}{\ln\left(\omega/\Lambda_{\text{QCD}} \right)}\right]^{\tilde{a}+1}. 
\end{align} 
Here, we have retained terms up to next-to-leading order in the small $|\vp|$ expansion.

%%%%%%%%%%%%%%%%%%%%

%\bibliographystyle{acm}
%\bibliographystyle{plain}
\bibliographystyle{unsrt}
\bibliography{reference_DS,reference_PG}

\end{document}